\documentclass[12pt]{article}
\usepackage{epsfig}
\usepackage{pstricks,rotating, graphicx}
\usepackage{a4}
\usepackage{latexsym}
\usepackage{cite}

\usepackage{color}
\usepackage{colordvi}
\usepackage[hang]{subfigure}

\graphicspath{{Pics/}}
\DeclareGraphicsExtensions{.eps,.ps}

\usepackage[width=16.5cm, left=2.2cm,top=2.5cm,height=24.cm]{geometry}

\usepackage{pslatex}
\usepackage{txfonts}
\usepackage[latin1]{inputenc}
\usepackage[T1]{fontenc}

\newcommand{\eps}{\varepsilon}
\newcommand{\gev}{\,\textrm{GeV}}
\newcommand{\tev}{\,\textrm{TeV}}
\newcommand{\csec}[1]{Sec.~\ref{#1}}
\newcommand{\ce}[1]{Eq.~(\ref{#1})}
\newcommand{\ced}[2]{Eqs.~(\ref{#1})--(\ref{#2})}
\newcommand{\cf}[1]{{Fig.~\ref{#1}}}

\newcommand{\cfa}[2]{{Figs.~\ref{#1} and \ref{#2}}}
\newcommand{\cfd}[2]{{Figs.~\ref{#1}--\ref{#2}}}
\newcommand{\ct}[1]{{Tab.~\ref{#1}}}
\newcommand{\ctd}[2]{{Tabs.~\ref{#1}--\ref{#2}}}

\newcommand{\re}[1]{\;\textrm{Re} \left( #1 \right)}
\newcommand{\madloop}{\textsc{MadLoop}}
\newcommand{\madfks}{\textsc{MadFKS}}
\newcommand{\madgraph}{\textsc{MadGraph}}
\newcommand{\mathematica}{\textsc{Mathematica}}
\newcommand{\feyncalc}{\textsc{FeynCalc}}

\def\muf{{\mu^{}_f}}
\def\mufs{{\mu^{\,2}_f}}
\def\mur{{\mu^{}_r}}
\def\murs{{\mu^{\,2}_r}}

\def\Pnsp#1{P_{\rm ns}^{(#1),+}}
\def\Pnsm#1{P_{\rm ns}^{(#1),-}}
\def\Pps#1{P_{\rm ps}^{(#1)}}
\def\Pqq#1{P_{\rm qq}^{(#1)}}
\def\Pgq#1{P_{\rm gq}^{(#1)}}
\def\Pqg#1{P_{\rm qg}^{(#1)}}
\def\Pgg#1{P_{\rm gg}^{(#1)}}

\def\b#1{\beta_{#1}}

\def\ciqo{c_{i,\rm q}^{(1)}}
\def\cinpt{c_{i,\rm ns}^{(2),+}}

\def\cipt{c_{i,\rm ps}^{(2)}}
\def\ciqt{c_{i,\rm q}^{(2)}}
\def\cigo{c_{i,\rm g}^{(1)}}
\def\cigt{c_{i,\rm g}^{(2)}}

\def\ctqo{c_{3,\rm q}^{(1)}}

\def\ctnmt{c_{3,\rm ns}^{(2),-}}

\def\hmass{m_H}
\def\wmass{M_W}
\def\zmass{M_Z}
\def\vmass{M_V}
\def\bmass{m_b}
\def\tmass{m_t}

\begin{document}

\begin{titlepage}
\noindent
DESY 11-153 \\
CP3-11-28 \\
LPN 11-51 \\
SFB/CPP-11-50 \\
September 2011 \\
\vspace{1.3cm}

\begin{center}
\Large{\bf 
Vector boson fusion at NNLO in QCD: SM Higgs and beyond
}\\
\vspace{1.5cm}
\large
Paolo~Bolzoni$^{a}$, Fabio~Maltoni$^{b}$, Sven-Olaf~Moch$^{c}$ and Marco~Zaro$^{b}$\\
\vspace{1.2cm}
\normalsize
{\it $^a$II.~Institut f\"ur Theoretische Physik, Universit\"at Hamburg\\
\vspace{0.1cm}
Luruper Chaussee 149, D--22761 Hamburg, Germany}\\
\vspace{0.5cm}
{\it $^{b}$Center for Cosmology, Particle Physics and Phenomenology (CP3), \\
Universit\'e Catholique de Louvain, \\ Chemin du Cyclotron 2, B--1348 Louvain-la-Neuve, Belgium}\\
\vspace{0.5cm}
{\it $^c$Deutsches Elektronensynchrotron DESY \\
\vspace{0.1cm}
Platanenallee 6, D--15738 Zeuthen, Germany}\\

\vspace{1.8cm}

\large
{\bf Abstract}
\vspace{-0.2cm}
\end{center}
Weak vector boson fusion provides a unique channel to directly probe the mechanism of electroweak symmetry 
breaking at hadron colliders.  We present a method that allows to calculate total cross sections to 
next-to-next-to-leading order (NNLO) in QCD for an arbitrary $V^* V^* \to X$  process, the so-called structure 
function approach. By discussing  the case of Higgs production in detail, we estimate several classes of previously neglected contributions
 and we argue that such method is accurate at a precision level 
well above the typical residual scale and PDF uncertainties at NNLO.  Predictions  
for cross sections at the Tevatron and the LHC are presented for a variety of cases:  the Standard Model Higgs (including anomalous couplings), neutral and charged scalars in extended  Higgs sectors and (fermiophobic) vector resonance production. 
Further results can be easily obtained through the public use of  the {\sc VBF@NNLO} code.
\vfill
\end{titlepage}

%\newpage
%\tableofcontents

\newpage
%%
%% ---------------------------------------------------------------------------
%%
\section{Introduction}
\label{sec:Intro}

The central theme of the physics program at the Large Hadron Collider (LHC) is 
the search for the Standard Model (SM) Higgs boson and more in general
the elucidation of the mechanism of electroweak symmetry breaking.
The three dominant production mechanisms for the SM Higgs boson are,
(in order of importance) gluon-gluon fusion via a top-quark loop, vector-boson fusion (VBF)
and Higgs-Strahlung, {\it i.e.} associated production with $W$ and $Z$-bosons 
(see e.g.~\cite{Djouadi:2005gi,Harlander:2007zz}). In extensions of the SM with a richer
Higgs sector, such as in supersymmetry, or in strongly interacting light Higgs scenarios~\cite{Giudice:2007fh}, the relative
importance of the various channels might depend on the details or parameters of the model.
In any case, VBF  remains of primary importance, being the channel where longitudinal vector
boson scattering gives rise to violation of unitarity  at around 1 TeV, if no other particle or interaction
is present than what it is currently known from experiments.

Precise knowledge of the expected rates for the Higgs boson production processes is an essential 
prerequisite for any experimental search and, after discovery, will be a crucial input to 
proceed to accurate measurements.
At a hadron collider such as the LHC, precision predictions need to include 
higher-order radiative corrections which usually implies 
the next-to-next-to-leading order (NNLO) in Quantum Chromodynamics (QCD) 
and the next-to-leading order (NLO) as far as electroweak corrections are concerned.
If accounted for, these higher-order quantum corrections generally stabilize the theoretical predictions 
through an apparent convergence of the perturbative expansion 
and a substantially reduced dependence on the choice of the factorization and renormalization scales.
For inclusive Higgs boson production in gluon-gluon fusion and the $WH$ and $ZH$ mode, the exact NNLO corrections in QCD 
are available~\cite{Harlander:2002wh,Anastasiou:2002yz,Ravindran:2003um,Brein:2003wg} 
and have been shown to be very important (see e.g. \cite{Dittmaier:2011ti}).

Higgs boson production via VBF is the mechanism with the second largest rate in the SM
and offers a clean experimental signature with the presence of at least two jets 
in the forward/backward rapidity region~\cite{Cahn:1983ip,Kane:1984bb,Kleiss:1986xp} 
and a variety of Higgs boson decay modes to be searched for~\cite{Rainwater:1997dg, Rainwater:1998kj,Kauer:2000hi,Mangano:2002wn} . 
In extensions of the
SM which  feature scalar or vector state(s) with reduced couplings to fermions, VBF can become
the leading production mechanism.
VBF is a pure electroweak process at leading order (LO)  
and it acquires corrections at NLO in QCD~\cite{Han:1992hr,Figy:2003nv} 
as well as in the electroweak sector~\cite{Ciccolini:2007jr,Ciccolini:2007ec} 
leading to a typical accuracy for the total cross-section in the $5-10\%$ range.
Recently, the NNLO QCD corrections for the VBF process have been computed~\cite{Bolzoni:2010xr} 
in the so-called structure function approach~\cite{Han:1992hr}, which builds upon
the approximate, although very accurate, 
factorization of the QCD corrections between the two parton lines associated with the colliding hadrons.
At NNLO the results are very stable with respect to variations of the renormalization and factorization scales, $\mur$ and $\muf$,
and can take full advantage of modern sets of precise parton distribution functions (PDFs) at the same accuracy.
The small theoretical uncertainty for the inclusive rate due to missing higher order QCD corrections 
as well as the PDF uncertainties are estimated to be at the $2$\% level each for a wide range of Higgs boson masses.

The purpose of the present article is two-fold. 
First, it provides an extensive documentation of the NNLO computation of Refs.~\cite{Bolzoni:2010xr,Bolzoni:2010as} 
and a broad phenomenological study for the LHC at various center-of-mass energies and for the Tevatron.
Second, it exploits the universality of the structure function approach 
to describe the production of any color-neutral final state from the fusion of vector bosons 
and applies it to a number of new physics models, e.g. with an extended Higgs sector, 
or to vector resonance production (see e.g.,~\cite{Csaki:2003dt,Csaki:2003zu}).

The outline is as follows: In \csec{sec:SetStage} we define the VBF process as a signal, discuss
carefully its contributions and give arguments for the ultimate theoretical 
precision one could possibly aim at in predictions of its rates.
\csec{sec:QftAtHO} is devoted to a discussion of the structure function approach at NNLO in QCD 
along with detailed computations of neglected contributions as a means of estimating the accuracy.
In particular, we study the non-factorizable diagrams and the ones involving heavy quarks, 
all of which are not accounted for in the structure function approach.
Phenomenological results for inclusive rates at the LHC and Tevatron are presented in \csec{sec:Phenomenology}.
Extensions of the structure function approach to Higgs boson production via VBF 
in specific classes of models beyond the SM (see e.g.,~\cite{Zaro:2010fc} for charged Higgs bosons) 
as well as modifications due to anomalous couplings ($WWH$, $ZZH$, etc.) and vector resonances are discussed in \csec{sec:BSMHiggs}. 
Finally, we conclude in \csec{sec:Conclusions} 
and document some technical aspects and tables with VBF cross sections in the Appendices.

%%
%% ----------------------------------------------------------------------------
%%
\renewcommand{\thefigure}{\thesection.\arabic{figure}}
\setcounter{figure}{0}
\section{Setting the stage}
\label{sec:SetStage}

Processes in collider physics are always defined on simple conventions typically based
on leading-order Feynman diagrams. While this normally poses no problems,
it might lead to ambiguities when decays of resonant states and/or higher order effects are included. 
Consider, as a simple example, the class of processes which involve only
the electroweak coupling $\alpha_{EW}$ at the leading order,
such as Drell-Yan (with the decay into two jets), single-top production and
Higgs boson production via coupling with a vector boson. 
Most of such processes can be easily defined at leading order considering the corresponding
resonant intermediate or final states, while some, such as VBF are (quantum-mechanically) ambiguous already at the leading order:
$pp \to Hjj$ with vector bosons in the $t$-channel can interfere with $pp \to HV^{(*)} \to Hjj$,
{\it i.e.}, with Higgs associated production with a vector boson then decaying into two jets. 
Such interference, however, is quite small everywhere in the phase space and 
it can formally be reduced to zero by just taking the narrow width limit. 
This suggests that considering the two processes distinct is a handy approximation, at least at the leading
order.  For all of the processes in this class, {\it i.e.}, single-top and Higgs electroweak production, aiming at a better
precision by including higher-order QCD effects creates further ambiguities as it opens 
up more possibilities for interferences (one notable example is  $tW$ at NLO overlapping with $t\bar t$ production)
and the reliability of the approximations made has to be carefully assessed.

In general two complementary approaches can be followed.
The first is to consider all interferences  exactly, and introduce a gauge-invariant scheme to properly handle the width effects.  
This can be consistently done  at NLO order in QCD and electroweak corrections, following well-known and
established techniques, such as the use of the complex-mass scheme~\cite{Argyres:1995ym}. 
This is the path followed for instance in Ref.~\cite{Ciccolini:2007ec} 
for the calculation of QCD and electroweak effects in the order $\alpha_{EW}^3$ process $pp \to Hjj$.
The advantage of this approach is that all interference effects are correctly
taken into account in any region of phase space, including where tight cuts might create significant enhancements. 
This is the only way to proceed when interference effects are similar to or larger than the corrections from higher orders. 
However, when such effects are small, it offers several drawbacks. 
The first is the unnecessary complexity of the calculation itself. 
The second  is that the operational separation between the two processes, which can be quite useful at the practical level, 
for example in experimental analyses, is lost. 
In this context, even the definition of
signal and background might not be meaningful and the distinction possibly leads to confusion. 
In this case another approach can be followed. Use a simple  process definition and systematically 
check the impact of higher-order corrections as well as those from interferences to set the ultimate practical precision
that can be achieved. In this section we argue that for vector-boson fusion this procedure is sound and
can lead to a definition which is unambiguous  to better than 1\%, {\it i.e.} more than sufficient for all practical applications at hadron colliders.

\begin{figure}[ht!]
\centering
\includegraphics[scale=0.5]{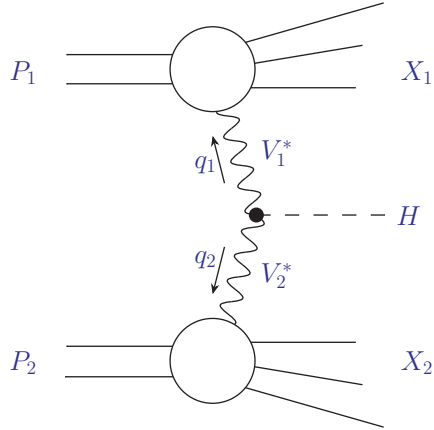}
\caption{\small
  \label{fig:vbf}
Higgs production via the VBF process.
}
\end{figure}

In short, we define VBF  as the Higgs production for vanishing quark masses, through direct coupling to vector bosons
in the $t$-channel,  and with no color exchange between the two colliding hadrons,
{\it i.e.}, all the processes that can be represented by the diagram in \cf{fig:vbf},
where no heavy-quark loop is to be included in the blobs while additional vector bosons in a color singlet state might
appear at one or more loops (not shown).

The definition above, while academic in nature, fits extremely well with what is looked for and associated to in the current
experimental searches. For example, it excludes the interfering effects with $s$-channel associated production, but includes
the additional exchange of two gluons  in a color singlet state in the
$t$-channel. These two effects, as we will argue in the following, are tiny
and can be neglected in the numerical evaluation of the total cross section. 
However, it is useful to keep in mind that our definition put them on a different ground. 
The main reason/motivation is that among the most important
characteristics exploited in the experimental analyses  is the presence of two
high-invariant mass forward-backward jets and the absence of radiation in the central region. 
Both these effects are present in the case of the color-singlet component of double-gluon exchange in the $t$-channel, 
while they are not typical of the $s$-channel contribution. 

The definition given above allows us to systematically classify processes for
Higgs boson production as VBF and non-VBF. In the latter we include also possible interferences between amplitudes belonging to the VBF class and those in the non-VBF one.\\

\noindent ``VBF'' processes:
\begin{itemize}
\item 
  Factorizable contributions in QCD, see \cf{fig:vbf}. This class is evaluated exactly in this work for massless quarks.
  It provides the bulk of all QCD corrections up to order $\alpha_s^2$ to a precision better than 1\%.

\item Non-factorizable contributions in QCD. This class starts at order $\alpha_s^2/N_c^2$. 
  It is estimated in \csec{sec:dia-pentagon} to contribute less than 1\% to the total VBF cross section.
  
\item Electroweak corrections to diagrams in \cf{fig:vbf}. 
  These are relevant corrections which have been calculated in Ref.~\cite{Ciccolini:2007ec}. A combination  with  
  the NNLO QCD ones calculated in this work has been reported in  Ref.~\cite{Dittmaier:2011ti}.
\end{itemize}

\noindent ``Non-VBF'' processes:
\begin{itemize}
\item Single-quark line contributions, as calculated in
  Ref.~\cite{Harlander:2008xn}. 
  These effects are  smaller than 1\% in differential cross sections.
  
\item Interferences in VBF itself known at NLO in QCD and electroweak, as
  calculated in Ref.~\cite{Ciccolini:2007ec}.  These effects can be calculated at LO and are found to be very small.
  
\item Interferences between VBF and associated $WH$ and $WZ$ production at NLO in QCD and electroweak, as calculated in Ref.~\cite{Ciccolini:2007ec}. These effects can be easily calculated at LO and are found to be very small.

\item Interferences with the top-loop mediated Higgs production,
  Refs.~\cite{Andersen:2006ag,Andersen:2007mp} 
  and contributions calculated in \csec{sec:dia-tbloop}. These effects are found to contribute less than 1\% to the total cross section. 

\item
  $t$-channel vector boson production in presence of heavy-quark loops (triangles and boxes), see \csec{sec:dia-tbloop}.
  These effects are estimated to contribute less than 1\% to the total cross section.
\end{itemize}

We conclude this section by mentioning that the same kind of difficulties/ambiguities arise when the Higgs width becomes non-negligible, {\it i.e.} 
for large Higgs masses, let us say $m_H > 500$ GeV. Apart from the need of a Breit-Wigner smearing of cross sections calculated in the narrow-width limit that can be easily implemented, interference effects with ``background'' processes can become important and need to be accounted for. Such effects have been considered in full generality at LO in studies of vector boson scattering~\cite{Ballestrero:2007xq,Ballestrero:2009vw} and, for leptonic final states of the vector bosons, in EW production of $V_1V_2 j j$ at NLO in QCD~\cite{Jager:2006cp,Jager:2006zc,Bozzi:2007ur,Arnold:2011wj}.

%%
%% ----------------------------------------------------------------------------
%%
\renewcommand{\thetable}{\thesection.\arabic{table}}
\setcounter{table}{0}
\renewcommand{\theequation}{\thesection.\arabic{equation}}
\setcounter{equation}{0}
\renewcommand{\thefigure}{\thesection.\arabic{figure}}
\setcounter{figure}{0}
\section{VBF at higher orders}
\label{sec:QftAtHO}

Let us discuss the radiative corrections to the VBF process with emphasis on the QCD contributions.
As mentioned above, at LO the Higgs production in VBF proceeds purely through electroweak interactions (see \cf{fig:vbflo}), 
with the cross section for $pp \to Hjj$ being of order $\alpha_{EW}^3$.
\begin{figure}[t!]
\centering
\includegraphics[scale=0.333]{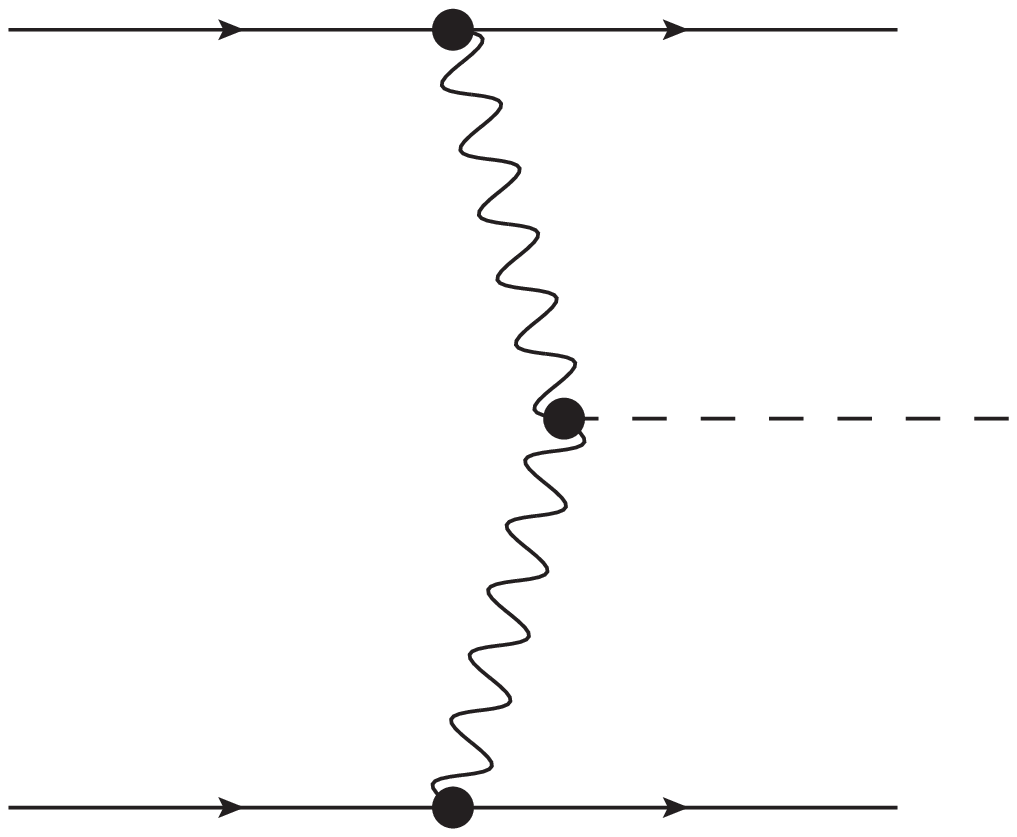}
\caption{\small
  \label{fig:vbflo}
Higgs production via the VBF process at LO in QCD.
}
\end{figure}

In the QCD improved parton model, higher order corrections arise.
All NLO QCD contributions to the VBF cross section (order $\alpha_{EW}^3 \alpha_s$) 
can be treated exactly in the structure function approach, which we 
discuss in detail and extend to NNLO  below in \csec{sec:SFapproach}.
At NNLO, {\it i.e.} at order $\alpha_{EW}^3 \alpha_s^2$, certain other contributions arise, 
which we estimate.
These are diagrams involving gluon exchange between the two quark lines (cf. \csec{sec:dia-pentagon}),
and diagrams involving closed heavy-quark loops (cf. \csec{sec:dia-tbloop}).
We also comment briefly on electroweak corrections at one-loop in \csec{sec:NLO-EW}.

\subsection{Structure function approach}
\label{sec:SFapproach}

The structure function approach is based on the observation that to a very good approximation 
the VBF process can be described as a double deep-inelastic scattering process (DIS), 
see \cf{fig:vbf}, 
where two (virtual) vector-bosons $V_i$ (independently) emitted from
the hadronic initial states fuse into a Higgs boson. 
This approximation builds on the absence (or smallness) of the QCD interference between 
the two inclusive final states $X_1$ and $X_2$. 
In this case the total cross section is given as a product 
of the matrix element $\mathcal M^{\mu\rho}$ for VBF, 
{\it i.e.}, $V_1^\mu V_2^\rho \to H$, which in the SM reads 
\begin{equation}
  \label{eq:vbf-Mmunu-SM}
  \mathcal M^{\mu\nu} = 2 \left(\sqrt 2 G_F\right)^{1/2} M_{V_i}^2 g^{\mu\nu}
  \, ,
\end{equation}
and of the DIS hadronic tensor $W_{\mu \nu}$\,:
\begin{eqnarray}
  \label{eq:disapproach}
d\sigma &=& \frac{1}{2 S} 2 G_F^2 M^2_{V_1} M^2_{V_2} \frac{1}{\left(Q^2_1 +M^2_{V_1}\right)^2}
 \frac{1}{\left(Q^2_2 +M^2_{V_2}\right)^2} W_{\mu\nu} \left(x_1, Q^2_1\right)
 \mathcal M^{\mu\rho} \mathcal {M^*}^{\nu\sigma} W_{\rho\sigma}\left(x_2,Q^2_2\right)\times 
\nonumber
\\
&&
 \times
 \frac{d^3 P_{X_1}}{\left(2\pi\right)^3 2 E_{X_1}}  \frac{d^3 P_{X_2}}{\left(2\pi\right)^3 2 E_{X_2}} ds_1 ds_2 \frac{d^3 P_H}{\left(2\pi\right)^3 2 E_H}
 \left(2\pi\right)^4 \delta^4\left( P_1+P_2-P_{X_1} -P_{X_2}-P_{H} \right) \, .
\end{eqnarray}
Here $G_F$ is Fermi's constant and $\sqrt{S}$ is the center-of-mass energy of the collider.
$Q^2_i=-q_i^2$, $x_i=Q_i^2/(2P_i\cdot q_i)$ are the usual DIS variables, 
$s_i=(P_i+q_i)^2$ are the invariant masses of the $i$-th proton remnant, 
and $M_{V_i}$ denote the vector-boson masses, see \cf{fig:vbf}.
The three-particle phase space $dPS$ of the VBF process is given in the second line of \ce{eq:disapproach}. 
It is discussed in detail in \csec{sec:PhaseSpace}.

Higgs production in VBF requires the hadronic tensor $W_{\mu \nu}$ 
for DIS neutral and charged current reactions, {\it i.e.}, the scattering off a $Z$ as well as off a $W^\pm$-boson.
It is commonly expressed in terms of the standard DIS structure functions $F_i(x,Q^2)$ with $i=1,2,3$.
{\it i.e.} $F_i^V$ with $i=1,2,3$ and $V \in \{Z,W^\pm\}$ and we employ the particle data group (PDG) 
conventions~\cite{Nakamura:2010pdg}.
Thus, 
\begin{eqnarray}
  \label{eq:disWmunu}
W_{\mu\nu} \left(x_i, Q_i^2\right) %\, = \,}}
&=& 
\left( - g_{\mu \nu} + \frac{q_{i,\, \mu} q_{i,\, \nu}}{q_i^2} \right)\, F_{1}(x_i,Q_i^2) 
+ \frac{ {\hat P}_{i,\, \mu}{\hat P}_{i,\, \nu} }{P_i \cdot q_i}\, F_{2}(x_i,Q_i^2)
+ {\rm{i}} \epsilon_{\mu\nu\alpha\beta} \frac{P_i^\alpha q_i^\beta}{2 P_i \cdot q_i} F_{3}(x_i,Q_i^2)
\, ,
\qquad
\end{eqnarray}
where $\epsilon_{\mu\nu\alpha\beta}$ is the completely antisymmetric tensor
and the momentum ${\hat P}_{i}$ reads
\begin{equation}
  \label{eq:def-Phat}
  {\hat P}_{i,\, \mu} = P_{i,\, \mu} - \frac{P_i \cdot q_i}{q_i^2}\, q_{i,\, \mu}
\, .
\end{equation}
\begin{figure}[t!]
\centering
\includegraphics[scale=0.333]{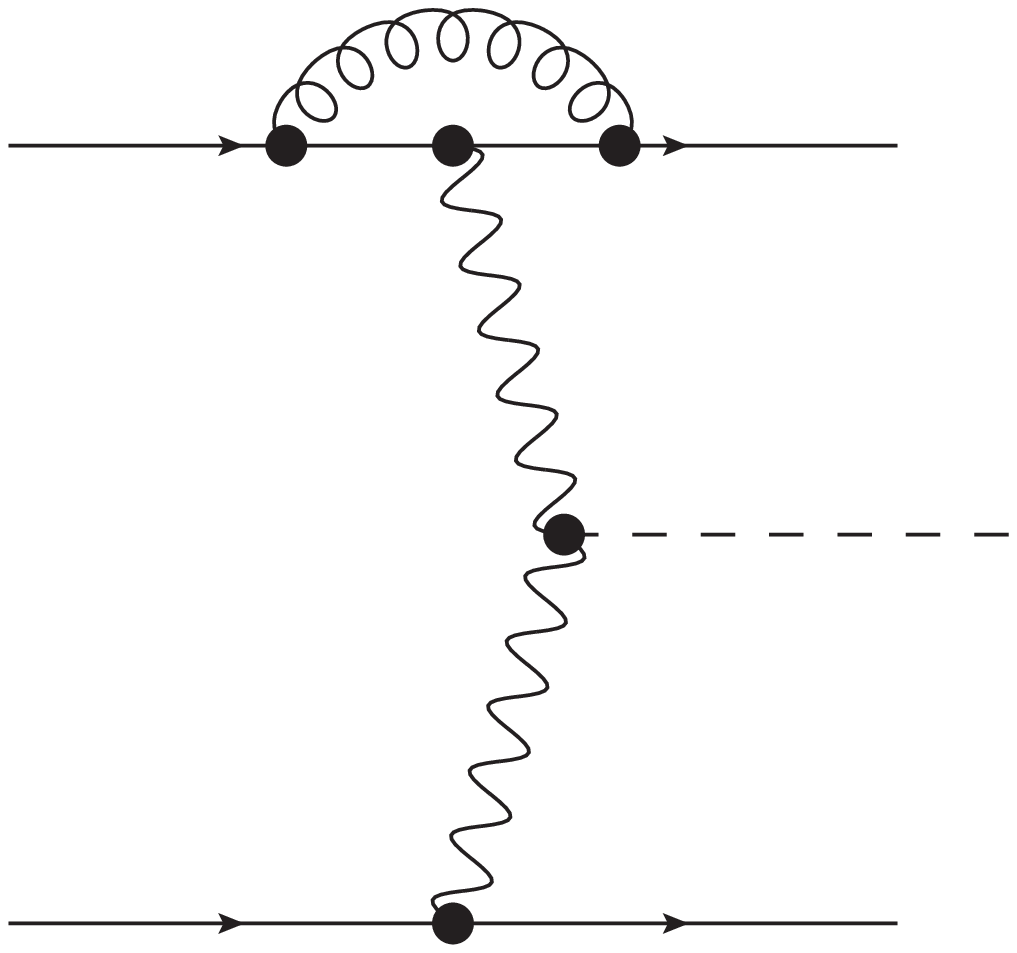}
\hspace{1cm}
\includegraphics[scale=0.333]{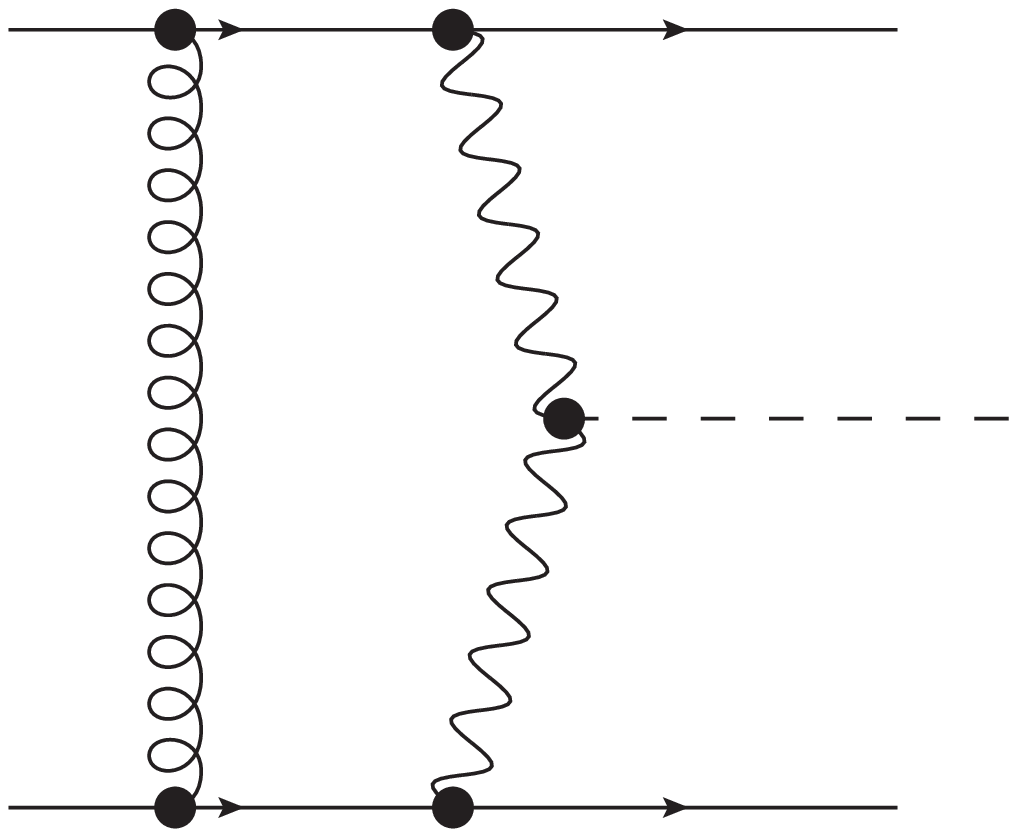}
\caption{\small
  \label{fig:vbfnlo}
Higgs production via the VBF process at NLO in QCD.
}
\end{figure}
The factorization underlying \ce{eq:disapproach} does not hold exactly already at LO,
because interference can occur either between identical final state quarks ({\it i.e.}, $uu\to H uu$) or between
processes where either a $W$ or a $Z$ can be exchanged ({\it i.e.}, $ud\to Hud$).
However, at LO, these contributions can be easily computed and they have been included in our results.
On the other hand, simple arguments of kinematics (based on the behavior of the propagators in the matrix element~\cite{Dicus:1985zg})
show that such contributions are heavily suppressed already at LO and contribute to the total cross section well below the 1\% level,
a fact that has been confirmed by a complete calculation even through NLO~\cite{Ciccolini:2007ec}.
Apart from these interference effects, the factorization of \ce{eq:disapproach} is still exact at NLO.
This is due to color conservation: QCD corrections to the upper quark line 
are independent from those of the lower line. \cf{fig:vbfnlo}~(left) shows a
sample diagram accounted for by the structure function approach, while the
diagram \cf{fig:vbfnlo}~(right) vanishes at NLO due to color conservation, 
{\it i.e.}, Tr($t^a)=0$ for generators $t^a$ of the color SU$(N_c)$ gauge group.

The evaluation of \ce{eq:vbf-Mmunu-SM} and \ce{eq:disWmunu} leads to the explicit
result for the squared hadronic tensor in \ce{eq:disapproach} in terms of the
DIS structure functions~\cite{Han:1992hr} (see also the review~\cite{Djouadi:2005gi})\,:
\begin{eqnarray}
 \label{eq:WMMW-SM}
{\lefteqn{
 W_{\mu\nu} \left(x_1, Q^2_1\right)\mathcal M^{\mu\rho} \mathcal {M^*}^{\nu\sigma} W_{\rho\sigma}\left(x_2,Q^2_2\right) 
\,=\, 4 \sqrt 2 G_F M_{V_i}^4 \, \times 
}} 
\nonumber\\
&\times&
	 \left\{ F_1 \left(x_1, Q^2_1\right) F_1 \left(x_2,Q^2_2\right) \left[2+\frac{ (q_1\cdot q_2)^2}{q_1^2 q_2^2}\right]+ \right. 
\nonumber\\
& &
	 +\frac{ F_1 \left(x_1, Q^2_1\right) F_2 \left(x_2,Q^2_2\right) }{P_2\cdot q_2}
 \left[  \frac{ (P_2 \cdot q_2)^2 }{q_2^2} +\frac{1}{q_1^2}\left(P_2\cdot q_1-\frac{P_2 \cdot q_2}{q_2^2} q_1 \cdot q_2\right)^2\right] 
\nonumber\\
& &
	 +\frac{ F_2 \left(x_1, Q^2_1\right) F_1 \left(x_2,Q^2_2\right) }{P_1\cdot q_1}
 \left[  \frac{ (P_1 \cdot q_1)^2 }{q_1^2} +\frac{1}{q_2^2}\left(P_1\cdot q_2-\frac{P_1 \cdot q_1}{q_1^2} q_1 \cdot q_2\right)^2\right] 
\nonumber\\
& &
 	+\frac{ F_2 \left(x_1, Q^2_1\right) F_2 \left(x_2,Q^2_2\right) }{(P_1\cdot q_1) (P_2\cdot q_2)} 
\, \times
\nonumber\\
& &
\hspace*{5mm} \times
 \left(P_1\cdot P_2 - \frac{(P_1\cdot q_1)  (P_2\cdot q_1)}{q_1^2} - \frac{(P_1\cdot q_2) (P_2\cdot q_2)}{q_2^2} 
   +\frac{(P_1\cdot q_1) (P_2\cdot q_2) (q_1\cdot q_2)}{q_1^2 q_2^2} \right)^2 
\nonumber\\
& &
 	\left. +\frac{ F_3 \left(x_1, Q^2_1\right) F_3 \left(x_2,Q^2_2\right) }{2 (P_1\cdot q_1) (P_2\cdot q_2)} 
          \biggl((P_1\cdot P_2) (q_1 \cdot q_2) - (P_1 \cdot q_2) (P_2\cdot q_1) \biggr)
 \right\} \, .
\end{eqnarray}

At this stage it remains to insert the DIS structure functions $F_i^V$ with $i=1,2,3$ and $V \in \{Z,W^\pm\}$.
At NLO in QCD, explicit expression have been given in Ref.~\cite{Han:1992hr} using the results of Ref.~\cite{Bardeen:1978yd}.
For the necessary generalization beyond NLO, let us briefly review the basic formulae.
QCD factorization allows to express the structure functions as convolutions of the PDFs in the proton 
and the short-distance Wilson coefficient functions $C_{i}$.
The gluon PDF at the factorization scale $\muf$ is denoted by $g(x,\muf)$ and 
the quark (or anti-quark) PDF by $q_{i}(x,\muf)$ (or ${\bar q}_{i}(x,\muf)$) for a specific quark flavor $i$.
The latter PDFs appear in the following combinations, 
\begin{eqnarray}
  \label{eq:pdf-s-v}
  &q_{\rm s} \,=\, \sum\limits_{i=1}^{n_f}\, \biggl(q_i+{\bar q}_i \biggr)
  \, ,
  \qquad\qquad
  &q^{\rm v}_{\rm ns} \,=\, \sum\limits_{i=1}^{n_f}\, \biggl(q_i-{\bar q}_i \biggr)
  \, ,
\\
  \label{eq:pdf-pm}
  &q^+_{{\rm ns},i} \,=\, \biggl(q_i + {\bar q}_i \biggr) - q_{\rm s} 
\, ,
\qquad\qquad
%  \label{eq:minus-pdf}
  &q^-_{{\rm ns},i} \,=\, \biggl(q_i - {\bar q}_i \biggr) - q^{\rm v}_{\rm ns} 
  \, ,
\end{eqnarray}
as the singlet distribution $q_{\rm s}$, the (non-singlet) valence distribution $q^{\rm v}_{\rm ns}$ 
as well as flavor asymmetries of $q^\pm_{{\rm ns},i}$. 
All of them are subject to well-defined transformation properties under the flavor isospin, see
e.g. \cite{Moch:2004pa,Vogt:2004ns}.

\bigskip

For the neutral current $Z$-boson exchange the DIS structure functions $F_i^Z$ can be written as follows:
\begin{eqnarray}
  \label{eq:disncZ-12}
  {\lefteqn{
      F_i^Z(x,Q^2) \,=\, f_i(x)\, \int\limits_0^1\, dz\, \int\limits_0^1\, dy \,
      \delta(x-yz)\, \sum\limits_{j=1}^{n_f}\, 
      \left( v_j^2 + a_j^2 \right)\, \times }}
  \\
  \nonumber
  & &
  \times\, \biggl\{ 
  q^+_{{\rm ns},j}(y,\muf)\, C^+_{i,\rm ns}(z,Q,\mur,\muf) + q_{\rm s}(y,\muf) \, C_{i,\rm q}(z,Q,\mur,\muf) + g(y,\muf) \, C_{i,\rm g}(z,Q,\mur,\muf)
  \biggr\}
  \, ,
\\
  \label{eq:disncZ-3}
  {\lefteqn{
      F_3^Z(x,Q^2) \,=\, \int\limits_0^1\, dz\, \int\limits_0^1\, dy \,
      \delta(x-yz)\, \sum\limits_{i=1}^{n_f}\, 
      2 \, v_i\, a_i \, \times }}
  \\
  \nonumber
  & &
  \times\, \biggl\{ 
  q^-_{{\rm ns},i}(y,\muf)\, C^-_{3,\rm ns}(z,Q,\mur,\muf) + q^{\rm v}_{\rm ns}(y,\muf)\, C^{\rm v}_{3,\rm ns}(z,Q,\mur,\muf) 
  \biggr\}
  \, ,
\end{eqnarray}
where $i=1,2$ and the pre-factors in \ce{eq:disncZ-12} are $f_1(x) = 1/2$, $f_2(x) = x$. 
The vector- and axial-vector coupling constants $v_i$ and $a_i$ in
\ce{eq:disncZ-12} are given by
\begin{eqnarray}
  \label{eq:va-coupl-Z12}
  v_i^2 + a_i^2 &=& \left\{ 
    \begin{array}[h]{cc}
      {1 \over 4} + \left({1 \over 2} - {4 \over 3} \sin^2 \theta_w \right)^2 & u\mbox{-type\, quarks}
      \, ,
      \\[2ex]
      {1 \over 4} + \left({1 \over 2} - {2 \over 3} \sin^2 \theta_w \right)^2 & d\mbox{-type\, quarks}
      \, ,
    \end{array}
    \right.
\end{eqnarray}
and, likewise, in \ce{eq:disncZ-3},
\begin{eqnarray}
  \label{eq:va-coupl-Z3}
  2 v_i a_i &=& \left\{ 
    \begin{array}[h]{cc}
      {1 \over 2} - {4 \over 3} \sin^2 \theta_w & u\mbox{-type\, quarks}
      \, ,
      \\[2ex]
      {1 \over 2} - {2 \over 3} \sin^2 \theta_w & d\mbox{-type\, quarks}
      \, .
    \end{array}
    \right.
\end{eqnarray}

The coefficient functions $C_{i}$ in \ced{eq:disncZ-12}{eq:disncZ-3} parameterize the hard partonic scattering process.
They depend only on the scaling variable $x$, and on dimensionless ratios of $Q^2$,
$\muf$ and the renormalization scale $\mur$. 
The perturbative expansion of $C_{i}$ in the strong coupling $\alpha_s$ 
up to two loops reads in the non-singlet sector,
\begin{eqnarray}
  \label{eq:ns-coeff}
  C^+_{i,\rm ns}(x) &=& 
  \delta(1-x) 
  + a_s \biggl\{ 
  \ciqo + L_M \Pqq0
  \biggr\} 
\\
& &
  + a^2_s \biggl\{ 
  \cinpt 
  + L_M \biggl( \Pnsp1 + \ciqo \* (\Pqq0 - \b0) \biggr) 
  + L_M^2 \biggl( {1 \over 2} \* \Pqq0 \* ( \Pqq0 - \b0) \biggr) 
\nonumber
\\
& &
  + L_R \, \b0 \* \ciqo + L_R L_M \, \b0 \* \Pqq0 
  \biggr\} 
  \, , 
\nonumber
\\
  \label{eq:3-coeff}
  C^-_{3,\rm ns}(x) &=& 
  \delta(1-x) 
  + a_s \biggl\{ 
  \ctqo + L_M \Pqq0
  \biggr\} 
\\
& &
  + a^2_s \biggl\{ 
  \ctnmt 
  + L_M \biggl( \Pnsm1 + \ctqo \* (\Pqq0 - \b0) \biggr) 
  + L_M^2 \biggl( {1 \over 2} \* \Pqq0 \* ( \Pqq0 - \b0) \biggr) 
\nonumber
\\
& &
  + L_R \, \b0 \* \ctqo + L_R L_M \, \b0 \* \Pqq0 
  \biggr\} 
  \, ,\nonumber 
\end{eqnarray}
where $a_s = \alpha_s(\mur)/(4 \pi)$ and $i=1,2$ in \ce{eq:ns-coeff}.
The complete scale dependence, 
{\it i.e.} the towers of logarithms in $L_M = \ln(Q^2 / \mufs)$ and $L_R = \ln(\murs / \mufs)$ 
(keeping $\mur \neq \muf$), has been derived by renormalization group methods
(see, e.g. \cite{vanNeerven:2000uj})
in terms of splitting functions $P_{ij}^{(l)}$ and the coefficients of the QCD
beta function, $\beta_l$.
In our normalization of the expansion parameter, $a_s = \alpha_s/(4 \pi)$, 
the conventions for the running coupling are
\begin{eqnarray}
  \label{eq:arun}
  \frac{d}{d \ln \mu^2}\: \frac{\alpha_s}{4\pi} \:\: \equiv \:\: 
  \frac{d\,a_s}{d \ln \mu^2} \:\: = \:\: 
  - \beta_0\, a_s^2 - \ldots 
  \:\: ,
  \qquad\qquad
  \b0 \,=\, {11 \over 3} C_A - { 2\over 3} n_f
  \, ,
\end{eqnarray}
with $\beta_0$ the usual expansion coefficient of the QCD beta function,
$C_A=3$ and $n_f$ the number of light flavors.

Note, that the valence coefficient function $C^{\rm v}_{3,\rm ns}$ in \ce{eq:disncZ-3} 
is defined as $C^{\rm v}_{3,\rm ns} = C^-_{3,\rm ns} + C^{\rm s}_{3,\rm ns}$. 
However, we have $C^{\rm s}_{3,\rm ns} \neq 0$ starting at three-loop order only,  
so that \ce{eq:disncZ-3} suffices with $C^{\rm v}_{3,\rm ns} = C^-_{3,\rm ns}$ up to NNLO.
In the singlet sector we have
\begin{eqnarray}    
  \label{eq:q-coeff}
  C_{i,\rm q}(x) &=& 
  \delta(1-x) 
  + a_s \biggl\{ 
  \ciqo + L_M \Pqq0
  \biggr\} 
\\
& &
  + a^2_s \biggl\{ 
  \ciqt 
  + L_M \biggl( \Pqq1 + \ciqo \* (\Pqq0 - \b0) + \cigo \* \Pgq0 \biggr) 
  + L_M^2 \biggl( {1 \over 2} \* \Pqq0 \* ( \Pqq0 - \b0) + {1 \over 2} \Pqg0 \* \Pgq0 \biggr) 
\nonumber
\\
& &
  + L_R \, \b0 \* \ciqo + L_R L_M \, \b0 \* \Pqq0 
  \biggr\} 
  \, , 
\nonumber
\\
\label{eq:g-coeff}
  C_{i,\rm g}(x) &=& 
  a_s \biggl\{ 
  \cigo + L_M \, \Pqg0
  \biggr\}
\\
& &
  + a^2_s \biggl\{ 
  \cigt 
  + L_M \biggl( \Pqg1 + \ciqo \* \Pqg0 + \cigo \* (\Pgg0 - \b0) \biggr) 
  + L_M^2 \biggl( {1 \over 2} \Pqq0 \* \Pqg0 + {1 \over 2} \* \Pqg0 \* ( \Pgg0 - \b0) \biggr) 
\nonumber
\\
& &
  + L_R \, \b0 \* \cigo + L_R L_M \, \b0 \* \Pqg0 
  \biggr\}
  \, ,
\nonumber
\end{eqnarray}
where again $i=1,2$ in \ce{eq:q-coeff}. 
The quark-singlet contribution contains the so-called pure-singlet part, $C_{i,\rm q} = C^+_{i,\rm ns} + C_{i,\rm ps}$, 
{\it i.e.} $\Pqq1 = \Pnsp1 + \Pps1$ and $\ciqt = \cinpt + \cipt$ in \ce{eq:q-coeff}.
Note, that starting at two-loop order we have $C_{i,\rm ps} \neq 0$. 
The DIS coefficient functions $c_{i,\rm k}^{(l)}$ are known to NNLO from Refs.~\cite{vanNeerven:1991nn,Zijlstra:1992qd,Zijlstra:1992kj,Moch:1999eb}, 
likewise, NNLO evolution of the PDFs has been determined in Refs.~\cite{Moch:2004pa,Vogt:2004mw} 
and even the hard corrections at order $\alpha_s^3$ are available~\cite{Vermaseren:2005qc,Moch:2008fj}.
Accurate parametrizations of all coefficient functions in \ced{eq:ns-coeff}{eq:g-coeff} 
can be taken e.g. from Refs.~\cite{Vermaseren:2005qc, Moch:2008fj} 
and the splitting functions $P_{ij}^{(l)}$ are given e.g. in Refs.~\cite{Moch:2004pa,Vogt:2004mw}~\footnote{
\label{footnote:nf-coeff}
Note, that with the conventions of Refs.~\cite{Moch:2004pa,Vogt:2004mw,Vermaseren:2005qc, Moch:2008fj} 
both the pure-singlet and the gluon coefficient functions as well as the 
the splitting functions $\Pqg0$ and $\Pqg1$ in \ced{eq:ns-coeff}{eq:g-coeff} 
need to be divided by a factor $2 n_f$ to account for the contribution of {\bf one}
individual quark flavor (not the anti-quark).
}.
All products in \ced{eq:ns-coeff}{eq:g-coeff} are understood as Mellin convolutions. 
They can be easily evaluated in terms of harmonic polylogarithms $H_{\vec{m}}(x)/(1\pm x)$ up to weight 4, 
see~\cite{Remiddi:1999ew}, and for their numerical evaluation we have used the {\sc Fortran} package~\cite{Gehrmann:2001pz}.

\bigskip

For the charged current case with $W^\pm$-boson exchange the DIS structure functions 
$F_i^{W^\pm}$ are given by,
\begin{eqnarray}
  \label{eq:disccWm-12}
  {\lefteqn{
  F_i^{W^-}(x,Q^2) \,=\, \frac{1}{2}\, f_i(x)\, \int\limits_0^1\, dz\, \int\limits_0^1\, dy \,
  \delta(x-yz)\, \frac{1}{n_f}
  \sum\limits_{j=1}^{n_f}\, 
  \left( v_j^2 + a_j^2 \right)\, \times }}
  \\
  \nonumber
  & &
  \times 
  \left\{ 
    \delta q^-_{\rm ns}(y,\muf)\, C^-_{i,\rm ns}(z,Q,\mur,\muf)    
    + q_{\rm s}(y,\muf) \, C_{i,\rm q}(z,Q,\mur,\muf) + g(y,\muf) \, C_{i,\rm g}(z,Q,\mur,\muf)
  \right\}
  \, ,
  \\
  \label{eq:disccWm-3}
  {\lefteqn{
  F_3^{W^-}(x,Q^2) \,=\, \frac{1}{2}\, \int\limits_0^1\, dz\, \int\limits_0^1\, dy \,
  \delta(x-yz)\,  \frac{1}{n_f}
  \sum\limits_{i=1}^{n_f}\, 
  2 \, v_i\, a_i \, \times }}
  \\
  \nonumber
  & &
  \times 
  \left\{ 
    \delta q^+_{\rm ns}(y,\muf)\, C^+_{3,\rm ns}(z,Q,\mur,\muf) 
    + q^{\rm v}_{\rm ns}(y,\muf) \, C^{\rm v}_{3,\rm ns}(z,Q,\mur,\muf) 
  \right\}
  \, ,
\end{eqnarray}
where, as above, $C_{i,\rm{q}}=C_{i,\rm{ns}}^++C_{i,\rm{ps}}$ and, also, 
$C^{\rm v}_{3,\rm ns} = C^-_{i,\rm ns}$ up to two-loop order.
The asymmetry $\delta q^\pm_{\rm ns}$ parametrizes the iso-triplet component of the proton, 
{\it i.e.} $u \neq d$ and so on. It is defined as 
\begin{eqnarray}
  \label{eq:du-pdf}
  \delta q^\pm_{\rm ns} &=& \sum\limits_{i \in u\rm{-type}} \sum\limits_{j \in d\rm{-type}} \, 
\left\{
  \biggl(q_i\pm{\bar q}_i \biggr) - \biggl(q_j\pm{\bar q}_j \biggr)
\right\}
  \, .
\end{eqnarray}
Its numerical impact is expected to be small though. 
The respective results for $F_i^{W^+}$ are obtained from \ced{eq:disccWm-12}{eq:disccWm-3} 
with the simple replacement $\delta q^\pm_{\rm ns} \to - \delta q^\pm_{\rm ns}$.

The vector- and axial-vector coupling constants $v_i$ and $a_i$ 
are given by
\begin{equation}
  \label{eq:va-coupl-W}
  v_i = a_i = {1 \over \sqrt{2}}
  \, .
\end{equation}
The coefficient functions in \ced{eq:disccWm-12}{eq:disccWm-3} including their 
dependence on the factorization and the renormalization scales 
can be obtained from \ced{eq:ns-coeff}{eq:g-coeff} with the help of the 
following simple substitutions $c_{i,\rm{ns}}^{(2),+} \leftrightarrow c_{i,\rm{ns}}^{(2),-}$,
$c_{3,\rm{ns}}^{(2),-}\leftrightarrow c_{3,\rm{ns}}^{(2),+}$ and
$P_{\rm{ns}}^{(1),+}\leftrightarrow P_{\rm{ns}}^{(1),-}$ and so on.
Again, all expressions for the coefficient and splitting functions are given in Refs.~\cite{Vermaseren:2005qc, Moch:2008fj} 
and~\cite{Moch:2004pa,Vogt:2004mw}, respectively.

\bigskip

\ce{eq:disapproach} with the explicit expressions for the DIS structure functions 
inserted provides the backbone of our NNLO QCD predictions for Higgs production in VBF. 
However, as emphasized above, the underlying factorization is not exact beyond
NLO and therefore, the non-factorizable corrections need to be estimated.
This will be done in the following.

\subsection{Non-factorizable contributions}
\label{sec:dia-pentagon}
In order to assess the quality of the factorization approach, we now 
estimate the size of the non-factorizable contributions, {\it i.e.} those coming from diagrams involving the 
exchange of gluons between the two quark lines and not included in the structure function approach.
Neglecting interferences between $t$- and $u$-channel diagrams, which are kinematically
suppressed, this class of diagrams vanishes at NLO because of color conservation, but contributes at NNLO for the first time. 
Here, the notion ``class of diagrams'' refers to a gauge invariant subset of the diagrams that contributes to a certain process. 
Examples of non-factorizable diagrams are shown in \cf{fig:dble-pent}.
\begin{figure}[t!]
\centering
\includegraphics[scale=0.3]{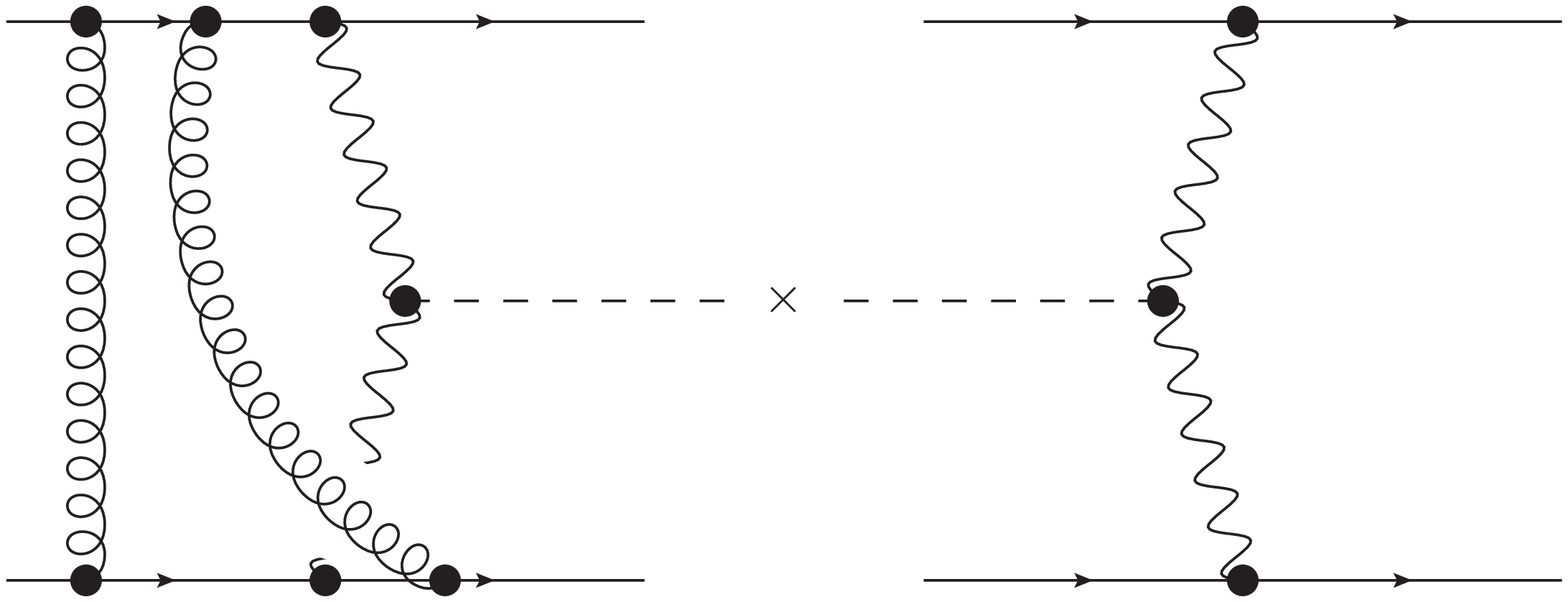}
\hspace*{10mm}     
\includegraphics[scale=0.3]{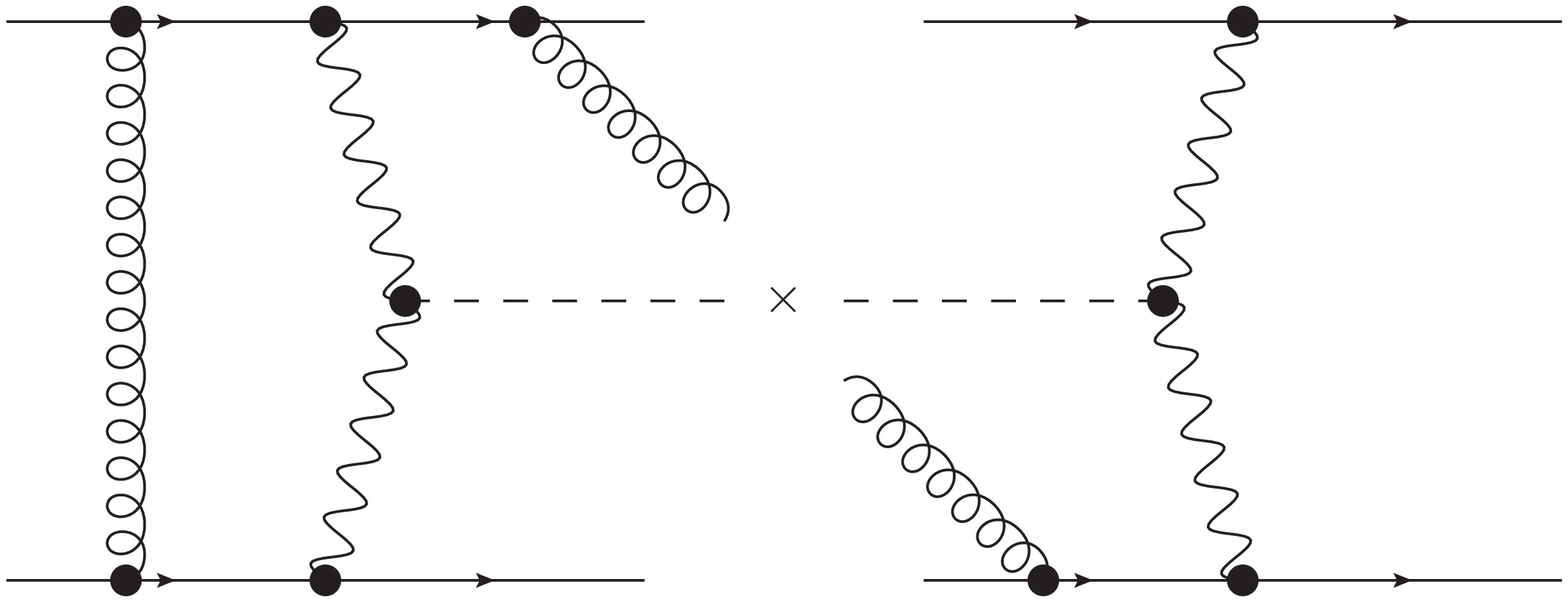} 
\\[2ex]
\caption{\small
  \label{fig:dble-pent}
Examples of squared matrix elements contributing at NNLO to VBF involving a double gluon exchange between the two quark lines.
}
\end{figure}
\begin{figure}[t!]
\centering
\includegraphics[scale=0.500]{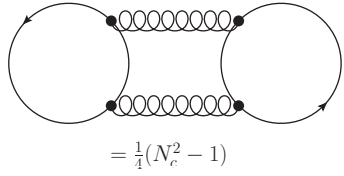}
\\[4ex]
\caption{\small
  \label{fig:col-penta}
  Color configurations associated to non-factorizable double-gluon exchange corrections to VBF at NNLO. 
}
\end{figure}
\begin{figure}[t!]
\centering
\includegraphics[scale=0.500]{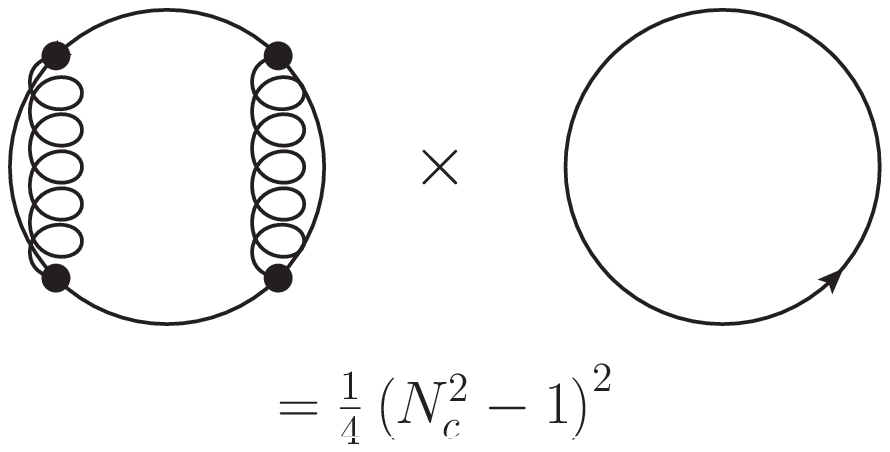}
\hspace*{ 7mm}
\includegraphics[scale=0.500]{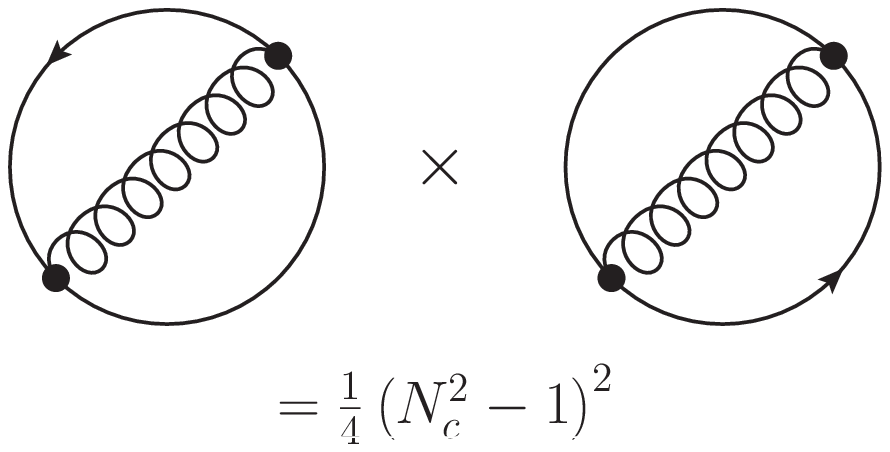}
\hspace*{ 7mm}
\includegraphics[scale=0.500]{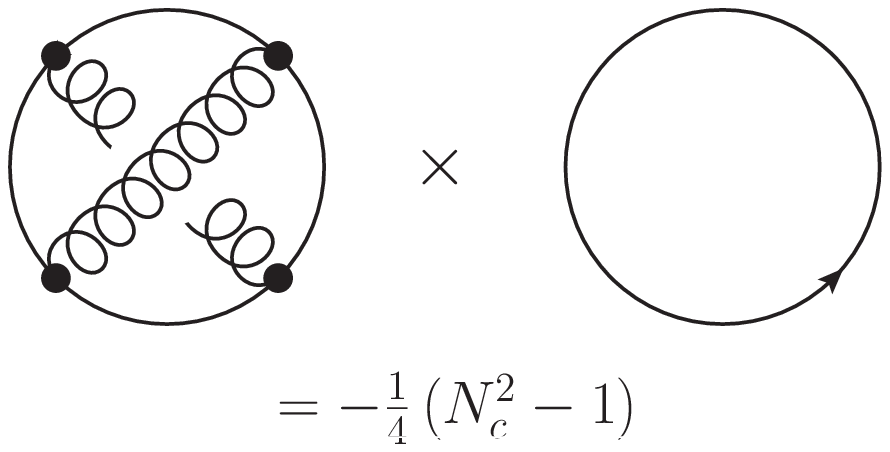} 
\\[4ex]
\caption{\small
  \label{fig:col-fact}
  Color configurations associated to factorizable corrections to VBF at NNLO. }
\end{figure}
Before we present a detailed numerical estimate of the size of the non-factorizable contributions,
we briefly recall two general arguments from Refs.~\cite{Bolzoni:2010xr,Bolzoni:2010as} justifying their omission.

The first argument is based on the study of the associated color factors.
The possible color configurations for  factorizable and non-factorizable corrections are shown
in \cfa{fig:col-penta}{fig:col-fact}, respectively, together with the associated color factors. 
The leading color factor of the factorizable corrections is $(N_c^2 -1)^2/4=16$, 
while we have $(N_c^2-1)/4=2$ (for $ N_c = 3$) for the double gluon exchange consisting of a double color-traces.
Hence the non-factorizable corrections are suppressed by a factor $O(1/N_c^2)$ with respect to the leading factorizable ones. 

The second argument is based on the kinematical dependence of diagrams like those shown in \cf{fig:dble-pent}.
Such contributions, see e.g., \cf{fig:dble-pent} (right), 
come from the interference of diagrams with  one or two gluons radiated by the upper quark line with diagrams 
where gluons are radiated by the lower line.  Angular ordering in gluon emission, however, 
leads to radiation  close to the quark from which it is emitted,  and since these quarks tend to be very forward (or backward) due to the exchange
of a spin-1 particle in the $t$-channel, there is generally very little overlap in phase space. 
Those arguments have already been used in~\cite{Figy:2007kv} to justify neglecting real-virtual double gluon exchange diagrams 
in the computation of NLO QCD corrections for Higgs in VBF in association with three jets.

We now corroborate these considerations with a quantitative analysis. Let us express the total cross section at order $\alpha_s^2$
as 
\begin{eqnarray}
&&\sigma_{NNLO} = \sigma_0  \left( 1 +  \alpha_s  \Delta_1 +  \alpha_s^2 \Delta_2 \right)\,,
 \end{eqnarray}
with $\Delta_2  =  \Delta^{fact}_{2}+ \Delta^{non-fact}_{2}$. As already said, $ \Delta_1$ receives  contributions 
mainly from factorizable corrections. The non-factorizable ones that are not kinematically suppressed are exactly zero due to color while
interference effects between amplitudes with identical quark lines in the final state are highly suppressed (and not included beyond LO in our approach).

The exact calculation of $\Delta^{non-fact}_{2}$ being out of reach, we  estimate  the ratio of the non-factorizable contributions, $\Delta^{non-fact}_{2}$, vs. the factorizable ones, $\Delta^{fact}_{2}$, as follows
\begin{equation}
  \label{eq:Rnlo}
  R_2 \, \equiv \, \frac{\Delta^{non-fact}_{2}}{\Delta^{fact}_{2}}\,\simeq\, 
  \frac{1}{N_c^2-1} R_{1} \,\equiv\, \frac{1}{N_c^2-1}\, \frac{\Delta^{non-fact, \, {U(1)}}_{1} }{\Delta^{fact, \, {U(1)}}_{1} }
  \, ,
\end{equation}
where $\Delta^{fact, {U(1)}}_{1} = \Delta^{fact}_{1}/C_F$  and $\Delta^{non-fact, {U(1)}}_{1}$ denotes the ``would-be'' impact of the corrections coming from non-factorizable diagrams at NLO,
{\it i.e.}, the class of diagrams involving the exchange of one gluon between
the two quark lines, computed as if the color factor were non-vanishing. 
In other words, $\Delta^{non-fact, {U(1)}}_{1}$ can be thought of as  the correction due to the gauge invariant class of diagrams where an extra $U(1)$ gauge boson is exchanged between the two quark lines including real and virtual diagrams.
The $R_2 \simeq R_1$ approximation assumes, of course, that the ratios will not dramatically change in going from NLO to NNLO. As there is no  substantial difference in the kinematics and no non-Abelian vertices enter at NNLO in diagrams where two gluons are exchanged in a color singlet in the $t$-channel, we can conclude that \ce{eq:Rnlo} should provide a reasonable estimate.

To compute $\Delta^{non-fact}_{1}$ we need to account for the four diagrams 
shown in \cf{fig:penta-nlo} along with the analogous real emission terms. 
As our argument only needs to provide an estimate of the corrections and it is based 
on the kinematics we can slightly simplify the calculation of the tensor integrals by 
considering only vector couplings of the vector boson to the quarks, which eliminates all but the scalar five-point functions.
The latter can be reduced in terms of scalar four-point functions~\cite{vanNeerven:1983vr,Denner:2002ii,Fleischer:2010sq}
and then evaluated, together with the scalar four- and three-point functions coming from the
Passarino-Veltmann reduction, with the help of the \textsc{QcdLoop} package~\cite{Ellis:2007qk}. 
For completeness, we stress that the results for the virtual diagrams have been checked against the amplitude automatically generated by \madloop~\cite{Hirschi:2011pa}, 
where machine precision agreement has been found point by point in the phase space.

The combination of the virtual and the real emission part, with the
subtraction of the soft divergences (no collinear divergences occur in this class of diagrams), 
has been done via \madfks~\cite{Frederix:2009yq} that generates all the needed
counterterms  automatically and performs also the integration over phase space. 
In practice, the computation of the $\mathcal O(\alpha_s)$ part of the cross section that
enters in $\Delta^{non-fact}_{1}$,  has been obtained as the difference of the complete NLO and the corresponding Born cross sections, 
and special attention has been paid to controlling the uncertainty of the numerical integration of
the real emission contributions.
\begin{figure}[t!]
\centering
\includegraphics[scale=0.280]{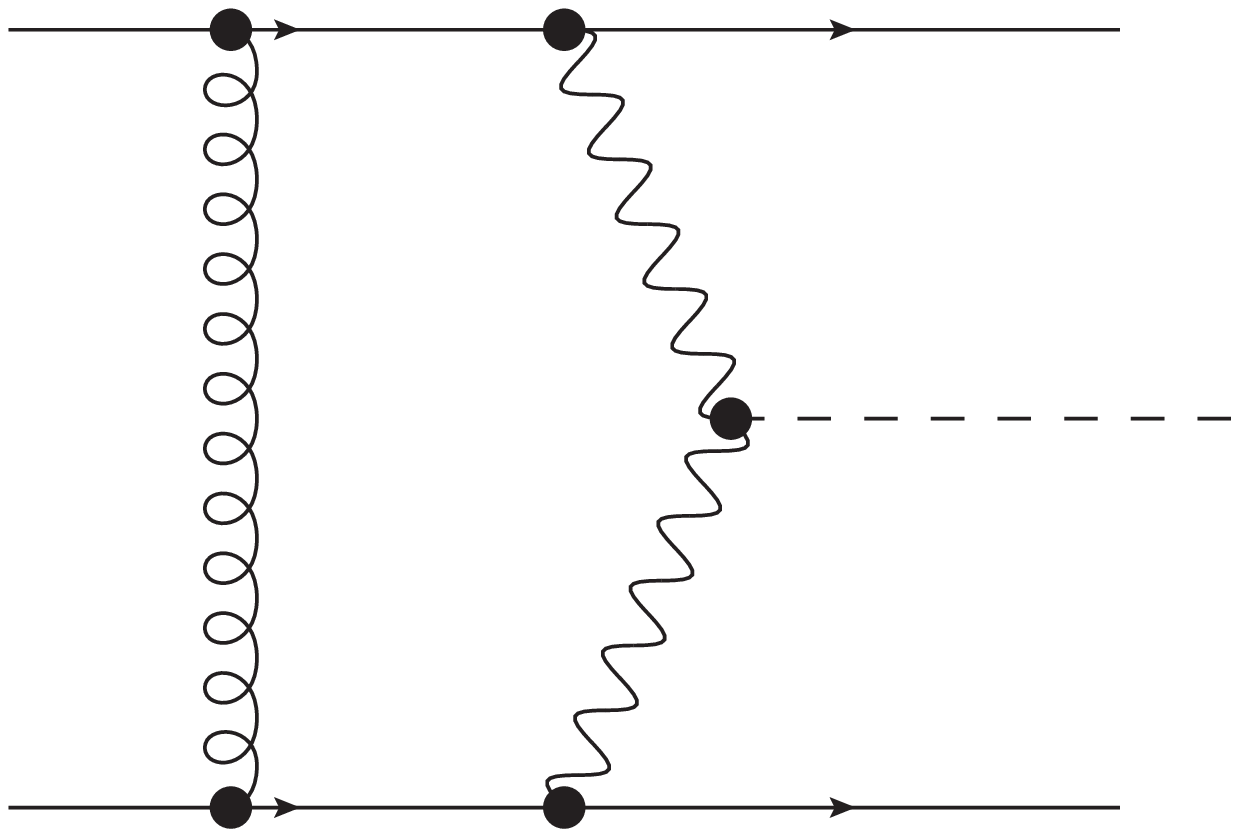}
\hspace*{3mm}
\includegraphics[scale=0.280]{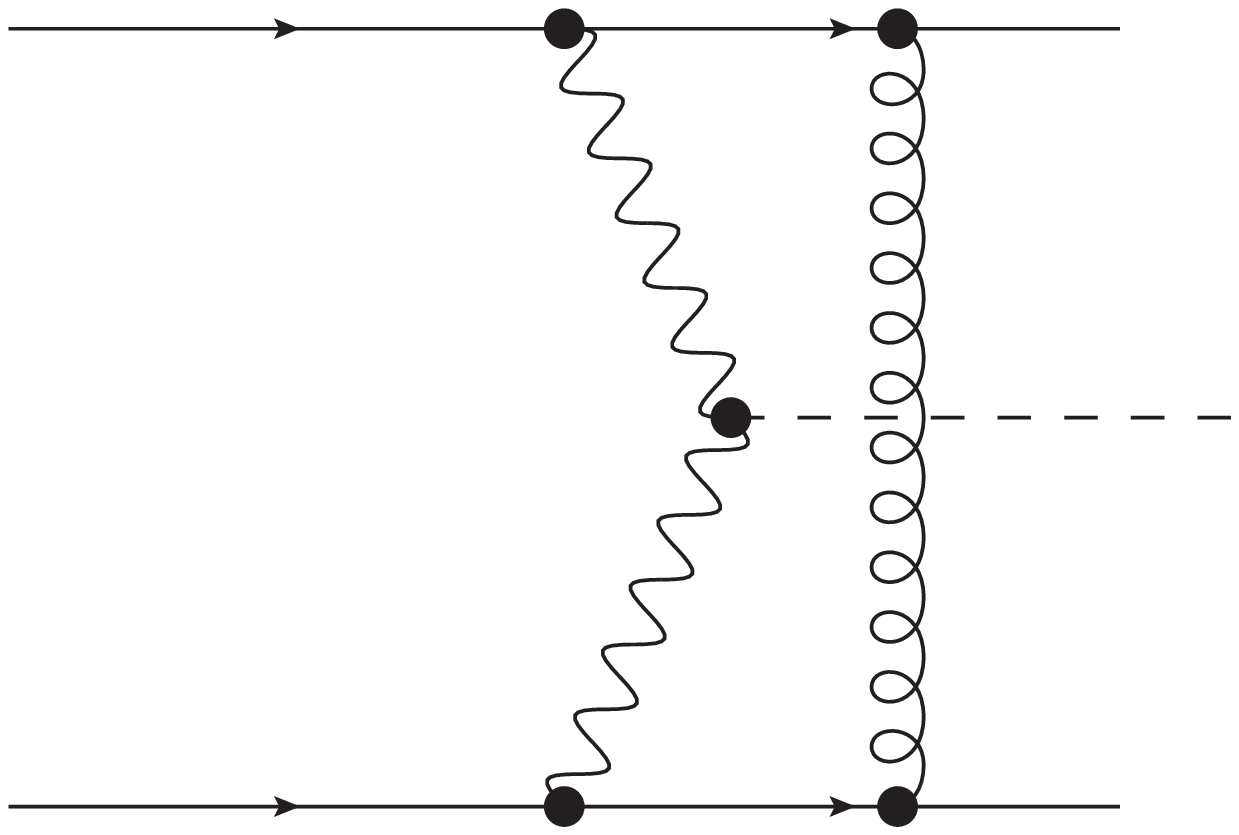} 
%\\[4ex]
\hspace*{3mm}
\includegraphics[scale=0.280]{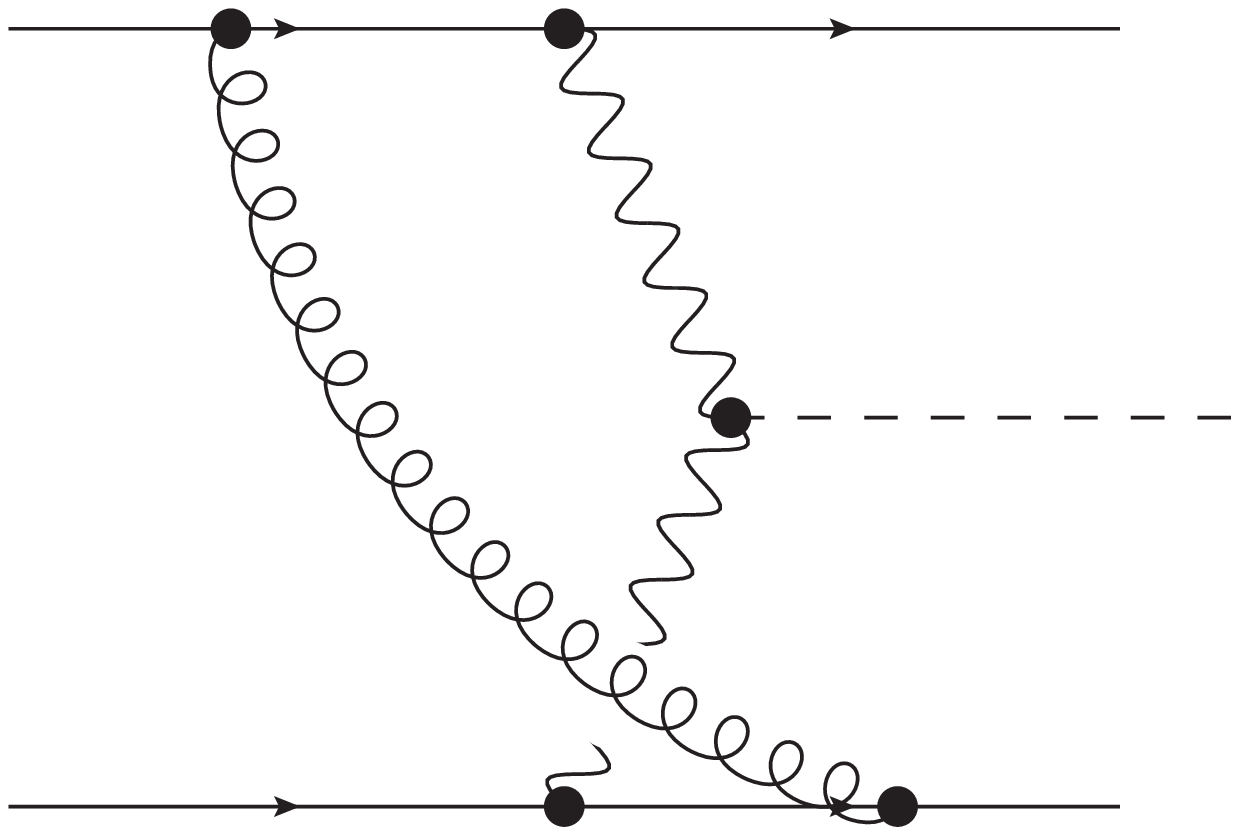} 
\hspace*{3mm}
\includegraphics[scale=0.280]{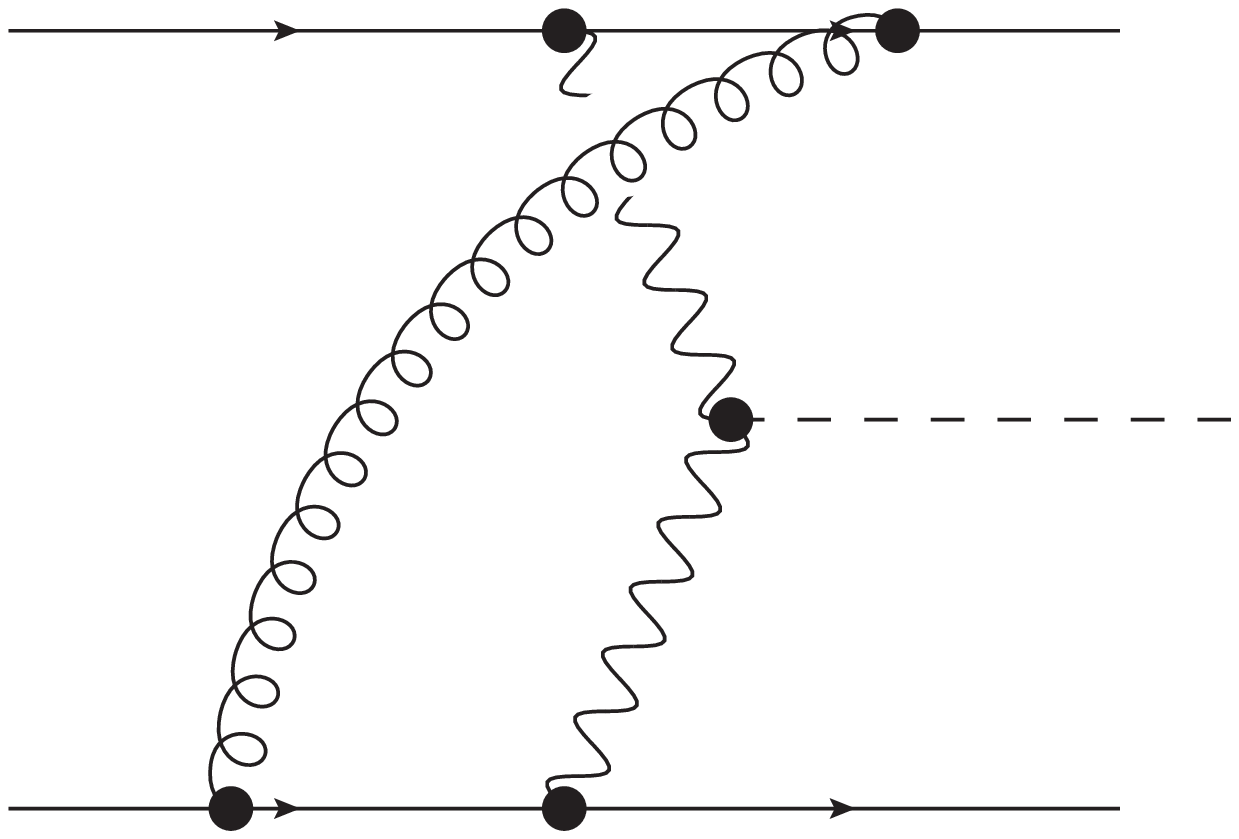} 
\\[4ex]
\caption{\small
  \label{fig:penta-nlo}
The four virtual topologies with one gluon exchange between the quark lines
that would contribute at NLO, barring the vanishing color factor. Those diagrams
represent the virtual part included in $\Delta^{non-fact}_{NLO}$.
}
\end{figure}

\begin{table}[ht!]
\begin{center}
\begin{tabular}{|c|cccc|}
\hline $ \hmass \, [\gev] $ & $\sigma_0 \,[\rm{pb}]$ & $ \Delta_{1}^{non-fact, \, U(1)} $ & $ \Delta_{1}^{fact, \, U(1)}$ & $ R_1 $ \\ \hline
 $ 100 $ & $ 3.06 \cdot 10^{-5} $ & $ 3.00 (4) \cdot 10^{-2} $ & $ 8.79 \cdot 10^{-2} $  & $ 0.34 $ \\
 $ 120 $ & $ 2.09 \cdot 10^{-5} $ & $ 2.90 (5) \cdot 10^{-2} $ & $ 9.22 \cdot 10^{-2} $  & $ 0.31 $ \\
 $ 150 $ & $ 1.19 \cdot 10^{-5} $ & $ 2.21 (5) \cdot 10^{-2} $ & $ 9.91 \cdot 10^{-2} $  & $ 0.22 $ \\
 $ 200 $ & $ 4.87 \cdot 10^{-6} $ & $ -3.2 (5) \cdot 10^{-3} $ & $ 1.12 \cdot 10^{-1} $  & $ -0.03 $ \\
 $ 250 $ & $ 2.04 \cdot 10^{-6} $ & $ 8 (4) \cdot 10^{-4} $ & $ 1.25 \cdot 10^{-1} $  & $ 0.01 $ \\
 $ 300 $ & $ 8.68 \cdot 10^{-7} $ & $ 2.7 (4) \cdot 10^{-3} $ & $ 1.39 \cdot 10^{-1} $  & $ 0.02 $ \\
\hline
\end{tabular}
\caption{\small
 Non-diagonal NLO QCD corrections to VBF at the Tevatron, $\sqrt S = 1.96 \tev$. Numbers have been computed ignoring the vanishing color factors of the diagrams in \cf{fig:penta-nlo}. The MRST2002~\cite{Martin:2002aw} NLO PDF set has been used. Renormalization and factorization scales have been set to $\wmass$. Integration errors, if relevant, are shown in parenthesis.}
\label{tab:penta-1960}
\end{center}
\end{table}

\begin{table}[ht!]
\begin{center}
\begin{tabular}{|c|cccc|}
\hline $ \hmass \, [\gev] $ & $\sigma_0 \,[\rm{pb}]$ & $ \Delta_{1}^{non-fact, \, U(1)} $ & $ \Delta_{1}^{fact, \, U(1)}$ & $ R_1 $ \\ \hline
 $ 100 $ & $ 2.04 \cdot 10^{-3} $ & $ 6.6 (3) \cdot 10^{-3} $ & $ 1.57 \cdot 10^{-2} $  & $ 0.42 $ \\
 $ 120 $ & $ 1.74 \cdot 10^{-3} $ & $ 5.4 (4) \cdot 10^{-3} $ & $ 1.54 \cdot 10^{-2} $  & $ 0.35 $ \\
 $ 150 $ & $ 1.39 \cdot 10^{-3} $ & $ 3.4 (2) \cdot 10^{-3} $ & $ 1.53 \cdot 10^{-2} $  & $ 0.22 $ \\
 $ 200 $ & $ 9.84 \cdot 10^{-4} $ & $ -3.6 (3) \cdot 10^{-3} $ & $ 1.59 \cdot 10^{-2} $  & $ -0.22 $ \\
 $ 250 $ & $ 7.12 \cdot 10^{-4} $ & $ -1.2 (2) \cdot 10^{-3} $ & $ 1.71 \cdot 10^{-2} $  & $ -0.07 $ \\
 $ 300 $ & $ 5.26 \cdot 10^{-4} $ & $ -4 (2) \cdot 10^{-4} $ & $ 1.85 \cdot 10^{-2} $  & $ -0.02 $ \\
 $ 400 $ & $ 3.00 \cdot 10^{-4} $ & $ 2 (2) \cdot 10^{-4} $ & $ 2.24 \cdot 10^{-2} $  & $ 0.01 $ \\
 $ 500 $ & $ 1.78 \cdot 10^{-4} $ & $ 1 (2) \cdot 10^{-4} $ & $ 2.69 \cdot 10^{-2} $  & $ 0.00 $ \\
 $ 650 $ & $ 8.66 \cdot 10^{-5} $ & $ 5 (2) \cdot 10^{-4} $ & $ 3.45 \cdot 10^{-2} $  & $ 0.01 $ \\
 $ 800 $ & $ 4.41 \cdot 10^{-5} $ & $ 2 (2) \cdot 10^{-4} $ & $ 4.28 \cdot 10^{-2} $  & $ 0.00 $ \\
 $ 1000 $ & $ 1.88 \cdot 10^{-5} $ & $ 5 (2) \cdot 10^{-4} $ & $ 5.46 \cdot 10^{-2} $  & $ 0.01 $ \\
\hline
\end{tabular}
\caption{\small
 Non-diagonal NLO QCD corrections to VBF at the LHC, $\sqrt S = 7 \tev$. Numbers have been computed ignoring the vanishing color factors of the diagrams in \cf{fig:penta-nlo}. The MRST2002~\cite{Martin:2002aw} NLO PDF set has been used. Renormalization and factorization scales have been set to $\wmass$. Integration errors, if relevant, are shown in parenthesis.}
\label{tab:penta-7000}
\end{center}
\end{table}

\begin{table}[ht!]
\begin{center}
\begin{tabular}{|c|cccc|}
\hline $ \hmass \, [\gev] $ & $\sigma_0 \,[\rm{pb}]$ & $ \Delta_{1}^{non-fact, \, U(1)} $ & $ \Delta_{1}^{fact, \, U(1)}$ & $ R_1 $ \\ \hline
 $ 100 $ & $ 5.63 \cdot 10^{-3} $ & $ 4.0 (4) \cdot 10^{-3} $ & $ 6.34 \cdot 10^{-3} $  & $ 0.63 $ \\
 $ 120 $ & $ 4.97 \cdot 10^{-3} $ & $ 3.1 (3) \cdot 10^{-3} $ & $ 5.76 \cdot 10^{-3} $  & $ 0.54 $ \\
 $ 140 $ & $ 4.42 \cdot 10^{-3} $ & $ 2.8 (3) \cdot 10^{-3} $ & $ 4.94 \cdot 10^{-3} $  & $ 0.56 $ \\
 $ 150 $ & $ 4.17 \cdot 10^{-3} $ & $ 2.4 (3) \cdot 10^{-3} $ & $ 4.62 \cdot 10^{-3} $  & $ 0.51 $ \\
 $ 155 $ & $ 4.06 \cdot 10^{-3} $ & $ 1.7 (3) \cdot 10^{-3} $ & $ 4.51 \cdot 10^{-3} $  & $ 0.37 $ \\
 $ 160 $ & $ 3.95 \cdot 10^{-3} $ & $ 1.5 (3) \cdot 10^{-3} $ & $ 4.39 \cdot 10^{-3} $  & $ 0.33 $ \\
 $ 165 $ & $ 3.84 \cdot 10^{-3} $ & $ 8 (2) \cdot 10^{-4} $ & $ 4.28 \cdot 10^{-3} $  & $ 0.18 $ \\
 $ 170 $ & $ 3.74 \cdot 10^{-3} $ & $ 10 (3) \cdot 10^{-5} $ & $ 4.20 \cdot 10^{-3} $  & $ 0.02 $ \\
 $ 175 $ & $ 3.64 \cdot 10^{-3} $ & $ -8 (3) \cdot 10^{-4} $ & $ 4.08 \cdot 10^{-3} $  & $ -0.21 $ \\
 $ 180 $ & $ 3.54 \cdot 10^{-3} $ & $ -1.3 (2) \cdot 10^{-3} $ & $ 3.87 \cdot 10^{-3} $  & $ -0.35 $ \\
 $ 185 $ & $ 3.45 \cdot 10^{-3} $ & $ -3.4 (4) \cdot 10^{-3} $ & $ 3.86 \cdot 10^{-3} $  & $ -0.89 $ \\
 $ 190 $ & $ 3.36 \cdot 10^{-3} $ & $ -2.8 (3) \cdot 10^{-3} $ & $ 3.82 \cdot 10^{-3} $  & $ -0.73 $ \\
 $ 195 $ & $ 3.27 \cdot 10^{-3} $ & $ -2.5 (3) \cdot 10^{-3} $ & $ 3.74 \cdot 10^{-3} $  & $ -0.67 $ \\
 $ 200 $ & $ 3.19 \cdot 10^{-3} $ & $ -2.6 (2) \cdot 10^{-3} $ & $ 3.74 \cdot 10^{-3} $  & $ -0.71 $ \\
 $ 210 $ & $ 3.03 \cdot 10^{-3} $ & $ -1.6 (3) \cdot 10^{-3} $ & $ 3.62 \cdot 10^{-3} $  & $ -0.43 $ \\
 $ 230 $ & $ 2.75 \cdot 10^{-3} $ & $ -7 (5) \cdot 10^{-4} $ & $ 3.40 \cdot 10^{-3} $  & $ -0.21 $ \\
 $ 250 $ & $ 2.50 \cdot 10^{-3} $ & $ -9 (2) \cdot 10^{-4} $ & $ 3.32 \cdot 10^{-3} $  & $ -0.26 $ \\
 $ 300 $ & $ 1.99 \cdot 10^{-3} $ & $ -5 (2) \cdot 10^{-4} $ & $ 3.21 \cdot 10^{-3} $  & $ -0.14 $ \\
 $ 400 $ & $ 1.32 \cdot 10^{-3} $ & $ 2 (2) \cdot 10^{-4} $ & $ 3.98 \cdot 10^{-3} $  & $ 0.05 $ \\
 $ 500 $ & $ 9.17 \cdot 10^{-4} $ & $ 3 (2) \cdot 10^{-4} $ & $ 5.32 \cdot 10^{-3} $  & $ 0.06 $ \\
 $ 650 $ & $ 5.59 \cdot 10^{-4} $ & $ 6 (2) \cdot 10^{-4} $ & $ 7.98 \cdot 10^{-3} $  & $ 0.07 $ \\
 $ 800 $ & $ 3.57 \cdot 10^{-4} $ & $ 2 (2) \cdot 10^{-4} $ & $ 1.11 \cdot 10^{-2} $  & $ 0.01 $ \\
 $ 1000 $ & $ 2.06 \cdot 10^{-4} $ & $ 3 (1) \cdot 10^{-4} $ & $ 1.63 \cdot 10^{-2} $  & $ 0.02 $ \\
\hline
\end{tabular}
\caption{\small
 Non-diagonal NLO QCD corrections to VBF at the LHC, $\sqrt S = 14 \tev$. Numbers have been computed ignoring the vanishing color factors of the diagrams in \cf{fig:penta-nlo}. The MRST2002~\cite{Martin:2002aw} NLO PDF set has been used. Renormalization and factorization scales have been set to $\wmass$. Integration errors, if relevant, are shown in parenthesis.}
\label{tab:penta-14000}
\end{center}
\end{table}

In \ctd{tab:penta-1960}{tab:penta-14000} we present the results for the 
non-factorizable corrections, $\Delta^{non-fact}_{1}$, at the Tevatron and at the LHC, 
compared with the quark-initiated factorizable ones, $\Delta^{fact}_{1}$, 
and we have used the same color factor of the Born term. For the sake of simplicity, 
we have focused on the $ud\to udH$ channel, with $Z$-boson exchange and, 
as mentioned above, we have considered only the vector coupling of the $Z$-boson to the quarks.
Interestingly, the numbers in \ctd{tab:penta-1960}{tab:penta-14000} display a sudden change of sign for 
$\Delta^{non-fact}_{1}$ at around $\hmass =180\gev = 2\,\zmass$\,, {\it i.e.}, the threshold of the $h \to ZZ$ process, 
which is due to the use of the zero-width approximation for the $Z$-bosons in the loop propagators, 
cf. also Refs.~\cite{Bredenstein:2006rh,Bredenstein:2006ha}. A consistent inclusion of $Z$-boson width effects, which
is beyond the scope of our analysis, would regularize this behavior.

The results of \ctd{tab:penta-1960}{tab:penta-14000} show an $R_{1}$ always well below unity.
Once the $O(1/N_c^2)$ color suppression in \ce{eq:Rnlo} is taken into account, 
one finds an upper bound on $R_{2} < 10\%$. 
As the impact of $\Delta^{fact}_{2} $ on the total cross-section is at the $1\%$ level, 
we estimate the contribution of the omitted non-factorizable corrections  at most at the per-mil level, hence negligible in our scheme.
Finally, note that $\Delta^{non-fact}_{1}$ decreases with increasing Higgs
boson masses $\hmass$, so that, as expected from the fact that the Higgs boson
acts as a ``kinematical de-correlator'' between the two jets,
the size of the non-factorizable corrections becomes totally negligible for $\hmass > 300$ GeV.

\subsection{Contributions from heavy-quark loops}
\label{sec:dia-tbloop}
Diagrams with heavy-quark loops can also provide contributions at NNLO in QCD that are not included in the structure function approach. 
Following our definition of VBF given in \csec{sec:SetStage} these contributions are classified in the strict sense as ``non-VBF'' processes.
However, given that such effects are genuinely new at NNLO and, 
moreover, that they have not been subject to extensive consideration in the
context of VBF in the literature before, cf. the previous discussions in Refs.~\cite{Bolzoni:2010as,Zaro:2010zz},
they deserve detailed study here.

\begin{figure}[ht!]
\centering
\includegraphics[scale=0.350]{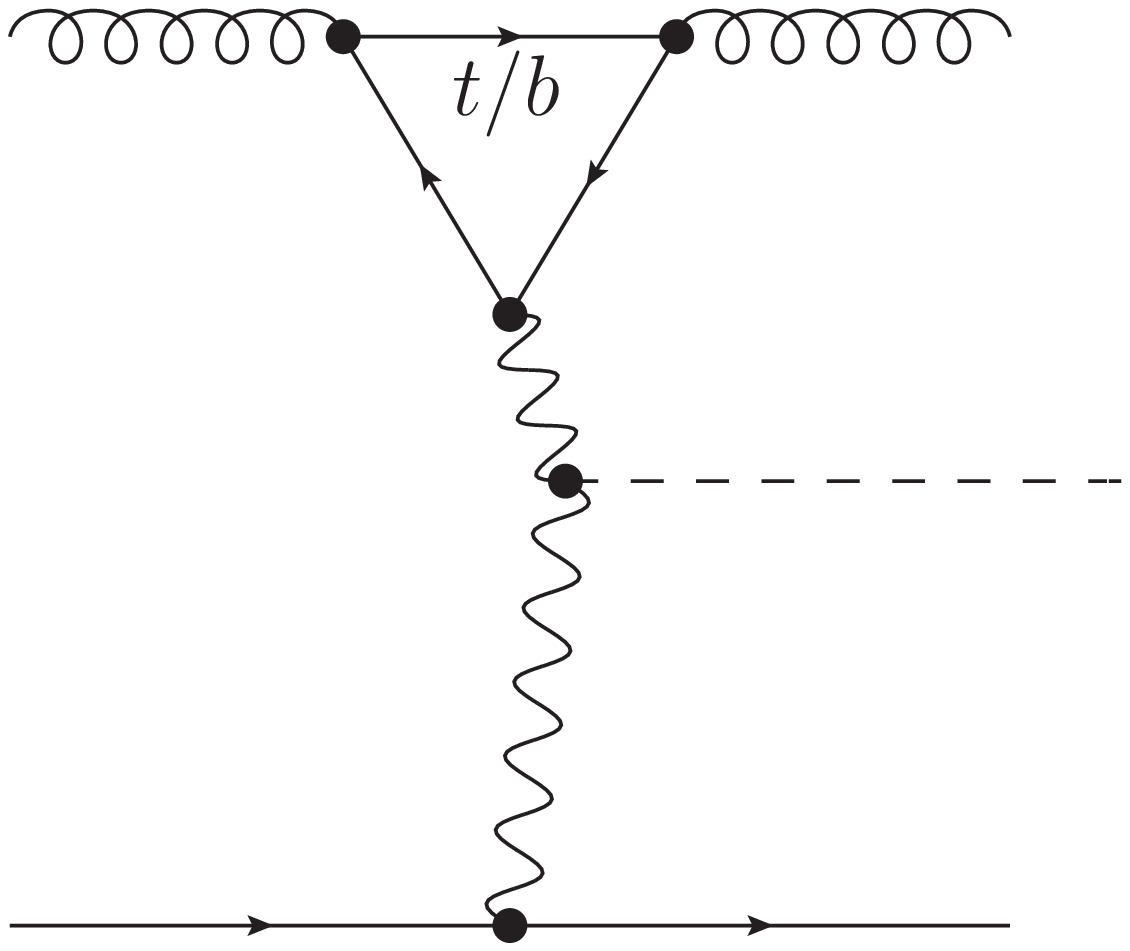}
\hspace*{5mm}
\includegraphics[scale=0.350]{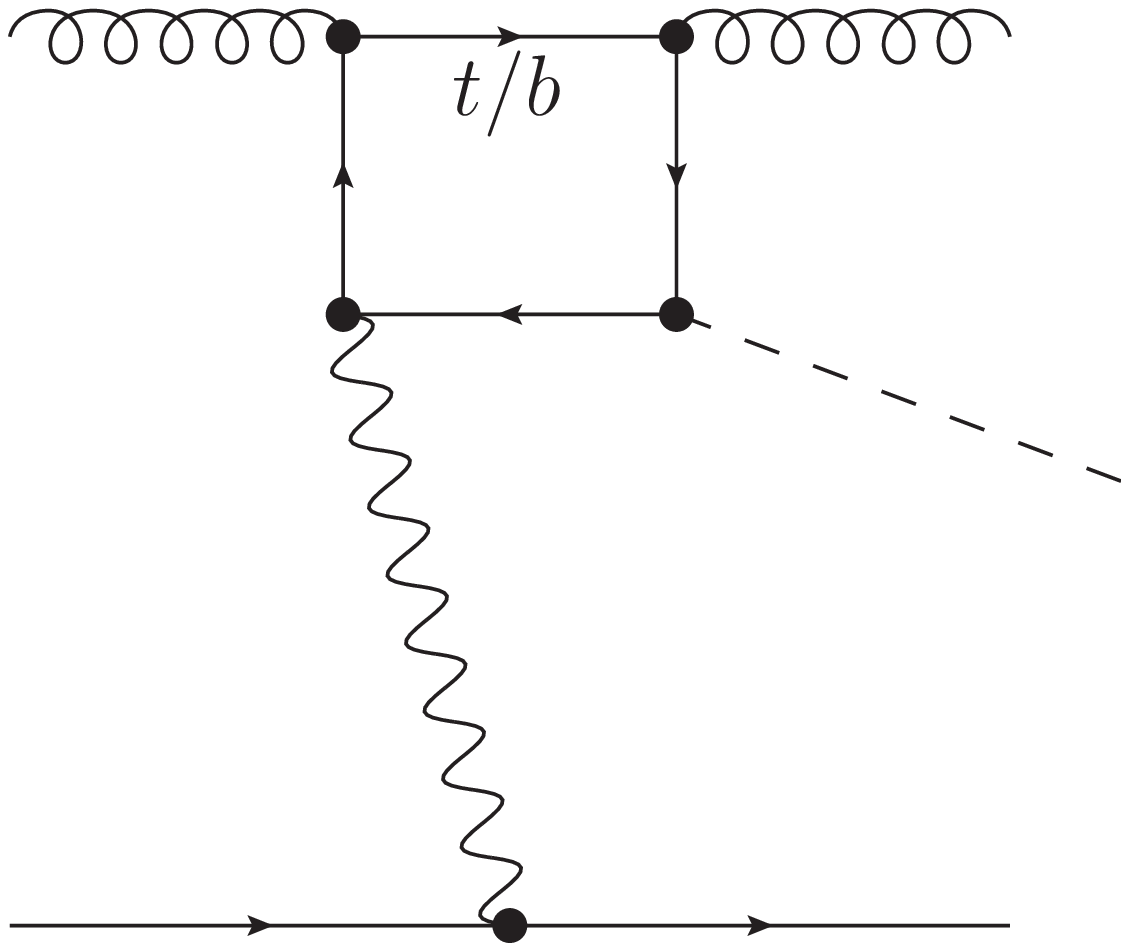}
\caption{\small
  \label{fig:heavyquarks}
Next-to-next-to-leading order QCD corrections due to heavy-quarks $(t/b)$ loops: pure one-loop diagrams contributing through their modulo squared.
}
\end{figure}

\begin{figure}[ht!]
\centering
\includegraphics[scale=0.350]{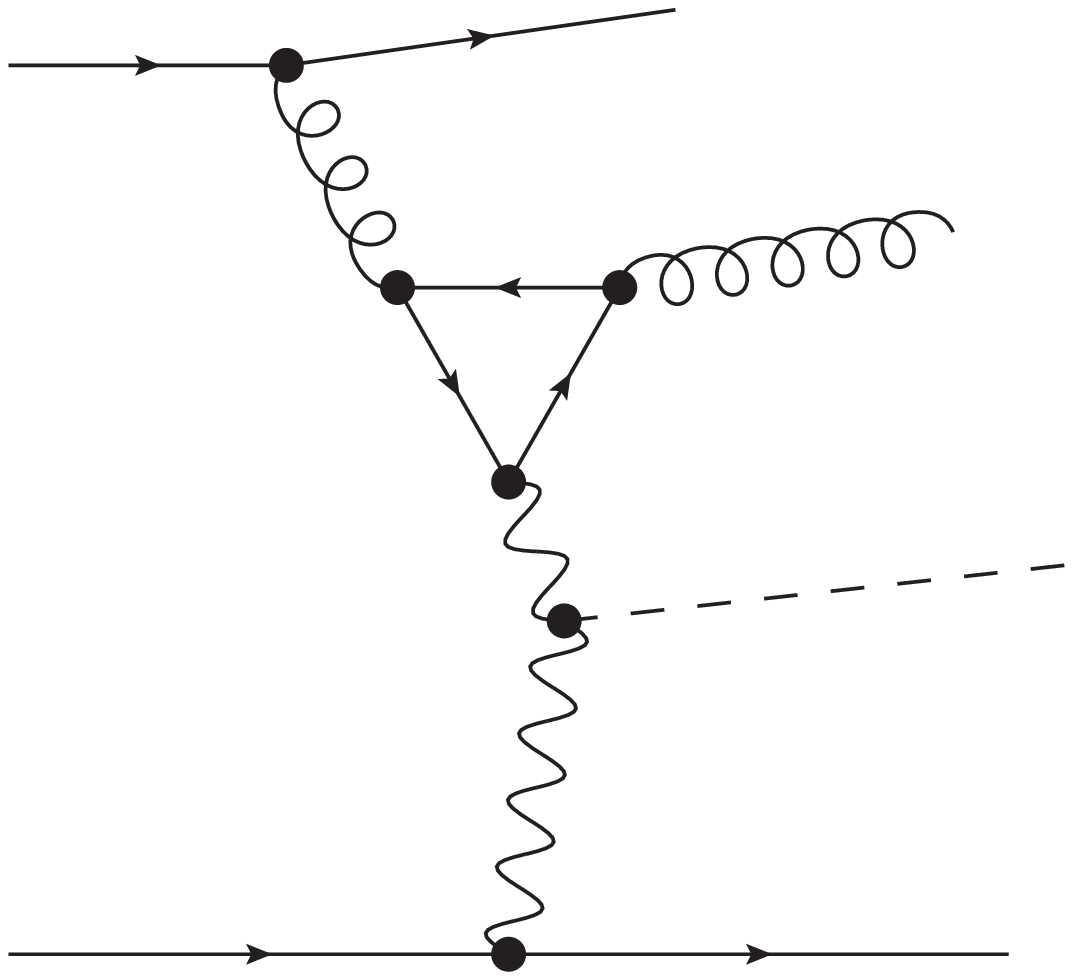}
\hspace*{5mm}
\includegraphics[scale=0.350]{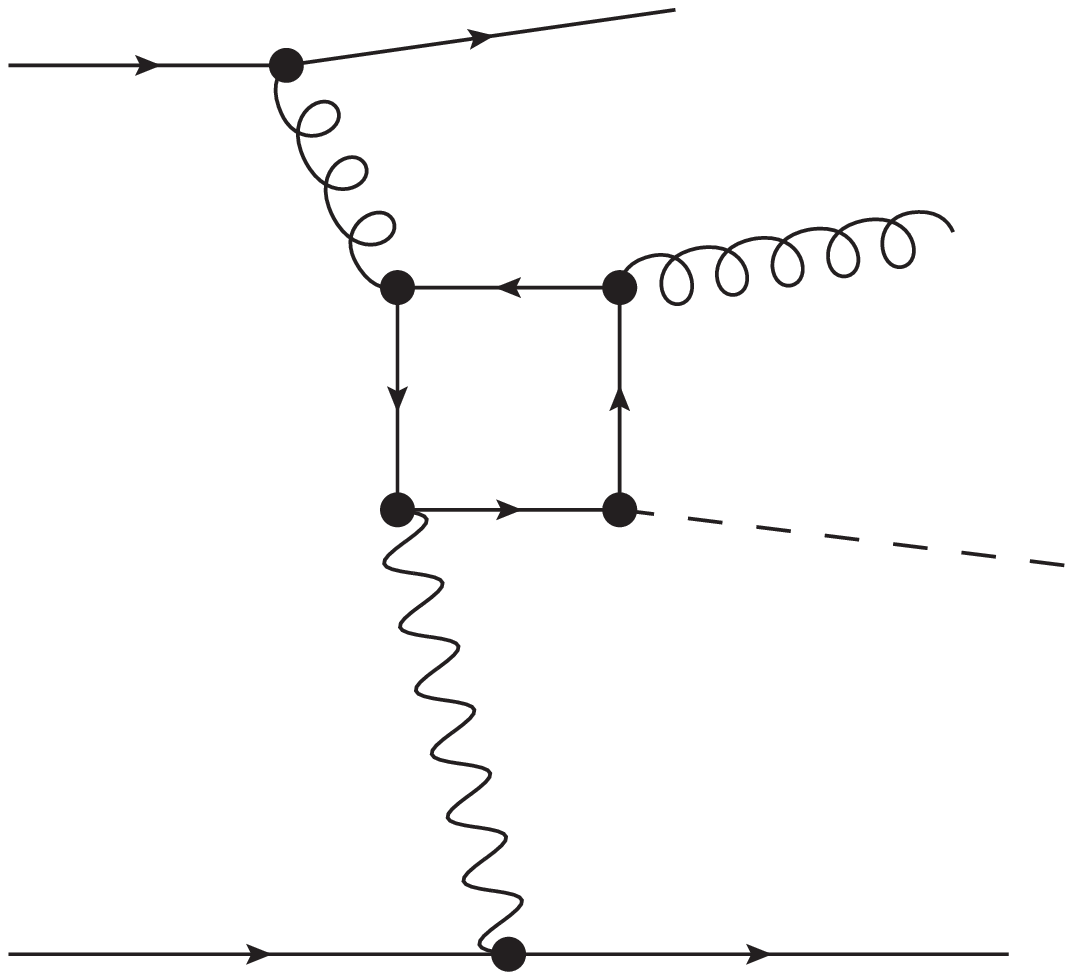}
\caption{\small
  \label{fig:heavyquarks-tri-glu}
Next-to-next-to-leading order QCD corrections due to heavy-quarks  $(t/b)$ loops: one-loop plus extra parton diagrams interfering with VBF NLO real corrections.
}
\end{figure}

\begin{figure}[ht!]
\centering
\includegraphics[scale=0.350]{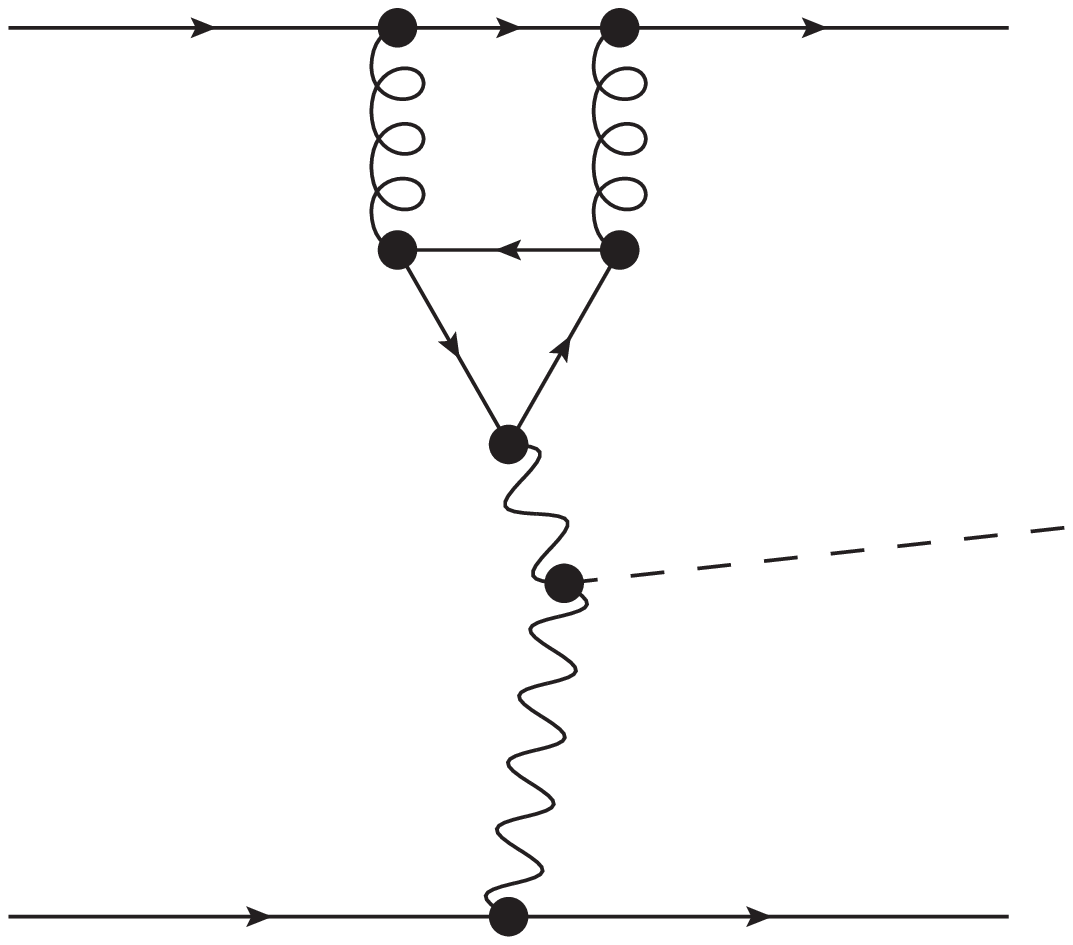}
\hspace*{5mm}
\includegraphics[scale=0.350]{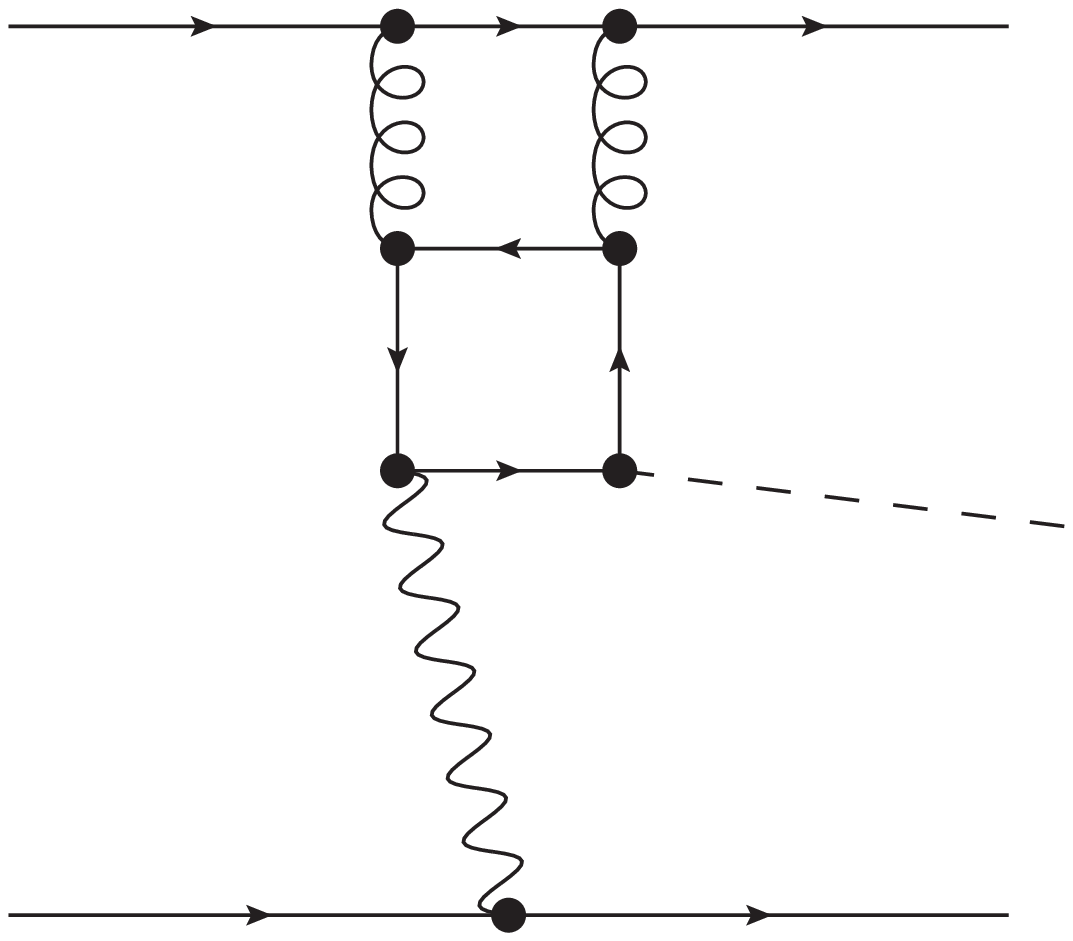}
\caption{\small
  \label{fig:heavyquarks-tri2loop}
Next-to-next-to-leading order QCD corrections due to heavy-quarks $(t/b)$ loops two-loop diagrams interfering with VBF LO diagrams.
}
\end{figure}

We distinguish three different classes of contributions:
the modulo squared of one-loop diagrams with no extra radiation requiring a quark-gluon initial state, Fig.~\ref{fig:heavyquarks};
the interference of one-loop diagrams with an extra parton in the final state, Fig.~\ref{fig:heavyquarks-tri-glu},
with the VBF real tree-level diagrams;
and the interference of  two-loop diagrams, Fig.~\ref{fig:heavyquarks-tri2loop}, with  VBF diagrams at the Born level.
Each of the three classes of loop diagrams has no soft/collinear divergencies and is gauge invariant, thus can be treated independently. 

Such contributions appear only for neutral weak currents. Moreover, in both boxes and triangles only the axial coupling of the $Z$-boson to 
the quarks survives, so that a mass-degenerate quark doublet gives zero contribution. Therefore only the top and bottom quarks need to be considered.

\subsubsection{Quark-gluon initiated contributions via the square of one-loop diagrams}
\label{sec:oneloophq}
Let us start with  the diagrams shown in \cf{fig:heavyquarks}.
A simple estimate~\cite{Bolzoni:2010xr} obtained in the limit $\bmass\rightarrow 0,\,\, \tmass\rightarrow \infty$, 
where only the contribution from the triangle is parametrically relevant, associates to this class 
an effect of less than one per-mil of the total cross section~\cite{Bolzoni:2010as,Zaro:2010zz}. By itself this
result is non-trivial given that the contributions of the diagrams in \cf{fig:heavyquarks} 
are proportional to the quark-gluon parton luminosity, which is potentially
large, especially at LHC. Thus, based on considering the triangle alone, 
it can be argued that contributions from the heavy-quark loops in \cf{fig:heavyquarks} 
can be safely neglected. 

We now investigate to which extent  the conclusion above is confirmed by a complete calculation 
of these contributions. 
Note that the corresponding one-loop diagrams  are long known~\cite{Kniehl:1990zu,Kniehl:1990iva} (see also~\cite{Brein:2003wg}), although in a different kinematic regime ({\it i.e.}, time-like) for the $Z$-boson.  Our results confirm previous findings and extend
them for the first time to the $t$-channel regime.
\begin{figure}[t!]
\centering
\includegraphics[scale=0.40]{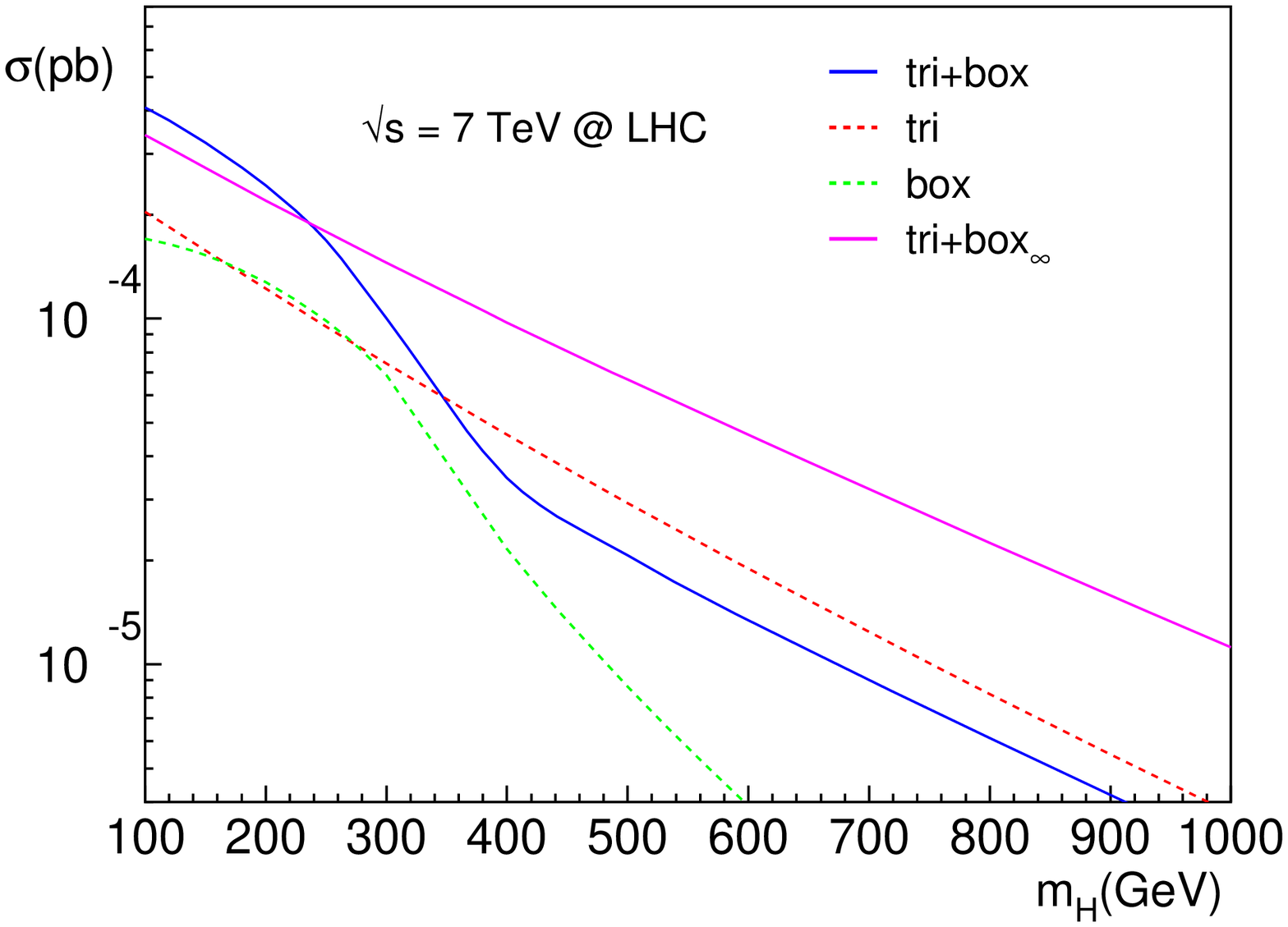}
\includegraphics[scale=0.40]{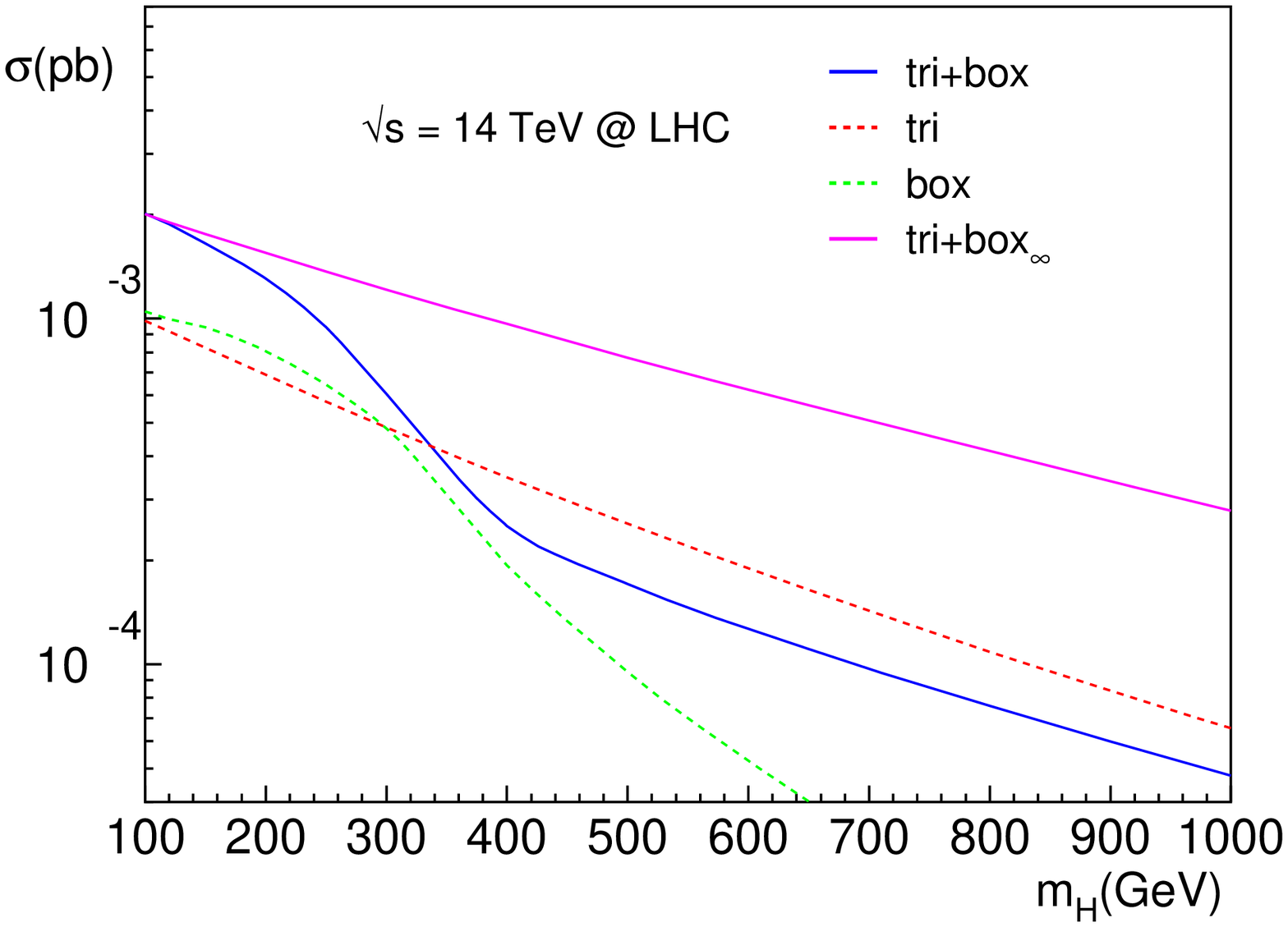}\\[4ex]
\includegraphics[scale=0.40]{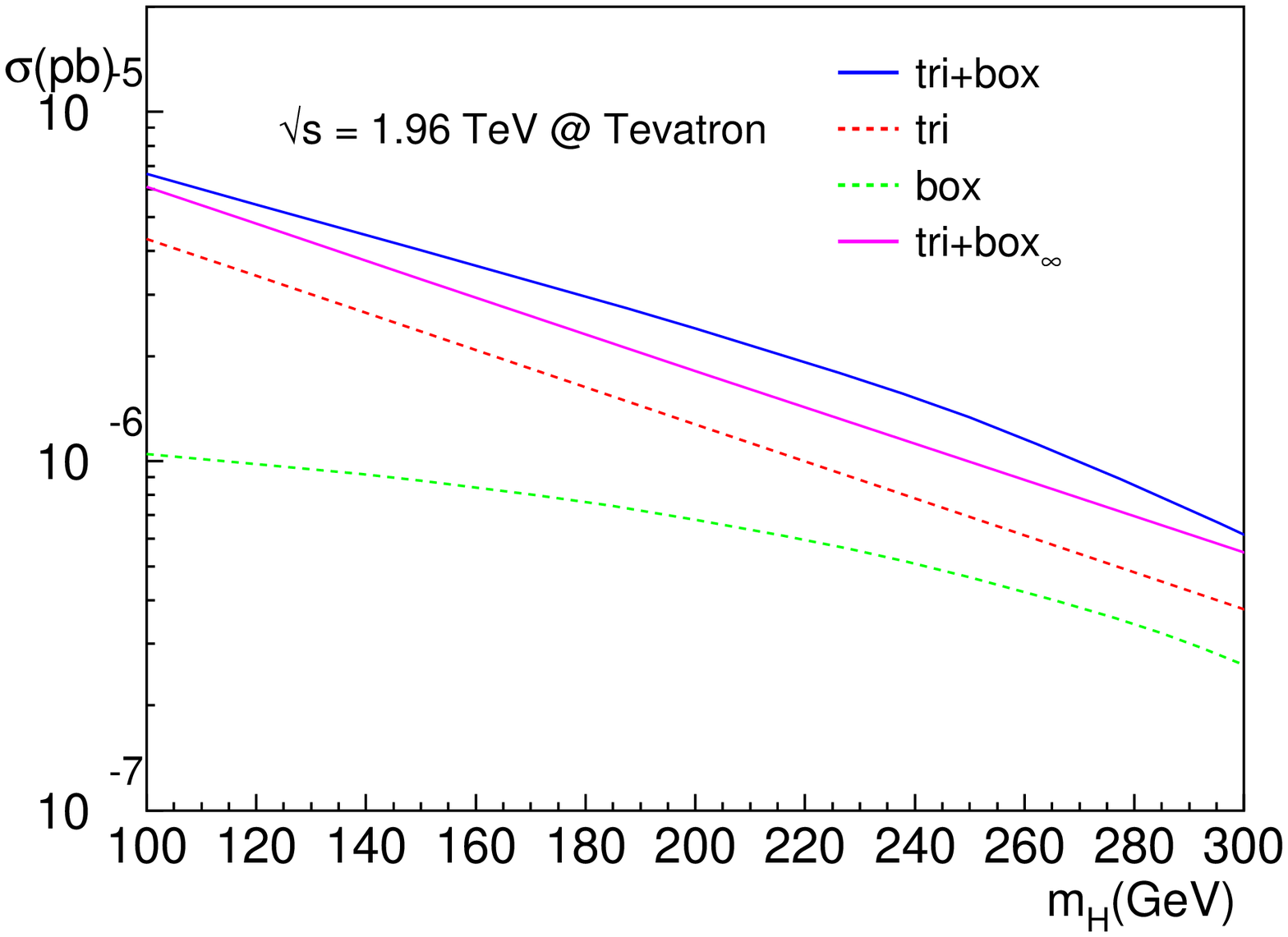}
\caption{\small 
Contribution of the heavy-quarks (t/b) loops to the total NNLO
VBF cross section, at the LHC with $\sqrt S =7\tev $ (top-left), $\sqrt S =14\tev$ (top-right) and at the Tevatron, $\sqrt S =1.96\tev$. 
Numbers are computed with the MSTW 2008~\cite{Martin:2009iq} NNLO PDF set. 
The renormalization and factorization scales have been set to $\wmass$.}
\label{fig:hqxsection}
\end{figure}  
\begin{table}[ht!]
\begin{center}
\begin{tabular}{|c|ccccc|}
\hline $ \hmass \,[\gev] $ & $ \sigma_{\rm{LO}} $ & $ \sigma_{ \rm{tri}+\rm{box} } $ & $ \sigma_{ \rm{tri} } $ & $ \sigma_{ \rm{box} } $ & $ \sigma_{ \rm{tri}+\rm{box} } ^{\infty} $ \\ \hline
$  100 $ & $ 1.49               $ & $ 4.09 \cdot 10^{-4} $ & $ 2.04 \cdot 10^{-4} $ & $ 1.70 \cdot 10^{-4} $ & $ 3.40 \cdot 10^{-4} $ \\
$  120 $ & $ 1.22               $ & $ 3.75 \cdot 10^{-4} $ & $ 1.84 \cdot 10^{-4} $ & $ 1.65 \cdot 10^{-4} $ & $ 3.12 \cdot 10^{-4} $ \\
$  150 $ & $ 9.19 \cdot 10^{-1} $ & $ 3.23 \cdot 10^{-4} $ & $ 1.58 \cdot 10^{-4} $ & $ 1.53 \cdot 10^{-4} $ & $ 2.73 \cdot 10^{-4} $ \\
$  200 $ & $ 6.01 \cdot 10^{-1} $ & $ 2.43 \cdot 10^{-4} $ & $ 1.22 \cdot 10^{-4} $ & $ 1.27 \cdot 10^{-4} $ & $ 2.20 \cdot 10^{-4} $ \\
$  250 $ & $ 4.09 \cdot 10^{-1} $ & $ 1.69 \cdot 10^{-4} $ & $ 9.49 \cdot 10^{-5} $ & $ 9.87 \cdot 10^{-5} $ & $ 1.78 \cdot 10^{-4} $ \\
$  300 $ & $ 2.87 \cdot 10^{-1} $ & $ 1.00 \cdot 10^{-4} $ & $ 7.42 \cdot 10^{-5} $ & $ 6.85 \cdot 10^{-5} $ & $ 1.45 \cdot 10^{-4} $ \\
$  400 $ & $ 1.52 \cdot 10^{-1} $ & $ 3.47 \cdot 10^{-5} $ & $ 4.62 \cdot 10^{-5} $ & $ 2.16 \cdot 10^{-5} $ & $ 9.76 \cdot 10^{-5} $ \\
$  500 $ & $ 8.58 \cdot 10^{-2} $ & $ 2.07 \cdot 10^{-5} $ & $ 2.93 \cdot 10^{-5} $ & $ 8.64 \cdot 10^{-6} $ & $ 6.67 \cdot 10^{-5} $ \\
$  650 $ & $ 3.95 \cdot 10^{-2} $ & $ 1.10 \cdot 10^{-5} $ & $ 1.53 \cdot 10^{-5} $ & $ 2.75 \cdot 10^{-6} $ & $ 3.85 \cdot 10^{-5} $ \\
$  800 $ & $ 1.93 \cdot 10^{-2} $ & $ 6.12 \cdot 10^{-6} $ & $ 8.20 \cdot 10^{-6} $ & $ 1.00 \cdot 10^{-6} $ & $ 2.25 \cdot 10^{-5} $ \\
$ 1000 $ & $ 8.00 \cdot 10^{-3} $ & $ 2.89 \cdot 10^{-6} $ & $ 3.71 \cdot 10^{-6} $ & $ 2.97 \cdot 10^{-7} $ & $ 1.12 \cdot 10^{-5} $ \\
\hline
\end{tabular}
\caption{\small
Values of the contributions to the total NNLO VBF cross-section due to the heavy-quark loop diagrams shown in \cf{fig:heavyquarks} at the LHC, $\sqrt S = 7 \tev$. The MSTW2008~\cite{Martin:2009iq} NNLO PDF set has been used.  The LO cross-section, computed with LO PDFs, is also shown for comparison. Renormalization and factorization scales have been set to $\wmass$. Integration errors are below the 1\% level. Cross-sections are in pb.}
\label{tab:tbloops-7000}
\end{center}
\end{table}

\begin{table}[ht!]
\begin{center}
\begin{tabular}{|c|ccccc|}
\hline $ \hmass \,[\gev] $ & $ \sigma_{\rm{LO}} $ & $ \sigma_{ \rm{tri}+\rm{box} } $ & $ \sigma_{ \rm{tri} } $ & $ \sigma_{ \rm{box} } $ & $ \sigma_{ \rm{tri}+\rm{box} } ^{\infty} $ \\ \hline
$  100 $ & $ 5.08               $ & $ 2.02 \cdot 10^{-3} $ & $ 9.86 \cdot 10^{-4} $ & $ 1.05 \cdot 10^{-3} $ & $ 2.01 \cdot 10^{-3} $ \\
$  120 $ & $ 4.29               $ & $ 1.88 \cdot 10^{-3} $ & $ 9.18 \cdot 10^{-4} $ & $ 9.99 \cdot 10^{-4} $ & $ 1.90 \cdot 10^{-3} $ \\
$  150 $ & $ 3.40               $ & $ 1.65 \cdot 10^{-3} $ & $ 8.24 \cdot 10^{-4} $ & $ 9.44 \cdot 10^{-4} $ & $ 1.76 \cdot 10^{-3} $ \\
$  200 $ & $ 2.40               $ & $ 1.31 \cdot 10^{-3} $ & $ 6.88 \cdot 10^{-4} $ & $ 8.06 \cdot 10^{-4} $ & $ 1.55 \cdot 10^{-3} $ \\
$  250 $ & $ 1.76               $ & $ 9.44 \cdot 10^{-4} $ & $ 5.76 \cdot 10^{-4} $ & $ 6.46 \cdot 10^{-4} $ & $ 1.37 \cdot 10^{-3} $ \\
$  300 $ & $ 1.33               $ & $ 6.05 \cdot 10^{-4} $ & $ 4.85 \cdot 10^{-4} $ & $ 4.82 \cdot 10^{-4} $ & $ 1.21 \cdot 10^{-3} $ \\
$  400 $ & $ 8.09 \cdot 10^{-1} $ & $ 2.52 \cdot 10^{-4} $ & $ 3.48 \cdot 10^{-4} $ & $ 1.94 \cdot 10^{-4} $ & $ 9.66 \cdot 10^{-4} $ \\
$  500 $ & $ 5.25 \cdot 10^{-1} $ & $ 1.71 \cdot 10^{-4} $ & $ 2.55 \cdot 10^{-4} $ & $ 9.52 \cdot 10^{-5} $ & $ 7.72 \cdot 10^{-4} $ \\
$  650 $ & $ 2.97 \cdot 10^{-1} $ & $ 1.11 \cdot 10^{-4} $ & $ 1.64 \cdot 10^{-4} $ & $ 4.02 \cdot 10^{-5} $ & $ 5.62 \cdot 10^{-4} $ \\
$  800 $ & $ 1.80 \cdot 10^{-1} $ & $ 7.59 \cdot 10^{-5} $ & $ 1.09 \cdot 10^{-4} $ & $ 1.93 \cdot 10^{-5} $ & $ 4.14 \cdot 10^{-4} $ \\
$ 1000 $ & $ 9.82 \cdot 10^{-2} $ & $ 4.76 \cdot 10^{-5} $ & $ 6.53 \cdot 10^{-5} $ & $ 8.05 \cdot 10^{-6} $ & $ 2.79 \cdot 10^{-4} $ \\
\hline
\end{tabular}
\caption{\small
Values of the contributions to the total NNLO VBF cross-section due to the heavy-quark loop diagrams shown in \cf{fig:heavyquarks} at the LHC, $\sqrt S = 14 \tev$. The MSTW2008~\cite{Martin:2009iq} NNLO PDF set has been used.  The LO cross-section, computed with LO PDFs, is also shown for comparison. Renormalization and factorization scales have been set to $\wmass$. Integration errors are below the 1\% level. Cross-sections are in pb.}
\label{tab:tbloops-14000}
\end{center}
\end{table}

The contribution of these diagrams to Higgs boson production in VBF are shown in \cf{fig:hqxsection}, 
where the sum of the triangle and the box, 
the triangle and the box alone and the limit $\tmass\rightarrow \infty$ 
at $\sqrt{S} = 7\,\rm{TeV}$ and $14\,\rm{TeV}$ for the LHC
and $\sqrt{S} = 1.96\,\rm{TeV}$ for the Tevatron have been plotted.
The pole mass values $\bmass=4.62\gev$, $\tmass=174.3\gev$ have been used.
The numbers for the LHC corresponding to the various scenarios in \cf{fig:hqxsection} 
are reported in \ctd{tab:tbloops-7000}{tab:tbloops-14000}. 
From these numbers we can see that the limit $\tmass\rightarrow \infty$, $\bmass \rightarrow 0$ is a good estimate of the 
upper bound of the integrated cross section.
In this case we verified that, as reported also in~\cite{Kniehl:1990zu}, the contribution of the box goes to zero.
Moreover, specially for large $\hmass$, the triangle alone is also an upper bound, while for small 
$\hmass$ it approximates the total cross-section within a factor of 2.
Therefore, given our aim to asses the importance of these contributions relative to the VBF cross section, 
it gives a reasonable estimate of the exact value.

Our results have been obtained by two independent computations of the squared amplitude, and also agree
point-by-point in phase space with the results of a code automatically  generated by \madloop~\cite{Hirschi:2011pa}.
The upshot is that the exact computation corroborates the findings of~\cite{Bolzoni:2010xr}.
The impact of the diagrams in \cf{fig:heavyquarks} 
on the VBF Higgs production cross section is always below the per-mil level of
the Born contribution, and therefore can be safely neglected. 

The exact result for the VBF cross section  in the quark-gluon channel at NNLO allows for a comparison with the Higgs-Strahlung case. 
In this case a sizable destructive interference between the triangle and the box diagrams 
takes place~\cite{Kniehl:1990iva,Brein:2003wg}, while this does not happen for VBF (see also the plots in \cf{fig:hqxsection}). 
It is instructive to find the origin of such a different behavior of the triangle and the box diagrams contributions 
in an $s$-channel process~\cite{Kniehl:1990iva} compared to a $t$-channel one (cf. \cf{fig:hqxsection}).
To this aim  it is useful to  define the relative phase angle $\phi$ between the triangle and the box amplitudes 
$\mathcal{M}_{\textrm{tri}}$ and $\mathcal{M}_{\textrm{box}}$ by
\begin{equation}
\label{eq:interference}
|\mathcal{M}_{\textrm{tri}}+\mathcal{M}_{\textrm{box}}|^2=|\mathcal{M}_{\textrm{tri}}|^2+
|\mathcal{M}_{\textrm{box}}|^2+2|\mathcal{M}_{\textrm{tri}}||\mathcal{M}_{\textrm{box}}|\cos\,\phi
\, .
\end{equation}
Recall that in the $s$-channel case in the large-$\tmass$ limit 
$\mathcal{M}_{\textrm{tri}}$ and $\mathcal{M}_{\textrm{box}}$ are real and opposite in sign 
thus implying $\phi\sim \pi$~\cite{Kniehl:1990iva}.
This remains true for the crossed amplitudes in the $t$-channel case of \cf{fig:hqxsection} 
because, due to unitarity, the asymptotic behavior in the large-$\tmass$ limit remains the same.
However, for finite top-quark masses and away from large-$\tmass$ limit,  both
$\mathcal{M}_{\textrm{tri}}$ and $\mathcal{M}_{\textrm{box}}$ 
acquire an imaginary part above the threshold $\sqrt{S} \simeq 2\tmass$ in the $s$-channel process.
This is due to the gluon-gluon scattering process opening to real top-quark pair-production.
Thus, $\mathcal{M}_{\textrm{tri}}$ and $\mathcal{M}_{\textrm{box}}$ become  complex numbers 
pointing in opposite directions in the complex plane within a tolerance of less than $50^\circ$
resulting in a strong destructive interference between them, see~\cite{Kniehl:1990iva}.
The different findings in the VBF case can now be easily understood: in 
the $t$-channel process of \cf{fig:heavyquarks} only $\mathcal{M}_{\textrm{box}}$ 
can develop an imaginary part due to an intermediate real top-quark pair while 
$\mathcal{M}_{\textrm{tri}}$ remains real. As a net result $\mathcal{M}_{\textrm{tri}}$ and $\mathcal{M}_{\textrm{box}}$
are likely to be almost orthogonal in the complex plane thus suppressing the interference term
in \ce{eq:interference}.

Finally, compared to the Higgs-Strahlung process, 
{\it i.e.}, the associated production of Higgs and a $Z$-boson in the gluon-fusion channel, 
the NNLO corrections in VBF from the heavy quarks are by far less important~\cite{Brein:2003wg}.
This difference has also a simple physical explanation, because in the Higgs-Strahlung this contribution 
proceeds through a $gg$ initiated $s$-channel, whereas VBF has the vector boson in the $t$-channel.
The $s$-channel propagator enhances the small $x$-region where the gluon luminosity steeply rises. 
In the crossed case, {\it i.e.}, in VBF, the dominant contribution to the cross
section comes from effective parton momentum fractions $\langle x \rangle \sim 10^{-2}$, where the gluon luminosity is not large yet.

\subsubsection{One-loop plus extra real parton diagrams}
\label{sec:triglu}
Let us now discuss the contributions coming from  diagrams in \cf{fig:heavyquarks-tri-glu} in the $qq$-channel and the crossed $qg$ one.
These diagrams feature one initial (final) state on-shell gluon  attached to the heavy-quark loop, together with 
the $Z$-boson and a time-like (space-like) off-shell gluon. We estimate the overall contribution of this class of diagrams by computing
only the triangles, yet  keeping the $\tmass$ and $\bmass$ finite. This choice is justified by the fact that triangles are parametrically
leading, especially for large values of $\hmass$ and, as shown in \csec{sec:oneloophq}, provide an approximate estimate for the total 
contribution, although sufficiently accurate for our purposes. We also verified that the limit $\tmass \to \infty$, 
which we expect to be close to the upper bound for the cross-section, is
indeed quite close ($\simeq 20 \% $) to the values we have computed.
Note that triangle diagrams as in \cf{fig:heavyquarks-tri-glu} have already been computed before 
in the context of $Z$-boson decay into hadrons~\cite{Kniehl:1989qu} or hadro-production~\cite{Gonsalves:1991qn} in a different kinematic regime.
As an additional simplification, we will consider only the parton channel with an extra gluon in the final state,
{\it i.e.}, the reaction $qq' \to qq'Hg$,  the other parton channels having  similar or smaller impact.
This latter assumption is supported by the previous studies of $Z$-boson hadro-production~\cite{Gonsalves:1991qn},
where it was shown that the entire class of diagrams in \cf{fig:heavyquarks-tri-glu} 
is heavily suppressed with respect to the two-loop ones of~\cf{fig:heavyquarks-tri2loop} 
considered below in \csec{sec:anomalyhq}.

We have computed the necessary expressions  for the interference of the one-loop diagrams with the tree-level ones 
analytically using \mathematica~and \feyncalc~\cite{Mertig:1990an} and have performed 
the phase-space integration using \madgraph~\cite{Maltoni:2002qb,Alwall:2007st}.
The respective numbers for the contribution to the VBF total cross-section 
of the process $qq'\to Hqq'g$ due to the heavy-quark triangle in \cf{fig:heavyquarks-tri-glu}
at the LHC, at  $\sqrt S = 7 \tev$ and $14 \tev$ are given in \ct{tab:tbloops-tri-glu}.
Again, we found these contributions to be totally negligible compared to the LO VBF cross-section, 
and also about one order of magnitude smaller than the ones studied in \csec{sec:oneloophq}.
\begin{table}[t!]
  \begin{minipage}[b]{0.5\linewidth}\centering
    \begin{tabular}{|c|cc|}
\hline $ \hmass\, [\gev] $ & $ \sigma_{LO} $ & $  \sigma_{\rm{tri}}^{ 1 \, \rm{loop+glu}} $ \\ \hline
$  100 $ & $ 1.49               $ & $ 5.10 \cdot 10^{-5} $ \\
$  120 $ & $ 1.22               $ & $ 4.02 \cdot 10^{-5} $ \\
$  150 $ & $ 9.19 \cdot 10^{-1} $ & $ 2.89 \cdot 10^{-5} $ \\
$  200 $ & $ 6.01 \cdot 10^{-1} $ & $ 1.74 \cdot 10^{-5} $ \\
$  250 $ & $ 4.09 \cdot 10^{-1} $ & $ 1.11 \cdot 10^{-5} $ \\
$  300 $ & $ 2.87 \cdot 10^{-1} $ & $ 7.41 \cdot 10^{-6} $ \\
$  400 $ & $ 1.52 \cdot 10^{-1} $ & $ 3.53 \cdot 10^{-6} $ \\
$  500 $ & $ 8.58 \cdot 10^{-2} $ & $ 1.84 \cdot 10^{-6} $ \\
$  650 $ & $ 3.95 \cdot 10^{-2} $ & $ 7.72 \cdot 10^{-7} $ \\
$  800 $ & $ 1.93 \cdot 10^{-2} $ & $ 3.52 \cdot 10^{-7} $ \\
$ 1000 $ & $ 8.00 \cdot 10^{-3} $ & $ 1.34 \cdot 10^{-7} $ \\
\hline
\end{tabular}

  \end{minipage}
  \begin{minipage}[b]{0.5\linewidth}\centering
    \begin{tabular}{|c|cc|}
\hline $ \hmass\, [\gev] $ & $ \sigma_{LO} $ & $  \sigma_{\rm{tri}}^{ 1 \, \rm{loop+glu}} $ \\ \hline
$  100 $ & $ 5.08               $ & $ 2.66 \cdot 10^{-4} $ \\
$  120 $ & $ 4.29               $ & $ 2.20 \cdot 10^{-4} $ \\
$  150 $ & $ 3.40               $ & $ 1.67 \cdot 10^{-4} $ \\
$  200 $ & $ 2.40               $ & $ 1.11 \cdot 10^{-4} $ \\
$  250 $ & $ 1.76               $ & $ 7.62 \cdot 10^{-5} $ \\
$  300 $ & $ 1.33               $ & $ 5.48 \cdot 10^{-5} $ \\
$  400 $ & $ 8.09 \cdot 10^{-1} $ & $ 3.02 \cdot 10^{-5} $ \\
$  500 $ & $ 5.25 \cdot 10^{-1} $ & $ 1.82 \cdot 10^{-5} $ \\
$  650 $ & $ 2.97 \cdot 10^{-1} $ & $ 9.21 \cdot 10^{-6} $ \\
$  800 $ & $ 1.80 \cdot 10^{-1} $ & $ 5.12 \cdot 10^{-6} $ \\
$ 1000 $ & $ 9.82 \cdot 10^{-2} $ & $ 2.54 \cdot 10^{-6} $ \\
\hline
\end{tabular}

  \end{minipage}
\caption{\small
Values of the contributions to the total NNLO VBF cross-section due to the heavy-quark triangle plus gluon emission diagrams shown in \cf{fig:heavyquarks-tri-glu} at the LHC, at  $\sqrt S = 7 \tev$ (left) and $\sqrt S = 14 \tev$ (right). The MSTW2008~\cite{Martin:2009iq} NNLO PDF set has been used.  The LO cross-section, computed with LO PDFs, is also shown for comparison. Renormalization and factorization scales have been set to $\wmass$. Integration errors are below the 1\% level. Cross-sections are in pb.}
\label{tab:tbloops-tri-glu}
\end{table}

\subsubsection{Two-loop diagrams}
\label{sec:anomalyhq}

Finally, we consider the  two-loop contributions of~\cf{fig:heavyquarks-tri2loop} 
where the heavy-quark loop is attached via two gluons  to the light-quark
line originating from one of the protons. These diagrams interfere with the Born VBF amplitudes.
The effective coupling to the light-quark line singles out the iso-triplet
component of the proton in the squared matrix elements, 
since only the axial part of the $Z$-boson coupling contributes for
non-degenerate heavy-quarks in the loop.
That is to say, this class of diagrams is proportional 
to the (non-singlet) distribution $\delta q^-_{{\rm ns}}$ of \ce{eq:du-pdf}, which is generally small.

In analogy to the previous \csec{sec:triglu}, we  aim at an estimate 
of the size of the contributions of~\cf{fig:heavyquarks-tri2loop} rather than the exact result.
To that end, we will restrict ourselves to the two-loop triangle diagrams.
This choice is, of course, also driven by the fact that the two-loop double box 
in \cf{fig:heavyquarks-tri2loop} is currently unknown for the required VBF
kinematical configuration and its computation would be  a tremendous task in itself.

The two-loop triangles for the $q{\bar q}Z$ vertex 
have been computed in \cite{Kniehl:1989qu}
for the case $\bmass=0$ and $\tmass \neq 0$, see also~\cite{Gonsalves:1991qn}.
Rather compact results in terms of harmonic polylogarithms~\cite{Remiddi:1999ew} 
(see~\cite{Gehrmann:2001pz} for numerical routines) for all
kinematic configurations have been obtained in~\cite{Bernreuther:2005rw} 
and we use the latter expressions to compute the numbers shown 
in \ct{tab:tbloops-tri2loop} for the LHC at  $\sqrt S = 7 \tev$ and $14 \tev$.
Again, a comparison of the cross section numbers in \ct{tab:tbloops-tri2loop}
for the reaction $qq'\to Hqq'$ at NNLO in QCD mediated by the two-loop triangle
with the LO cross-section shows that such contributions are below the per-mil level.\\

In ending this discussion of heavy-quark loop contributions we briefly remark that, of course, 
the diagrams shown in \cfd{fig:heavyquarks}{fig:heavyquarks-tri2loop},
also contribute in heavy-quark DIS, if the full neutral-current reactions are considered, 
{\it i.e.}, both $\gamma$ and $Z$-boson exchange at high $Q^2$.
Currently available DIS data on heavy-quark production, however, 
is usually taken at $Q^2$ values where these contributions are not relevant. 
Their existence is an issue, though, to be recalled in the definition of
variable-flavor number schemes, see e.g.,~\cite{Alekhin:2009ni}.
\begin{table}[ht]
  \begin{minipage}[b]{0.5\linewidth}\centering
    \begin{tabular}{|c|cc|}
\hline $ \hmass\, [\gev] $ & $ \sigma_{LO} $ & $  \sigma_{\rm{tri}}^{ 2\, \rm{loop}} $ \\ \hline
$  100 $ & $ 1.49               $ & $ 8.38 \cdot 10^{-4} $ \\
$  120 $ & $ 1.22               $ & $ 7.08 \cdot 10^{-4} $ \\
$  150 $ & $ 9.19 \cdot 10^{-1} $ & $ 5.60 \cdot 10^{-4} $ \\
$  200 $ & $ 6.01 \cdot 10^{-1} $ & $ 3.90 \cdot 10^{-4} $ \\
$  250 $ & $ 4.09 \cdot 10^{-1} $ & $ 2.81 \cdot 10^{-4} $ \\
$  300 $ & $ 2.87 \cdot 10^{-1} $ & $ 2.06 \cdot 10^{-4} $ \\
$  400 $ & $ 1.52 \cdot 10^{-1} $ & $ 1.18 \cdot 10^{-4} $ \\
$  500 $ & $ 8.58 \cdot 10^{-2} $ & $ 7.08 \cdot 10^{-5} $ \\
$  650 $ & $ 3.95 \cdot 10^{-2} $ & $ 3.52 \cdot 10^{-5} $ \\
$  800 $ & $ 1.93 \cdot 10^{-2} $ & $ 1.84 \cdot 10^{-5} $ \\
$ 1000 $ & $ 8.00 \cdot 10^{-3} $ & $ 8.18 \cdot 10^{-6} $ \\
\hline
\end{tabular}

  \end{minipage}
  \begin{minipage}[b]{0.5\linewidth}\centering
    \begin{tabular}{|c|cc|}
\hline $ \hmass\, [\gev] $ & $ \sigma_{LO} $ & $  \sigma_{\rm{tri}}^{ 2\, \rm{loop}} $ \\ \hline
$  100 $ & $ 5.08               $ & $ 2.30 \cdot 10^{-3} $ \\
$  120 $ & $ 4.29               $ & $ 2.01 \cdot 10^{-3} $ \\
$  150 $ & $ 3.40               $ & $ 1.67 \cdot 10^{-3} $ \\
$  200 $ & $ 2.40               $ & $ 1.25 \cdot 10^{-3} $ \\
$  250 $ & $ 1.76               $ & $ 9.63 \cdot 10^{-4} $ \\
$  300 $ & $ 1.33               $ & $ 7.60 \cdot 10^{-4} $ \\
$  400 $ & $ 8.09 \cdot 10^{-1} $ & $ 4.99 \cdot 10^{-4} $ \\
$  500 $ & $ 5.25 \cdot 10^{-1} $ & $ 3.44 \cdot 10^{-4} $ \\
$  650 $ & $ 2.97 \cdot 10^{-1} $ & $ 2.10 \cdot 10^{-4} $ \\
$  800 $ & $ 1.80 \cdot 10^{-1} $ & $ 1.35 \cdot 10^{-4} $ \\
$ 1000 $ & $ 9.82 \cdot 10^{-2} $ & $ 7.91 \cdot 10^{-5} $ \\
\hline
\end{tabular}

  \end{minipage}
\caption{\small
Values of the contributions to the total NNLO VBF cross-section due to the two-loop trinagle diagram shown in \cf{fig:heavyquarks-tri2loop} at the LHC, at  $\sqrt S = 7 \tev$ (left) and $\sqrt S = 14 \tev$ (right). The MSTW2008~\cite{Martin:2009iq} NNLO PDF set has been used.The LO cross-section, computed with LO PDFs, is also shown for comparison. Renormalization and factorization scales have been set to $\wmass$. Integration errors are below the 1\% level. Cross-sections are in pb.}
\label{tab:tbloops-tri2loop}
\end{table}

\subsection{One-loop electroweak corrections}
\label{sec:NLO-EW}

We briefly discuss the electroweak corrections at one-loop.
The combined strong and electroweak NLO corrections to Higgs production in VBF
have been computed in~\cite{Ciccolini:2007jr,Ciccolini:2007ec} and can be
obtained via the program {\sc HAWK}~\cite{hawk:2010}. 

In absence of a full calculation up to corrections of order $\alpha_s^2 \alpha_{EW}$ with respect to the Born amplitude, which
is currently beyond capabilities, a combination of EW and NNLO QCD corrections is possible, yet formally subject to ambiguities.
A pragmatic way to proceed is to follow two different approaches, multiplicative or additive, and the corresponding differences used to assess the impact
of neglected terms. Using a compact notation, we define
\begin{equation}
\sigma^{QCD}_{NNLO} = \sigma_0 +  \alpha_s \sigma_1^{QCD}+  \alpha_s^2 \sigma_2^{QCD} 
\end{equation}
the total cross section at NNLO in QCD,
\begin{equation}
\sigma^{EW} = \sigma_0 +  \alpha_{EW} \sigma_1^{EW} 
\end{equation}
the total cross sections including NLO EW corrections,
and $\sigma_{NLO+EW}$ the result of the full calculation at NLO in QCD and EW of Refs.~\cite{Ciccolini:2007jr,Ciccolini:2007ec}.

The additive scheme amounts to simply define
\begin{eqnarray}
&&\sigma_{NNLO+EW} \equiv \sigma_{NLO+EW} +  \alpha_s^2 \sigma_2^{QCD}
\, ,
 \end{eqnarray}
${\it i.e.}$, to add the missing $\alpha_s^2$ terms to the full NLO+EW calculation. In this scheme no assumption on the factorization of 
QCD and EW corrections is made and only terms that are known are included. It demands, however,  the two terms to be evaluated  with
exactly the same EW ($G_F,\zmass,\wmass,\dots$) and QCD (scales, $\alpha_s$, and PDFs) parameters, something possibly error-prone when different codes are used.

The multiplicative scheme amounts to assuming that the QCD and EW corrections factorize to a very
good approximation at NLO as well as at NNLO. In this case one can define
\begin{eqnarray}
&&\sigma_{NNLO+EW} \equiv \sigma^{QCD}_{NNLO} \left( 1 + \frac{\sigma_1^{EW}}{\sigma_0} \right)
\, . 
\end{eqnarray}
This is a very handy approximation: it implies that the EW corrections can be evaluated independently as effects such as scale 
and PDF choices are mostly canceled in the ratio $\sigma_1^{EW}/{\sigma_0}$. Note, however, that it implies the inclusion of  
unknown higher order terms in the results.  This is the approach followed in Ref.~\cite{Dittmaier:2011ti} and
the results relevant for Higgs production at the LHC can be found there. We have explicitly  checked for a few values of the Higgs mass, that the
differences between the two schemes are very small and totally negligible at the LHC.

%%
%% ---------------------------------------------------------------------------
%%
\renewcommand{\thetable}{\thesection.\arabic{table}}
\setcounter{table}{0}
\renewcommand{\thefigure}{\thesection.\arabic{figure}}
\setcounter{figure}{0}
\section{VBF production in the Standard Model}
\label{sec:Phenomenology}

We are now in a position to present an extensive phenomenological analysis 
for the VBF production mechanism at the LHC at the center of mass energies 
of $\sqrt{S}=7 \tev$ and $\sqrt{S}=14 \tev$ and the Tevatron, $\sqrt{S}=1.96 \tev$, 
employing the structure function approach up to the NNLO in QCD.
This is a significant extension of our previous studies in Refs.~\cite{Bolzoni:2010xr,Bolzoni:2010as}.
For the numerical results we use the following values for the electroweak parameters (see also \cite{Dittmaier:2011ti}):
The masses of the vector bosons are $\wmass = 80.398 \gev$ and $\zmass = 91.1876 \gev $, 
Fermi's constant and the weak mixing angle are taken to be 
$G_F = 1.16637 \cdot 10^{-5}$ and $\sin^2 \theta_w = 0.23119$. 
The widths of the vector bosons, $\Gamma_W$ and $\Gamma_Z$, 
have been set to zero for simplicity, their effect being of order $10^{-3}$ or less on the total rate.
Moreover, we provide numbers for all PDFs currently available at NNLO accuracy in QCD, 
ABKM~\cite{Alekhin:2009ni,Alekhin:2010iu}, 
HERAPDF1.5~\cite{herapdf:2009wt,herapdfgrid:2010}, 
JR09~\cite{JimenezDelgado:2008hf,JimenezDelgado:2009tv}, 
MSTW2008~\cite{Martin:2009iq} and 
NNPDF2.1~\cite{Ball:2011uy}
always using the default value of the strong coupling $\alpha_s$ required by the respective set.
A cut of $1 \gev$ has been used to regulate 
the phase space integration over $Q$ which extends to vanishing $Q^2$ (see \csec{sec:PhaseSpace}). 
In the numerical evaluation we have checked explicitly that the results remain unchanged upon variations 
of this cut on $Q$, a fact readily understood by realizing that the effective
$\langle Q \rangle$ in VBF is much larger, typically $\langle Q \rangle \simeq 20 \gev$,
see the discussion in~\cite{Bolzoni:2010xr}.

All results presented here can also be obtained through our publicly usable {\sc VBF@NNLO} code~\cite{vbfnnlo:2010}.

\subsection{Cross section predictions for the Tevatron}
\label{sec:TevCrossSections}

Let us start off with the Tevatron, where 
the cross section is roughly $0.1$~pb for a Higgs boson with mass $\hmass=100 \gev$ 
and steeply falling as a function of $\hmass$.
An exhaustive list of cross-sections, for different values of the Higgs boson mass 
in the range $\hmass \in [90,300] \gev$ at LO, NLO and NNLO in QCD
are given in \ctd{tab:table-tev-a09-Sc1}{tab:table-tev-nnn-Sc1}.

The central values have been obtained by setting 
the factorization and the renormalization scales $\mu_r = \mu_f =Q$, 
where $Q$ is the virtuality of the vector bosons which fuse into the Higgs, cf. \ce{eq:disapproach}.
Our results in \ctd{tab:table-tev-a09-Sc1}{tab:table-tev-nnn-Sc1} show 
that the NLO corrections are always positive and not too large of the order 
of 1-2\%, while the NNLO corrections are typically small of the order 
of $\pm$1\% depending on the chosen PDF set.

The theoretical uncertainty of the predictions has been determined by varying
the scales independently in a large range $\mu_r, \mu_f\in [Q/4,4Q]$.
Here we find markedly a clear improvement due to the NNLO corrections computed.
While we observe, e.g. for a Higgs mass of $\hmass=100 \gev$, variations of 
order $\pm$20-30\% for $\sigma_{LO}$, 
this reduces to $\pm$5-10\% for $\sigma_{NLO}$, 
and to $\pm$2\% for $\sigma_{NNLO}$.
The scale uncertainty increases slightly for larger Higgs masses, e.g. for $\hmass=250 \gev$, we find 
order $\pm$30-50\%, $\pm$10-12\% and to $\pm$3\% 
at LO, NLO and NNLO, respectively, see also \cf{fig:tev-mh}.
Of course, other scale choices are possible relating $\mu_r,\mu_f$ 
to the Higgs boson mass $\hmass$ or the $W$-boson mass $\wmass$, see \cf{fig:tev-scale} for a comparison.
As discussed in~\cite{Bolzoni:2010xr} the present choice, 
{\it i.e.}, relating $\mu_r$ and $\mu_f$ to $Q$ turns out to be the most natural one 
with the point of minimal sensitivity being $\mu_r, \mu_f \simeq Q$,
as it exhibits the best pattern of apparent convergence of the perturbative expansion.
Considering the size of the neglected contributions discussed at length above in \csec{sec:QftAtHO} 
the residual theoretical uncertainty due to perturbative QCD corrections at higher orders 
at the Tevatron is of the order 2-3\%.

The PDF dependence is clearly the dominating source of uncertainty at the
Tevatron, the individual PDFs report an uncertainity in the PDFs at NNLO (sometimes
combined with the one in the strong coupling $\alpha_s$) of the order $\pm$1-3\%.
Moreover the central values obtained from the various NNLO PDFs differ by
roughly 5\% at low Higgs masses, e.g. $\hmass=100 \gev$, which is increasing towards larger Higgs
masses, e.g., to 10\% at $\hmass=250 \gev$, see also \cf{fig:tev-pdfs}.
This is due to the larger values of effective $\langle x \rangle $ at which the parton luminosities are probed.
In summary, thus, the uncertainty due to the non-perturbative parameters (PDFs, $\alpha_s$)
can be estimated to be of the order 5\% for a light Higgs boson at the Tevatron.

%%%%%%%%% Tevatron %%%%%%%%%%%%%%%%%%%
%
\begin{figure}[ht!]
\centering
\includegraphics[scale=0.65]{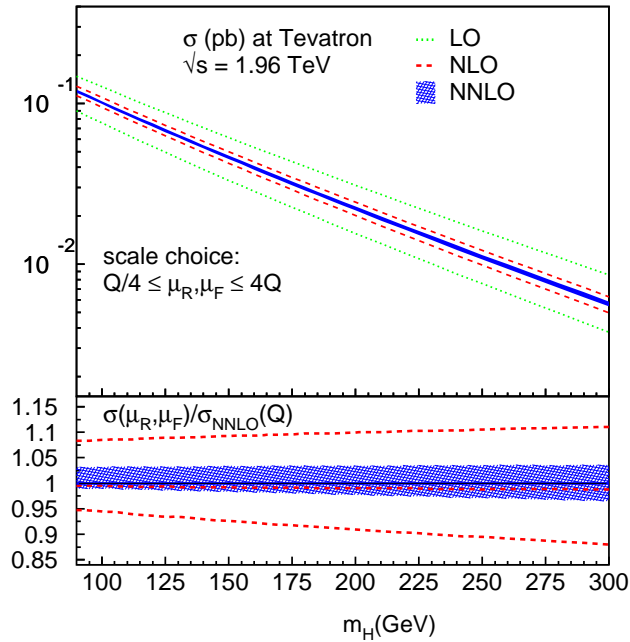}
\caption{\small
  \label{fig:tev-mh}
The total VBF cross sections at the Tevatron, $\sqrt S = 1.96 \tev$, at LO, NLO and NNLO in QCD
with the scale uncertainity from the variation $\mu_r,\mu_f \in [Q/4,4Q]$.
The MSTW2008~\cite{Martin:2009iq} PDF set (68\% CL) has been used. Numbers are in pb.
}
\end{figure}
\begin{figure}[ht!]
\centering
\includegraphics[scale=0.425]{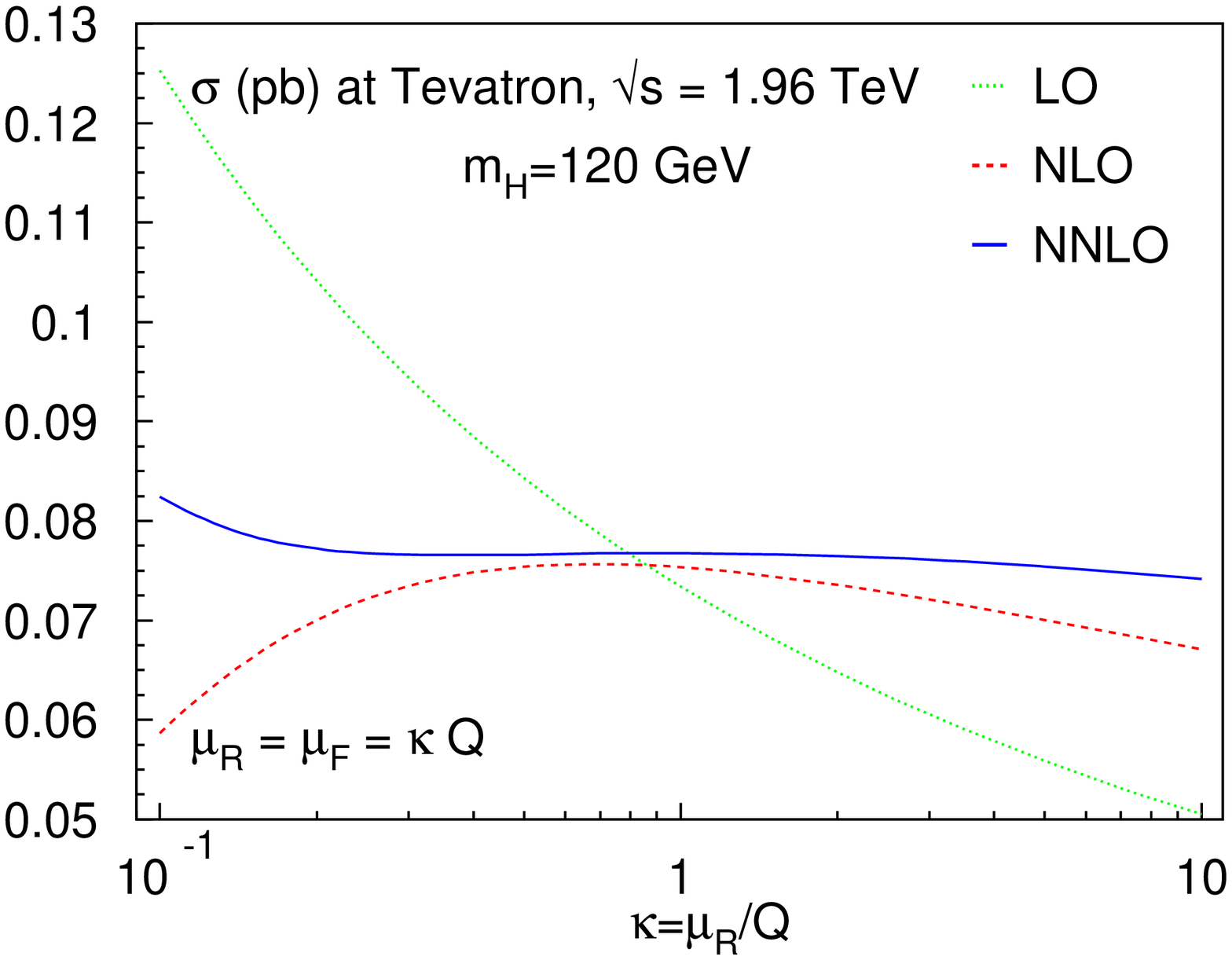}
\hspace*{3mm}
\includegraphics[scale=0.425]{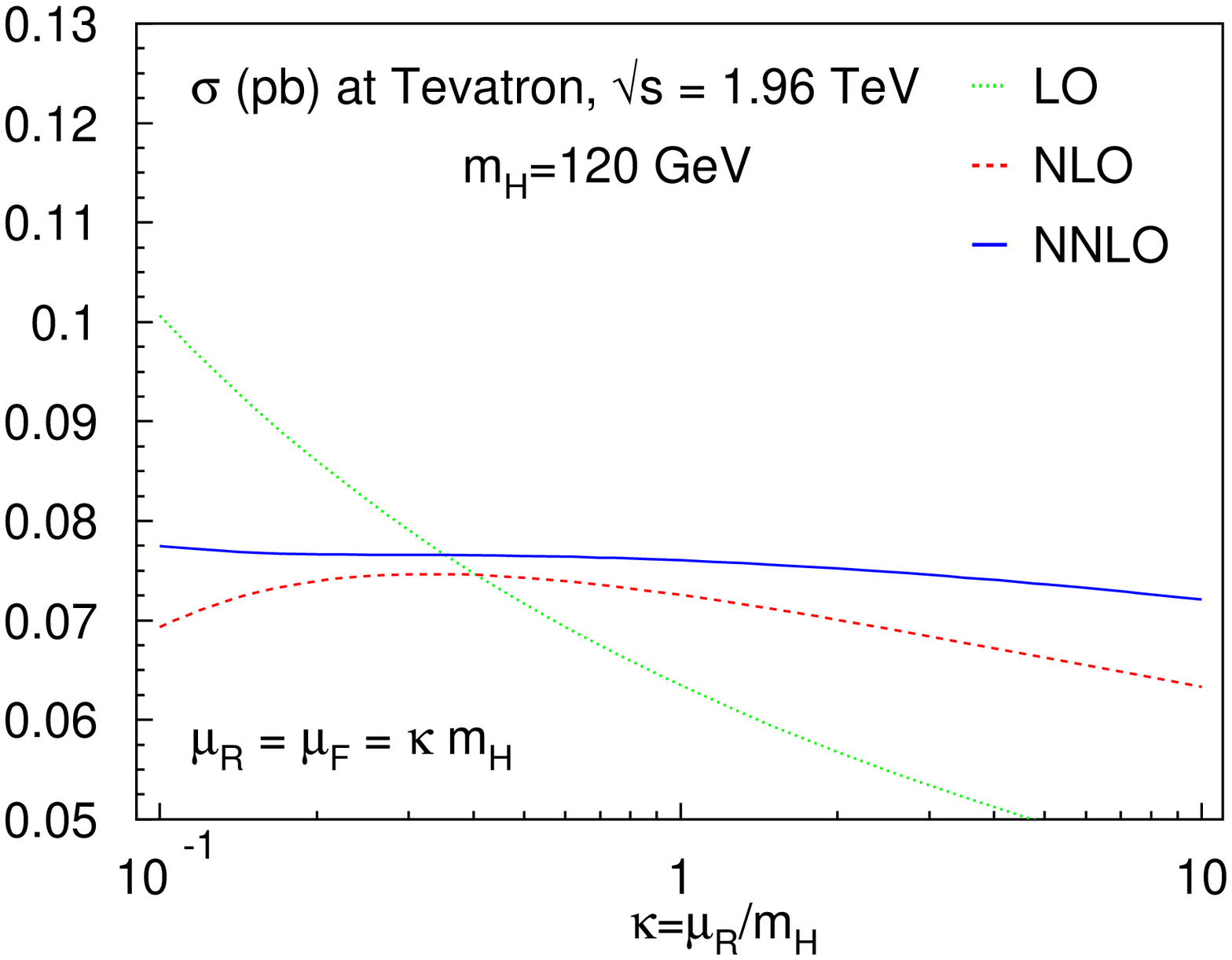}
\caption{\small
  \label{fig:tev-scale}
Scale dependence of the VBF cross sections at the Tevatron, $\sqrt S = 1.96 \tev$ 
at LO, NLO and NNLO in QCD for $\hmass = 120 \gev$ and the choice 
$\mu_r = \mu_f = \kappa Q$ (left) 
and
$\mu_r = \mu_f = \kappa \hmass$ (right).
The ABKM~\cite{Alekhin:2009ni} PDF set has been used. Numbers are in pb.
}
\end{figure}
\begin{figure}[ht!]
\centering
\includegraphics[scale=0.42]{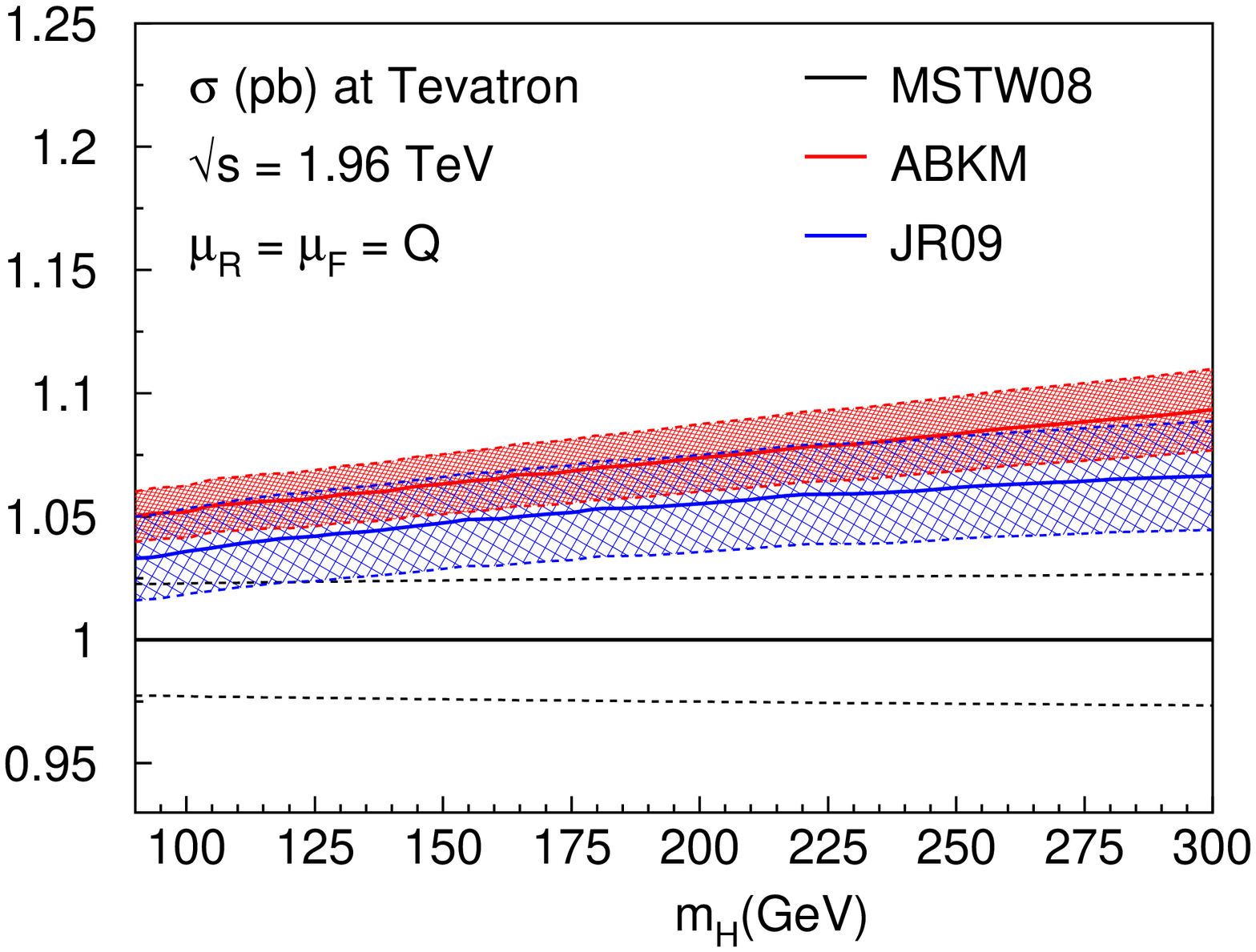}
\hspace*{3mm}
\includegraphics[scale=0.42]{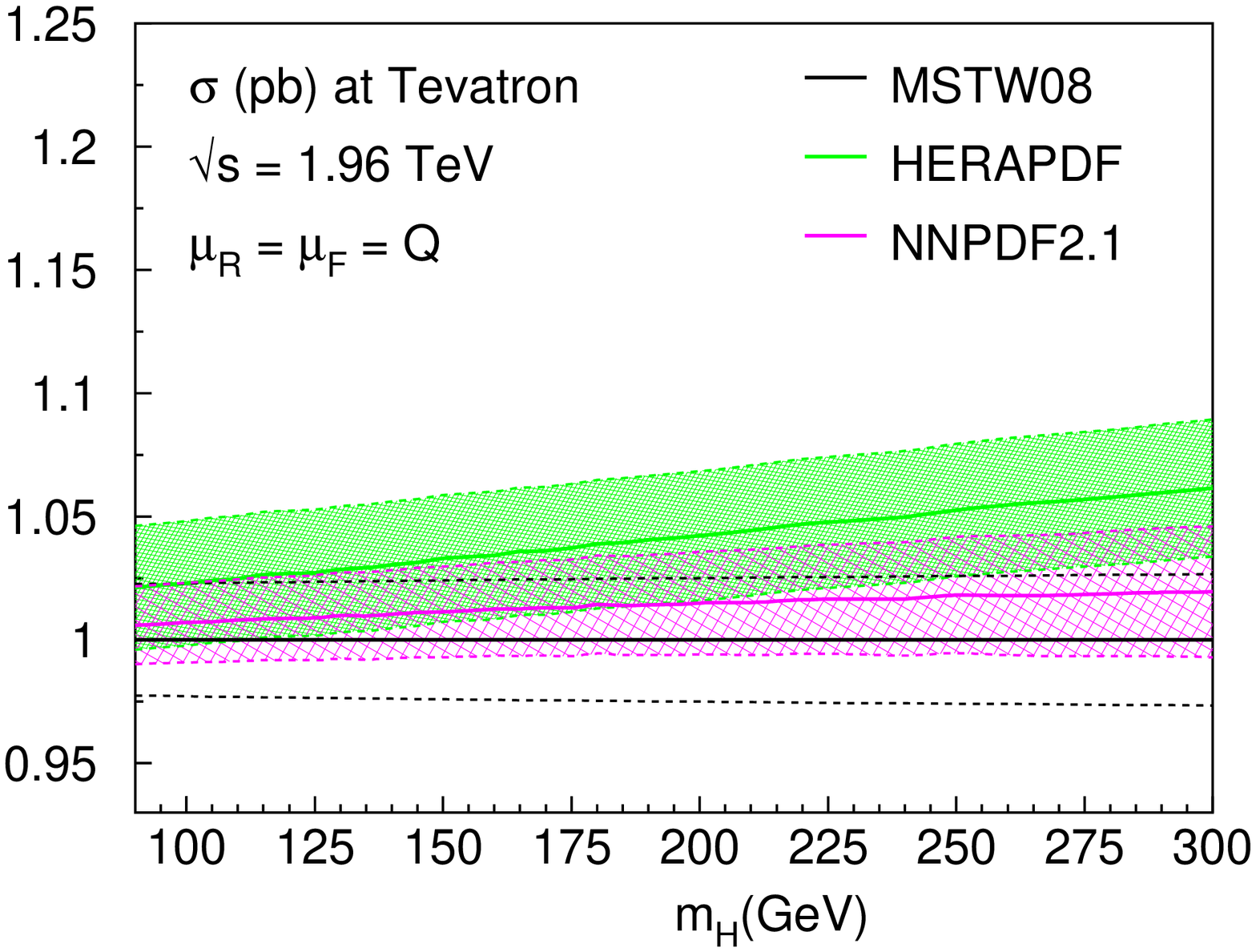}
\caption{\small
  \label{fig:tev-pdfs}
The PDF uncertainity of the VBF cross sections at the Tevatron, $\sqrt S = 1.96 \tev$, 
at NNLO in QCD for PDF sets of 
ABKM~\cite{Alekhin:2009ni},
HERAPDF1.5~\cite{herapdf:2009wt,herapdfgrid:2010}, 
JR09~\cite{JimenezDelgado:2008hf,JimenezDelgado:2009tv},
MSTW2008~\cite{Martin:2009iq} (68 \% CL), and NNPDF~\cite{Ball:2011uy}.
All results have been normalized to the best fit of MSTW2008.
}
\end{figure}

%%%%%%%%% LHC7 %%%%%%%%%%%%%%%%%%%
%
\begin{figure}[ht!]
\centering
\includegraphics[scale=0.65]{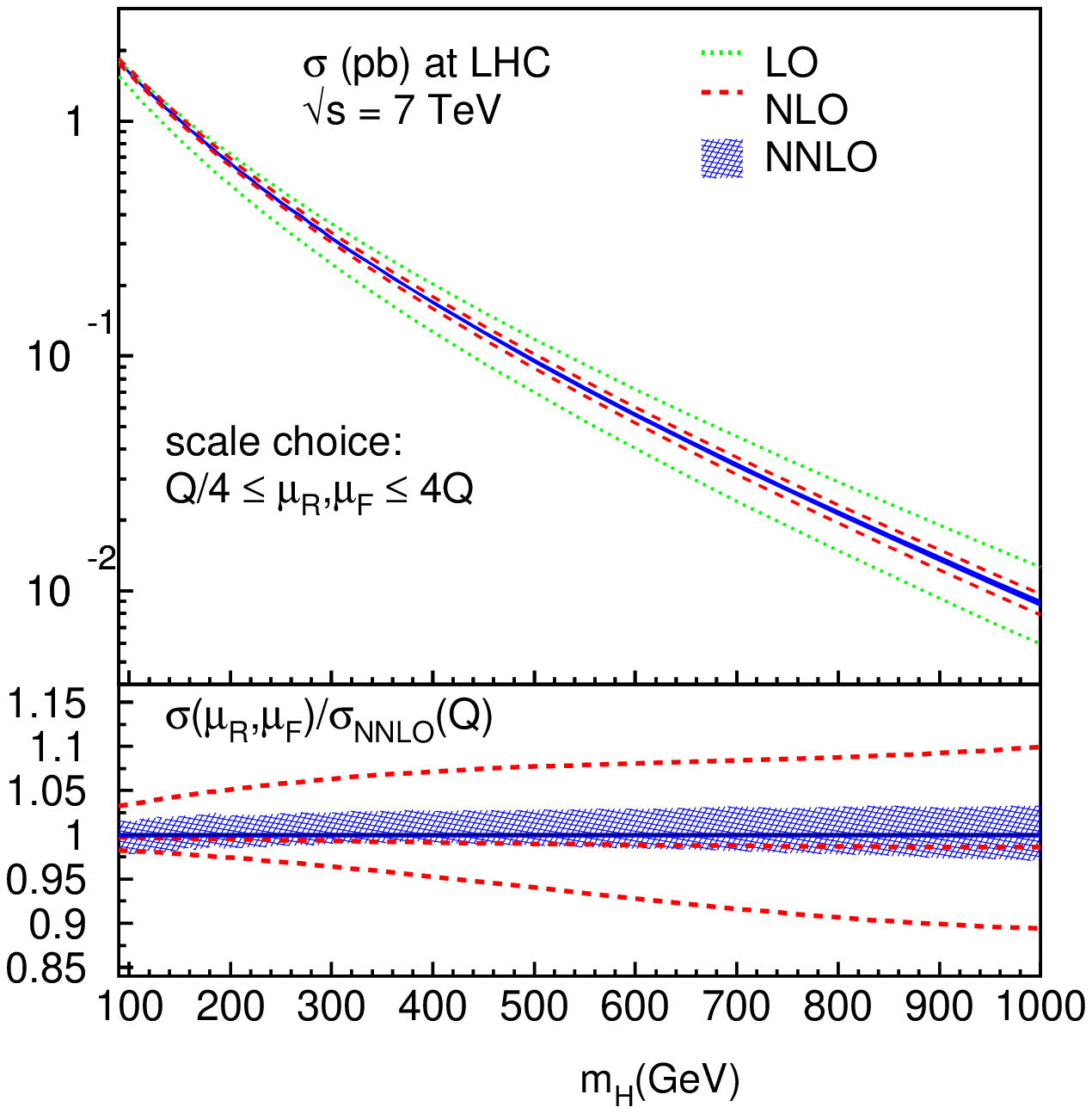}
\caption{\small
  \label{fig:lhc7-mh}
The total VBF cross sections at the LHC, $\sqrt S = 7 \tev$, at LO, NLO and NNLO in QCD
with the scale uncertainity from the variation $\mu_r,\mu_f \in [Q/4,4Q]$.
The MSTW2008~\cite{Martin:2009iq} PDF set (68\% CL) has been used. Numbers are in pb.
}
\end{figure}
\begin{figure}[ht!]
\centering
\includegraphics[scale=0.425]{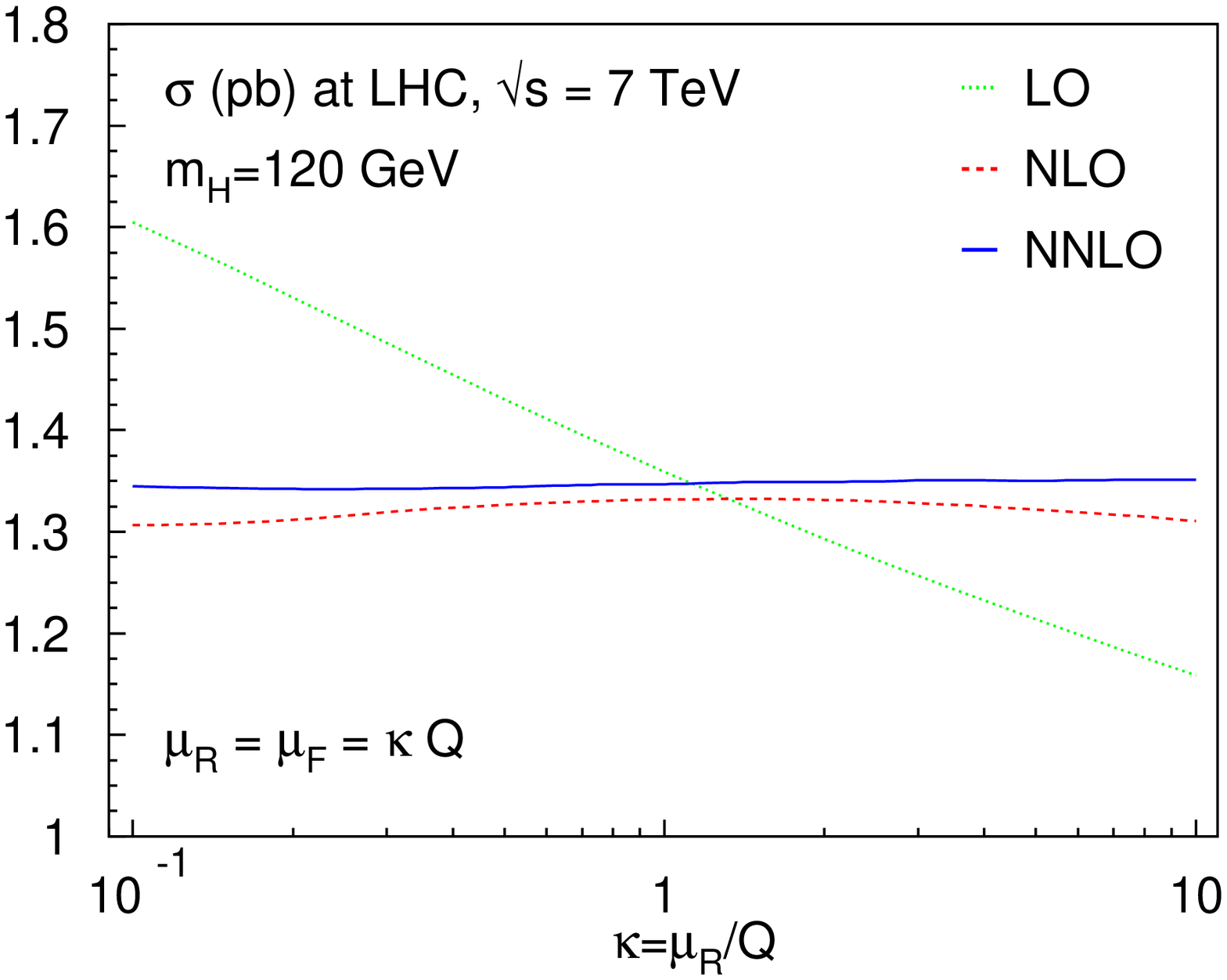}
\hspace*{3mm}
\includegraphics[scale=0.425]{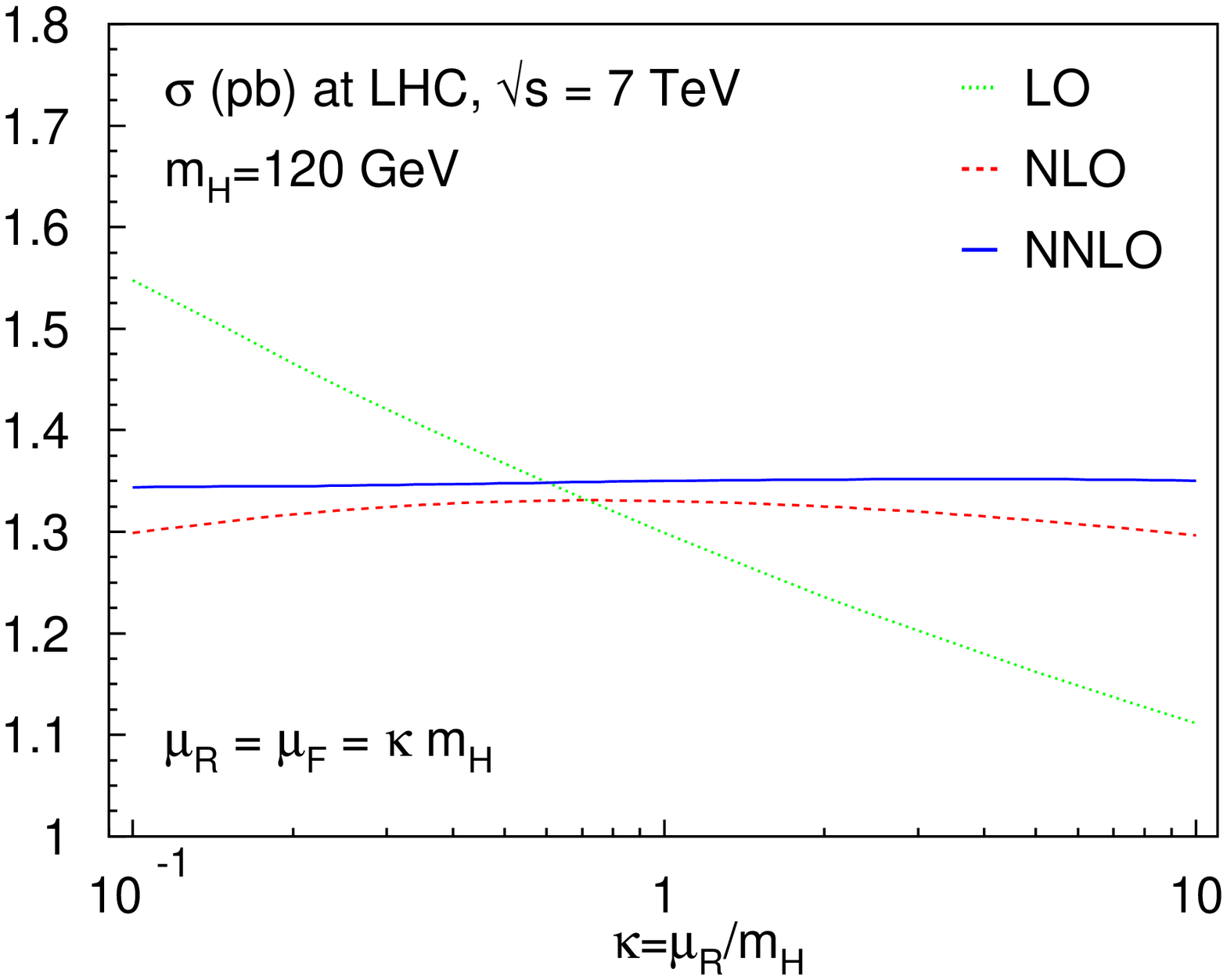}
\caption{\small
  \label{fig:lhc7-scale}
Scale dependence of the VBF cross sections at the LHC, $\sqrt S = 7 \tev$, 
at LO, NLO and NNLO in QCD for $\hmass = 120 \gev$ and the choice 
$\mu_r = \mu_f = \kappa Q$ (left) 
and
$\mu_r = \mu_f = \kappa \hmass$ (right).
The ABKM~\cite{Alekhin:2009ni} PDF set has been used. Numbers are in pb.
}
\end{figure}
\begin{figure}[ht!]
\centering
\includegraphics[scale=0.42]{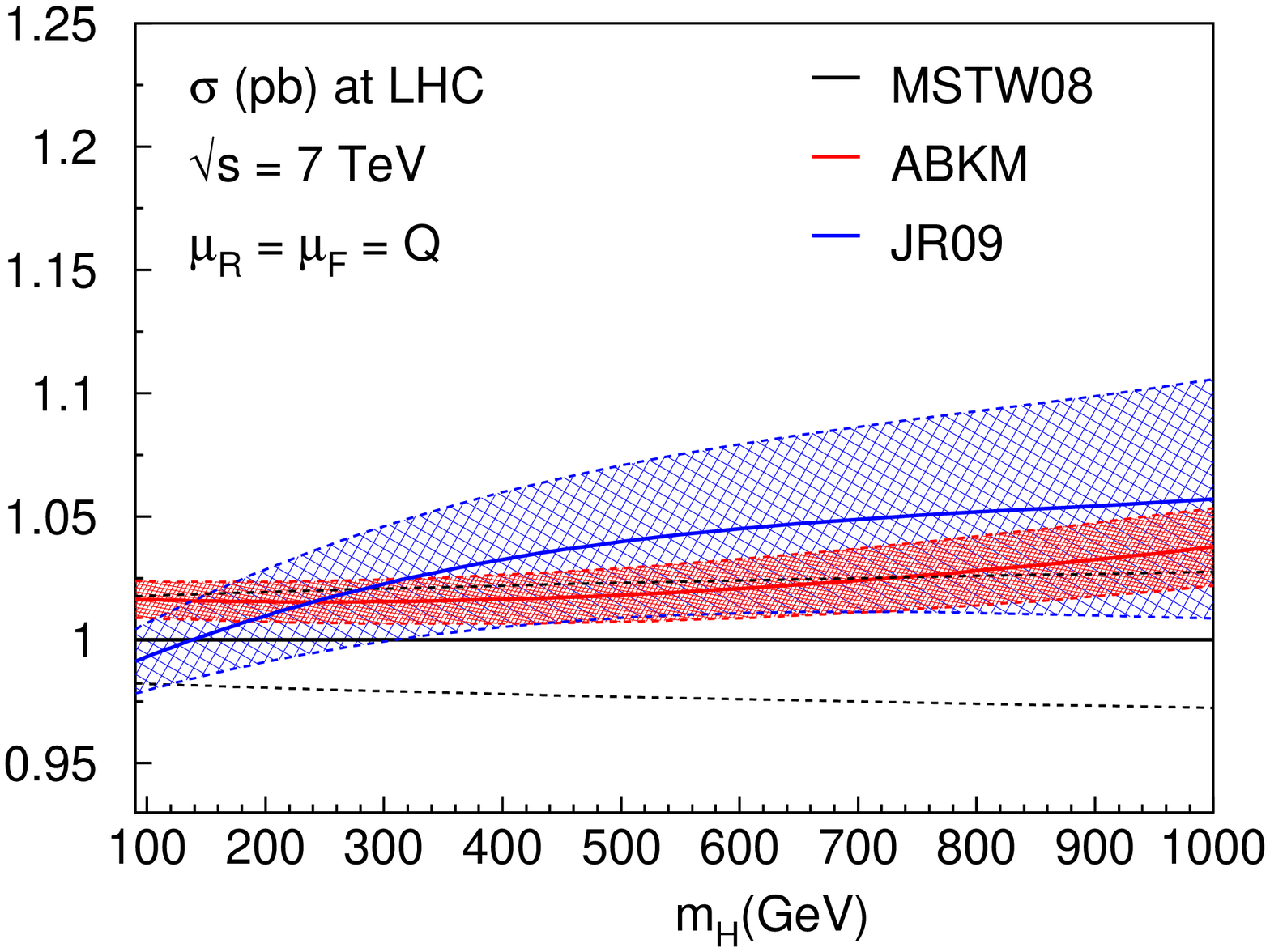}
\hspace*{3mm}
\includegraphics[scale=0.42]{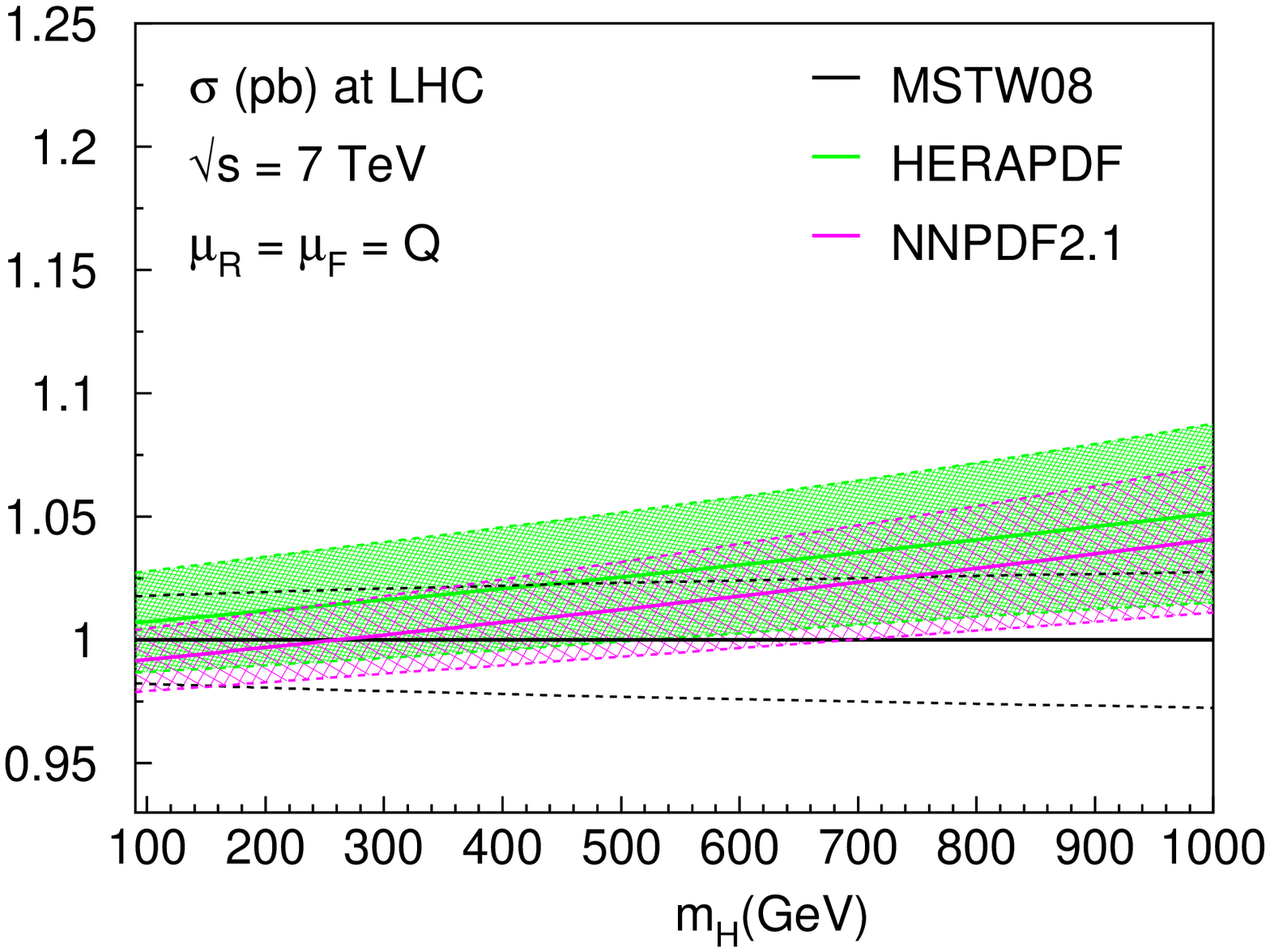}
\caption{\small
  \label{fig:lhc7-pdfs}
The PDF uncertainity of the VBF cross sections at the LHC, $\sqrt S = 7 \tev$, 
at NNLO in QCD for PDF sets of 
ABKM~\cite{Alekhin:2009ni},
HERAPDF1.5~\cite{herapdf:2009wt,herapdfgrid:2010}, 
JR09~\cite{JimenezDelgado:2008hf,JimenezDelgado:2009tv},
MSTW2008~\cite{Martin:2009iq} (68 \% CL), and NNPDF~\cite{Ball:2011uy}.
All results have been normalized to the best fit of MSTW2008.
}
\end{figure}

%%%%%%%%% LHC14 %%%%%%%%%%%%%%%%%%%
%
\begin{figure}[ht!]
\centering
\includegraphics[scale=0.65]{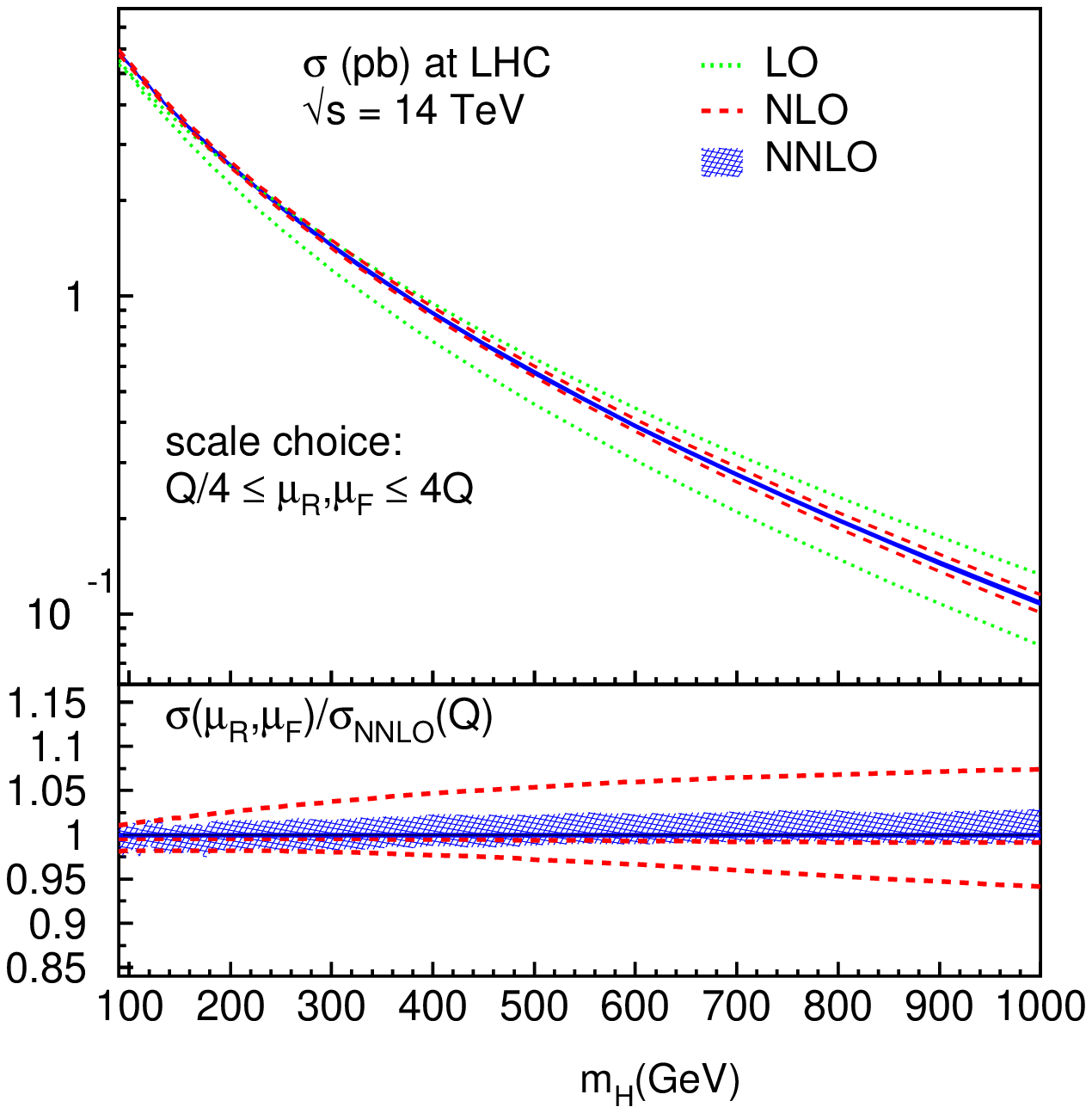}
\caption{\small
  \label{fig:lhc14-mh}
The total VBF cross sections at the LHC, $\sqrt S = 14 \tev$, at LO, NLO and NNLO in QCD
with the scale uncertainity from the variation $\mu_r,\mu_f \in [Q/4,4Q]$.
The MSTW2008~\cite{Martin:2009iq} PDF set (68\% CL) has been used. Numbers are in pb.
}
\end{figure}
\begin{figure}[ht!]
\centering
\includegraphics[scale=0.425]{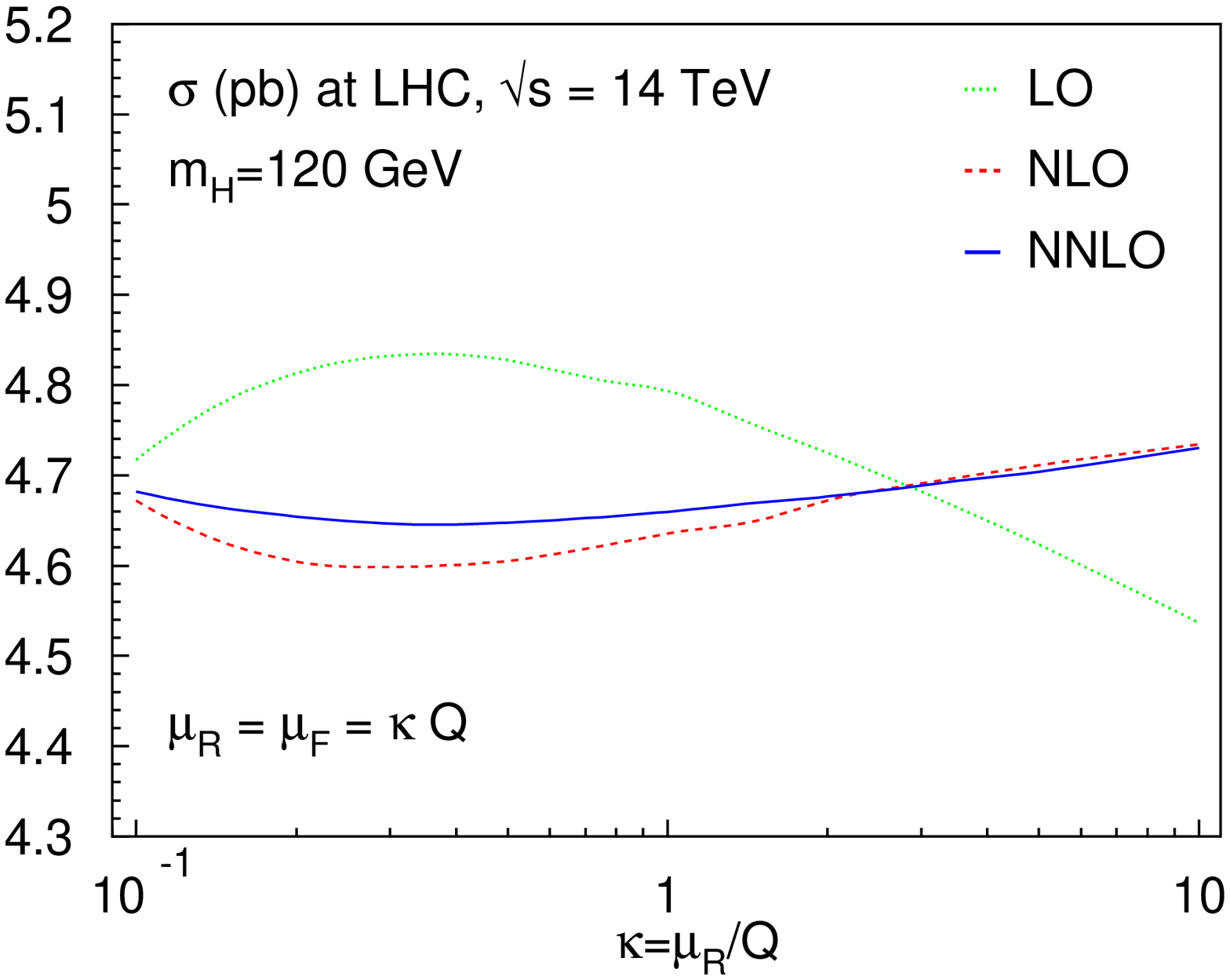}
\hspace*{3mm}
\includegraphics[scale=0.425]{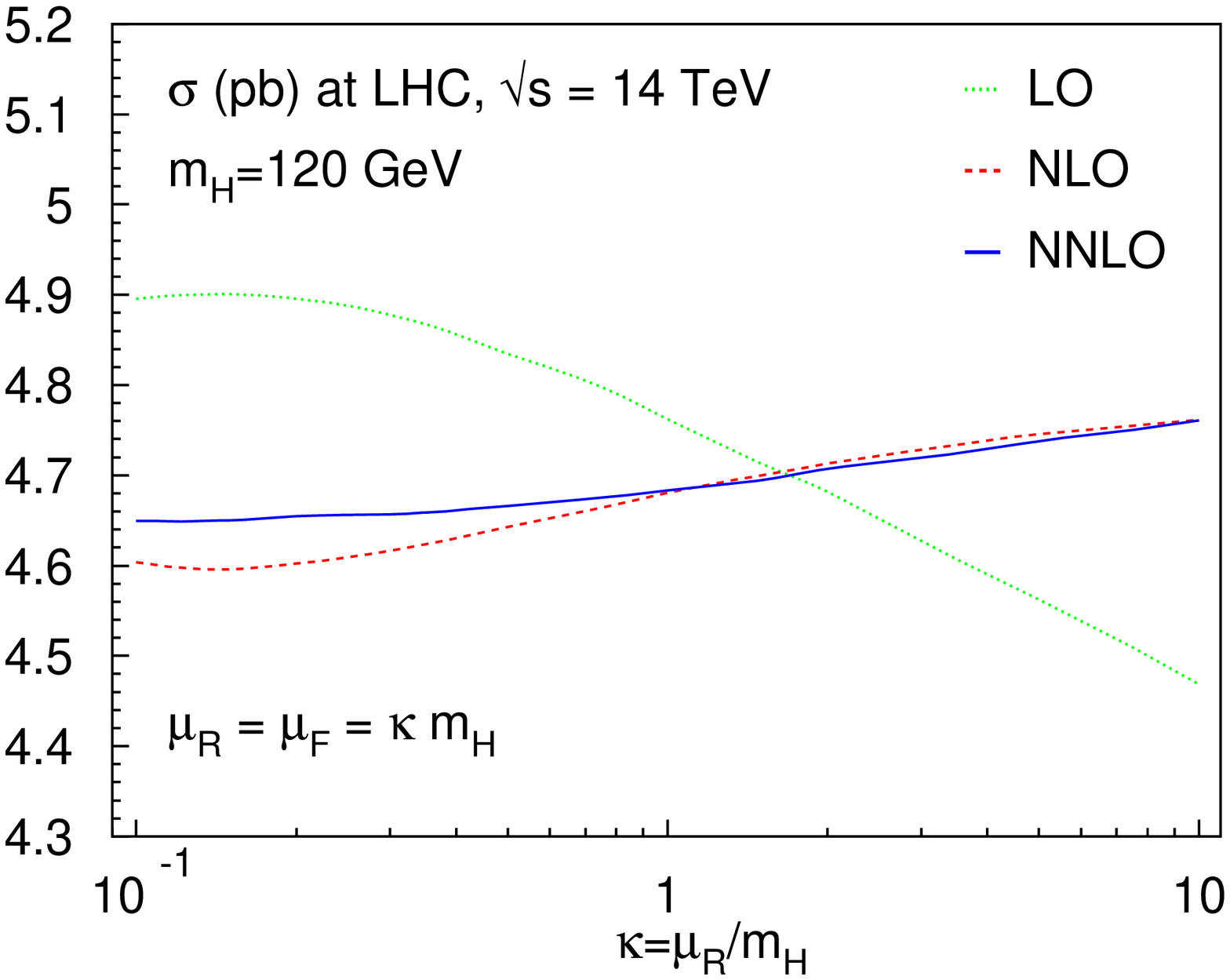}
\caption{\small
  \label{fig:lhc14-scale}
Scale dependence of the VBF cross sections at the LHC, $\sqrt S = 14 \tev$, 
at LO, NLO and NNLO in QCD for $\hmass = 120 \gev$ and the choice 
$\mu_r = \mu_f = \kappa Q$ (left) 
and
$\mu_r = \mu_f = \kappa \hmass$ (right).
The ABKM~\cite{Alekhin:2009ni} PDF set has been used. Numbers are in pb.
}
\end{figure}
\begin{figure}[ht!]
\centering
\includegraphics[scale=0.42]{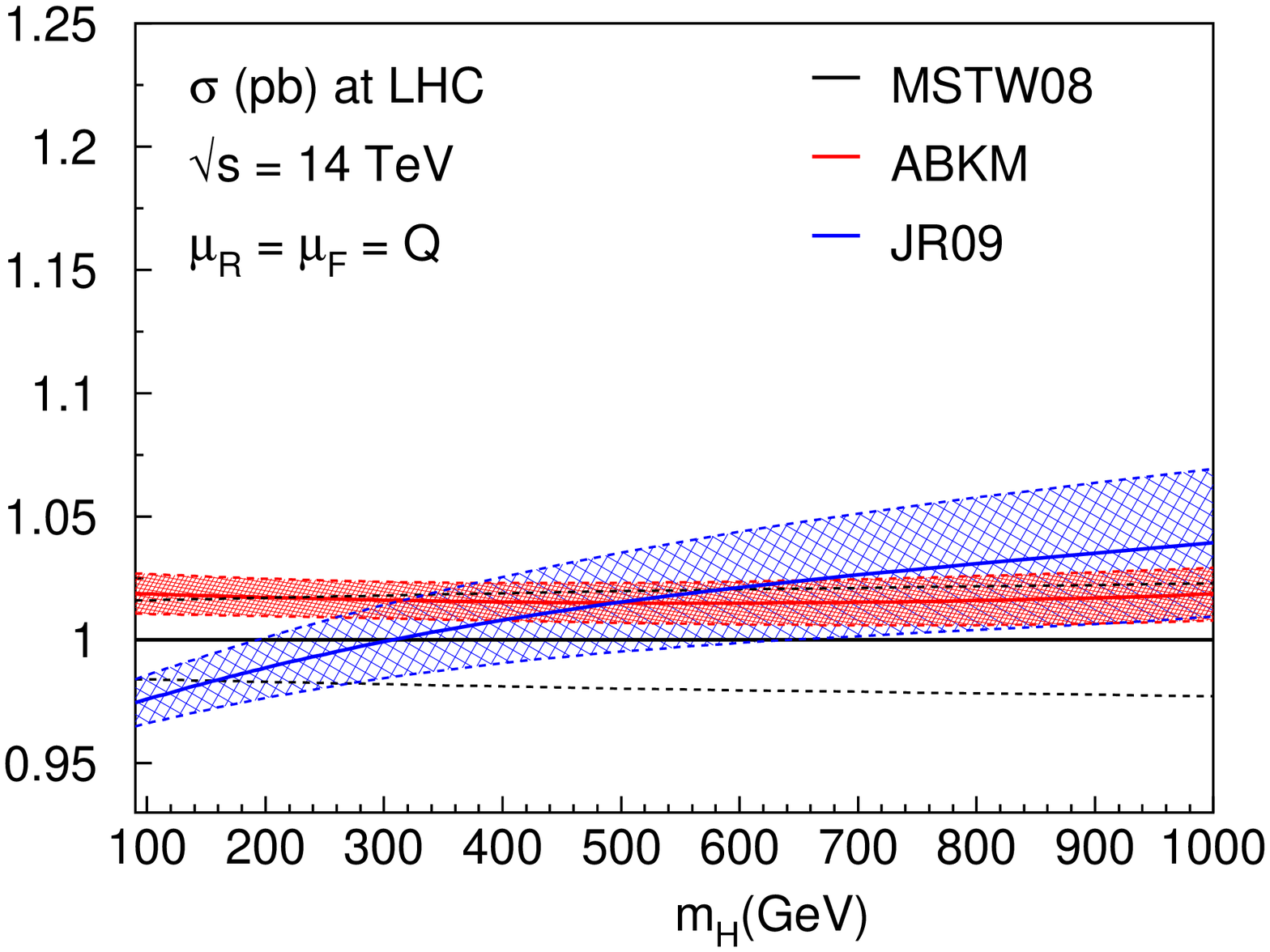}
\hspace*{3mm}
\includegraphics[scale=0.42]{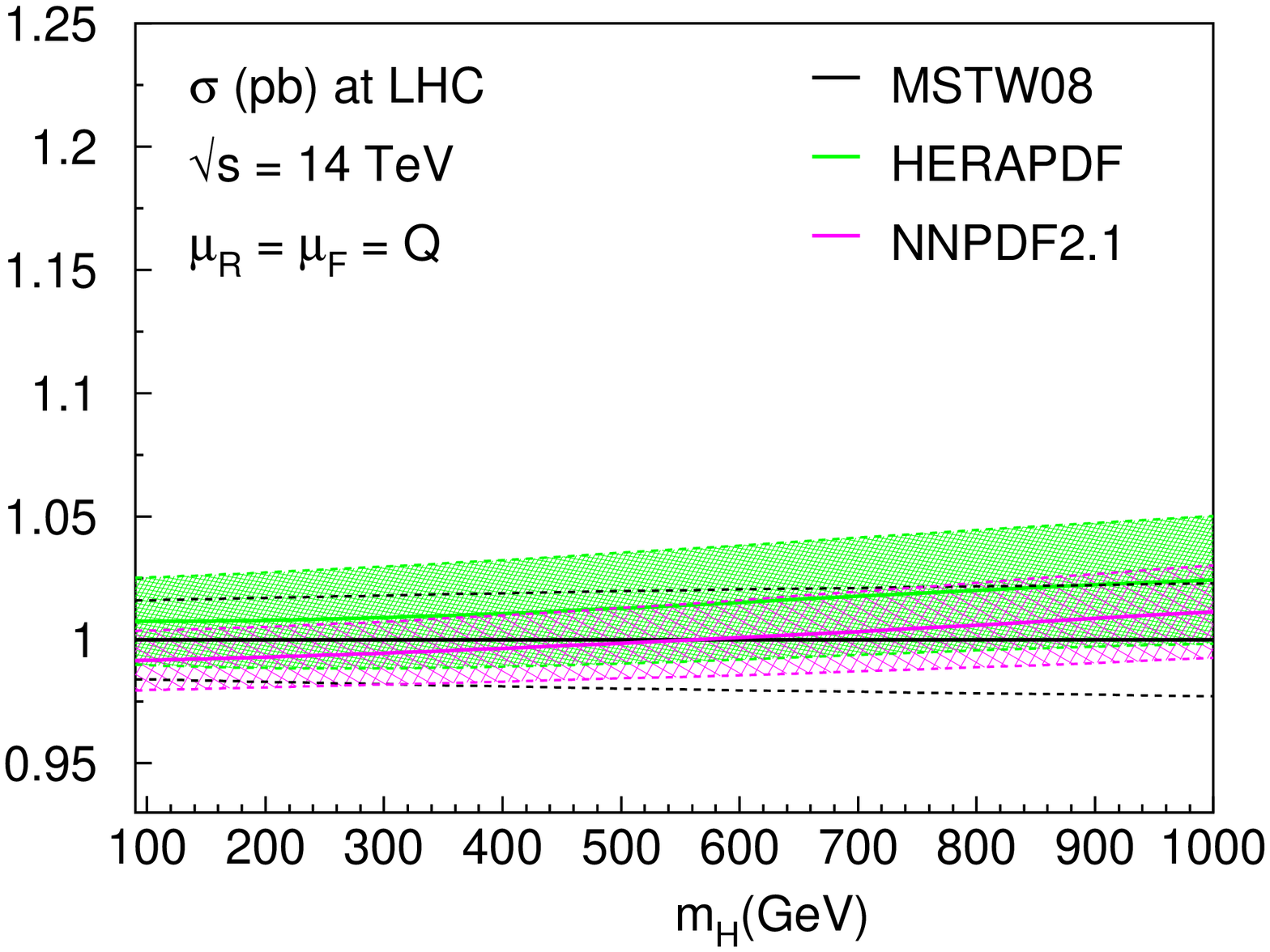}
\caption{\small
  \label{fig:lhc14-pdfs}
The PDF uncertainity of the VBF cross sections at the LHC, $\sqrt S = 14 \tev$, 
at NNLO in QCD for PDF sets of 
ABKM~\cite{Alekhin:2009ni},
HERAPDF1.5~\cite{herapdf:2009wt,herapdfgrid:2010}, 
JR09~\cite{JimenezDelgado:2008hf,JimenezDelgado:2009tv},
MSTW2008~\cite{Martin:2009iq} (68 \% CL), and NNPDF~\cite{Ball:2011uy}.
All results have been normalized to the best fit of MSTW2008.
}
\end{figure}

\subsection{Cross section predictions for the LHC}
\label{sec:LHCCrossSections}

Let us now present the numbers for the current and foreseen center-of-mass energies of the LHC.
For a Higgs boson with a mass $\hmass=100 \gev$ the cross section is roughly
$1.6 (5.5)$~pb at $\sqrt{S}=7(14) \tev$ 
and, again, steeply falling as a function of $\hmass$, e.g. to $0.1(0.5)$~pb at $\hmass=500 \gev$. 
This is illustrated in \cfa{fig:lhc7-mh}{fig:lhc14-mh}.
The complete listings of cross-sections at LO, NLO and NNLO in QCD for Higgs boson masses 
in the range $\hmass \in [90,1000] \gev$ for $\mu_r = \mu_f =Q$ 
are given in \ctd{tab:table-lhc7-a09-Sc1}{tab:table-lhc7-nnn-Sc1} ($\sqrt{S}=7 \tev$)
and in \ctd{tab:table-lhc14-a09-Sc1}{tab:table-lhc14-nnn-Sc1} ($\sqrt{S}=14 \tev$), respectively.

In analogy to the preceeding discussion in \csec{sec:TevCrossSections} 
our results in \ctd{tab:table-lhc7-a09-Sc1}{tab:table-lhc7-nnn-Sc1} demonstrate 
a very good apparent convergence of the perturbative expansion, 
{\it i.e.}, the NLO corrections of the order of 5\%, 
while the NNLO corrections are of the order of $\pm$1\%, only.
The NNLO results display also an impressive perturbative stability with respect to scale variations, 
see \cfa{fig:lhc7-scale}{fig:lhc14-scale} 
also for a comparison of the scale choice relating $\mu_r$ and $\mu_f$ to $Q$ 
and to $\hmass$, respectively. 
The preference to low scales is obvious from those plots, again 
with the point of minimal sensitivity being of the order $\mu_r, \mu_f \simeq Q$.
Given that all the perturbative QCD corrections not accounted for in the structure function
approach are very small and negligible, as discussed in \csec{sec:QftAtHO},
we can estimate the residual theoretical uncertainty due to uncalculated
higher orders in QCD at the LHC to be of order 2\% for the entire Higgs mass
range up to $\hmass = 1 \tev$.

Let us also comment on the uncertainties coming from the PDFs (and the
associated value of $\alpha_s$ as provided by the respective PDF set).
All PDF sets under consideration are displayed in \cfa{fig:lhc7-pdfs}{fig:lhc14-pdfs}.
In comparison to the Tevatron, we see that over a large range of Higgs masses,  $\hmass \simeq 100 \dots 300 \gev$, 
there are rather small differences between these sets only, because quark PDFs
are well constrained in the relevant $x$-region.
Generally, the PDF uncertainties are larger for the lower running energy of the LHC.
The plots in \cfa{fig:lhc7-pdfs}{fig:lhc14-pdfs} show that an almost constant $2\%$ PDF uncertainty can be
associated to the cross section for the LHC for a light Higgs boson. 
This deteriorates towards larger Higgs masses, e.g., the uncertainty being of order $\pm$10-12\% for $\hmass = 1 \tev$.

In summary, our results demonstrate that at NNLO in QCD the theoretical uncertainty is reduced to
about $2\%$ reaching the same level of ambiguity at which the Higgs production
signal via VBF can be defined phenomenologically, as discussed in \csec{sec:SetStage}.
Moreover, the study of the available NNLO PDF sets shows that all
non-perturbative parameters (PDFs and $\alpha_s$) needed for VBF precision predictions are well under control.
For larger Higgs masses, the PDF uncertainty is the dominating.

%\cleardoublepage
%\newpage

%%
%% ----------------------------------------------------------------------------
%%
\renewcommand{\theequation}{\thesection.\arabic{equation}}
\setcounter{equation}{0}
\renewcommand{\thefigure}{\thesection.\arabic{figure}}
\setcounter{figure}{0}
\section{VBF production beyond the Standard Model}
\label{sec:BSMHiggs}

One of the most important features of the structure function approach resides in its universality: as long as the final state
$X$ in $V^* V^* \to X$ fusion process does not interact (strongly) with the quark lines, the cross section can be
factorized using \ce{eq:disapproach} and QCD corrections decouple from the nature of the fusion process. 
This property allows the computation of  production cross sections at NNLO in QCD for
a variety of processes relevant for new physics searches in a straightforward way.  
In this section we present results for  a few examples: Higgs production through anomalous couplings
which are relevant when new physics states are heavy and the effects enter only at loop level; 
extra neutral and charged scalar production in extended Higgs sectors, such as in  two Higgs doublet and triplet Higgs models;
neutral and charged  vector fermiophobic resonance production.

%%
%% ---------------------------------------------------------------------------
%%
\subsection{Anomalous $VVH$ couplings}
\label{sec:AnomalousCoup}

Due to its very peculiar signature with two forward tagging jets, whose 
angular correlations can also be studied and related to the Higgs CP
properties independently of the Higgs decay channel, VBF can be a powerful means to discover new physics residing at a scale $\Lambda$ and responsible for anomalous $VVH$ vertices~\cite{Plehn:2001nj}.
The most generic structure for the $VVH$ vertex which generalizes \ce{eq:vbf-Mmunu-SM}, has the form~\cite{Takubo:2010fe}:
\begin{equation}
	\Gamma^{\mu\nu}\left(q_1,q_2\right)=2 \vmass^2\left( \left(\sqrt 2 G_F\right)^{1/2} + \frac{a_1}{\Lambda}\right) g_{\mu\nu} +
	\frac{a_2}{\Lambda} \left(q_1\cdot q_2g^{\mu\nu}-q_2^{\mu}q_1^{\nu}\right) + \frac{a_3}{\Lambda} \eps^{\mu\nu\rho\sigma} q_{1\rho}q_{2\sigma},
	\label{eq:VVH-anom}
\end{equation}
where the effective coefficients $a_i$ are dimensionless and vanish in the SM.
The VBF cross-section including anomalous vertices is known at NLO in
QCD~\cite{Hankele:2006ma}, and has been included in the program {\sc VBFNLO}~\cite{Arnold:2008rz}.

As discussed in~\cite{Plehn:2001nj}, at high energies one expects the effective coefficients $a_i$ to become form factors, 
{\it i.e.} to display a dependence on the scattering energy such that  unitarity would be preserved. 
However, since at the LHC the typical scales in a VBF process can be still considered small with respect to a new physics
scale of the order of 1 TeV, we will present numbers for constant values of $a_i$ and  $\Lambda = 500$ GeV. With the vertex in  \ce{eq:VVH-anom}, 
the expression corresponding to \ce{eq:WMMW-SM} becomes a bit more involved than its SM counterpart and it is given in the Appendix, ~\csec{sec:WMMW-anom}. 

Since $a_1$ only changes the cross-section by an overall factor, we set it to 0. We therefore plot 
in \cf{fig:anom-xsect} the total cross-sections at the LHC for different values of $a_2$ and $a_3$. 
Uncertainties at NNLO are again found to be of the order of a few percent.

\begin{figure}[h!]
	\begin{center}
\includegraphics[width=0.45\textwidth]{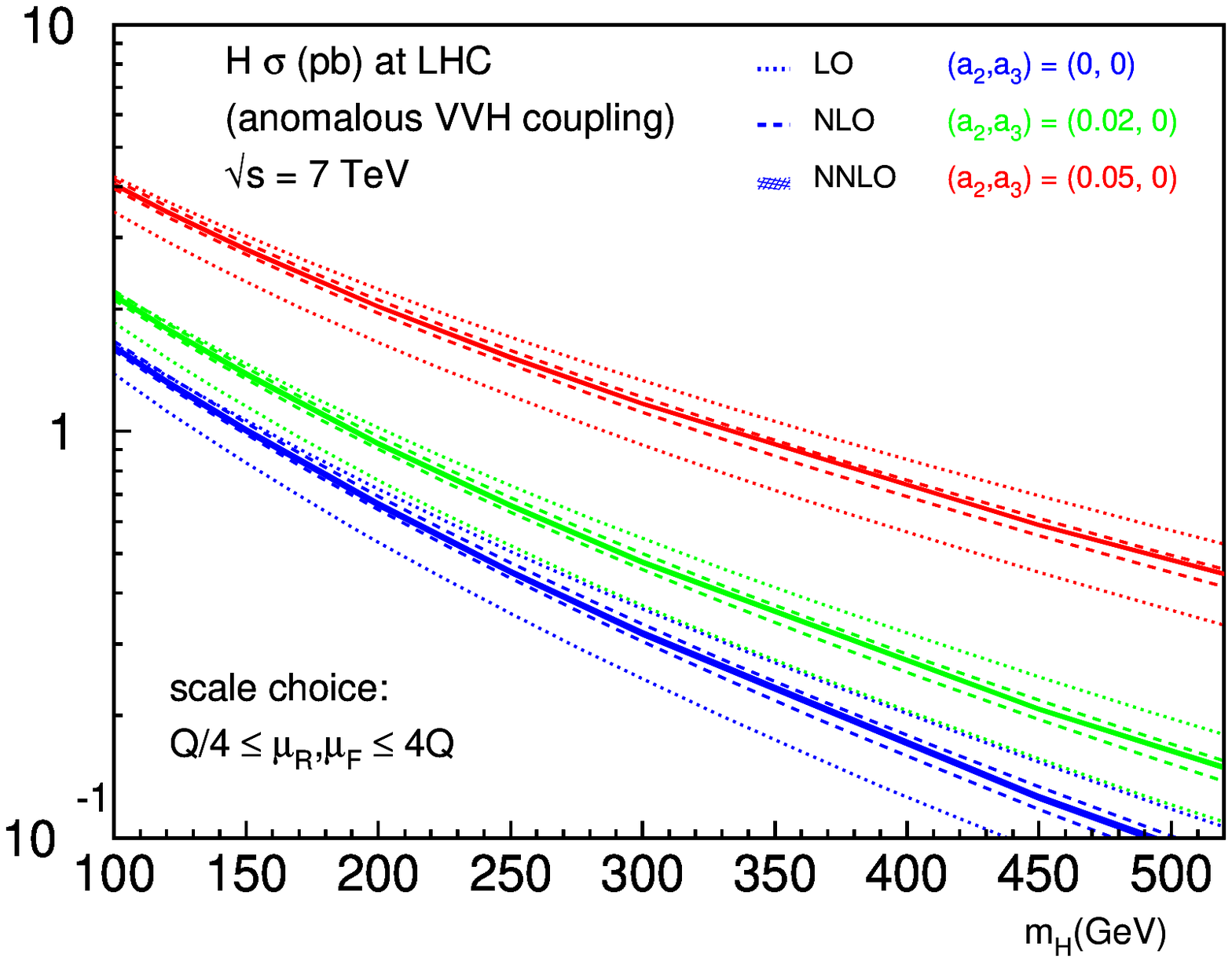}
\hspace{0.5cm}
\includegraphics[width=0.45\textwidth]{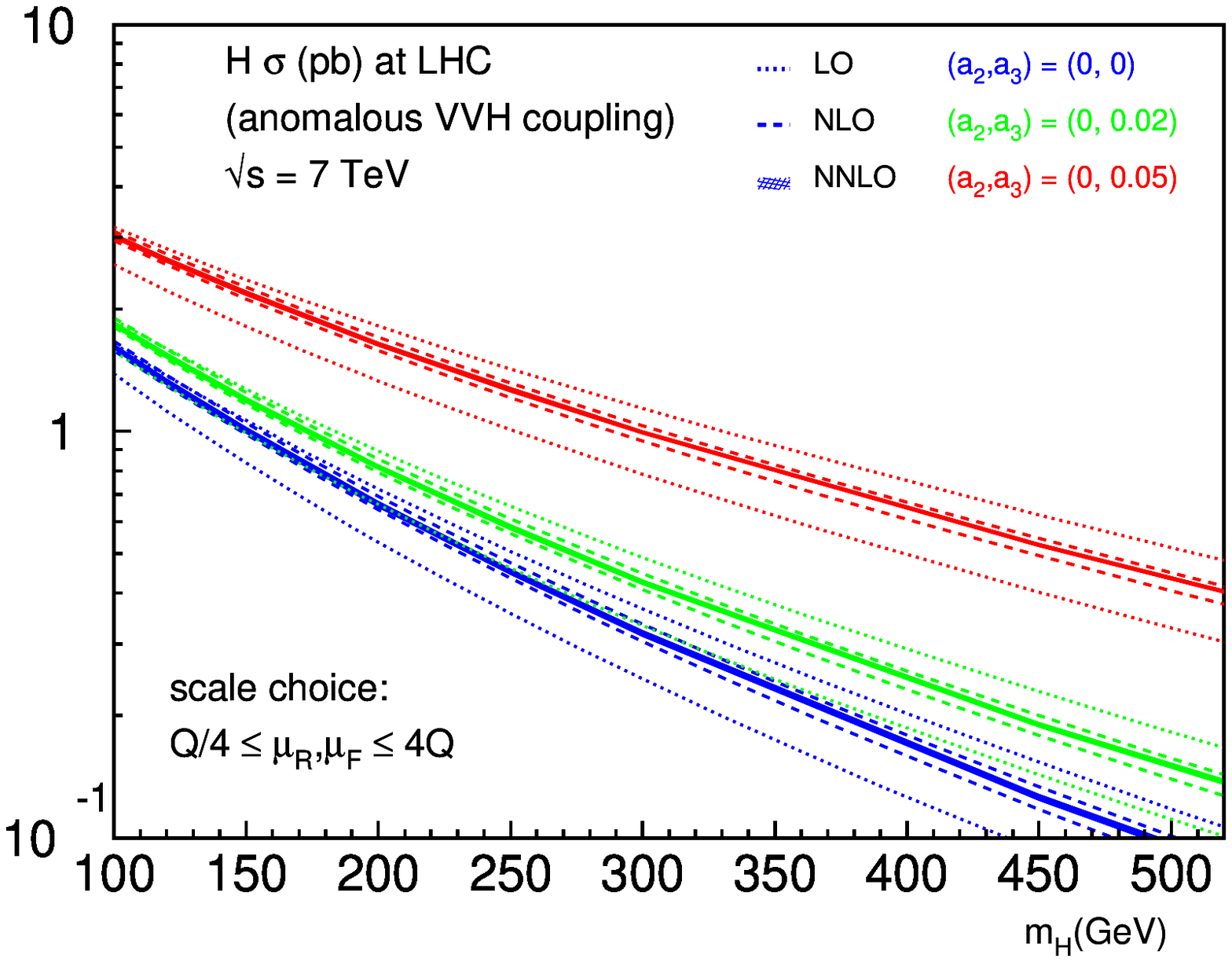}
\\[4ex]
\includegraphics[width=0.45\textwidth]{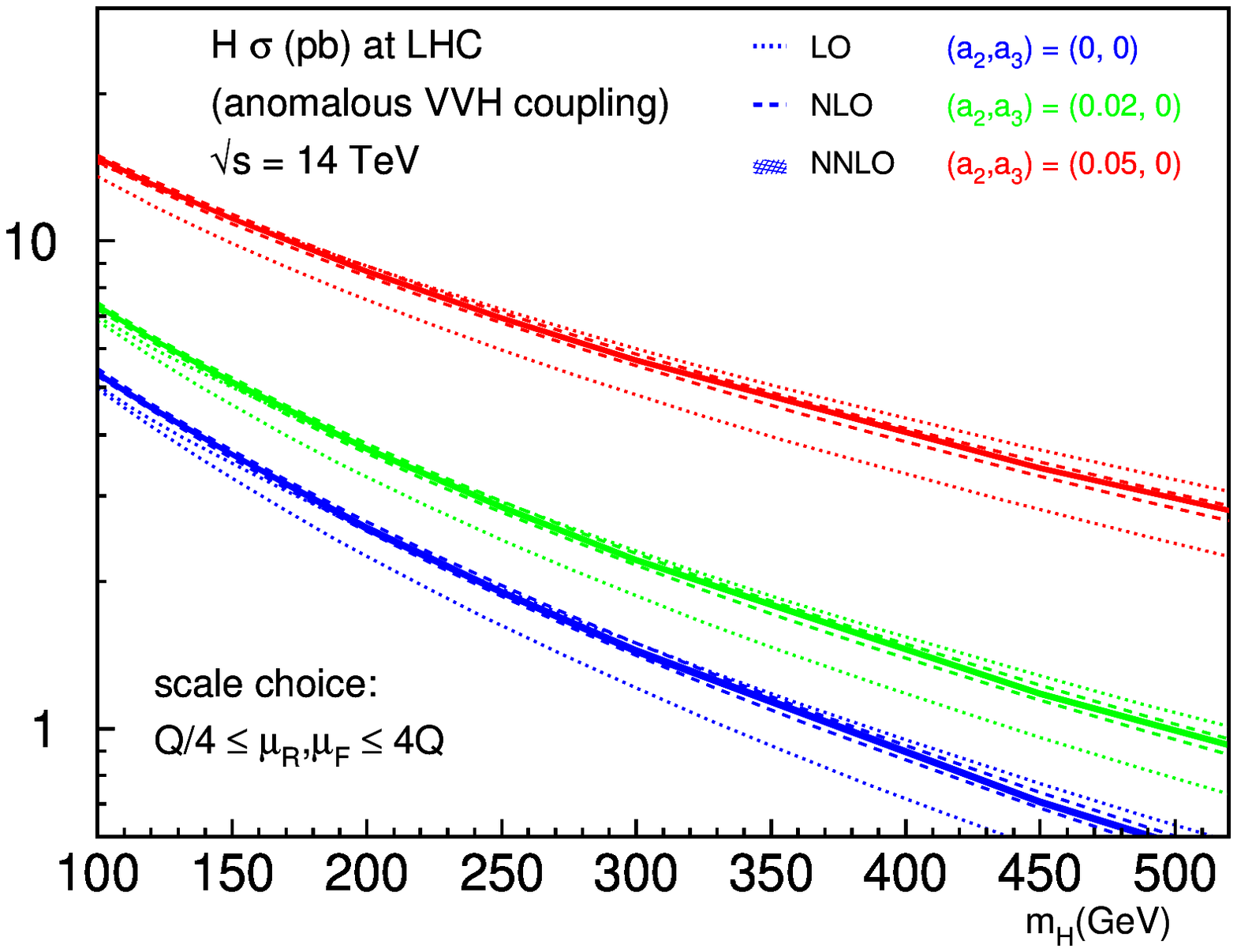}
\hspace{0.5cm}
\includegraphics[width=0.45\textwidth]{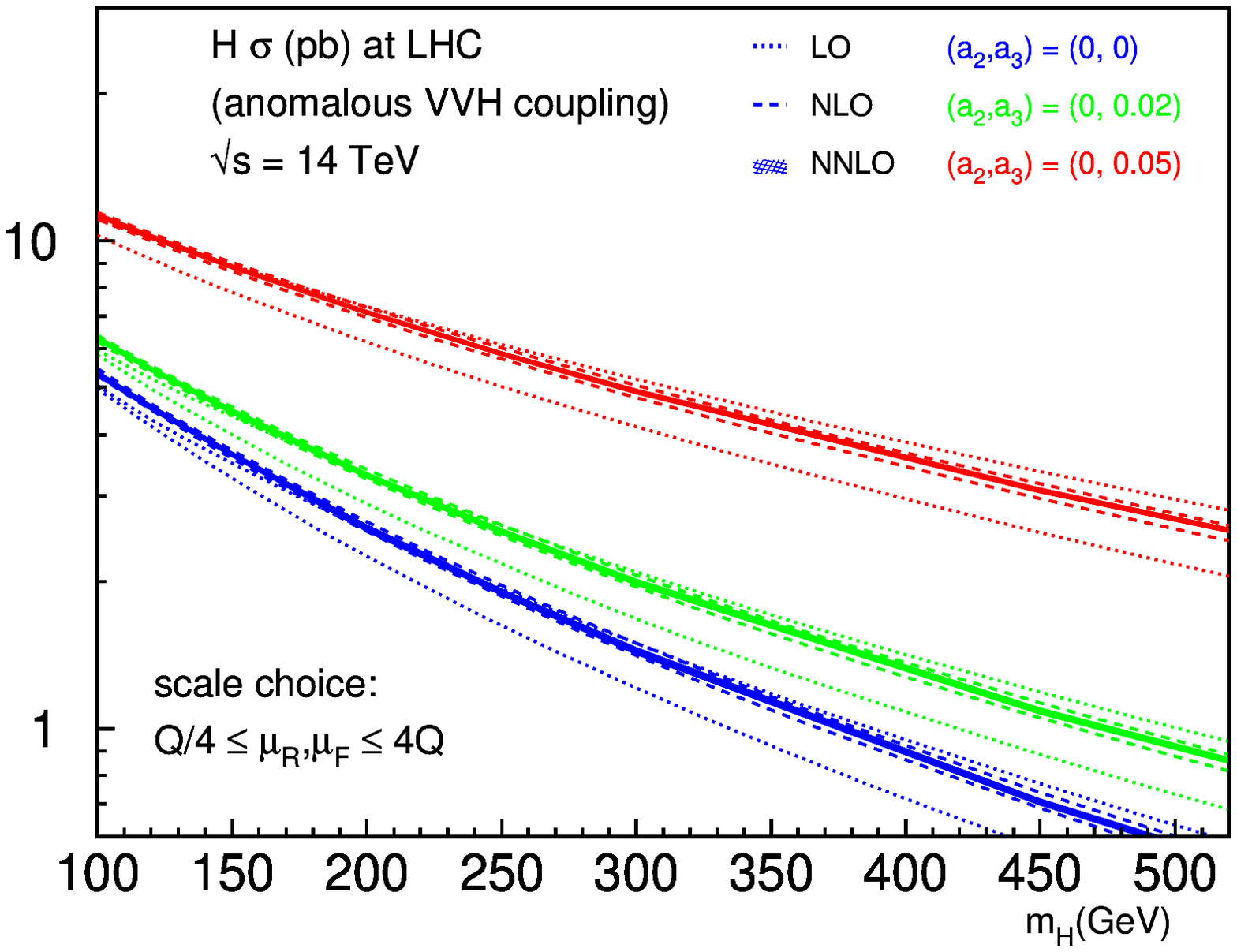}

\caption{\small
\label{fig:anom-xsect}
The total cross-section for Higgs production
via VBF at the LHC, with $\sqrt S = 7\tev$ (top) and $\sqrt S = 14\tev$ (bottom). The values used for the parameters $a_2$, $a_3$ are written on the plots. We have assumed $\Lambda= 500\gev$. The MSTW 2008~\cite{Martin:2009iq}
PDF set has been used. The uncertainty bands are obtained from the variation of
the renormalization and factorization scale in the interval $Q/4<\mu_f,\mu_r<4Q$, 
where $Q$ is the virtuality of the vector boson.}
\end{center}
\end{figure}

\subsection{New neutral and charged Higgs bosons}
\label{sec:ChargedHiggs}
Many extension of the SM feature an Higgs sector with an extra
${\rm SU}(2)$ doublet and/or triplet, that lead to a particle spectrum including several neutral Higgs
bosons, as well as single- and, possibly, double-charged scalars. Because of the particular signature
with forward jets, VBF can be a really powerful search channel also for these states.

In a generic two Higgs doublet model (2HDM), 
such as that in the  Minimal Supersymmetric Standard Model (MSSM)~\cite{Hollik:2008xn}, 
the production rate of a CP-even light or heavy Higgs scalar  via VBF is equal to the SM one times overall
factors~\cite{heinemeyer:1998np,degrassi:2002fi,frank:2006yh}. Therefore, to obtain the total
cross-section including QCD NNLO corrections for the $h^0$ or $H^0$ production, one
just needs to rescale the SM results given in~\csec{sec:Phenomenology}. 

Pseudo-scalar states do not couple to vector bosons at the tree level,
therefore VBF is not really relevant in this case. On the other hand, it is of great phenomenological 
interest to study the production of
charged Higgs bosons. As we will briefly see, there exist a class of models in
which charged Higgs boson production via VBF can be observed.
\begin{figure}[h!]
	\begin{center}
\includegraphics[width=0.45\textwidth]{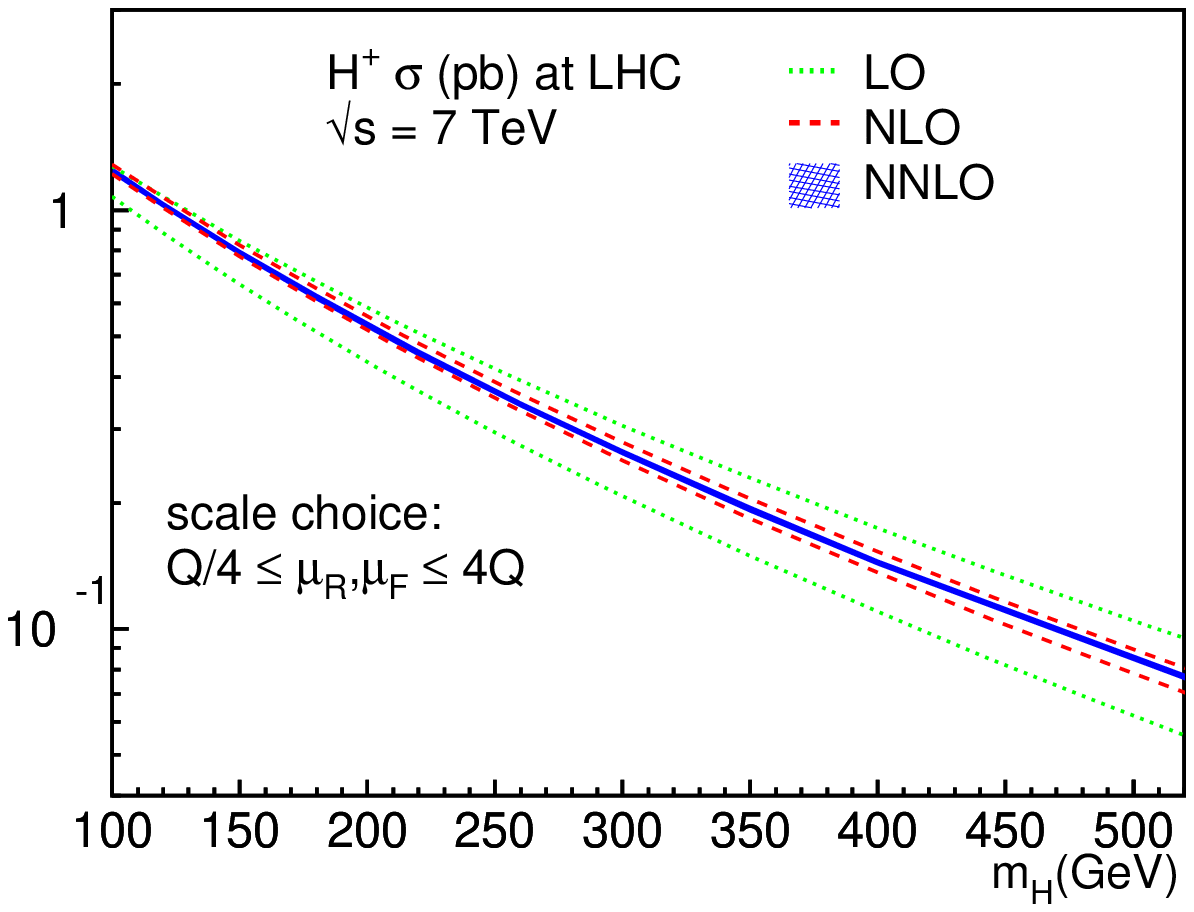}
\hspace{0.5cm}
\includegraphics[width=0.45\textwidth]{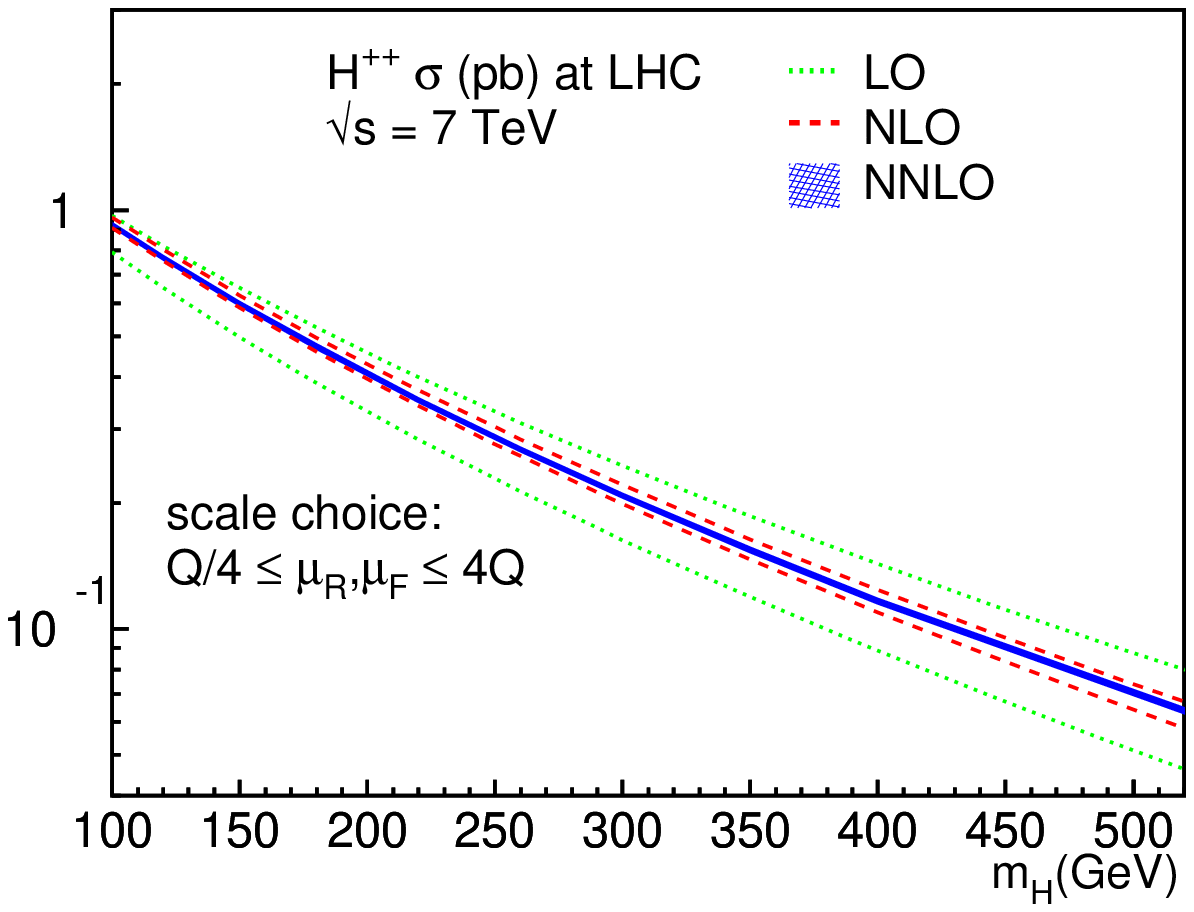}
\\[4ex]
\includegraphics[width=0.45\textwidth]{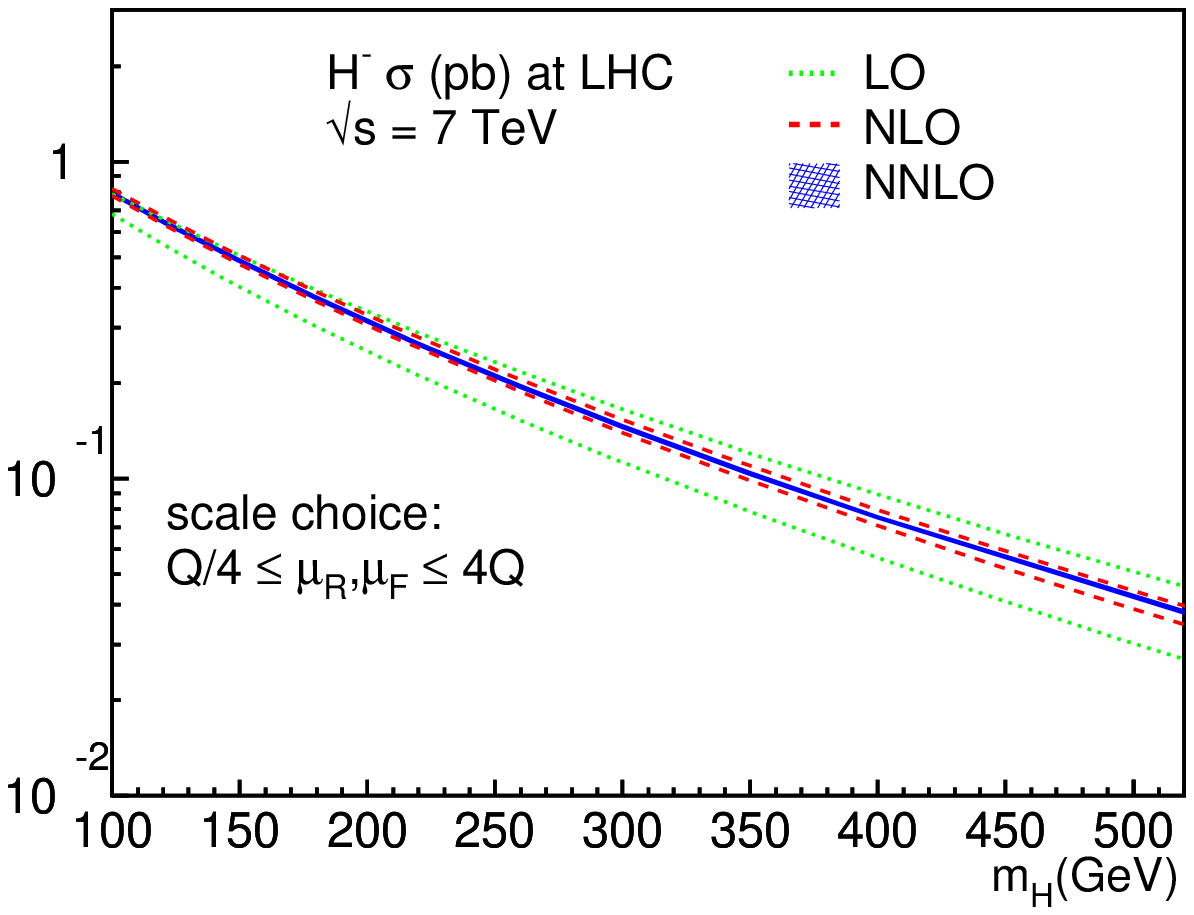}
\hspace{0.5cm}
\includegraphics[width=0.45\textwidth]{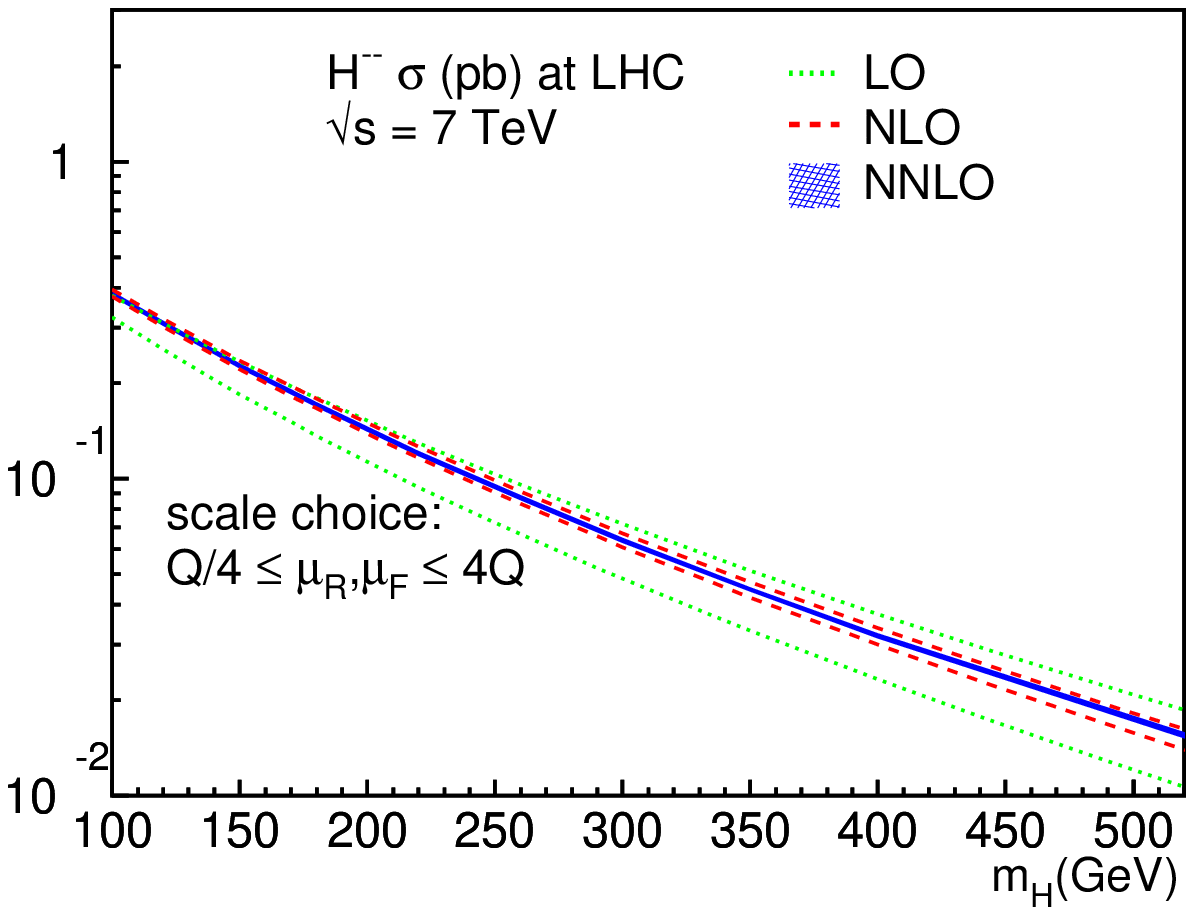}

\caption{\small
\label{fig:charged-xsect}
The total cross-section for charged Higgs production
via VBF at the LHC, with $\sqrt S = 7\tev$. The MSTW 2008~\cite{Martin:2009iq}
PDF set has been used. The uncertainty bands are obtained from the variation of
the renormalization and factorization scale in the interval $Q/4<\mu_f,\mu_r<4Q$, 
where $Q$ is the virtuality of the vector boson. Plots refer to: $H^+$
production (top-left), $H^{++}$ production (top-right), $H^-$ production
(bottom-left), $H^{--}$ production (bottom-right).
}
\end{center}
\end{figure}

\begin{figure}[h!]
	\begin{center}
\includegraphics[width=0.40\textwidth]{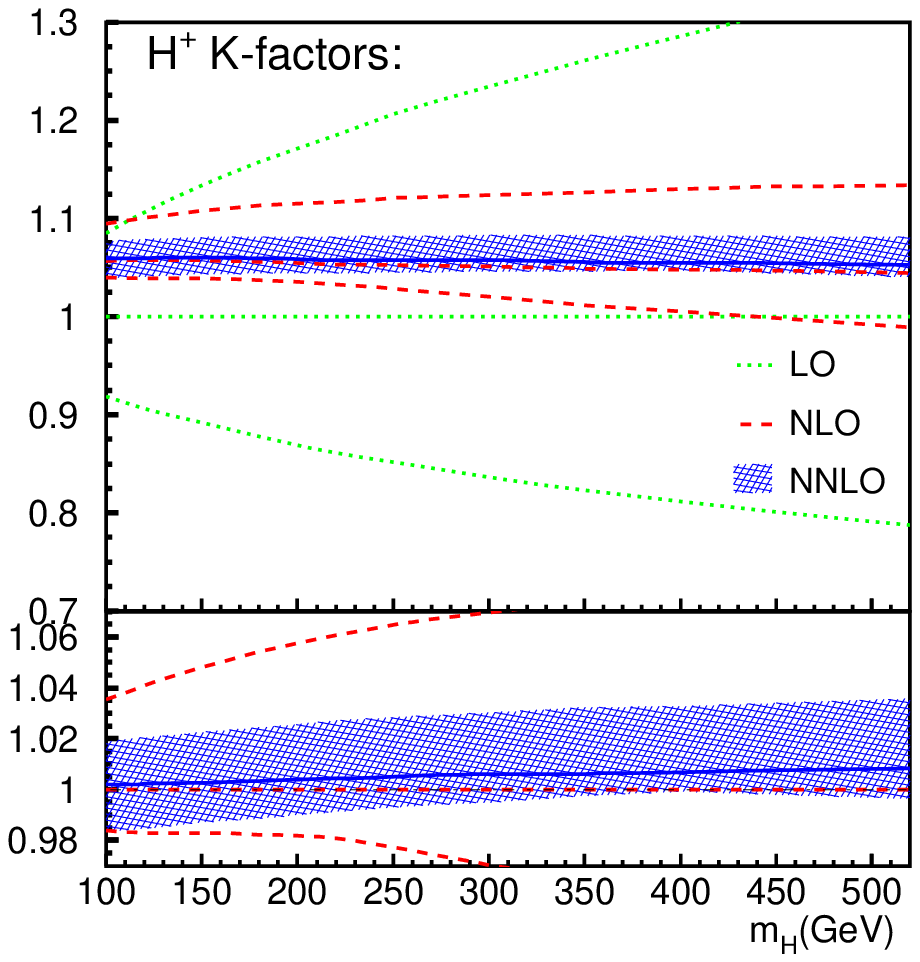}
\hspace{0.5cm}
\includegraphics[width=0.40\textwidth]{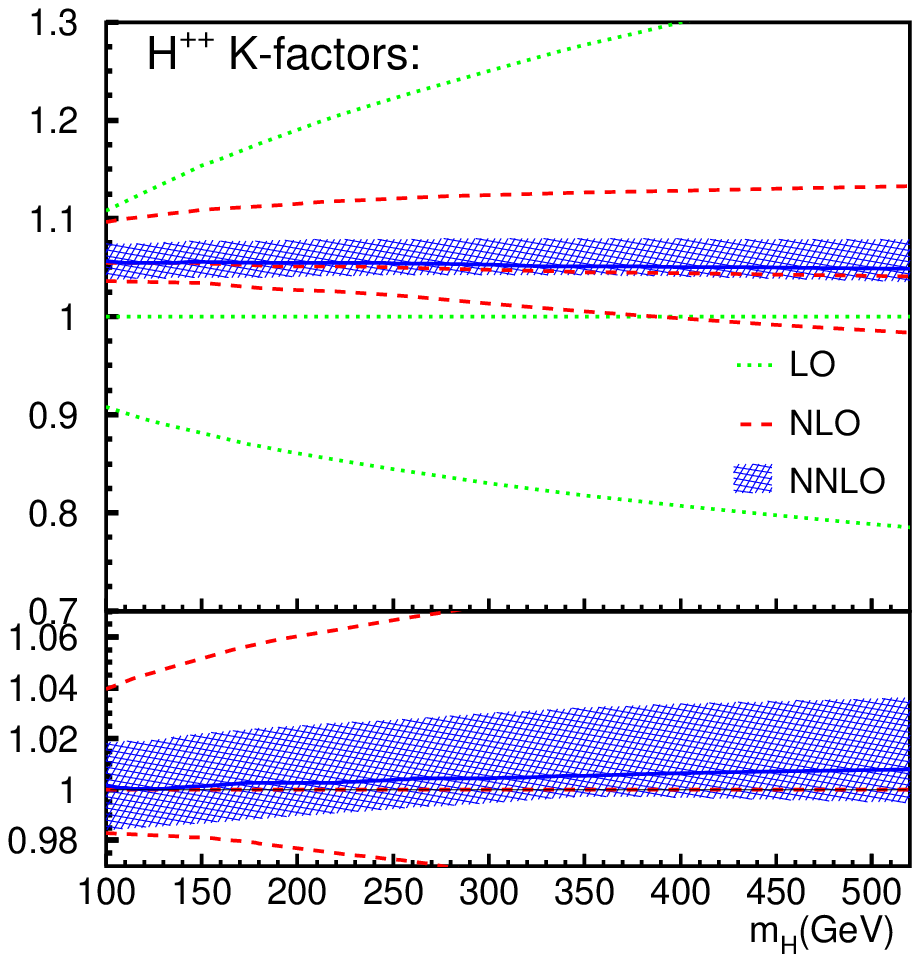}
\\[4ex]
\includegraphics[width=0.40\textwidth]{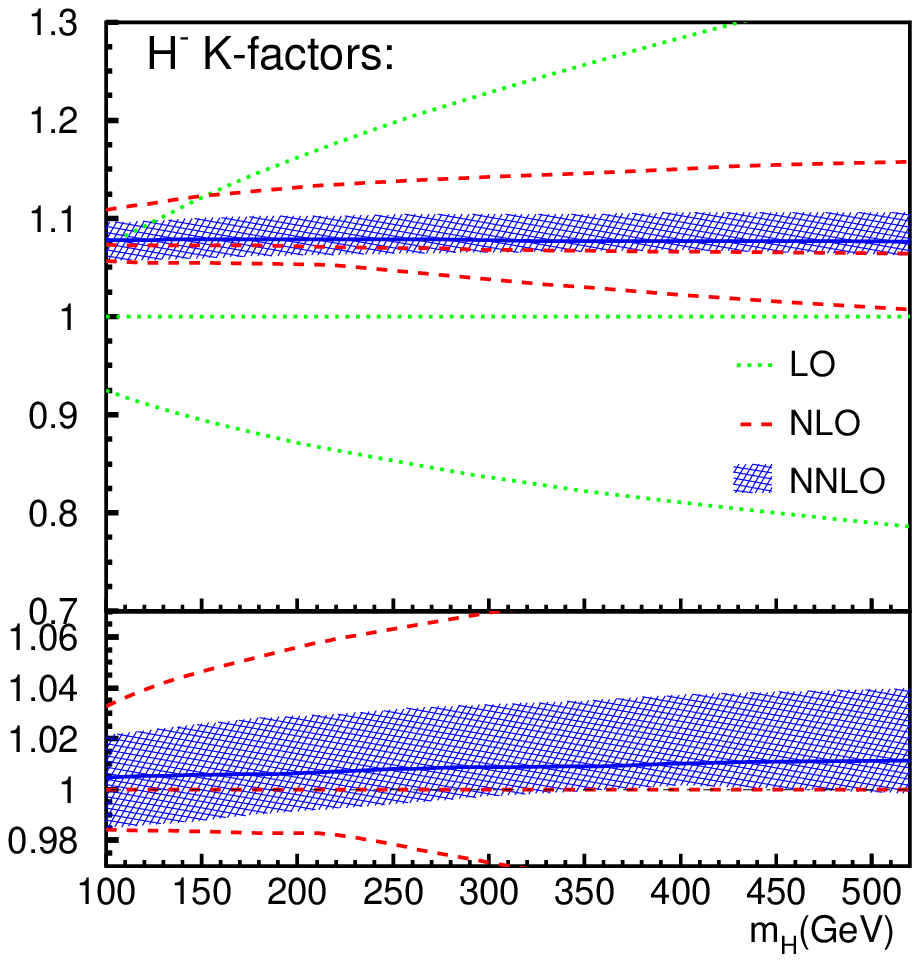}
\hspace{0.5cm}
\includegraphics[width=0.40\textwidth]{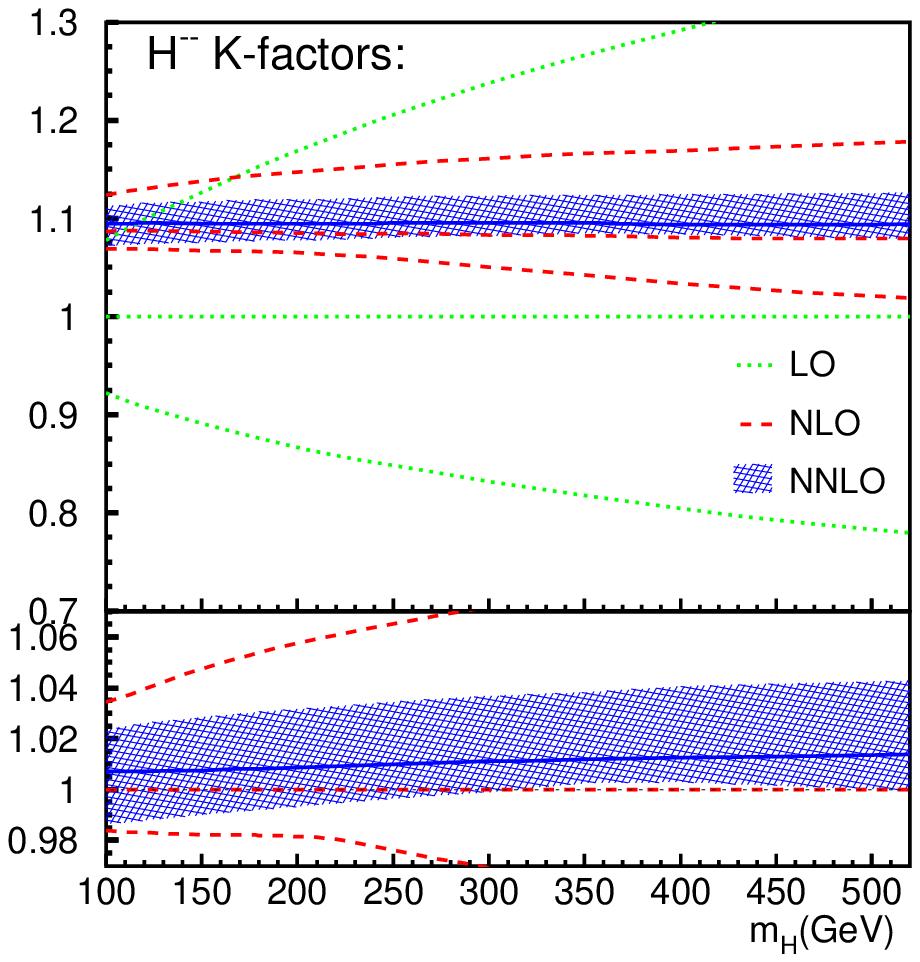}

\caption{\small
\label{fig:charged-kfact}
The (N)NLO/LO (upper inlay) and NNLO/NLO (lower inlay) $K$-factors for charged Higgs production
via VBF at the LHC, with $\sqrt S = 7\tev$. The MSTW 2008~\cite{Martin:2009iq}
PDF set has been used. The uncertainty bands are obtained from the variation of
the renormalization and factorization scale in the interval $Q/4<\mu_f,\mu_r<4Q$, 
where $Q$ is the virtuality of the vector boson. 
Plots refer to: $H^+$ production (top-left), $H^{++}$ production (top-right),
$H^-$ production (bottom-left), $H^{--}$ production (bottom-right).
}
\end{center}
\end{figure}

\begin{figure}[h!]
	\begin{center}
\includegraphics[width=0.40\textwidth]{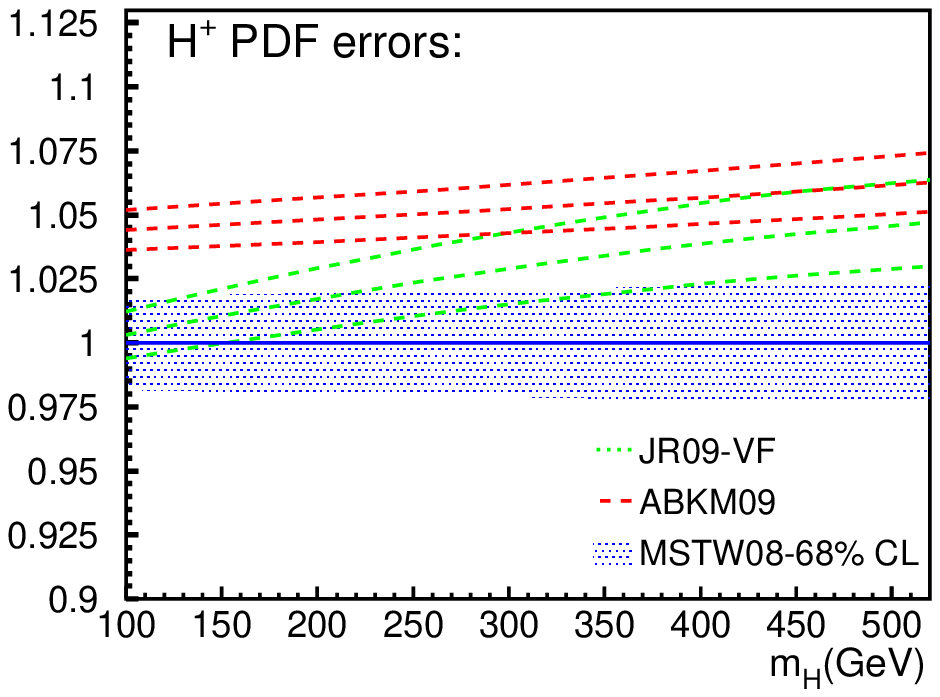}
\hspace{0.5cm}
\includegraphics[width=0.40\textwidth]{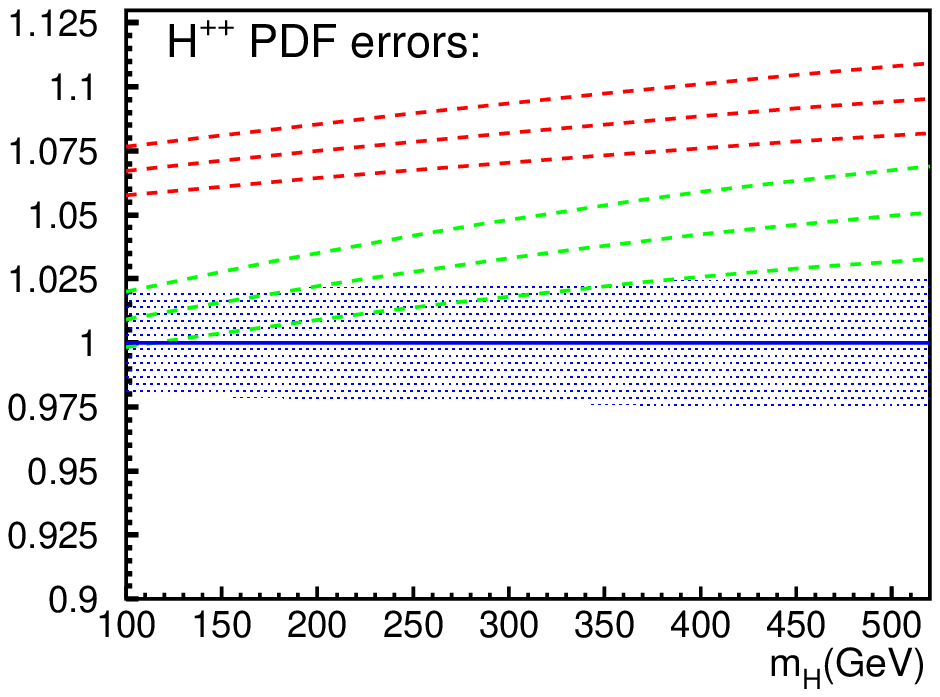}
\\[4ex]
\includegraphics[width=0.40\textwidth]{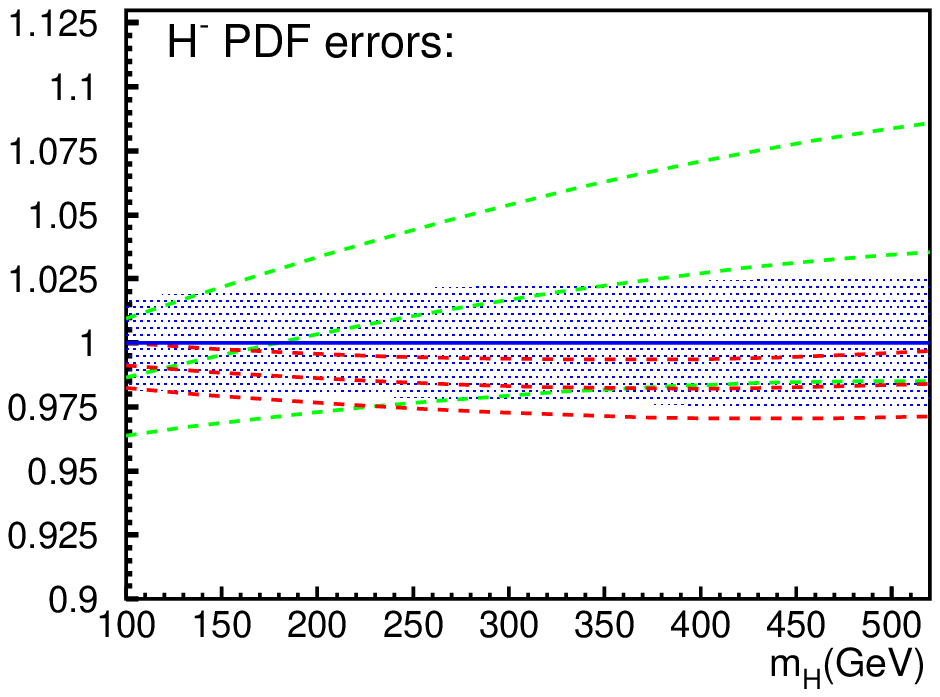}
\hspace{0.5cm}
\includegraphics[width=0.40\textwidth]{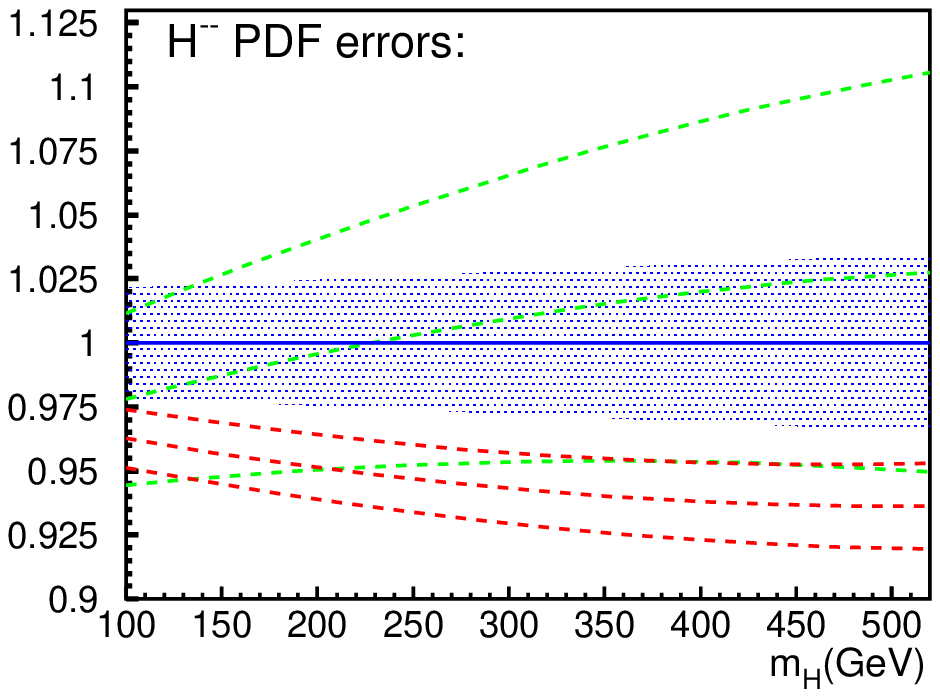}

\caption{\small
\label{fig:charged-pdf}
The PDF uncertainties to the NNLO total cross section for charged Higgs boson
production in VBF at the LHC, with $\sqrt S = 7\tev$. Plots are normalized to
the MSTW 2008~\cite{Martin:2009iq} central value. The errors for the MSTW set
are at 68\% CL. The plots show also the central value and error band for the
JR09-VF NNLO set~\cite{JimenezDelgado:2008hf, JimenezDelgado:2009tv} and for
the ABKM09 NNLO set~\cite{Alekhin:2009ni, Alekhin:2010iu}. Plots refer to:
$H^+$ production (top-left), $H^{++}$ production (top-right), $H^-$ production
(bottom-left), $H^{--}$ production (bottom-right)
}
\end{center}
\end{figure}

In models where the extra Higgs bosons are included in ${\rm SU}(2)$
doublets, like the MSSM or the 2HDM, isospin conservation forbids (at the tree
level) the appearance of $W^{\pm}H^{\mp}Z$ and $W^{\pm}H^{\mp}\gamma$ vertices,
while they can be loop-induced~\cite{Kanemura:1997ej}. However, in this case the
corresponding total cross-section for VBF is totally negligible~\cite{Asakawa:2006gm}, leaving the associated production of a charged
Higgs boson and a top-quark from a bottom-gluon initial state~\cite{Berger:2003sm,Weydert:2009vr} 
as the only phenomenologically relevant production channel. 

This situation changes when models with isospin triplets in the
Higgs-sector are considered. Many versions of such models exist in literature, with a very
interesting particle content including double-charged Higgs bosons.  In many of such models
the coupling of the charged Higgs bosons to gauge bosons is allowed with a
strength proportional to the ratio ${v'}/{v}$, where $v'$ is the triplet
vacuum expectation value (vev) and $v$ the SM one. Even though $v'$ is constrained
by electroweak data, in particular by the $\rho$ parameter of the SM, to be at most
a few GeV, this bound can be loosened by including more triplets~\cite{Georgi:1985nv}. In this case, the production of single- and double-charged Higgs bosons via 
VBF can be relevant at colliders~\cite{Gunion:1989ci,Vega:1989tt,Asakawa:2006gm}.
The Lorentz structure of the $VVH$ vertex is, 
like the SM one in \ce{eq:vbf-Mmunu-SM}, proportional to the metric tensor (see, e.g., the Feynman rules given in Ref.~\cite{Godfrey:2010qb}),
and it can be cast in the generic form
\begin{equation}
\Gamma^{\mu\nu}_{V_iV_jH}=2 \left(\sqrt 2 G_F\right)^{1/2} m_i m_j \; F_{ij}\left(-i g^{\mu\nu}\right),
\label{eq:charged-vertex}
\end{equation}
where the dimensionless constant $F_{ij}$ depends on the particular model (in
the SM the non-vanishing ones are $F_{ZZ}=F_{W^{\pm}W^{\mp}}=1$).
Because of the rich variety of available models, and in order to be as much general as possible, 
we will present numbers with $\left|F_{ij}\right|=1$.

In~\cf{fig:charged-xsect} we show the total cross section for $H^{\pm},H^{\pm\pm}$ production via VBF at the LHC with $\sqrt S = 7\tev$.
The plots clearly display the excellent convergence of the perturbative series and the reduction of theoretical
uncertainties obtained by the inclusion of higher order corrections, as it  can be also be  appreciated by
inspecting~\cf{fig:charged-kfact}, where the $K$-factors are shown. In
particular we find that the NNLO $K$-factor for the nominal scale choice 
($\mu_r=\mu_f=Q$) differs from one for less than 1\% in almost the whole mass range shown. 
From \cf{fig:charged-pdf} it is clear that the uncertainty 
coming from the PDFs is sizable,  in particular for large values of the charged Higgs masses, 
where the PDFs are probed at larger values of $x$,  {\it i.e.}, in a region, where the PDF uncertainties generally increase.

\subsection{Production of fermiophobic vector boson resonances}
\label{sec:VectorReson}
In many extensions of the SM, additional vector bosons can appear in the
particle spectrum, together with or as an alternative to the Higgs boson, see
e.g.~\cite{Csaki:2003dt,Csaki:2003zu,Chivukula:2006cg,He:2007ge}.
The existence of heavy vector bosons, however, is severely constrained by electroweak
precision data and direct searches and masses are pushed to the multi-TeV scale. 
This holds unless their couplings to (light) fermions are for some reason suppressed and 
only couplings to the SM weak vector bosons take place. 
In such models, where the new heavy vector resonances are dubbed ``fermiophobic'',
VBF can become the dominant production channel.

While a detailed phenomenological analysis is beyond our scope,  we 
just limit ourselves to  presenting a few results 
motivated by warped scenarios~\cite{Agashe:2007ki,Agashe:2008jb} 
for the production of a neutral and a single-charged vector resonance, 
which couples to SM vector bosons via a trilinear vertex of the form
\begin{equation}
	\Gamma^{\mu_1 \mu_2 \mu_3}\left(p_1, p_2, p_3\right) = g_{123} \left[ g^{\mu_1 \mu_2}\left(p_1^{\mu_3}-p_2^{\mu_3}\right) - 
	g^{\mu_2 \mu_3}\left(p_2^{\mu_1}-p_3^{\mu_1}\right) -
        g^{\mu_3 \mu_1}\left(p_3^{\mu_2}-p_1^{\mu_2}\right) \right]. \label{VectorVert}
\end{equation}
This particular choice assumes that the new vector bosons are explicit or ``hidden'' gauge vectors  of a new sector.
Different vertices are also possible  (see e.g.,~\cite{DaRold:2005zs,Falkowski:2011ua}) and can be easily implemented in our calculation.

The expression for the total cross-section in terms of structure functions and particle momenta is given in \csec{sec:WMMW-vector}.
The results we present are for neutral and charged vector boson production, which we
call $W'^{\pm}$ or $Z'$ and which we assume  not to be coupled to photons. 
(This is in fact not a real limitation as in practice forward jets are required entailing
de facto a minimum $Q^2$ for the exchanged particle in the $t$-channel 
and such a cut could also be effectively included in the structure function approach). 
For the sake of simplicity, we also assume that no $ZZZ'$ vertex exists 
in our model and that in the $WZW'$ and $WWZ'$ vertices the coupling constants
are equal to those appearing in the respective $WWZ$ SM vertex:
\begin{equation}
  g_{WWZ'} = g_{WZW'} = g_{WWZ} = g_w \cos \theta_w .
\end{equation}
The corresponding results are shown in~\cf{fig:vector-xsect}.
\begin{figure}[h!]
	\begin{center}
\includegraphics[width=0.45\textwidth]{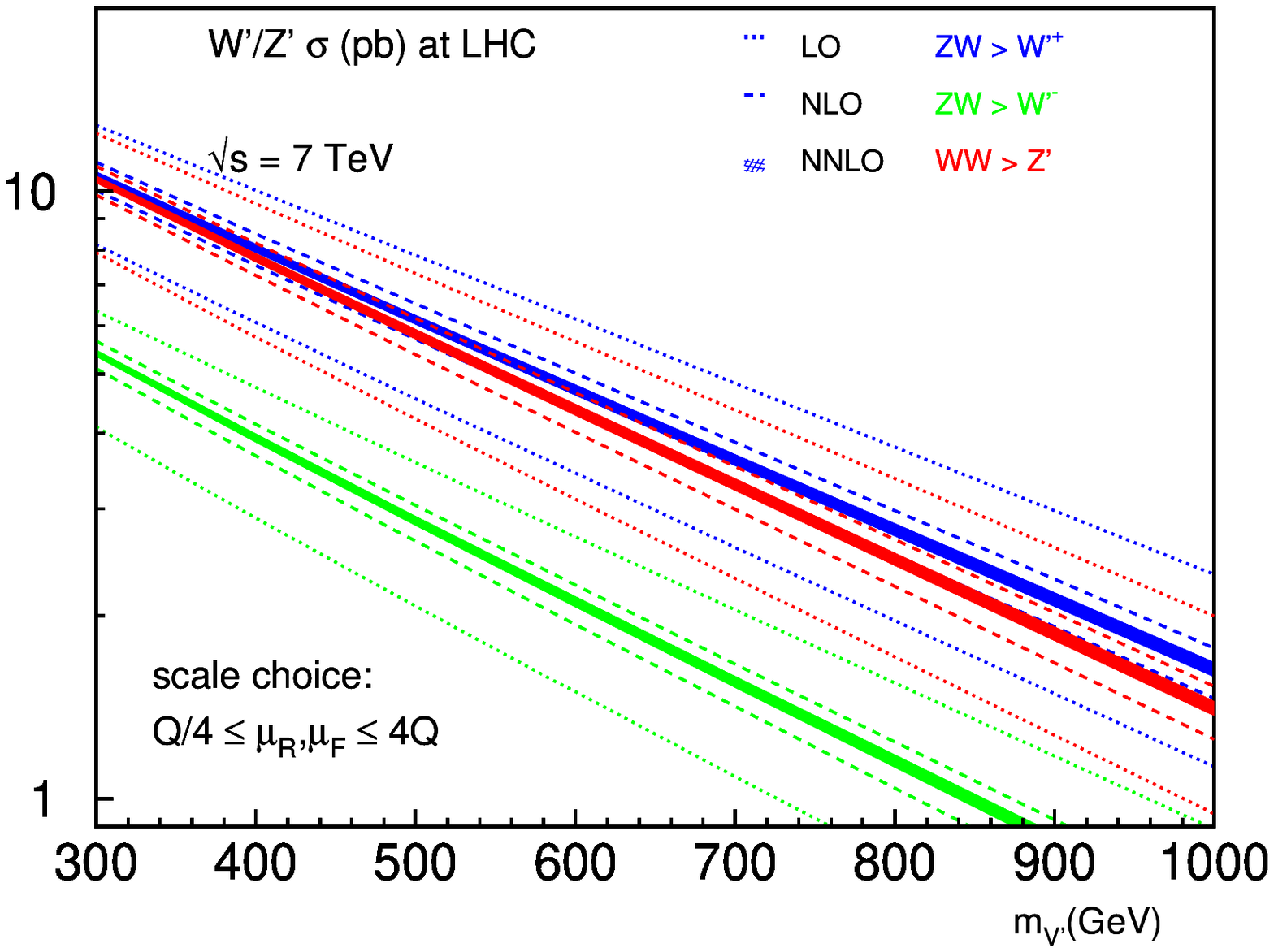}
\hspace{0.5cm}
\includegraphics[width=0.45\textwidth]{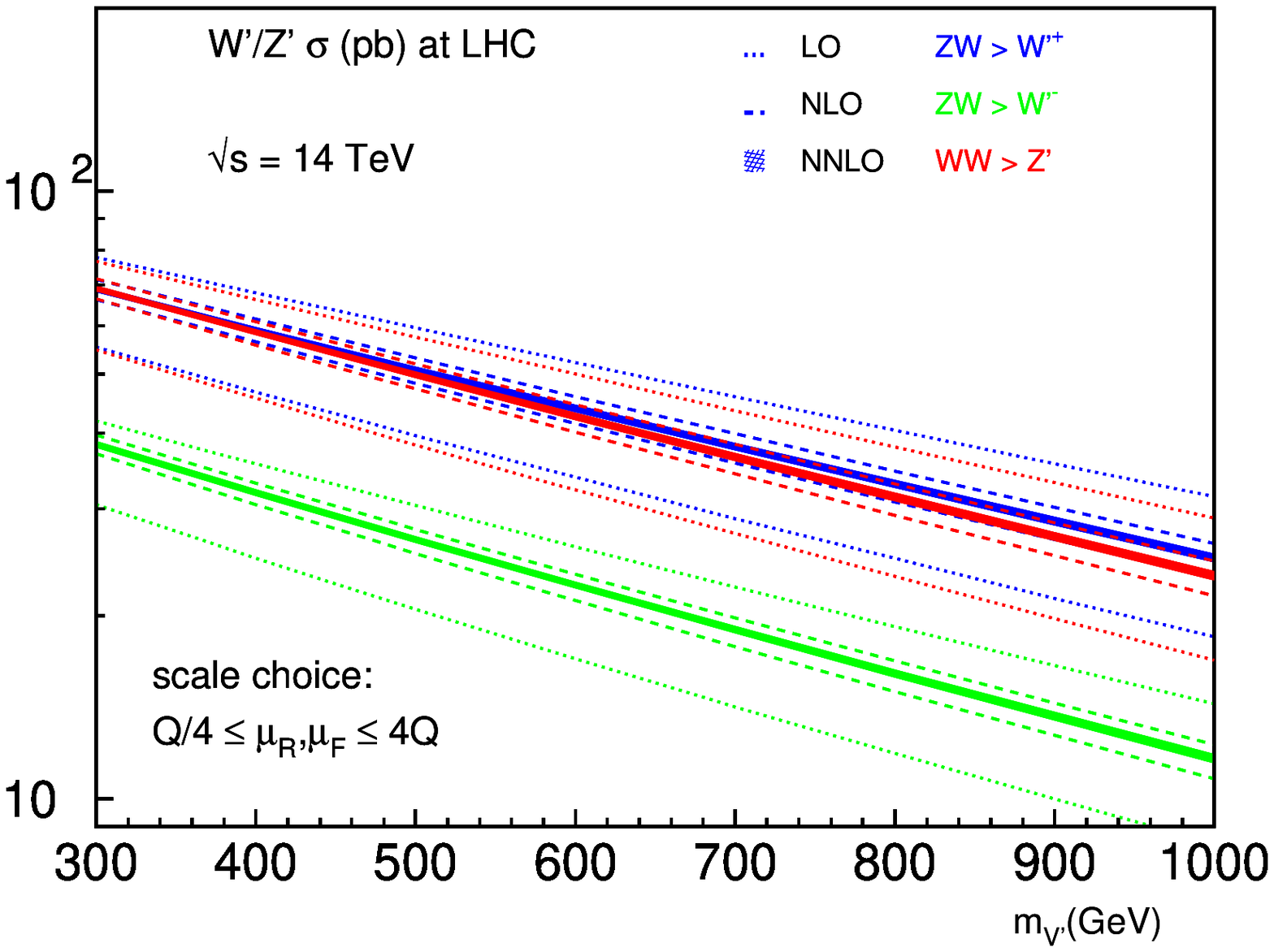}
\caption{\small
\label{fig:vector-xsect}
The total cross-section for a vector resonance ($Z'$ or $W'$)
via VBF at the LHC, $\sqrt S = 7\tev$ (top) and $\sqrt S = 14\tev$ (bottom). 
The trilinear coupling between vector bosons is assumed to be equal to the one for $ZWW$ in the SM. 
The MSTW 2008~\cite{Martin:2009iq} PDF set has been used. 
The uncertainty bands are obtained from the variation of
the renormalization and factorization scale in the interval $Q/4<\mu_f,\mu_r<4Q$, where $Q$ is the virtuality of the vector boson.}
\end{center}
\end{figure}
%

%%
%% ---------------------------------------------------------------------------
%%
\newpage
\section{Conclusions}
\label{sec:Conclusions}
The computation of NNLO in QCD predictions for production cross sections at
the LHC is in general a formidable, but nevertheless strongly motivated task. 
To date only a handful of very simple processes (with $2\to 1$ kinematics at the Born level) are known at this order: 
Drell-Yan, Higgs production in gluon-gluon fusion, and $V^* \to VH$. 
In this work we have argued that
VBF processes provide a notable exception, being the only class of processes beyond one-body final states at the Born level
(actually starting from a genuine three-body one) whose total cross section can be calculated at NNLO through the use of currently 
available techniques and results. We have shown that the key reason
for this unique possibility is that the structure function approach and the factorization approximation on which it is based, 
work extremely well up to order $\alpha_s^2$, corrections being kinematically and parametrically (by $\alpha_s^2/N_c^2$) 
suppressed and de facto negligible. Considering also contributions coming from  processes involving virtual heavy quarks, 
neglected in the past,  we find that  the residual theoretical uncertainty  from higher order QCD corrections is at the $2$\% 
level over a wide range of Higgs boson masses. With the PDF uncertainties
being also of the same order, one concludes that cross sections 
for this class of processes are among the most precise rate predictions available in LHC phenomenology. 
In this respect, our results strongly encourage the efforts towards the calculation of the differential NNLO rates in DIS, 
which is  now in sight~\cite{Daleo:2009yj}. 
A fully exclusive DIS computation at NNLO would put VBF on par with gluon-gluon fusion, and $V^* \to VH$ 
(see e.g., \cite{Ferrera:2011bk} for recent progress).
It would allow to mimic experimental selection cuts on the forward jets 
and to estimate non-trivial effects such as central jet-veto efficiencies, for the first time at NNLO.

Another very pleasant feature of the structure function approach is its
universality: any weak boson fusion process to an arbitrary
$n$-body final state $X_n$ (of particles not or very weakly interacting with the
quark lines), {\it i.e.}, $V^* V^* \to X_n$, can be easily computed at the NNLO. 
As first simple applications we have considered fusion of $V^* V^*$ to one-particle BSM final
states:  neutral scalar production with the full set of anomalous
vector-vector-scalar couplings, single-charged and double-charged scalars,
neutral and charged heavy vector resonances. Other models and scattering
processes, such as two-body (scalar-scalar, scalar-vector, vector-vector, and
fermion-anti-fermion) final states can be easily implemented. In fact, thanks
to its modularity, the structure function approach to VBF cross sections via
{\sc VBF@NNLO} could be automatized using tools like
{\sc FeynRules}~\cite{Christensen:2008py} and \madgraph\ 5~\cite{Alwall:2011uj} for
matrix element evaluation in arbitrary new physics models. NNLO results  could
then be used to provide the best normalization to fully exclusive NLO-accurate
event samples automatically generated via, for example,
{\sc aMC@NLO}~\cite{Ossola:2007ax,Frederix:2009yq,Hirschi:2011pa}. 
Work in this direction is in progress.
In the meantime, all results presented in this article can also be easily obtained through 
the public use of our {\sc VBF@NNLO} code~\cite{vbfnnlo:2010}.

%%
%% ---------------------------------------------------------------------
%%
\subsection*{Acknowledgments}
We thank R.~Frederix, S.~Frixione and V.~Hirschi for their help 
in using \madfks~\cite{Frederix:2009yq} and \madloop~\cite{Hirschi:2011pa}, 
and A.~Akeroyd, G.~Bozzi, C.~Grojean, B.~Kniehl and A.~Vicini for discussions. 
The Feynman diagrams in this article have been prepared using 
{\sc Jaxodraw}~\cite{Binosi:2003yf} and {\sc Axodraw}~\cite{Vermaseren:1994je}.
This work is partially supported by the 
Belgian Federal Office for Scientific, Technical and Cultural Affairs through Interuniversity Attraction Pole No. P6/11, 
by the Deutsche Forschungsgemeinschaft in Sonderforschungs\-be\-reich/Transregio~9 
and by the European Commission through contract PITN-GA-2010-264564 ({\it LHCPhenoNet}). 

\bigskip

\appendix
%%
%% ---------------------------------------------------------------------
%%
\renewcommand{\theequation}{\ref{sec:appA}.\arabic{equation}}
\setcounter{equation}{0}
\section{Useful formulae}
\label{sec:appA}

\subsection{The VBF phase space}

\label{sec:PhaseSpace}
In this appendix we briefly document the parameterization for the phase space of the VBF process. 
We will first take the most general case, which is the production of an $n$-particles final state via VBF with momenta $K_1$, \ldots , $K_n$, then we will specialize to the case of one particle.\\
In the structure function approach we can consider the proton remnants as massive particles, and integrate over their masses, which we label $s_1$, $s_2$.\\
We will call $P_1$, $P_2$ the momenta of the incoming protons, and $P_{X_1}$, $P_{X_2}$ the momenta of the proton remnants.
The choice of kinematical variables is guided by the requirement to resemble the DIS kinematics as much as possible.
We work in the hadronic center-of-mass reference frame, with
\begin{equation}
P_1=\frac{1}{2}\left(\sqrt{S},0,0,\sqrt{S}\right)
\, , \qquad
P_2=\frac{1}{2}\left(\sqrt{S},0,0,-\sqrt{S}\right) 
\, .
\end{equation}
Then, the Lorentz invariant phase space for this process is (cf. \ce{eq:disapproach}),
\begin{eqnarray}
\label{eq:PS}
dPS \,=\, \prod_{i=1,2}ds_i\; \frac{d^4 P_{X_i}}{(2\pi)^4}\; 2\pi \delta\left(P_{X_i}^2-s_i\right) \;
dPS_n\left(K_1, \ldots, K_n\right)\;
(2\pi)^4 \delta^4\left(P_1+P_2-P_{X_1}-P_{X_2}-\sum_{j=1,n} K_j\right)
\, . 
\nonumber \\
\end{eqnarray}
In \ce{eq:PS} we have separated the phase space of the proton remnants and the one of the particles produced via VBF. 
In order to solve the kinematics, let $q_i$ be the momentum exchanged by each of the two protons, with the convention, 
as for DIS, that the direction of $q_i$ is incoming with respect to the proton vertex 
(see \cf{fig:vbf}),
\begin{equation}
q_i=P_{X_i}-P_i
\, .
\end{equation}
We parameterize $q_i$ in terms of the two light-like momenta $P_i$ and the
transverse components ${\bf q_i^{\bot}}$:
\begin{equation}
  q_1 = \frac{2 P_1 \cdot q_1}{2 P_1\cdot P_2} P_2 + \frac{2 P_2 \cdot q_1}{2 P_1\cdot P_2} P_1 +{\bf q_1^{\bot}}
\, ,
\end{equation}
\begin{equation}
  q_2 = \frac{2 P_2 \cdot q_2}{2 P_1\cdot P_2} P_1 + \frac{2 P_1 \cdot q_2}{2 P_1\cdot P_2} P_2 +{\bf q_2^{\bot}}
\, .
\end{equation}
The Bjorken scaling variables $x_i$ for the DIS process are given by the scalar products $P_i\cdot q_i$ 
\begin{equation}
2 P_i\cdot q_i = \frac{-q_i ^2}{x_i}= \frac{Q_i^2}{x_i}
\, ,
\end{equation}
where $Q_i^2=-q_i^2$. In analogy one can also define variables $y_i$ via the relation
\begin{equation}
2 P_2\cdot q_1= \frac{Q_1^2}{y_1}
\, ,\qquad 
2 P_1\cdot q_2= \frac{Q_2^2}{y_2}
\, ,
\end{equation}
so that 
\begin{equation}
  y_i=-\frac{Q_i^4}{x_iS(Q_i^2- {q_i^{\bot}}^2)}
\, , 
\end{equation}
with $ q_i^{\bot}=\sqrt{\left| {\bf q_i^{\bot}}^2\right|}$.\\
Then, the integration measure can be expressed as
\begin{equation}
d^4 q_i = dQ_i^2\;d^2 {\bf q_i ^{\bot}}\; \frac{dx_i}{2x_i}
\, ,
\end{equation}
and
\begin{equation}
P_{X_i}^2 = Q_i^2\left(\frac{1}{x_i}-1\right)
\, .
\end{equation}
The phase space \ce{eq:PS} reduces now to the following form,
\begin{equation}
\label{eq:PS-v2}
dPS \,=\, 
\frac{1}{(2\pi)^2}\prod_{i=1,2} dQ^2_i\;d^2 {\bf q_i ^{\bot}}\; \frac{dx_i}{2x_i}\;
dPS_n\left(K_1, \ldots, K_n\right)\;
(2\pi)^4 \delta^4\left(q_1 + q_2 + \sum _{j=1,n} K_j\right)
\, ,
\end{equation}
and the integrations on the transverse components can be cast in polar coordinates,
\begin{equation}
d^2 {\bf q_1 ^{\bot}}\; d^2 {\bf q_2 ^{\bot}} \,=\,
{ q_1 ^{\bot}} \;d { q_1 ^{\bot}} d\varphi_1 \;{ q_2 ^{\bot}}\; d{ q_2 ^{\bot}} d\varphi_2
=2\pi\;{ q_1 ^{\bot}} \;d { q_1 ^{\bot}} \;{ q_2 ^{\bot}}\; d{ q_2^{\bot}} d\varphi_{12}
\, .
\end{equation}
The momentum-conservation condition imposed by the Dirac delta-function in \ce{eq:PS-v2} allows to write a relation between the proton remnants variables and the total energy of the VBF products:
\begin{eqnarray}
  S_{VBF} &=& \left( K_1 + \ldots + K_n \right)^2 = \left(q_1 + q_2\right)^2 = \nonumber\\
  &=&
  \frac{1}{Q_1^2 Q_2^2} Sx_1 x_2 \left(Q_1^2 -{ q_1^{\bot}}^2\right) 
\left(Q_2^2 -{ q_2^{\bot}}^2\right)+ \frac{Q_1^2 Q_2^2}{S x_1 x_2} -Q_1^2 -Q_2^2 -2 { q_1^{\bot}}{ q_2^{\bot}} \cos\varphi_{12} 
\, .
\label{eq:svbf}
\end{eqnarray}
Using this equation one can generate the phase space of the VBF-produced
particles once the proton remnants variables are fixed, 
simply in the same way as for the decay of a particle.\\

We will finally go through the details of the case in which only one particle with mass $m$ and momentum $K$ is produced via VBF. In this case we simply have
\begin{equation}
  \label{eq:PS1}
  dPS_1\left(K\right) = \frac{d^4 K }{\left(2\pi\right)^4} \; 2\pi \delta\left(K^2- m^2\right)\, ,
\end{equation}
and the $d^4 K$ integration can be eliminated using the momentum-conservation Dirac delta-function, so that we just need to solve the mass-shell condition imposed by the delta-function in \ce{eq:PS1} (which is nothing but \ce{eq:svbf} with $S_{VBF} = m^2$):\\
\begin{equation}
  m^2 = \frac{1}{Q_1^2 Q_2^2} Sx_1 x_2 \left(Q_1^2 -{ q_1^{\bot}}^2\right) 
\left(Q_2^2 -{ q_2^{\bot}}^2\right)+ \frac{Q_1^2 Q_2^2}{S x_1 x_2} -Q_1^2 -Q_2^2 -2 { q_1^{\bot}}{ q_2^{\bot}} \cos\varphi_{12} 
\, .
\label{eq:MH}
\end{equation}
This equation can, of course, be easily solved for $\cos\varphi_{12}$, but in this case it would lead to an (integrable) singularity in the Jacobian, and therefore to numerical problems.
Therefore, our choice has been to write \ce{eq:MH} as a second degree equation for $q_2^{\bot}$:
\begin{equation}
A{q_2^{\bot} }^2+ B { q_2^{\bot}} +C \,=\, 0 
\, , \label{eq:qt2}
\end{equation}
with
\begin{eqnarray}
A&=&\frac{ S x_1 x_2 \left({ q_1^{\bot}}^2-Q_1^2\right) }{Q_1^2 Q_2^2} 
\, ,\\
B&=&-2 q_1^{\bot} \cos\varphi_{12}
\, ,\\
C&=&\frac{Q_1^2 Q_2^2}{ S x_1 x_2 } -\frac{S x_1 x_2 q_1^{\bot}}{Q_1^2}+S
x_1 x_2 -m^2-Q_1^2-Q_2^2
\, .
\end{eqnarray}
Since two solutions of $ q_2^{\bot}$ exist for each set of parameters $x_1$, $x_2$, $Q_1^2$, $Q_2^2$, $ q_1^{\bot}$, $\varphi$, 
two phase space points are evaluated. 
We require ``physical'' phase space points to have positive $ q_2^{\bot}$ 
and the number of rejected points can be greatly reduced if one asks
\begin{equation}
0<Q_2^2<S x_1 x_2 
\frac{\left(Q_1^2 -{ q_1^{\bot}}^2 \right)  \left[ Q_1^4 -Q_1^2 \left(S x_1 x_2-m^2\right) +{ q_1^{\bot}}^2  S x_1 x_2 \right]}
{Q_1^2 \left[ Q_1^4 -Q_1^2\left(  {  q_1^{\bot}}^2 \sin^2 \varphi_{12} +S x_1 x_2\right)+ { q_1^{\bot}}^2 S x_1 x_2 \right] }
\, ,
\label{eq:intq2}
\end{equation}
which corresponds to the positivity of the discriminant of~\ce{eq:qt2}. 
The other parameters are generated in the following ranges:
\begin{eqnarray}
&&
	x_1 \cdot x_2 \,\in\, \left[\frac{m^2}{S},1\right] \, , \qquad
	\log x_1 \,\in\, \left[\log \frac{m^2}{S},0\right]\, , \qquad
	Q_1^2 \,\in\, \left[Q_0^2,\left(\sqrt S - m \right)^2\right]\, , \qquad
\nonumber\\
&&
	 q_1^{\bot} \,\in\, \left[0, Q_1\right]\, , \qquad
	\varphi \,\in\, \left[0,2\pi\right]\, ,
	\label{eq:intpar}
\end{eqnarray}
where $Q_0$ is a technical cut set to $1\gev$ which prevents sampling PDFs and $\alpha_s$ at too low scales. 
It also sets the lower bound for $Q_2$ and we have checked the independence of the cross section results on value of $Q_0$.
The differential cross-section from~\ce{eq:disapproach}, multiplied by the Jacobian corresponding 
to~\ced{eq:intq2}{eq:intpar}, is integrated using {\sc VEGAS}~\cite{Lepage:1977sw}.

\subsection{The VBF cross-section with anomalous $VVH$ couplings}
\label{sec:WMMW-anom}
We report here the formula corresponding to \ce{eq:WMMW-SM} for the vertex in \ce{eq:VVH-anom}. We rewrite it as:
\begin{equation}
	\Gamma^{\mu\nu}\left(q_1,q_2\right)=A_1 g_{\mu\nu} +
	A_2 \left(q_1\cdot q_2g^{\mu\nu}-q_2^{\mu}q_1^{\nu}\right) + A_3 \eps^{\mu\nu\rho\sigma} q_{1\rho}q_{2\sigma},
\end{equation}
where we have set
\begin{eqnarray}
	A_1&=&2 \vmass^2\left( \left(\sqrt 2 G_F\right)^{1/2} + \frac{a_1}{\Lambda}\right),\nonumber\\ 
	A_2&=& \frac{a_2}{\Lambda}, \nonumber\\
	A_3&=& \frac{a_3}{\Lambda}.
\end{eqnarray}
\ce{eq:WMMW-SM}, with the replacement $\mathcal M^{\mu\nu}=\Gamma^{\mu\nu}\left(q_1,q_2\right)$ has now the form:
\begin{eqnarray}
W_{\mu\nu} \left(x_1, Q^2_1\right)\mathcal M^{\mu\rho} \mathcal {M^*}^{\nu\sigma} W_{\rho\sigma}\left(x_2,Q^2_2\right)=
\sum_{i,j=1}^3 C_{ij}F_i \left(x_1, Q^2_1\right) F_j \left(x_2,Q^2_2\right). \label{eq:WMMW-cij}
\end{eqnarray}
Defining
\begin{equation}
  q_1\cdot q_2 = q_{12}, \qquad P_a \cdot q_b = P_{ab},
\end{equation}
the non-vanishing $C_{ij}$ read
%{
%\input{eq_anomalous}
%}
\begin{eqnarray}
	C_{11}&=& \;\left(2+\frac{ {\; {q_{12}}} ^2}{\; {q_1}^2 \; {q_2}^2}\right)\left|A_1 \right|^2
+ 6 \; {q_{12}}\re{A_1 A_2^*} 
+\;\left(\; {q_1}^2 \; {q_2}^2+ 2 {\; {q_{12}}}^2\right)\left|A_2\right|^2\nonumber\\
&&+2 \;\left({\; {q_{12}}}^2-\; {q_1}^2 \; {q_2}^2\right) \left|A_3 \right|^2,
	\\
	C_{12}&=&\frac{1}{\; {P_{22}}}
 \;\left[  \frac{ {\; {P_{22}}}^2 }{\; {q_2}^2} +\frac{1}{\; \; {q_1}^2}\;\left(\; {P_{21}}-\frac{\; {P_{22}}}{\; {q_2}^2} \; {q_{12}}\right)^2\right] \left|A_1 \right|^2
 +2 \frac{\; {P_{22}} \; {q_{12}}}{\; {q_2}^2}\re{A_1 A_2^*}\nonumber\\
	&& +\; {P_{21}} \;\left(2  \; {q_{12}} -\frac{ {\; {P_{21}}} \; {q_2}^2}{\; {P_{22}}}\right)\left|A_2 \right|^2
+\;\left(2 \; {P_{21}} \; {q_{12}} -\frac{ {\; {P_{21}}}^2 \; {q_2}^2}{\; {P_{22}}} -\; {P_{22}} \; \; {q_1}^2\right)\left|A_3 \right|^2,
	\\
	C_{21}&=&\frac{1}{\; {P_{11}}}
 \;\left[  \frac{ {\; {P_{11}}}^2 }{\; \; {q_1}^2} +\frac{1}{\; {q_2}^2}\;\left(\; {P_{12}}-\frac{\; {P_{11}}}{\; \; {q_1}^2} \; {q_{12}}\right)^2\right] \left|A_1 \right|^2
 +2 \frac{\; {P_{11}} \; {q_{12}}}{\; \; {q_1}^2}\re{A_1 A_2^*}\nonumber\\
	&& + \;\left(2 \; {P_{12}} \; {q_{12}} -\frac{ {\; {P_{12}}}^2 \; \; {q_1}^2}{\; {P_{11}}}\right)\left|A_2 \right|^2
+\;\left(2 \; {P_{12}} \; {q_{12}} -\frac{ {\; {P_{12}}}^2 \; \; {q_1}^2}{\; {P_{11}}} -\; {P_{11}} \; {q_2}^2\right)\left|A_3 \right|^2,
	\\
	C_{22}&=&\frac{1}{\; {P_{11}} \; {P_{22}}} 
 \;\left(\; \frac{s}{2} - \frac{\; {P_{11}} \; {P_{21}}}{\; \; {q_1}^2} - \frac{\; {P_{12}} \; {P_{22}}}{\; {q_2}^2} +\frac{\; {P_{11}} \; {P_{22}} \; {q_{12}}}{\; {q_1}^2 \; {q_2}^2} \right)^2 \left|A_1 \right|^2  \nonumber\\
 	&&- 2\;\left(\; {P_{12}} \; {P_{21}}- \; {q_{12}} \; \frac{s}{2}\right)
	\;\left( \frac{\; {q_{12}}}{\; {q_1}^2 \; {q_2}^2} +\frac{\; s}{2\; {P_{11}} \; {P_{22}}}- \frac{\; {P_{21}}}{\; {P_{22}} \; \; {q_1}^2} - \frac{\; {P_{12}}}{\; {P_{11}} \; {q_2}^2} \right)\re{A_1 A_2^*}\nonumber\\
	&&+2 \;\left( \frac{\; {q_{12}}}{\; {q_1}^2 \; {q_2}^2} +\frac{\; s }{2\; {P_{11}} \; {P_{22}}}- \frac{\; {P_{21}}}{\; {P_{22}} \; \; {q_1}^2} - \frac{\; {P_{12}}}{\; {P_{11}} \; {q_2}^2} \right)\; \eps_{\mu\nu\rho\sigma} P_1^{\mu}P_2^{\nu}\; \; {q_1}^{\rho}\; {q_2}^{\sigma}\;\re{A_1 A_3^*}\nonumber\\
	&&+\frac{\;\left(\; {q_{12}} \; s - 2\; {P_{12}} \; {P_{21}}\right)^2}{4\; {P_{11}} \; {P_{22}}}\left|A_2 \right|^2
	-\frac{\; {P_{12}} \; {P_{21}}-\; {q_{12}} \; s}{\; {P_{11}} \; {P_{22}}}\; \eps_{\mu\nu\rho\sigma} P_1^{\mu}P_2^{\nu}\; \; {q_1}^{\rho}\; {q_2}^{\sigma}\;\re{A_2 A_3^*}\nonumber\\
	&&+\;\left\{s \;\left(\; {q_{12}} - \frac{\; {P_{21}} \; {q_2}^2}{\; {P_{22}}} - \frac{\; {P_{12}} \; \; {q_1}^2}{\; {P_{11}}}
+\frac{\; {P_{12}} \; {P_{21}} \; {q_{12}}}{\; {P_{11}} \; {P_{22}}}\right)
\right.\nonumber\\
	&&\left.
+\frac{1}{\; {P_{11}} \; {P_{22}}}\;\left[\; \frac{s^2}{4} \;\left( \; {q_1}^2 \; {q_2}^2- \; {q_{12}}^{2}\right)- 
\; {P_{12}}^2 \; {P_{21}}^2
\right]-\; {P_{11}} \; {P_{22}} + 2 \; {P_{12}} \; {P_{21}}
	\right\} \left|A_3 \right|^2,
	\\
	C_{33}&=&\frac{\; {q_{12}} \; s - 2\; {P_{12}} \; {P_{21}}}{2\; {P_{11}} \; {P_{22}}} \left|A_1 \right|^2
        +\frac{1}{2}\;\left[\frac{\; s}{2\; {P_{11}} \; {P_{22}}} \;\left( \; {q_1}^2 \; {q_2}^2+ {\; {q_{12}}}^2\right)
+\; {q_{12}} -\frac{\; {P_{12}} \; \; {q_1}^2}{\; {P_{11}}} -\frac{\; {P_{21}} \; {q_2}^2}{\; {P_{22}}} \right.
\nonumber\\
	&&\left.-
\frac{\; {P_{12}} \; {P_{21}} \; {q_{12}}}{\; {P_{11}} \; {P_{22}}}\right]\re{A_1 A_2^*}
+\frac{\; {q_{12}}}{2\; {P_{11}} \; {P_{22}}}\; \eps_{\mu\nu\rho\sigma} P_1^{\mu}P_2^{\nu}\; \; {q_1}^{\rho}\; {q_2}^{\sigma}\;\re{A_1 A_3^*}\nonumber\\
	&&+
\frac{\; {q_{12}}}{2}\;\left(\; {q_{12}} -\frac{\; {P_{12}} \; \; {q_1}^2}{\; {P_{11}}} -\frac{\; {P_{21}} \; {q_2}^2}{\; {P_{22}}}
+\frac{\;s \; {q_1}^{2} \; {q_2}^{2} \;}{2\; {P_{11}} \; {P_{22}}}\right)\left|A_2 \right|^2
\nonumber\\
&&+
\frac{\; {q_1}^2 \; {q_2}^2}{2\; {P_{11}} \; {P_{22}}}\; \eps_{\mu\nu\rho\sigma} P_1^{\mu}P_2^{\nu}\; \; {q_1}^{\rho}\; {q_2}^{\sigma}\;\re{A_2 A_3^*}\nonumber\\
	&&
+\frac{\; {q_{12}}}{2}\;\left(\; {q_{12}} -\frac{\; {P_{12}} \; \; {q_1}^2}{\; {P_{11}}} -\frac{\; {P_{21}} \; {q_2}^2}{\; {P_{22}}}
+\frac{\; {P_{12}} \; {P_{21}} \; {q_1}^2 \; {q_2}^2}{\; {P_{11}} \; {P_{22}} \; {q_{12}}}\right)
\left|A_3 \right|^2
	.
\end{eqnarray}

\subsection{The VBF cross-section for a vector resonance}
\label{sec:WMMW-vector}
Finally, we give the formula for the total cross-section for the production of a vector resonance in the structure function approach.\\
We assume the usual gauge and Lorentz invariant tri-linear vertex:
\begin{equation}
	\Gamma^{\mu_1 \mu_2 \mu_3}\left(p_1, p_2, p_3\right) = g_{123} \left[ g^{\mu_1 \mu_2}\left(p_1^{\mu_3}-p_2^{\mu_3}\right) + 
	g^{\mu_2 \mu_3}\left(p_2^{\mu_1}-p_3^{\mu_1}\right) +
	g^{\mu_3 \mu_1}\left(p_3^{\mu_2}-p_1^{\mu_2}\right) \right],
\end{equation}
where all the momenta are taken to flow outside the vertex. Since in the
processes that we have considered the coupling $g_{123}$ 
can always be factorized from the total cross-section, we will set it to unity.\\
As in \csec{sec:WMMW-anom}, with the extra care to sum over the polarizations of the produced vector, we have
\begin{eqnarray}
W_{\mu\nu} \left(x_1, Q^2_1\right) \sum_\lambda \left(\mathcal M^{\mu\rho}_{\lambda} \mathcal {M^*}^{\nu\sigma}_{\lambda}\right) W_{\rho\sigma}\left(x_2,Q^2_2\right)=
\sum_{i,j=1}^3 C_{ij}F_i \left(x_1, Q^2_1\right) F_j \left(x_2,Q^2_2\right),
\end{eqnarray}
where we have set
\begin{equation}
	\mathcal M^{\mu\nu}_{\lambda} =  \Gamma^{\mu \nu}_ {\rho}\left(q_1, q_2, -q_1-q_2\right) \epsilon_{\lambda}^{\rho}\left(-q_1-q_2\right),
\end{equation}
and the non-vanishing coefficients $C_{ij}$ are
%{
%\input{eq_vector}
%}
\begin{eqnarray}
	C_{11} &=&\frac{1}{\; {q_1}^2 \; {q_2}^2} \;\left[ 4 \; {q_1}^2 \; {q_2}^2  \; {q_{12}} -14 {\;  {q_1}^4} \; {q_2}^2 -14 \; \; {q_1}^2 {\; {q_2}^4}  +
   11 \; \; {q_1}^2 {\; {q_{12}}} ^2+ 11 q_2 ^2 {\; {q_{12}}} ^2+ 2 {\; {q_{12}}} ^3  \phantom{\frac{1}{m_{V'}^2}} \right. \nonumber\\
  && + \left. \frac{1}{m_{V'}^2} \;\left(\; \; {q_1}^2- \; {q_2}^2\right)^2 \;\left(2 \; {q_1}^2 \; {q_2}^2 + {\; {q_{12}}}^2\right)\right] , \\
 %%%%%%%%%%%%%%%
  C_{12} &=& \frac{-1}{\; {P_{22}} \; \; {q_1}^2 {\; {q_2}^4} } \;\left[ 
  -11 {\; {P_{21}}}^2 \; \; {q_1}^2 {\; {q_2}^4}-2 {\; {P_{21}}}^2 \; {q_{12}} {\; {q_2}^4}+{P_{21}}^2 {\; {q_2}^6}+22 \; {P_{21}} \; {P_{22}} \; \; {q_1}^2 \; {q_{12}} \; {q_2}^2  \phantom{\frac{1}{m_{V'}^2}} \right.\nonumber \\
  &&  +4 \; {P_{21}} \; {P_{22}} {\; {q_{12}}}^2 \; {q_2}^2-2 \; {P_{21}} \; {P_{22}} \; {q_{12}} {\; {q_2}^4}+{P_{22}}^2 {\; \; {q_1}^4} \; {q_2}^2-11 {P_{22}}^2 \; \; {q_1}^2 {\; {q_{12}}}^2\nonumber \\
  &&-2 {\; {P_{22}}}^2 \; \; {q_1}^2 \; {q_{12}} \; {q_2}^2+5 {P_{22}}^2 \; \; {q_1}^2 {\; {q_2}^4}-2 {P_{22}}^2 {\; {q_{12}}}^3-3 {\; {P_{22}}}^2 \; {q_{12}}^2 \; {q_2}^2\nonumber \\
  &&\left.-\frac{1}{m_{V'}^2} \;\left(\; \; {q_1}^2- \; {q_2}^2\right)^2
  \;\left({P_2q_1}^2 {\; {q_2}^2}^2-2 \; {P_{21}} \; {P_{22}} \; {q_{12}} \; {q_2}^2+{\; {P_{22}}}^2 \; {q_1}^2 \; {q_2}^2+{\; {P_{22}}}^2 \; {q_{12}}^2\right) 
  \right] , 
\nonumber\\
&& \\
 %%%%%%%%%%%%%%%
	C_{21} &=& C_{12} \;\left(1 \leftrightarrow 2\right)  , \\
 %%%%%%%%%%%%%%%
 C_{22} &=&  \frac{1}{\; {P_{11}} \; {P_{22}} {\; \; {q_1}^4} {\; {q_2}^4} } \;\left\{
 3 {P_{11}}^2 {\; {P_{21}}}^2 \; \; {q_1}^2 {\; {q_2}^4}+2 {\; {P_{11}}}^2 {\; {P_{21}}}^2 \; {q_{12}} {\; {q_2}^4}-{\; {P_{11}}}^2 {\; {P_{21}}}^2 {\; {q_2}^6} 
 %\phantom{\frac{s^2}{4}}
\right.\nonumber \\
 && -6 {\; {P_{11}}}^2 \; {P_{21}} \; {P_{22}} \; \; {q_1}^2 \; {q_{12}} \; {q_2}^2-4 {\; {P_{11}}}^2 \; {P_{21}} \; {P_{22}} {\; {q_{12}}}^2 \; {q_2}^2+2 {\; {P_{11}}}^2 \; {P_{21}} \; {P_{22}} \; {q_{12}} {\; {q_2}^4}\nonumber \\
 && +3 {\; {P_{11}}}^2 {\; {P_{22}}}^2 \; \; {q_1}^2 {\; {q_{12}}}^2+2 {\; {P_{11}}}^2 {\; {P_{22}}}^2 {\; {q_{12}}}^3+3 {\; {P_{11}}}^2 {\; {P_{22}}}^2 {\; {q_{12}}}^2 \; {q_2}^2\nonumber \\
 && -2 \; {P_{11}} \; {P_{12}} \; {P_{21}} \; {P_{22}} {\; \; {q_1}^4} \; {q_2}^2+4 \; {P_{11}} \; {P_{12}} \; {P_{21}} \; {P_{22}} \; \; {q_1}^2 \; {q_{12}} \; {q_2}^2-2 \; {P_{11}} \; {P_{12}} \; {P_{21}} \; {P_{22}} \; \; {q_1}^2 {\; {q_2}^4}\nonumber \\
 && +2 \; {P_{11}} \; {P_{12}} {\; {P_{22}}}^2 {\; \; {q_1}^4} \; {q_{12}}-4 \; {P_{11}} \; {P_{12}} {\; {P_{22}}}^2 \; \; {q_1}^2 {\; {q_{12}}}^2-6 \; {P_{11}} \; {P_{12}} {\; {P_{22}}}^2 \; \; {q_1}^2 \; {q_{12}} \; {q_2}^2\nonumber \\
 && -{\; {P_{12}}}^2 {\; {P_{22}}}^2 {\; \; {q_1}^6}+2 {\; {P_{12}}}^2 {\; {P_{22}}}^2 {\; \; {q_1}^4} \; {q_{12}}+3 {\; {P_{12}}}^2 {\; {P_{22}}}^2 {\; \; {q_1}^4} \; {q_2}^2 \nonumber \\
 && +s\;    
\; {q_1}^2 \; {q_2}^2 \;\left(q_1 - q_2\right)^2 \;\left(\; {P_{11}} \; {P_{21}} \; {q_2}^2 - \; {P_{11}} \; {P_{22}} \; {q_{12}} + \; {P_{12}} \; {P_{22}} \; \; {q_1}^2 \right)
-\frac{s^2}{4}{\; \; {q_1}^4} {\; {q_2}^4} \;\left(q_1 -  q_2\right)^2 \nonumber \\
&&+ \frac{1}{4\; m_{V'}^2}\;\left(\; {q_1}^2- \; {q_2}^2\right)^2
 \;\left(2 \; {P_{11}} \; {P_{21}} \; {q_2}^2-2 \; {P_{11}} \; {P_{22}} \; {q_{12}}+ 2 \; {P_{12}} \; {P_{22}} \; {q_1}^2 \left. -s \; {q_1}^2 \; {q_2}^2 \right)^2 
\right\}
 , 
\nonumber\\
&& \\
 %%%%%%%%%%%%%%%
	C_{33} &=& \frac{1}{ 2 \; {P_{11}} \; {P_{22}}}  \;\left[
-8 \; {P_{11}} \; {P_{21}} \; {q_2}^2+8 \; {P_{11}} \; {P_{22}} \; {q_{12}}+\; {P_{12}} \; {P_{21}} \; \; {q_1}^2+6 \; {P_{12}} \; {P_{21}} \; {q_{12}}
+\phantom{\frac{1}{m_{V'}^2}}\right.\nonumber \\
&& \; {P_{12}} \; {P_{21}} \; {q_2}^2-8 \; {P_{12}} \; {P_{22}} \; \; {q_1}^2	
-\frac{s}{2}\;\left(
\; \; {q_1}^2 \; {q_{12}}-8 \; {q_1}^2 \; {q_2}^2+6 {\; {q_{12}}}^2+\; {q_{12}} \; {q_2}^2\right)\nonumber\\
&&\left. +\frac{1}{m_{V'}^2}\;\left(\; \; {q_1}^2- \; {q_2}^2\right)^2\;\left( - \; {P_{12}} \; {P_{21}} +\frac{s}{2} \; {q_{12}} \right) \right].
\end{eqnarray}
%

%%
%% ---------------------------------------------------------------------
%%
\renewcommand{\theequation}{\ref{sec:appB}.\arabic{table}}
\setcounter{table}{0}
\section{VBF cross sections}
\label{sec:appB}

We present the VBF cross sections for the Tevatron in \ctd{tab:table-tev-a09-Sc1}{tab:table-tev-nnn-Sc1},
and for the LHC in \ctd{tab:table-lhc7-a09-Sc1}{tab:table-lhc7-nnn-Sc1} 
and in \ctd{tab:table-lhc14-a09-Sc1}{tab:table-lhc14-nnn-Sc1}, respectively.
The PDF sets used and all other parameters are given in the table captions.

%***************** 1st table ************************ 

\begin{table}[tb!]
\begin{center}
\begin{tabular}{|c|c|c|c|}
\hline
$ \hmass \, [\gev] $ & $ \sigma_{LO} $ & $ \sigma_{NLO} $ & $ \sigma_{NNLO} $ \\
\hline
  90 &$1.206^{+0.363}_{-0.233}$$^{+0.012}_{-0.012}$$\cdot 10^{-1}$ &$1.220^{+0.103}_{-0.056}$$^{+0.012}_{-0.012}$$\cdot 10^{-1}$ &$1.241^{+0.034}_{-0.011}$$^{+0.012}_{-0.012}$$\cdot 10^{-1
}$          \\
  95 &$1.108^{+0.342}_{-0.218}$$^{+0.011}_{-0.011}$$\cdot 10^{-1}$ &$1.123^{+0.096}_{-0.053}$$^{+0.011}_{-0.011}$$\cdot 10^{-1}$ &$1.144^{+0.030}_{-0.012}$$^{+0.011}_{-0.011}$$\cdot 10^{-1
}$          \\
 100 &$1.019^{+0.322}_{-0.204}$$^{+0.010}_{-0.010}$$\cdot 10^{-1}$ &$1.035^{+0.091}_{-0.052}$$^{+0.011}_{-0.011}$$\cdot 10^{-1}$ &$1.054^{+0.029}_{-0.010}$$^{+0.011}_{-0.011}$$\cdot 10^{-1
}$          \\
 105 &$9.387^{+3.025}_{-1.906}$$^{+0.098}_{-0.098}$$\cdot 10^{-2}$ &$9.551^{+0.841}_{-0.497}$$^{+0.100}_{-0.100}$$\cdot 10^{-2}$ &$9.737^{+0.258}_{-0.113}$$^{+0.099}_{-0.099}$$\cdot 10^{-2
}$          \\
 110 &$8.652^{+2.847}_{-1.786}$$^{+0.092}_{-0.092}$$\cdot 10^{-2}$ &$8.814^{+0.788}_{-0.474}$$^{+0.093}_{-0.093}$$\cdot 10^{-2}$ &$8.990^{+0.247}_{-0.106}$$^{+0.093}_{-0.093}$$\cdot 10^{-2
}$          \\
 115 &$7.975^{+2.681}_{-1.666}$$^{+0.086}_{-0.086}$$\cdot 10^{-2}$ &$8.146^{+0.724}_{-0.456}$$^{+0.088}_{-0.088}$$\cdot 10^{-2}$ &$8.308^{+0.224}_{-0.107}$$^{+0.087}_{-0.087}$$\cdot 10^{-2
}$          \\
 120 &$7.360^{+2.519}_{-1.560}$$^{+0.081}_{-0.081}$$\cdot 10^{-2}$ &$7.533^{+0.681}_{-0.427}$$^{+0.082}_{-0.082}$$\cdot 10^{-2}$ &$7.679^{+0.214}_{-0.099}$$^{+0.082}_{-0.082}$$\cdot 10^{-2
}$          \\
\hline
\end{tabular}
\end{center}
\vspace*{-5mm}
\caption{
\small
\label{tab:table-tev-a09-Sc1}
Total VBF cross sections at the
Tevatron, $\sqrt S = 1.96 \tev$
at LO, NLO and NNLO in QCD. Errors shown are respectively scale and PDF uncertainities.  Scale uncertainities are evaluated by varying $\mu_r$ and $\mu_f$ in the interval
$\mu_r,\mu_f \in [Q/4,4Q]$.
The ABKM~\cite{Alekhin:2009ni} PDF set has been used.
Numbers are in pb.
}
\end{table}

%***************** 1st table ************************ 

\begin{table}[tb!]
\begin{center}
\begin{tabular}{|c|c|c|c|}
\hline
$ \hmass \, [\gev] $ & $ \sigma_{LO} $ & $ \sigma_{NLO} $ & $ \sigma_{NNLO} $ \\
\hline
  90 &$1.199^{+0.349}_{-0.226}$$^{+0.029}_{-0.029}$$\cdot 10^{-1}$ &$1.210^{+0.103}_{-0.053}$$^{+0.030}_{-0.030}$$\cdot 10^{-1}$ &$1.207^{+0.037}_{-0.010}$$^{+0.030}_{-0.030}$$\cdot 10^{-1
}$          \\
  95 &$1.103^{+0.329}_{-0.212}$$^{+0.026}_{-0.026}$$\cdot 10^{-1}$ &$1.113^{+0.098}_{-0.050}$$^{+0.027}_{-0.027}$$\cdot 10^{-1}$ &$1.112^{+0.033}_{-0.010}$$^{+0.027}_{-0.027}$$\cdot 10^{-1
}$          \\
 100 &$1.014^{+0.310}_{-0.198}$$^{+0.024}_{-0.024}$$\cdot 10^{-1}$ &$1.028^{+0.088}_{-0.050}$$^{+0.025}_{-0.025}$$\cdot 10^{-1}$ &$1.025^{+0.031}_{-0.009}$$^{+0.025}_{-0.025}$$\cdot 10^{-1
}$          \\
 105 &$9.332^{+2.923}_{-1.849}$$^{+0.225}_{-0.225}$$\cdot 10^{-2}$ &$9.467^{+0.838}_{-0.465}$$^{+0.234}_{-0.234}$$\cdot 10^{-2}$ &$9.458^{+0.293}_{-0.094}$$^{+0.234}_{-0.234}$$\cdot 10^{-2
}$          \\
 110 &$8.600^{+2.749}_{-1.733}$$^{+0.207}_{-0.207}$$\cdot 10^{-2}$ &$8.744^{+0.784}_{-0.442}$$^{+0.216}_{-0.216}$$\cdot 10^{-2}$ &$8.732^{+0.270}_{-0.088}$$^{+0.216}_{-0.216}$$\cdot 10^{-2
}$          \\
 115 &$7.930^{+2.593}_{-1.623}$$^{+0.191}_{-0.191}$$\cdot 10^{-2}$ &$8.076^{+0.741}_{-0.425}$$^{+0.200}_{-0.200}$$\cdot 10^{-2}$ &$8.074^{+0.247}_{-0.092}$$^{+0.200}_{-0.200}$$\cdot 10^{-2
}$          \\
 120 &$7.317^{+2.439}_{-1.519}$$^{+0.177}_{-0.177}$$\cdot 10^{-2}$ &$7.467^{+0.692}_{-0.406}$$^{+0.185}_{-0.185}$$\cdot 10^{-2}$ &$7.464^{+0.232}_{-0.087}$$^{+0.185}_{-0.185}$$\cdot 10^{-2
}$          \\
\hline
\end{tabular}
\end{center}
\vspace*{-5mm}
\caption{
\small
\label{tab:table-tev-h15-Sc1}
Total VBF cross sections at the
Tevatron, $\sqrt S = 1.96 \tev$
at LO, NLO and NNLO in QCD. Errors shown are respectively scale and PDF uncertainities.  Scale uncertainities are evaluated by varying $\mu_r$ and $\mu_f$ in the interval
$\mu_r,\mu_f \in [Q/4,4Q]$.
The HERAPDF1.5~\cite{herapdf:2009wt,herapdfgrid:2010} PDF set has been used.
Numbers are in pb.
}
\end{table}

%***************** 1st table ************************ 

\begin{table}[tb!]
\begin{center}
\begin{tabular}{|c|c|c|c|}
\hline
$ \hmass \, [\gev] $ & $ \sigma_{LO} $ & $ \sigma_{NLO} $ & $ \sigma_{NNLO} $ \\
\hline
  90 &$1.227^{+0.324}_{-0.219}$$^{+0.015}_{-0.015}$$\cdot 10^{-1}$ &$1.235^{+0.092}_{-0.045}$$^{+0.015}_{-0.015}$$\cdot 10^{-1}$ &$1.221^{+0.031}_{-0.008}$$^{+0.020}_{-0.020}$$\cdot 10^{-1
}$          \\
  95 &$1.129^{+0.306}_{-0.205}$$^{+0.014}_{-0.014}$$\cdot 10^{-1}$ &$1.140^{+0.085}_{-0.047}$$^{+0.014}_{-0.014}$$\cdot 10^{-1}$ &$1.125^{+0.029}_{-0.007}$$^{+0.019}_{-0.019}$$\cdot 10^{-1
}$          \\
 100 &$1.040^{+0.289}_{-0.192}$$^{+0.013}_{-0.013}$$\cdot 10^{-1}$ &$1.049^{+0.082}_{-0.043}$$^{+0.013}_{-0.013}$$\cdot 10^{-1}$ &$1.038^{+0.027}_{-0.008}$$^{+0.017}_{-0.017}$$\cdot 10^{-1
}$          \\
 105 &$9.581^{+2.730}_{-1.801}$$^{+0.124}_{-0.124}$$\cdot 10^{-2}$ &$9.699^{+0.751}_{-0.411}$$^{+0.123}_{-0.123}$$\cdot 10^{-2}$ &$9.580^{+0.248}_{-0.072}$$^{+0.162}_{-0.162}$$\cdot 10^{-2
}$          \\
 110 &$8.835^{+2.576}_{-1.691}$$^{+0.116}_{-0.116}$$\cdot 10^{-2}$ &$8.961^{+0.698}_{-0.401}$$^{+0.115}_{-0.115}$$\cdot 10^{-2}$ &$8.851^{+0.228}_{-0.078}$$^{+0.151}_{-0.151}$$\cdot 10^{-2
}$          \\
 115 &$8.152^{+2.433}_{-1.583}$$^{+0.109}_{-0.109}$$\cdot 10^{-2}$ &$8.290^{+0.650}_{-0.401}$$^{+0.108}_{-0.108}$$\cdot 10^{-2}$ &$8.182^{+0.213}_{-0.076}$$^{+0.140}_{-0.140}$$\cdot 10^{-2
}$          \\
 120 &$7.526^{+2.294}_{-1.485}$$^{+0.102}_{-0.102}$$\cdot 10^{-2}$ &$7.665^{+0.614}_{-0.367}$$^{+0.101}_{-0.101}$$\cdot 10^{-2}$ &$7.568^{+0.194}_{-0.073}$$^{+0.131}_{-0.131}$$\cdot 10^{-2
}$          \\
\hline
\end{tabular}
\end{center}
\vspace*{-5mm}
\caption{
\small
\label{tab:table-tev-gjr-Sc1}
Total VBF cross sections at the
Tevatron, $\sqrt S = 1.96 \tev$
at LO, NLO and NNLO in QCD. Errors shown are respectively scale and PDF uncertainities.  Scale uncertainities are evaluated by varying $\mu_r$ and $\mu_f$ in the interval
$\mu_r,\mu_f \in [Q/4,4Q]$.
The JR09~\cite{JimenezDelgado:2008hf,JimenezDelgado:2009tv} PDF set has been used.
Numbers are in pb.
}
\end{table}

%***************** 1st table ************************ 

\begin{table}[tb!]
\begin{center}
\begin{tabular}{|c|c|c|c|}
\hline
$ \hmass \, [\gev] $ & $ \sigma_{LO} $ & $ \sigma_{NLO} $ & $ \sigma_{NNLO} $ \\
\hline
  90 &$1.125^{+0.354}_{-0.227}$$^{+0.019}_{-0.019}$$\cdot 10^{-1}$ &$1.176^{+0.104}_{-0.055}$$^{+0.028}_{-0.028}$$\cdot 10^{-1}$ &$1.182^{+0.038}_{-0.011}$$^{+0.027}_{-0.027}$$\cdot 10^{-1
}$          \\
  95 &$1.035^{+0.335}_{-0.212}$$^{+0.018}_{-0.018}$$\cdot 10^{-1}$ &$1.083^{+0.096}_{-0.053}$$^{+0.026}_{-0.026}$$\cdot 10^{-1}$ &$1.088^{+0.035}_{-0.012}$$^{+0.025}_{-0.025}$$\cdot 10^{-1
}$          \\
 100 &$9.541^{+3.156}_{-1.991}$$^{+0.166}_{-0.166}$$\cdot 10^{-2}$ &$9.974^{+0.892}_{-0.511}$$^{+0.240}_{-0.240}$$\cdot 10^{-2}$ &$1.002^{+0.032}_{-0.011}$$^{+0.023}_{-0.023}$$\cdot 10^{-1
}$          \\
 105 &$8.794^{+2.984}_{-1.865}$$^{+0.154}_{-0.154}$$\cdot 10^{-2}$ &$9.181^{+0.848}_{-0.469}$$^{+0.222}_{-0.222}$$\cdot 10^{-2}$ &$9.235^{+0.298}_{-0.102}$$^{+0.213}_{-0.213}$$\cdot 10^{-2
}$          \\
 110 &$8.117^{+2.810}_{-1.752}$$^{+0.143}_{-0.143}$$\cdot 10^{-2}$ &$8.465^{+0.786}_{-0.453}$$^{+0.206}_{-0.206}$$\cdot 10^{-2}$ &$8.520^{+0.277}_{-0.104}$$^{+0.197}_{-0.197}$$\cdot 10^{-2
}$          \\
 115 &$7.494^{+2.652}_{-1.641}$$^{+0.133}_{-0.133}$$\cdot 10^{-2}$ &$7.813^{+0.739}_{-0.431}$$^{+0.191}_{-0.191}$$\cdot 10^{-2}$ &$7.866^{+0.257}_{-0.098}$$^{+0.183}_{-0.183}$$\cdot 10^{-2
}$          \\
 120 &$6.923^{+2.502}_{-1.536}$$^{+0.123}_{-0.123}$$\cdot 10^{-2}$ &$7.217^{+0.691}_{-0.414}$$^{+0.177}_{-0.177}$$\cdot 10^{-2}$ &$7.268^{+0.239}_{-0.095}$$^{+0.170}_{-0.170}$$\cdot 10^{-2
}$          \\
\hline
\end{tabular}
\end{center}
\vspace*{-5mm}
\caption{
\small
\label{tab:table-tev-m68-Sc1}
Total VBF cross sections at the
Tevatron, $\sqrt S = 1.96 \tev$
at LO, NLO and NNLO in QCD. Errors shown are respectively scale and PDF uncertainities.  Scale uncertainities are evaluated by varying $\mu_r$ and $\mu_f$ in the interval
$\mu_r,\mu_f \in [Q/4,4Q]$.
The MSTW2008~\cite{Martin:2009iq} PDF set (68\% CL) has been used.
Numbers are in pb.
}
\end{table}

%***************** 1st table ************************ 

\begin{table}[tb!]
\begin{center}
\begin{tabular}{|c|c|c|c|}
\hline
$ \hmass \, [\gev] $ & $ \sigma_{LO} $ & $ \sigma_{NLO} $ & $ \sigma_{NNLO} $ \\
\hline
  90 &$1.180^{+0.342}_{-0.224}$$^{+0.023}_{-0.023}$$\cdot 10^{-1}$ &$1.191^{+0.100}_{-0.053}$$^{+0.024}_{-0.024}$$\cdot 10^{-1}$ &$1.189^{+0.036}_{-0.012}$$^{+0.019}_{-0.019}$$\cdot 10^{-1
}$          \\
  95 &$1.085^{+0.323}_{-0.209}$$^{+0.022}_{-0.022}$$\cdot 10^{-1}$ &$1.096^{+0.094}_{-0.049}$$^{+0.022}_{-0.022}$$\cdot 10^{-1}$ &$1.095^{+0.034}_{-0.010}$$^{+0.017}_{-0.017}$$\cdot 10^{-1
}$          \\
 100 &$9.983^{+3.045}_{-1.962}$$^{+0.202}_{-0.202}$$\cdot 10^{-2}$ &$1.011^{+0.087}_{-0.048}$$^{+0.021}_{-0.021}$$\cdot 10^{-1}$ &$1.009^{+0.031}_{-0.009}$$^{+0.016}_{-0.016}$$\cdot 10^{-1
}$          \\
 105 &$9.190^{+2.875}_{-1.838}$$^{+0.188}_{-0.188}$$\cdot 10^{-2}$ &$9.324^{+0.828}_{-0.459}$$^{+0.194}_{-0.194}$$\cdot 10^{-2}$ &$9.303^{+0.293}_{-0.093}$$^{+0.151}_{-0.151}$$\cdot 10^{-2
}$          \\
 110 &$8.469^{+2.708}_{-1.721}$$^{+0.176}_{-0.176}$$\cdot 10^{-2}$ &$8.608^{+0.763}_{-0.440}$$^{+0.181}_{-0.181}$$\cdot 10^{-2}$ &$8.588^{+0.271}_{-0.095}$$^{+0.141}_{-0.141}$$\cdot 10^{-2
}$          \\
 115 &$7.805^{+2.550}_{-1.608}$$^{+0.164}_{-0.164}$$\cdot 10^{-2}$ &$7.951^{+0.713}_{-0.424}$$^{+0.170}_{-0.170}$$\cdot 10^{-2}$ &$7.933^{+0.252}_{-0.096}$$^{+0.132}_{-0.132}$$\cdot 10^{-2
}$          \\
 120 &$7.202^{+2.401}_{-1.509}$$^{+0.154}_{-0.154}$$\cdot 10^{-2}$ &$7.349^{+0.668}_{-0.399}$$^{+0.159}_{-0.159}$$\cdot 10^{-2}$ &$7.331^{+0.231}_{-0.092}$$^{+0.124}_{-0.124}$$\cdot 10^{-2
}$          \\
\hline
\end{tabular}
\end{center}
\vspace*{-5mm}
\caption{
\small
\label{tab:table-tev-nnn-Sc1}
Total VBF cross sections at the
Tevatron, $\sqrt S = 1.96 \tev$
at LO, NLO and NNLO in QCD. Errors shown are respectively scale and PDF uncertainities.  Scale uncertainities are evaluated by varying $\mu_r$ and $\mu_f$ in the interval
$\mu_r,\mu_f \in [Q/4,4Q]$.
The NNPDF2.1~\cite{Ball:2011uy} PDF set has been used.
Numbers are in pb.
}
\end{table}

%***************** 1st table ************************ 

\begin{table}[tb!]
\begin{footnotesize}
\begin{center}
\begin{tabular}{|c|c|c|c|}
\hline
$ \hmass \, [\gev] $ & $ \sigma_{LO} $ & $ \sigma_{NLO} $ & $ \sigma_{NNLO} $ \\
\hline
  90 &$1.856^{+0.140}_{-0.140}$$^{+0.013}_{-0.013}$$\phantom{\cdot 10^{-0}}$ &$1.797^{+0.062}_{-0.022}$$^{+0.013}_{-0.013}$$\phantom{\cdot 10^{-0}}$ &$1.820^{+0.021}_{-0.029}$$^{+0.014}_{-0.014}$$\phantom{\cdot 10^{-0}
}$          \\
  95 &$1.758^{+0.146}_{-0.138}$$^{+0.013}_{-0.013}$$\phantom{\cdot 10^{-0}}$ &$1.709^{+0.059}_{-0.025}$$^{+0.013}_{-0.013}$$\phantom{\cdot 10^{-0}}$ &$1.728^{+0.024}_{-0.027}$$^{+0.013}_{-0.013}$$\phantom{\cdot 10^{-0}
}$          \\
 100 &$1.667^{+0.146}_{-0.135}$$^{+0.012}_{-0.012}$$\phantom{\cdot 10^{-0}}$ &$1.622^{+0.059}_{-0.024}$$^{+0.012}_{-0.012}$$\phantom{\cdot 10^{-0}}$ &$1.643^{+0.022}_{-0.026}$$^{+0.012}_{-0.012}$$\phantom{\cdot 10^{-0}
}$          \\
 105 &$1.585^{+0.147}_{-0.133}$$^{+0.011}_{-0.011}$$\phantom{\cdot 10^{-0}}$ &$1.544^{+0.054}_{-0.026}$$^{+0.011}_{-0.011}$$\phantom{\cdot 10^{-0}}$ &$1.560^{+0.024}_{-0.021}$$^{+0.012}_{-0.012}$$\phantom{\cdot 10^{-0}
}$          \\
 110 &$1.506^{+0.148}_{-0.130}$$^{+0.011}_{-0.011}$$\phantom{\cdot 10^{-0}}$ &$1.466^{+0.058}_{-0.022}$$^{+0.011}_{-0.011}$$\phantom{\cdot 10^{-0}}$ &$1.483^{+0.026}_{-0.022}$$^{+0.011}_{-0.011}$$\phantom{\cdot 10^{-0}
}$          \\
 115 &$1.431^{+0.148}_{-0.129}$$^{+0.010}_{-0.010}$$\phantom{\cdot 10^{-0}}$ &$1.394^{+0.058}_{-0.018}$$^{+0.010}_{-0.010}$$\phantom{\cdot 10^{-0}}$ &$1.411^{+0.022}_{-0.014}$$^{+0.011}_{-0.011}$$\phantom{\cdot 10^{-0}
}$          \\
 120 &$1.363^{+0.145}_{-0.128}$$^{+0.010}_{-0.010}$$\phantom{\cdot 10^{-0}}$ &$1.330^{+0.054}_{-0.021}$$^{+0.010}_{-0.010}$$\phantom{\cdot 10^{-0}}$ &$1.345^{+0.022}_{-0.016}$$^{+0.010}_{-0.010}$$\phantom{\cdot 10^{-0}
}$          \\
\hline
\end{tabular}
\end{center}
\vspace*{-5mm}
\end{footnotesize}
\caption{
\small
\label{tab:table-lhc7-a09-Sc1}
Total VBF cross sections at the
LHC, $\sqrt S = 7 \tev$
at LO, NLO and NNLO in QCD. Errors shown are respectively scale and PDF uncertainities.  Scale uncertainities are evaluated by varying $\mu_r$ and $\mu_f$ in the interval
$\mu_r,\mu_f \in [Q/4,4Q]$.
The ABKM~\cite{Alekhin:2009ni} PDF set has been used.
Numbers are in pb.
}
\end{table}

%***************** 1st table ************************ 

\begin{table}[tb!]
\begin{footnotesize}
\begin{center}
\begin{tabular}{|c|c|c|c|}
\hline
$ \hmass \, [\gev] $ & $ \sigma_{LO} $ & $ \sigma_{NLO} $ & $ \sigma_{NNLO} $ \\
\hline
  90 &$1.870^{+0.135}_{-0.129}$$^{+0.036}_{-0.036}$$\phantom{\cdot 10^{-0}}$ &$1.810^{+0.063}_{-0.019}$$^{+0.036}_{-0.036}$$\phantom{\cdot 10^{-0}}$ &$1.801^{+0.026}_{-0.032}$$^{+0.036}_{-0.036}$$\phantom{\cdot 10^{-0}
}$          \\
  95 &$1.776^{+0.136}_{-0.129}$$^{+0.035}_{-0.035}$$\phantom{\cdot 10^{-0}}$ &$1.723^{+0.055}_{-0.024}$$^{+0.034}_{-0.034}$$\phantom{\cdot 10^{-0}}$ &$1.710^{+0.028}_{-0.026}$$^{+0.034}_{-0.034}$$\phantom{\cdot 10^{-0}
}$          \\
 100 &$1.686^{+0.136}_{-0.128}$$^{+0.033}_{-0.033}$$\phantom{\cdot 10^{-0}}$ &$1.637^{+0.054}_{-0.027}$$^{+0.033}_{-0.033}$$\phantom{\cdot 10^{-0}}$ &$1.627^{+0.023}_{-0.030}$$^{+0.033}_{-0.033}$$\phantom{\cdot 10^{-0}
}$          \\
 105 &$1.603^{+0.137}_{-0.128}$$^{+0.031}_{-0.031}$$\phantom{\cdot 10^{-0}}$ &$1.559^{+0.055}_{-0.025}$$^{+0.031}_{-0.031}$$\phantom{\cdot 10^{-0}}$ &$1.545^{+0.025}_{-0.023}$$^{+0.031}_{-0.031}$$\phantom{\cdot 10^{-0}
}$          \\
 110 &$1.524^{+0.140}_{-0.126}$$^{+0.030}_{-0.030}$$\phantom{\cdot 10^{-0}}$ &$1.484^{+0.055}_{-0.025}$$^{+0.030}_{-0.030}$$\phantom{\cdot 10^{-0}}$ &$1.473^{+0.024}_{-0.026}$$^{+0.030}_{-0.030}$$\phantom{\cdot 10^{-0}
}$          \\
 115 &$1.450^{+0.141}_{-0.123}$$^{+0.029}_{-0.029}$$\phantom{\cdot 10^{-0}}$ &$1.411^{+0.055}_{-0.020}$$^{+0.029}_{-0.029}$$\phantom{\cdot 10^{-0}}$ &$1.401^{+0.023}_{-0.022}$$^{+0.029}_{-0.029}$$\phantom{\cdot 10^{-0}
}$          \\
 120 &$1.380^{+0.139}_{-0.121}$$^{+0.028}_{-0.028}$$\phantom{\cdot 10^{-0}}$ &$1.346^{+0.053}_{-0.021}$$^{+0.028}_{-0.028}$$\phantom{\cdot 10^{-0}}$ &$1.335^{+0.025}_{-0.019}$$^{+0.027}_{-0.027}$$\phantom{\cdot 10^{-0}
}$          \\
\hline
\end{tabular}
\end{center}
\vspace*{-5mm}
\end{footnotesize}
\caption{
\small
\label{tab:table-lhc7-h15-Sc1}
Total VBF cross sections at the
LHC, $\sqrt S = 7 \tev$
at LO, NLO and NNLO in QCD. Errors shown are respectively scale and PDF uncertainities.  Scale uncertainities are evaluated by varying $\mu_r$ and $\mu_f$ in the interval
$\mu_r,\mu_f \in [Q/4,4Q]$.
The HERAPDF1.5~\cite{herapdf:2009wt,herapdfgrid:2010} PDF set has been used.
Numbers are in pb.
}
\end{table}

%***************** 1st table ************************ 

\begin{table}[tb!]
\begin{footnotesize}
\begin{center}
\begin{tabular}{|c|c|c|c|}
\hline
$ \hmass \, [\gev] $ & $ \sigma_{LO} $ & $ \sigma_{NLO} $ & $ \sigma_{NNLO} $ \\
\hline
  90 &$1.834^{+0.096}_{-0.111}$$^{+0.023}_{-0.023}$$\phantom{\cdot 10^{-0}}$ &$1.776^{+0.047}_{-0.024}$$^{+0.023}_{-0.023}$$\phantom{\cdot 10^{-0}}$ &$1.772^{+0.021}_{-0.026}$$^{+0.023}_{-0.023}$$\phantom{\cdot 10^{-0}
}$          \\
  95 &$1.739^{+0.102}_{-0.107}$$^{+0.022}_{-0.022}$$\phantom{\cdot 10^{-0}}$ &$1.688^{+0.044}_{-0.024}$$^{+0.022}_{-0.022}$$\phantom{\cdot 10^{-0}}$ &$1.684^{+0.019}_{-0.024}$$^{+0.023}_{-0.023}$$\phantom{\cdot 10^{-0}
}$          \\
 100 &$1.654^{+0.105}_{-0.107}$$^{+0.022}_{-0.022}$$\phantom{\cdot 10^{-0}}$ &$1.605^{+0.044}_{-0.021}$$^{+0.021}_{-0.021}$$\phantom{\cdot 10^{-0}}$ &$1.599^{+0.023}_{-0.017}$$^{+0.022}_{-0.022}$$\phantom{\cdot 10^{-0}
}$          \\
 105 &$1.572^{+0.108}_{-0.107}$$^{+0.021}_{-0.021}$$\phantom{\cdot 10^{-0}}$ &$1.530^{+0.045}_{-0.022}$$^{+0.021}_{-0.021}$$\phantom{\cdot 10^{-0}}$ &$1.526^{+0.019}_{-0.022}$$^{+0.021}_{-0.021}$$\phantom{\cdot 10^{-0}
}$          \\
 110 &$1.498^{+0.107}_{-0.108}$$^{+0.020}_{-0.020}$$\phantom{\cdot 10^{-0}}$ &$1.458^{+0.039}_{-0.023}$$^{+0.020}_{-0.020}$$\phantom{\cdot 10^{-0}}$ &$1.453^{+0.014}_{-0.020}$$^{+0.021}_{-0.021}$$\phantom{\cdot 10^{-0}
}$          \\
 115 &$1.428^{+0.105}_{-0.107}$$^{+0.019}_{-0.019}$$\phantom{\cdot 10^{-0}}$ &$1.390^{+0.043}_{-0.023}$$^{+0.019}_{-0.019}$$\phantom{\cdot 10^{-0}}$ &$1.388^{+0.016}_{-0.022}$$^{+0.020}_{-0.020}$$\phantom{\cdot 10^{-0}
}$          \\
 120 &$1.360^{+0.109}_{-0.103}$$^{+0.019}_{-0.019}$$\phantom{\cdot 10^{-0}}$ &$1.325^{+0.042}_{-0.020}$$^{+0.018}_{-0.018}$$\phantom{\cdot 10^{-0}}$ &$1.321^{+0.018}_{-0.018}$$^{+0.019}_{-0.019}$$\phantom{\cdot 10^{-0}
}$          \\
\hline
\end{tabular}
\end{center}
\vspace*{-5mm}
\end{footnotesize}
\caption{
\small
\label{tab:table-lhc7-gjr-Sc1}
Total VBF cross sections at the
LHC, $\sqrt S = 7 \tev$
at LO, NLO and NNLO in QCD. Errors shown are respectively scale and PDF uncertainities.  Scale uncertainities are evaluated by varying $\mu_r$ and $\mu_f$ in the interval
$\mu_r,\mu_f \in [Q/4,4Q]$.
The JR09~\cite{JimenezDelgado:2008hf,JimenezDelgado:2009tv} PDF set has been used.
Numbers are in pb.
}
\end{table}

%***************** 1st table ************************ 

\begin{table}[tb!]
\begin{footnotesize}
\begin{center}
\begin{tabular}{|c|c|c|c|}
\hline
$ \hmass \, [\gev] $ & $ \sigma_{LO} $ & $ \sigma_{NLO} $ & $ \sigma_{NNLO} $ \\
\hline
  90 &$1.674^{+0.127}_{-0.125}$$^{+0.021}_{-0.021}$$\phantom{\cdot 10^{-0}}$ &$1.782^{+0.059}_{-0.026}$$^{+0.035}_{-0.035}$$\phantom{\cdot 10^{-0}}$ &$1.789^{+0.028}_{-0.033}$$^{+0.032}_{-0.032}$$\phantom{\cdot 10^{-0}
}$          \\
  95 &$1.589^{+0.124}_{-0.123}$$^{+0.020}_{-0.020}$$\phantom{\cdot 10^{-0}}$ &$1.697^{+0.053}_{-0.033}$$^{+0.034}_{-0.034}$$\phantom{\cdot 10^{-0}}$ &$1.697^{+0.027}_{-0.026}$$^{+0.030}_{-0.030}$$\phantom{\cdot 10^{-0}
}$          \\
 100 &$1.510^{+0.129}_{-0.126}$$^{+0.019}_{-0.019}$$\phantom{\cdot 10^{-0}}$ &$1.608^{+0.059}_{-0.029}$$^{+0.032}_{-0.032}$$\phantom{\cdot 10^{-0}}$ &$1.616^{+0.025}_{-0.034}$$^{+0.029}_{-0.029}$$\phantom{\cdot 10^{-0}
}$          \\
 105 &$1.435^{+0.129}_{-0.122}$$^{+0.018}_{-0.018}$$\phantom{\cdot 10^{-0}}$ &$1.531^{+0.059}_{-0.028}$$^{+0.031}_{-0.031}$$\phantom{\cdot 10^{-0}}$ &$1.536^{+0.024}_{-0.033}$$^{+0.027}_{-0.027}$$\phantom{\cdot 10^{-0}
}$          \\
 110 &$1.366^{+0.130}_{-0.120}$$^{+0.017}_{-0.017}$$\phantom{\cdot 10^{-0}}$ &$1.456^{+0.055}_{-0.027}$$^{+0.029}_{-0.029}$$\phantom{\cdot 10^{-0}}$ &$1.460^{+0.025}_{-0.027}$$^{+0.026}_{-0.026}$$\phantom{\cdot 10^{-0}
}$          \\
 115 &$1.300^{+0.131}_{-0.118}$$^{+0.017}_{-0.017}$$\phantom{\cdot 10^{-0}}$ &$1.386^{+0.053}_{-0.022}$$^{+0.028}_{-0.028}$$\phantom{\cdot 10^{-0}}$ &$1.391^{+0.026}_{-0.028}$$^{+0.025}_{-0.025}$$\phantom{\cdot 10^{-0}
}$          \\
 120 &$1.239^{+0.132}_{-0.117}$$^{+0.016}_{-0.016}$$\phantom{\cdot 10^{-0}}$ &$1.320^{+0.055}_{-0.026}$$^{+0.027}_{-0.027}$$\phantom{\cdot 10^{-0}}$ &$1.324^{+0.025}_{-0.023}$$^{+0.024}_{-0.024}$$\phantom{\cdot 10^{-0}
}$          \\
 125 &$1.181^{+0.131}_{-0.116}$$^{+0.015}_{-0.015}$$\phantom{\cdot 10^{-0}}$ &$1.258^{+0.057}_{-0.020}$$^{+0.026}_{-0.026}$$\phantom{\cdot 10^{-0}}$ &$1.264^{+0.023}_{-0.023}$$^{+0.023}_{-0.023}$$\phantom{\cdot 10^{-0}
}$          \\
 130 &$1.127^{+0.130}_{-0.112}$$^{+0.015}_{-0.015}$$\phantom{\cdot 10^{-0}}$ &$1.200^{+0.053}_{-0.020}$$^{+0.024}_{-0.024}$$\phantom{\cdot 10^{-0}}$ &$1.203^{+0.026}_{-0.021}$$^{+0.022}_{-0.022}$$\phantom{\cdot 10^{-0}
}$          \\
 135 &$1.075^{+0.129}_{-0.110}$$^{+0.014}_{-0.014}$$\phantom{\cdot 10^{-0}}$ &$1.144^{+0.053}_{-0.018}$$^{+0.023}_{-0.023}$$\phantom{\cdot 10^{-0}}$ &$1.153^{+0.018}_{-0.023}$$^{+0.021}_{-0.021}$$\phantom{\cdot 10^{-0}
}$          \\
 140 &$1.027^{+0.129}_{-0.107}$$^{+0.014}_{-0.014}$$\phantom{\cdot 10^{-0}}$ &$1.094^{+0.051}_{-0.018}$$^{+0.022}_{-0.022}$$\phantom{\cdot 10^{-0}}$ &$1.099^{+0.024}_{-0.019}$$^{+0.020}_{-0.020}$$\phantom{\cdot 10^{-0}
}$          \\
 145 &$9.818^{+1.264}_{-1.063}$$^{+0.130}_{-0.130}$$\cdot 10^{-1}$ &$1.045^{+0.050}_{-0.018}$$^{+0.022}_{-0.022}$$\phantom{\cdot 10^{-0}}$ &$1.050^{+0.021}_{-0.021}$$^{+0.020}_{-0.020}$$\phantom{\cdot 10^{-0}
}$          \\
 150 &$9.383^{+1.246}_{-1.026}$$^{+0.125}_{-0.125}$$\cdot 10^{-1}$ &$9.998^{+0.486}_{-0.175}$$^{+0.207}_{-0.207}$$\cdot 10^{-1}$ &$1.003^{+0.021}_{-0.016}$$^{+0.019}_{-0.019}$$\phantom{\cdot 10^{-0}
}$          \\
 155 &$8.986^{+1.204}_{-1.026}$$^{+0.121}_{-0.121}$$\cdot 10^{-1}$ &$9.570^{+0.463}_{-0.185}$$^{+0.199}_{-0.199}$$\cdot 10^{-1}$ &$9.599^{+0.209}_{-0.154}$$^{+0.180}_{-0.180}$$\cdot 10^{-1
}$          \\
 160 &$8.599^{+1.224}_{-0.997}$$^{+0.116}_{-0.116}$$\cdot 10^{-1}$ &$9.161^{+0.450}_{-0.163}$$^{+0.191}_{-0.191}$$\cdot 10^{-1}$ &$9.176^{+0.183}_{-0.133}$$^{+0.173}_{-0.173}$$\cdot 10^{-1
}$          \\
 165 &$8.233^{+1.199}_{-0.973}$$^{+0.112}_{-0.112}$$\cdot 10^{-1}$ &$8.765^{+0.443}_{-0.143}$$^{+0.184}_{-0.184}$$\cdot 10^{-1}$ &$8.809^{+0.184}_{-0.150}$$^{+0.167}_{-0.167}$$\cdot 10^{-1
}$          \\
 170 &$7.886^{+1.182}_{-0.955}$$^{+0.108}_{-0.108}$$\cdot 10^{-1}$ &$8.394^{+0.439}_{-0.151}$$^{+0.176}_{-0.176}$$\cdot 10^{-1}$ &$8.411^{+0.202}_{-0.126}$$^{+0.160}_{-0.160}$$\cdot 10^{-1
}$          \\
 175 &$7.558^{+1.156}_{-0.931}$$^{+0.104}_{-0.104}$$\cdot 10^{-1}$ &$8.037^{+0.435}_{-0.127}$$^{+0.169}_{-0.169}$$\cdot 10^{-1}$ &$8.085^{+0.174}_{-0.129}$$^{+0.154}_{-0.154}$$\cdot 10^{-1
}$          \\
 180 &$7.244^{+1.148}_{-0.903}$$^{+0.100}_{-0.100}$$\cdot 10^{-1}$ &$7.713^{+0.420}_{-0.131}$$^{+0.163}_{-0.163}$$\cdot 10^{-1}$ &$7.732^{+0.187}_{-0.105}$$^{+0.148}_{-0.148}$$\cdot 10^{-1
}$          \\
 185 &$6.957^{+1.123}_{-0.887}$$^{+0.096}_{-0.096}$$\cdot 10^{-1}$ &$7.398^{+0.411}_{-0.123}$$^{+0.157}_{-0.157}$$\cdot 10^{-1}$ &$7.412^{+0.185}_{-0.090}$$^{+0.142}_{-0.142}$$\cdot 10^{-1
}$          \\
 190 &$6.667^{+1.106}_{-0.854}$$^{+0.093}_{-0.093}$$\cdot 10^{-1}$ &$7.105^{+0.384}_{-0.136}$$^{+0.151}_{-0.151}$$\cdot 10^{-1}$ &$7.127^{+0.166}_{-0.102}$$^{+0.137}_{-0.137}$$\cdot 10^{-1
}$          \\
 195 &$6.404^{+1.093}_{-0.833}$$^{+0.090}_{-0.090}$$\cdot 10^{-1}$ &$6.822^{+0.383}_{-0.135}$$^{+0.146}_{-0.146}$$\cdot 10^{-1}$ &$6.856^{+0.146}_{-0.114}$$^{+0.133}_{-0.133}$$\cdot 10^{-1
}$          \\
 200 &$6.156^{+1.062}_{-0.817}$$^{+0.087}_{-0.087}$$\cdot 10^{-1}$ &$6.543^{+0.385}_{-0.116}$$^{+0.140}_{-0.140}$$\cdot 10^{-1}$ &$6.580^{+0.143}_{-0.096}$$^{+0.128}_{-0.128}$$\cdot 10^{-1
}$          \\
 210 &$5.683^{+1.029}_{-0.767}$$^{+0.081}_{-0.081}$$\cdot 10^{-1}$ &$6.049^{+0.351}_{-0.121}$$^{+0.131}_{-0.131}$$\cdot 10^{-1}$ &$6.081^{+0.137}_{-0.090}$$^{+0.119}_{-0.119}$$\cdot 10^{-1
}$          \\
 220 &$5.256^{+0.986}_{-0.737}$$^{+0.075}_{-0.075}$$\cdot 10^{-1}$ &$5.593^{+0.336}_{-0.113}$$^{+0.121}_{-0.121}$$\cdot 10^{-1}$ &$5.623^{+0.126}_{-0.075}$$^{+0.111}_{-0.111}$$\cdot 10^{-1
}$          \\
 230 &$4.871^{+0.942}_{-0.695}$$^{+0.070}_{-0.070}$$\cdot 10^{-1}$ &$5.184^{+0.322}_{-0.109}$$^{+0.113}_{-0.113}$$\cdot 10^{-1}$ &$5.213^{+0.114}_{-0.075}$$^{+0.104}_{-0.104}$$\cdot 10^{-1
}$          \\
 240 &$4.523^{+0.905}_{-0.672}$$^{+0.066}_{-0.066}$$\cdot 10^{-1}$ &$4.811^{+0.295}_{-0.118}$$^{+0.106}_{-0.106}$$\cdot 10^{-1}$ &$4.834^{+0.110}_{-0.064}$$^{+0.097}_{-0.097}$$\cdot 10^{-1
}$          \\
 250 &$4.200^{+0.868}_{-0.633}$$^{+0.062}_{-0.062}$$\cdot 10^{-1}$ &$4.457^{+0.298}_{-0.101}$$^{+0.099}_{-0.099}$$\cdot 10^{-1}$ &$4.488^{+0.107}_{-0.055}$$^{+0.090}_{-0.090}$$\cdot 10^{-1
}$          \\
 260 &$3.907^{+0.830}_{-0.601}$$^{+0.058}_{-0.058}$$\cdot 10^{-1}$ &$4.147^{+0.272}_{-0.102}$$^{+0.092}_{-0.092}$$\cdot 10^{-1}$ &$4.175^{+0.098}_{-0.052}$$^{+0.085}_{-0.085}$$\cdot 10^{-1
}$          \\
 270 &$3.638^{+0.796}_{-0.574}$$^{+0.054}_{-0.054}$$\cdot 10^{-1}$ &$3.860^{+0.258}_{-0.101}$$^{+0.086}_{-0.086}$$\cdot 10^{-1}$ &$3.886^{+0.098}_{-0.044}$$^{+0.079}_{-0.079}$$\cdot 10^{-1
}$          \\
 280 &$3.391^{+0.762}_{-0.545}$$^{+0.051}_{-0.051}$$\cdot 10^{-1}$ &$3.594^{+0.247}_{-0.098}$$^{+0.081}_{-0.081}$$\cdot 10^{-1}$ &$3.623^{+0.090}_{-0.043}$$^{+0.075}_{-0.075}$$\cdot 10^{-1
}$          \\
 290 &$3.162^{+0.731}_{-0.516}$$^{+0.048}_{-0.048}$$\cdot 10^{-1}$ &$3.360^{+0.226}_{-0.101}$$^{+0.076}_{-0.076}$$\cdot 10^{-1}$ &$3.380^{+0.086}_{-0.036}$$^{+0.070}_{-0.070}$$\cdot 10^{-1
}$          \\
 300 &$2.956^{+0.698}_{-0.492}$$^{+0.045}_{-0.045}$$\cdot 10^{-1}$ &$3.137^{+0.217}_{-0.100}$$^{+0.071}_{-0.071}$$\cdot 10^{-1}$ &$3.156^{+0.078}_{-0.031}$$^{+0.066}_{-0.066}$$\cdot 10^{-1
}$          \\
 320 &$2.587^{+0.640}_{-0.445}$$^{+0.040}_{-0.040}$$\cdot 10^{-1}$ &$2.740^{+0.198}_{-0.088}$$^{+0.063}_{-0.063}$$\cdot 10^{-1}$ &$2.762^{+0.073}_{-0.027}$$^{+0.058}_{-0.058}$$\cdot 10^{-1
}$          \\
 340 &$2.268^{+0.589}_{-0.399}$$^{+0.035}_{-0.035}$$\cdot 10^{-1}$ &$2.405^{+0.180}_{-0.081}$$^{+0.056}_{-0.056}$$\cdot 10^{-1}$ &$2.424^{+0.066}_{-0.022}$$^{+0.052}_{-0.052}$$\cdot 10^{-1
}$          \\
 360 &$1.998^{+0.538}_{-0.362}$$^{+0.032}_{-0.032}$$\cdot 10^{-1}$ &$2.119^{+0.161}_{-0.078}$$^{+0.049}_{-0.049}$$\cdot 10^{-1}$ &$2.136^{+0.057}_{-0.018}$$^{+0.046}_{-0.046}$$\cdot 10^{-1
}$          \\
 380 &$1.766^{+0.494}_{-0.329}$$^{+0.028}_{-0.028}$$\cdot 10^{-1}$ &$1.870^{+0.147}_{-0.074}$$^{+0.044}_{-0.044}$$\cdot 10^{-1}$ &$1.886^{+0.052}_{-0.014}$$^{+0.041}_{-0.041}$$\cdot 10^{-1
}$          \\
 400 &$1.565^{+0.453}_{-0.299}$$^{+0.025}_{-0.025}$$\cdot 10^{-1}$ &$1.657^{+0.131}_{-0.070}$$^{+0.039}_{-0.039}$$\cdot 10^{-1}$ &$1.671^{+0.045}_{-0.014}$$^{+0.037}_{-0.037}$$\cdot 10^{-1
}$          \\
 450 &$1.169^{+0.367}_{-0.236}$$^{+0.020}_{-0.020}$$\cdot 10^{-1}$ &$1.235^{+0.103}_{-0.058}$$^{+0.030}_{-0.030}$$\cdot 10^{-1}$ &$1.248^{+0.035}_{-0.013}$$^{+0.028}_{-0.028}$$\cdot 10^{-1
}$          \\
 500 &$8.851^{+2.964}_{-1.876}$$^{+0.152}_{-0.152}$$\cdot 10^{-2}$ &$9.341^{+0.813}_{-0.485}$$^{+0.229}_{-0.229}$$\cdot 10^{-2}$ &$9.440^{+0.265}_{-0.110}$$^{+0.218}_{-0.218}$$\cdot 10^{-2
}$          \\
 550 &$6.770^{+2.425}_{-1.495}$$^{+0.120}_{-0.120}$$\cdot 10^{-2}$ &$7.143^{+0.645}_{-0.404}$$^{+0.178}_{-0.178}$$\cdot 10^{-2}$ &$7.225^{+0.209}_{-0.099}$$^{+0.170}_{-0.170}$$\cdot 10^{-2
}$          \\
 600 &$5.233^{+1.974}_{-1.203}$$^{+0.095}_{-0.095}$$\cdot 10^{-2}$ &$5.516^{+0.514}_{-0.339}$$^{+0.139}_{-0.139}$$\cdot 10^{-2}$ &$5.581^{+0.162}_{-0.086}$$^{+0.134}_{-0.134}$$\cdot 10^{-2
}$          \\
 650 &$4.076^{+1.621}_{-0.969}$$^{+0.076}_{-0.076}$$\cdot 10^{-2}$ &$4.296^{+0.411}_{-0.286}$$^{+0.110}_{-0.110}$$\cdot 10^{-2}$ &$4.345^{+0.131}_{-0.072}$$^{+0.106}_{-0.106}$$\cdot 10^{-2
}$          \\
 700 &$3.198^{+1.333}_{-0.785}$$^{+0.061}_{-0.061}$$\cdot 10^{-2}$ &$3.368^{+0.330}_{-0.239}$$^{+0.087}_{-0.087}$$\cdot 10^{-2}$ &$3.408^{+0.109}_{-0.062}$$^{+0.085}_{-0.085}$$\cdot 10^{-2
}$          \\
 750 &$2.526^{+1.096}_{-0.638}$$^{+0.050}_{-0.050}$$\cdot 10^{-2}$ &$2.658^{+0.268}_{-0.201}$$^{+0.070}_{-0.070}$$\cdot 10^{-2}$ &$2.692^{+0.081}_{-0.055}$$^{+0.068}_{-0.068}$$\cdot 10^{-2
}$          \\
 800 &$2.007^{+0.906}_{-0.522}$$^{+0.041}_{-0.041}$$\cdot 10^{-2}$ &$2.108^{+0.216}_{-0.168}$$^{+0.056}_{-0.056}$$\cdot 10^{-2}$ &$2.136^{+0.068}_{-0.047}$$^{+0.055}_{-0.055}$$\cdot 10^{-2
}$          \\
 850 &$1.599^{+0.753}_{-0.426}$$^{+0.033}_{-0.033}$$\cdot 10^{-2}$ &$1.680^{+0.179}_{-0.140}$$^{+0.045}_{-0.045}$$\cdot 10^{-2}$ &$1.702^{+0.056}_{-0.040}$$^{+0.045}_{-0.045}$$\cdot 10^{-2
}$          \\
 900 &$1.281^{+0.625}_{-0.349}$$^{+0.028}_{-0.028}$$\cdot 10^{-2}$ &$1.345^{+0.145}_{-0.118}$$^{+0.037}_{-0.037}$$\cdot 10^{-2}$ &$1.365^{+0.042}_{-0.036}$$^{+0.037}_{-0.037}$$\cdot 10^{-2
}$          \\
 950 &$1.030^{+0.521}_{-0.287}$$^{+0.023}_{-0.023}$$\cdot 10^{-2}$ &$1.081^{+0.120}_{-0.100}$$^{+0.030}_{-0.030}$$\cdot 10^{-2}$ &$1.096^{+0.035}_{-0.030}$$^{+0.030}_{-0.030}$$\cdot 10^{-2
}$          \\
1000 &$8.307^{+4.344}_{-2.363}$$^{+0.189}_{-0.189}$$\cdot 10^{-3}$ &$8.713^{+0.982}_{-0.834}$$^{+0.243}_{-0.243}$$\cdot 10^{-3}$ &$8.838^{+0.291}_{-0.259}$$^{+0.245}_{-0.245}$$\cdot 10^{-3
}$          \\
\hline
\end{tabular}
\end{center}
\vspace*{-5mm}
\end{footnotesize}
\caption{
\small
\label{tab:table-lhc7-m68-Sc1}
Total VBF cross sections at the
LHC, $\sqrt S = 7 \tev$
at LO, NLO and NNLO in QCD. Errors shown are respectively scale and PDF uncertainities.  Scale uncertainities are evaluated by varying $\mu_r$ and $\mu_f$ in the interval
$\mu_r,\mu_f \in [Q/4,4Q]$.
The MSTW2008~\cite{Martin:2009iq} PDF set (68\% CL) has been used.
Numbers are in pb.
}
\end{table}

%***************** 1st table ************************ 

\begin{table}[tb!]
\begin{footnotesize}
\begin{center}
\begin{tabular}{|c|c|c|c|}
\hline
$ \hmass \, [\gev] $ & $ \sigma_{LO} $ & $ \sigma_{NLO} $ & $ \sigma_{NNLO} $ \\
\hline
  90 &$1.818^{+0.112}_{-0.118}$$^{+0.026}_{-0.026}$$\phantom{\cdot 10^{-0}}$ &$1.760^{+0.056}_{-0.029}$$^{+0.026}_{-0.026}$$\phantom{\cdot 10^{-0}}$ &$1.774^{+0.029}_{-0.030}$$^{+0.023}_{-0.023}$$\phantom{\cdot 10^{-0}
}$          \\
  95 &$1.726^{+0.114}_{-0.118}$$^{+0.025}_{-0.025}$$\phantom{\cdot 10^{-0}}$ &$1.672^{+0.055}_{-0.026}$$^{+0.025}_{-0.025}$$\phantom{\cdot 10^{-0}}$ &$1.684^{+0.030}_{-0.031}$$^{+0.022}_{-0.022}$$\phantom{\cdot 10^{-0}
}$          \\
 100 &$1.640^{+0.118}_{-0.119}$$^{+0.024}_{-0.024}$$\phantom{\cdot 10^{-0}}$ &$1.591^{+0.052}_{-0.028}$$^{+0.024}_{-0.024}$$\phantom{\cdot 10^{-0}}$ &$1.603^{+0.028}_{-0.028}$$^{+0.021}_{-0.021}$$\phantom{\cdot 10^{-0}
}$          \\
 105 &$1.560^{+0.118}_{-0.118}$$^{+0.023}_{-0.023}$$\phantom{\cdot 10^{-0}}$ &$1.515^{+0.049}_{-0.029}$$^{+0.023}_{-0.023}$$\phantom{\cdot 10^{-0}}$ &$1.522^{+0.028}_{-0.027}$$^{+0.020}_{-0.020}$$\phantom{\cdot 10^{-0}
}$          \\
 110 &$1.482^{+0.121}_{-0.117}$$^{+0.022}_{-0.022}$$\phantom{\cdot 10^{-0}}$ &$1.441^{+0.051}_{-0.023}$$^{+0.022}_{-0.022}$$\phantom{\cdot 10^{-0}}$ &$1.448^{+0.027}_{-0.023}$$^{+0.019}_{-0.019}$$\phantom{\cdot 10^{-0}
}$          \\
 115 &$1.411^{+0.121}_{-0.116}$$^{+0.021}_{-0.021}$$\phantom{\cdot 10^{-0}}$ &$1.372^{+0.050}_{-0.021}$$^{+0.021}_{-0.021}$$\phantom{\cdot 10^{-0}}$ &$1.378^{+0.029}_{-0.024}$$^{+0.018}_{-0.018}$$\phantom{\cdot 10^{-0}
}$          \\
 120 &$1.344^{+0.126}_{-0.116}$$^{+0.020}_{-0.020}$$\phantom{\cdot 10^{-0}}$ &$1.308^{+0.049}_{-0.022}$$^{+0.020}_{-0.020}$$\phantom{\cdot 10^{-0}}$ &$1.318^{+0.021}_{-0.028}$$^{+0.017}_{-0.017}$$\phantom{\cdot 10^{-0}
}$          \\
 125 &$1.280^{+0.124}_{-0.111}$$^{+0.020}_{-0.020}$$\phantom{\cdot 10^{-0}}$ &$1.246^{+0.051}_{-0.018}$$^{+0.019}_{-0.019}$$\phantom{\cdot 10^{-0}}$ &$1.252^{+0.026}_{-0.019}$$^{+0.016}_{-0.016}$$\phantom{\cdot 10^{-0}
}$          \\
 130 &$1.221^{+0.122}_{-0.111}$$^{+0.019}_{-0.019}$$\phantom{\cdot 10^{-0}}$ &$1.189^{+0.050}_{-0.017}$$^{+0.019}_{-0.019}$$\phantom{\cdot 10^{-0}}$ &$1.195^{+0.027}_{-0.018}$$^{+0.016}_{-0.016}$$\phantom{\cdot 10^{-0}
}$          \\
 135 &$1.165^{+0.123}_{-0.109}$$^{+0.018}_{-0.018}$$\phantom{\cdot 10^{-0}}$ &$1.137^{+0.045}_{-0.019}$$^{+0.018}_{-0.018}$$\phantom{\cdot 10^{-0}}$ &$1.143^{+0.022}_{-0.017}$$^{+0.015}_{-0.015}$$\phantom{\cdot 10^{-0}
}$          \\
 140 &$1.112^{+0.122}_{-0.106}$$^{+0.017}_{-0.017}$$\phantom{\cdot 10^{-0}}$ &$1.087^{+0.045}_{-0.021}$$^{+0.017}_{-0.017}$$\phantom{\cdot 10^{-0}}$ &$1.093^{+0.021}_{-0.020}$$^{+0.015}_{-0.015}$$\phantom{\cdot 10^{-0}
}$          \\
 145 &$1.062^{+0.121}_{-0.103}$$^{+0.017}_{-0.017}$$\phantom{\cdot 10^{-0}}$ &$1.037^{+0.047}_{-0.016}$$^{+0.017}_{-0.017}$$\phantom{\cdot 10^{-0}}$ &$1.043^{+0.021}_{-0.015}$$^{+0.014}_{-0.014}$$\phantom{\cdot 10^{-0}
}$          \\
 150 &$1.015^{+0.120}_{-0.102}$$^{+0.016}_{-0.016}$$\phantom{\cdot 10^{-0}}$ &$9.924^{+0.453}_{-0.176}$$^{+0.160}_{-0.160}$$\cdot 10^{-1}$ &$9.990^{+0.199}_{-0.190}$$^{+0.134}_{-0.134}$$\cdot 10^{-1
}$          \\
 155 &$9.706^{+1.189}_{-0.999}$$^{+0.154}_{-0.154}$$\cdot 10^{-1}$ &$9.492^{+0.449}_{-0.143}$$^{+0.154}_{-0.154}$$\cdot 10^{-1}$ &$9.541^{+0.199}_{-0.146}$$^{+0.129}_{-0.129}$$\cdot 10^{-1
}$          \\
 160 &$9.283^{+1.160}_{-0.969}$$^{+0.148}_{-0.148}$$\cdot 10^{-1}$ &$9.086^{+0.440}_{-0.142}$$^{+0.148}_{-0.148}$$\cdot 10^{-1}$ &$9.146^{+0.169}_{-0.139}$$^{+0.124}_{-0.124}$$\cdot 10^{-1
}$          \\
 165 &$8.890^{+1.167}_{-0.949}$$^{+0.143}_{-0.143}$$\cdot 10^{-1}$ &$8.709^{+0.413}_{-0.140}$$^{+0.143}_{-0.143}$$\cdot 10^{-1}$ &$8.761^{+0.183}_{-0.148}$$^{+0.120}_{-0.120}$$\cdot 10^{-1
}$          \\
 170 &$8.514^{+1.143}_{-0.937}$$^{+0.137}_{-0.137}$$\cdot 10^{-1}$ &$8.341^{+0.416}_{-0.136}$$^{+0.138}_{-0.138}$$\cdot 10^{-1}$ &$8.390^{+0.174}_{-0.125}$$^{+0.115}_{-0.115}$$\cdot 10^{-1
}$          \\
 175 &$8.159^{+1.123}_{-0.929}$$^{+0.133}_{-0.133}$$\cdot 10^{-1}$ &$8.006^{+0.388}_{-0.138}$$^{+0.133}_{-0.133}$$\cdot 10^{-1}$ &$8.043^{+0.168}_{-0.124}$$^{+0.111}_{-0.111}$$\cdot 10^{-1
}$          \\
 180 &$7.831^{+1.096}_{-0.908}$$^{+0.128}_{-0.128}$$\cdot 10^{-1}$ &$7.679^{+0.394}_{-0.138}$$^{+0.128}_{-0.128}$$\cdot 10^{-1}$ &$7.707^{+0.181}_{-0.108}$$^{+0.107}_{-0.107}$$\cdot 10^{-1
}$          \\
 185 &$7.506^{+1.087}_{-0.879}$$^{+0.123}_{-0.123}$$\cdot 10^{-1}$ &$7.377^{+0.376}_{-0.131}$$^{+0.124}_{-0.124}$$\cdot 10^{-1}$ &$7.411^{+0.156}_{-0.113}$$^{+0.103}_{-0.103}$$\cdot 10^{-1
}$          \\
 190 &$7.204^{+1.064}_{-0.856}$$^{+0.119}_{-0.119}$$\cdot 10^{-1}$ &$7.065^{+0.377}_{-0.115}$$^{+0.120}_{-0.120}$$\cdot 10^{-1}$ &$7.118^{+0.154}_{-0.127}$$^{+0.100}_{-0.100}$$\cdot 10^{-1
}$          \\
 195 &$6.904^{+1.057}_{-0.824}$$^{+0.115}_{-0.115}$$\cdot 10^{-1}$ &$6.789^{+0.371}_{-0.109}$$^{+0.116}_{-0.116}$$\cdot 10^{-1}$ &$6.827^{+0.153}_{-0.094}$$^{+0.096}_{-0.096}$$\cdot 10^{-1
}$          \\
 200 &$6.640^{+1.027}_{-0.820}$$^{+0.111}_{-0.111}$$\cdot 10^{-1}$ &$6.526^{+0.344}_{-0.110}$$^{+0.112}_{-0.112}$$\cdot 10^{-1}$ &$6.557^{+0.156}_{-0.090}$$^{+0.093}_{-0.093}$$\cdot 10^{-1
}$          \\
 210 &$6.128^{+0.990}_{-0.776}$$^{+0.104}_{-0.104}$$\cdot 10^{-1}$ &$6.037^{+0.327}_{-0.108}$$^{+0.104}_{-0.104}$$\cdot 10^{-1}$ &$6.058^{+0.150}_{-0.086}$$^{+0.087}_{-0.087}$$\cdot 10^{-1
}$          \\
 220 &$5.656^{+0.960}_{-0.729}$$^{+0.097}_{-0.097}$$\cdot 10^{-1}$ &$5.589^{+0.312}_{-0.103}$$^{+0.098}_{-0.098}$$\cdot 10^{-1}$ &$5.612^{+0.122}_{-0.074}$$^{+0.081}_{-0.081}$$\cdot 10^{-1
}$          \\
 230 &$5.243^{+0.915}_{-0.703}$$^{+0.091}_{-0.091}$$\cdot 10^{-1}$ &$5.180^{+0.297}_{-0.101}$$^{+0.092}_{-0.092}$$\cdot 10^{-1}$ &$5.202^{+0.115}_{-0.071}$$^{+0.076}_{-0.076}$$\cdot 10^{-1
}$          \\
 240 &$4.856^{+0.884}_{-0.665}$$^{+0.085}_{-0.085}$$\cdot 10^{-1}$ &$4.809^{+0.279}_{-0.101}$$^{+0.086}_{-0.086}$$\cdot 10^{-1}$ &$4.828^{+0.111}_{-0.072}$$^{+0.071}_{-0.071}$$\cdot 10^{-1
}$          \\
 250 &$4.512^{+0.843}_{-0.633}$$^{+0.080}_{-0.080}$$\cdot 10^{-1}$ &$4.465^{+0.273}_{-0.092}$$^{+0.080}_{-0.080}$$\cdot 10^{-1}$ &$4.487^{+0.104}_{-0.057}$$^{+0.067}_{-0.067}$$\cdot 10^{-1
}$          \\
 260 &$4.188^{+0.813}_{-0.598}$$^{+0.075}_{-0.075}$$\cdot 10^{-1}$ &$4.154^{+0.254}_{-0.095}$$^{+0.076}_{-0.076}$$\cdot 10^{-1}$ &$4.176^{+0.097}_{-0.055}$$^{+0.063}_{-0.063}$$\cdot 10^{-1
}$          \\
 270 &$3.898^{+0.777}_{-0.570}$$^{+0.070}_{-0.070}$$\cdot 10^{-1}$ &$3.872^{+0.240}_{-0.095}$$^{+0.071}_{-0.071}$$\cdot 10^{-1}$ &$3.888^{+0.091}_{-0.054}$$^{+0.059}_{-0.059}$$\cdot 10^{-1
}$          \\
 280 &$3.631^{+0.743}_{-0.543}$$^{+0.066}_{-0.066}$$\cdot 10^{-1}$ &$3.614^{+0.225}_{-0.097}$$^{+0.067}_{-0.067}$$\cdot 10^{-1}$ &$3.627^{+0.084}_{-0.044}$$^{+0.056}_{-0.056}$$\cdot 10^{-1
}$          \\
 290 &$3.386^{+0.714}_{-0.514}$$^{+0.062}_{-0.062}$$\cdot 10^{-1}$ &$3.371^{+0.216}_{-0.088}$$^{+0.063}_{-0.063}$$\cdot 10^{-1}$ &$3.384^{+0.082}_{-0.037}$$^{+0.053}_{-0.053}$$\cdot 10^{-1
}$          \\
 300 &$3.158^{+0.688}_{-0.489}$$^{+0.058}_{-0.058}$$\cdot 10^{-1}$ &$3.146^{+0.210}_{-0.085}$$^{+0.059}_{-0.059}$$\cdot 10^{-1}$ &$3.163^{+0.075}_{-0.033}$$^{+0.050}_{-0.050}$$\cdot 10^{-1
}$          \\
 320 &$2.762^{+0.629}_{-0.443}$$^{+0.052}_{-0.052}$$\cdot 10^{-1}$ &$2.760^{+0.185}_{-0.084}$$^{+0.053}_{-0.053}$$\cdot 10^{-1}$ &$2.770^{+0.069}_{-0.030}$$^{+0.044}_{-0.044}$$\cdot 10^{-1
}$          \\
 340 &$2.422^{+0.575}_{-0.399}$$^{+0.047}_{-0.047}$$\cdot 10^{-1}$ &$2.423^{+0.171}_{-0.077}$$^{+0.047}_{-0.047}$$\cdot 10^{-1}$ &$2.435^{+0.062}_{-0.023}$$^{+0.040}_{-0.040}$$\cdot 10^{-1
}$          \\
 360 &$2.130^{+0.528}_{-0.364}$$^{+0.042}_{-0.042}$$\cdot 10^{-1}$ &$2.138^{+0.153}_{-0.075}$$^{+0.043}_{-0.043}$$\cdot 10^{-1}$ &$2.145^{+0.055}_{-0.017}$$^{+0.036}_{-0.036}$$\cdot 10^{-1
}$          \\
 380 &$1.879^{+0.481}_{-0.329}$$^{+0.037}_{-0.037}$$\cdot 10^{-1}$ &$1.891^{+0.139}_{-0.070}$$^{+0.038}_{-0.038}$$\cdot 10^{-1}$ &$1.897^{+0.050}_{-0.013}$$^{+0.032}_{-0.032}$$\cdot 10^{-1
}$          \\
 400 &$1.664^{+0.442}_{-0.299}$$^{+0.034}_{-0.034}$$\cdot 10^{-1}$ &$1.677^{+0.126}_{-0.066}$$^{+0.035}_{-0.035}$$\cdot 10^{-1}$ &$1.682^{+0.045}_{-0.012}$$^{+0.029}_{-0.029}$$\cdot 10^{-1
}$          \\
 450 &$1.239^{+0.355}_{-0.236}$$^{+0.026}_{-0.026}$$\cdot 10^{-1}$ &$1.256^{+0.098}_{-0.056}$$^{+0.027}_{-0.027}$$\cdot 10^{-1}$ &$1.259^{+0.033}_{-0.010}$$^{+0.023}_{-0.023}$$\cdot 10^{-1
}$          \\
 500 &$9.335^{+2.889}_{-1.852}$$^{+0.205}_{-0.205}$$\cdot 10^{-2}$ &$9.529^{+0.777}_{-0.469}$$^{+0.212}_{-0.212}$$\cdot 10^{-2}$ &$9.547^{+0.264}_{-0.086}$$^{+0.182}_{-0.182}$$\cdot 10^{-2
}$          \\
 550 &$7.133^{+2.336}_{-1.490}$$^{+0.162}_{-0.162}$$\cdot 10^{-2}$ &$7.313^{+0.616}_{-0.396}$$^{+0.169}_{-0.169}$$\cdot 10^{-2}$ &$7.330^{+0.197}_{-0.088}$$^{+0.146}_{-0.146}$$\cdot 10^{-2
}$          \\
 600 &$5.495^{+1.908}_{-1.189}$$^{+0.130}_{-0.130}$$\cdot 10^{-2}$ &$5.665^{+0.495}_{-0.331}$$^{+0.135}_{-0.135}$$\cdot 10^{-2}$ &$5.682^{+0.159}_{-0.081}$$^{+0.118}_{-0.118}$$\cdot 10^{-2
}$          \\
 650 &$4.268^{+1.559}_{-0.955}$$^{+0.105}_{-0.105}$$\cdot 10^{-2}$ &$4.428^{+0.396}_{-0.277}$$^{+0.110}_{-0.110}$$\cdot 10^{-2}$ &$4.434^{+0.125}_{-0.064}$$^{+0.096}_{-0.096}$$\cdot 10^{-2
}$          \\
 700 &$3.341^{+1.276}_{-0.772}$$^{+0.085}_{-0.085}$$\cdot 10^{-2}$ &$3.483^{+0.322}_{-0.235}$$^{+0.089}_{-0.089}$$\cdot 10^{-2}$ &$3.490^{+0.096}_{-0.061}$$^{+0.079}_{-0.079}$$\cdot 10^{-2
}$          \\
 750 &$2.631^{+1.051}_{-0.625}$$^{+0.070}_{-0.070}$$\cdot 10^{-2}$ &$2.758^{+0.261}_{-0.198}$$^{+0.073}_{-0.073}$$\cdot 10^{-2}$ &$2.763^{+0.082}_{-0.055}$$^{+0.065}_{-0.065}$$\cdot 10^{-2
}$          \\
 800 &$2.085^{+0.863}_{-0.510}$$^{+0.057}_{-0.057}$$\cdot 10^{-2}$ &$2.195^{+0.213}_{-0.167}$$^{+0.060}_{-0.060}$$\cdot 10^{-2}$ &$2.200^{+0.063}_{-0.046}$$^{+0.054}_{-0.054}$$\cdot 10^{-2
}$          \\
 850 &$1.660^{+0.712}_{-0.418}$$^{+0.047}_{-0.047}$$\cdot 10^{-2}$ &$1.755^{+0.176}_{-0.141}$$^{+0.050}_{-0.050}$$\cdot 10^{-2}$ &$1.758^{+0.052}_{-0.040}$$^{+0.045}_{-0.045}$$\cdot 10^{-2
}$          \\
 900 &$1.325^{+0.590}_{-0.340}$$^{+0.039}_{-0.039}$$\cdot 10^{-2}$ &$1.410^{+0.142}_{-0.119}$$^{+0.042}_{-0.042}$$\cdot 10^{-2}$ &$1.410^{+0.042}_{-0.034}$$^{+0.037}_{-0.037}$$\cdot 10^{-2
}$          \\
 950 &$1.063^{+0.489}_{-0.279}$$^{+0.033}_{-0.033}$$\cdot 10^{-2}$ &$1.137^{+0.117}_{-0.101}$$^{+0.035}_{-0.035}$$\cdot 10^{-2}$ &$1.138^{+0.034}_{-0.030}$$^{+0.031}_{-0.031}$$\cdot 10^{-2
}$          \\
1000 &$8.551^{+4.076}_{-2.295}$$^{+0.272}_{-0.272}$$\cdot 10^{-3}$ &$9.192^{+0.979}_{-0.850}$$^{+0.290}_{-0.290}$$\cdot 10^{-3}$ &$9.197^{+0.281}_{-0.254}$$^{+0.264}_{-0.264}$$\cdot 10^{-3
}$          \\
\hline
\end{tabular}
\end{center}
\vspace*{-5mm}
\end{footnotesize}
\caption{
\label{tab:table-lhc7-nnn-Sc1}
Total VBF cross sections at the
LHC, $\sqrt S = 7 \tev$
at LO, NLO and NNLO in QCD. Errors shown are respectively scale and PDF uncertainities.  Scale uncertainities are evaluated by varying $\mu_r$ and $\mu_f$ in the interval
$\mu_r,\mu_f \in [Q/4,4Q]$.
The NNPDF2.1~\cite{Ball:2011uy} PDF set has been used.
Numbers are in pb.
}
\end{table}

%***************** 1st table ************************ 

\begin{table}[tb!]
\begin{footnotesize}
\begin{center}
\begin{tabular}{|c|c|c|c|}
\hline
$ \hmass \, [\gev] $ & $ \sigma_{LO} $ & $ \sigma_{NLO} $ & $ \sigma_{NNLO} $ \\
\hline
  90 &$6.204^{+0.000}_{-0.130}$$^{+0.050}_{-0.050}$$\phantom{\cdot 10^{-0}}$ &$5.927^{+0.127}_{-0.080}$$^{+0.049}_{-0.049}$$\phantom{\cdot 10^{-0}}$ &$5.982^{+0.039}_{-0.112}$$^{+0.049}_{-0.049}$$\phantom{\cdot 10^{-0}
}$          \\
  95 &$5.920^{+0.000}_{-0.078}$$^{+0.047}_{-0.047}$$\phantom{\cdot 10^{-0}}$ &$5.698^{+0.081}_{-0.078}$$^{+0.047}_{-0.047}$$\phantom{\cdot 10^{-0}}$ &$5.724^{+0.063}_{-0.082}$$^{+0.047}_{-0.047}$$\phantom{\cdot 10^{-0}
}$          \\
 100 &$5.694^{+0.000}_{-0.117}$$^{+0.045}_{-0.045}$$\phantom{\cdot 10^{-0}}$ &$5.457^{+0.098}_{-0.089}$$^{+0.044}_{-0.044}$$\phantom{\cdot 10^{-0}}$ &$5.503^{+0.031}_{-0.109}$$^{+0.044}_{-0.044}$$\phantom{\cdot 10^{-0}
}$          \\
 105 &$5.425^{+0.022}_{-0.095}$$^{+0.042}_{-0.042}$$\phantom{\cdot 10^{-0}}$ &$5.220^{+0.093}_{-0.079}$$^{+0.042}_{-0.042}$$\phantom{\cdot 10^{-0}}$ &$5.261^{+0.044}_{-0.084}$$^{+0.042}_{-0.042}$$\phantom{\cdot 10^{-0}
}$          \\
 110 &$5.215^{+0.026}_{-0.124}$$^{+0.040}_{-0.040}$$\phantom{\cdot 10^{-0}}$ &$5.011^{+0.100}_{-0.063}$$^{+0.040}_{-0.040}$$\phantom{\cdot 10^{-0}}$ &$5.020^{+0.076}_{-0.045}$$^{+0.040}_{-0.040}$$\phantom{\cdot 10^{-0}
}$          \\
 115 &$5.003^{+0.020}_{-0.128}$$^{+0.038}_{-0.038}$$\phantom{\cdot 10^{-0}}$ &$4.825^{+0.073}_{-0.079}$$^{+0.038}_{-0.038}$$\phantom{\cdot 10^{-0}}$ &$4.854^{+0.026}_{-0.108}$$^{+0.038}_{-0.038}$$\phantom{\cdot 10^{-0}
}$          \\
 120 &$4.796^{+0.054}_{-0.135}$$^{+0.036}_{-0.036}$$\phantom{\cdot 10^{-0}}$ &$4.639^{+0.093}_{-0.069}$$^{+0.036}_{-0.036}$$\phantom{\cdot 10^{-0}}$ &$4.661^{+0.038}_{-0.074}$$^{+0.037}_{-0.037}$$\phantom{\cdot 10^{-0}
}$          \\
 125 &$4.612^{+0.050}_{-0.154}$$^{+0.035}_{-0.035}$$\phantom{\cdot 10^{-0}}$ &$4.443^{+0.092}_{-0.069}$$^{+0.035}_{-0.035}$$\phantom{\cdot 10^{-0}}$ &$4.485^{+0.031}_{-0.076}$$^{+0.035}_{-0.035}$$\phantom{\cdot 10^{-0}
}$          \\
 130 &$4.430^{+0.077}_{-0.151}$$^{+0.033}_{-0.033}$$\phantom{\cdot 10^{-0}}$ &$4.269^{+0.098}_{-0.083}$$^{+0.033}_{-0.033}$$\phantom{\cdot 10^{-0}}$ &$4.308^{+0.034}_{-0.071}$$^{+0.033}_{-0.033}$$\phantom{\cdot 10^{-0}
}$          \\
 135 &$4.265^{+0.079}_{-0.154}$$^{+0.032}_{-0.032}$$\phantom{\cdot 10^{-0}}$ &$4.110^{+0.086}_{-0.052}$$^{+0.032}_{-0.032}$$\phantom{\cdot 10^{-0}}$ &$4.145^{+0.039}_{-0.062}$$^{+0.032}_{-0.032}$$\phantom{\cdot 10^{-0}
}$          \\
 140 &$4.105^{+0.099}_{-0.165}$$^{+0.030}_{-0.030}$$\phantom{\cdot 10^{-0}}$ &$3.960^{+0.089}_{-0.053}$$^{+0.030}_{-0.030}$$\phantom{\cdot 10^{-0}}$ &$3.983^{+0.045}_{-0.049}$$^{+0.031}_{-0.031}$$\phantom{\cdot 10^{-0}
}$          \\
 145 &$3.948^{+0.109}_{-0.166}$$^{+0.029}_{-0.029}$$\phantom{\cdot 10^{-0}}$ &$3.816^{+0.083}_{-0.052}$$^{+0.029}_{-0.029}$$\phantom{\cdot 10^{-0}}$ &$3.839^{+0.053}_{-0.063}$$^{+0.029}_{-0.029}$$\phantom{\cdot 10^{-0}
}$          \\
 150 &$3.802^{+0.123}_{-0.166}$$^{+0.028}_{-0.028}$$\phantom{\cdot 10^{-0}}$ &$3.674^{+0.092}_{-0.051}$$^{+0.028}_{-0.028}$$\phantom{\cdot 10^{-0}}$ &$3.709^{+0.031}_{-0.068}$$^{+0.028}_{-0.028}$$\phantom{\cdot 10^{-0}
}$          \\
 155 &$3.662^{+0.128}_{-0.179}$$^{+0.027}_{-0.027}$$\phantom{\cdot 10^{-0}}$ &$3.544^{+0.089}_{-0.046}$$^{+0.027}_{-0.027}$$\phantom{\cdot 10^{-0}}$ &$3.571^{+0.044}_{-0.048}$$^{+0.027}_{-0.027}$$\phantom{\cdot 10^{-0}
}$          \\
 160 &$3.527^{+0.140}_{-0.174}$$^{+0.025}_{-0.025}$$\phantom{\cdot 10^{-0}}$ &$3.421^{+0.088}_{-0.046}$$^{+0.026}_{-0.026}$$\phantom{\cdot 10^{-0}}$ &$3.444^{+0.039}_{-0.051}$$^{+0.026}_{-0.026}$$\phantom{\cdot 10^{-0}
}$          \\
 165 &$3.406^{+0.140}_{-0.171}$$^{+0.024}_{-0.024}$$\phantom{\cdot 10^{-0}}$ &$3.299^{+0.084}_{-0.047}$$^{+0.025}_{-0.025}$$\phantom{\cdot 10^{-0}}$ &$3.348^{+0.020}_{-0.069}$$^{+0.025}_{-0.025}$$\phantom{\cdot 10^{-0}
}$          \\
 170 &$3.284^{+0.154}_{-0.172}$$^{+0.023}_{-0.023}$$\phantom{\cdot 10^{-0}}$ &$3.191^{+0.065}_{-0.048}$$^{+0.024}_{-0.024}$$\phantom{\cdot 10^{-0}}$ &$3.211^{+0.033}_{-0.049}$$^{+0.024}_{-0.024}$$\phantom{\cdot 10^{-0}
}$          \\
 175 &$3.173^{+0.150}_{-0.182}$$^{+0.023}_{-0.023}$$\phantom{\cdot 10^{-0}}$ &$3.078^{+0.082}_{-0.046}$$^{+0.023}_{-0.023}$$\phantom{\cdot 10^{-0}}$ &$3.103^{+0.032}_{-0.048}$$^{+0.023}_{-0.023}$$\phantom{\cdot 10^{-0}
}$          \\
 180 &$3.066^{+0.147}_{-0.175}$$^{+0.022}_{-0.022}$$\phantom{\cdot 10^{-0}}$ &$2.972^{+0.084}_{-0.029}$$^{+0.022}_{-0.022}$$\phantom{\cdot 10^{-0}}$ &$3.000^{+0.038}_{-0.043}$$^{+0.022}_{-0.022}$$\phantom{\cdot 10^{-0}
}$          \\
 185 &$2.963^{+0.160}_{-0.176}$$^{+0.021}_{-0.021}$$\phantom{\cdot 10^{-0}}$ &$2.871^{+0.077}_{-0.036}$$^{+0.021}_{-0.021}$$\phantom{\cdot 10^{-0}}$ &$2.905^{+0.027}_{-0.041}$$^{+0.022}_{-0.022}$$\phantom{\cdot 10^{-0}
}$          \\
 190 &$2.868^{+0.163}_{-0.183}$$^{+0.020}_{-0.020}$$\phantom{\cdot 10^{-0}}$ &$2.776^{+0.080}_{-0.040}$$^{+0.020}_{-0.020}$$\phantom{\cdot 10^{-0}}$ &$2.802^{+0.042}_{-0.034}$$^{+0.021}_{-0.021}$$\phantom{\cdot 10^{-0}
}$          \\
 195 &$2.767^{+0.171}_{-0.177}$$^{+0.020}_{-0.020}$$\phantom{\cdot 10^{-0}}$ &$2.686^{+0.082}_{-0.036}$$^{+0.020}_{-0.020}$$\phantom{\cdot 10^{-0}}$ &$2.715^{+0.030}_{-0.037}$$^{+0.020}_{-0.020}$$\phantom{\cdot 10^{-0}
}$          \\
 200 &$2.680^{+0.167}_{-0.175}$$^{+0.019}_{-0.019}$$\phantom{\cdot 10^{-0}}$ &$2.604^{+0.080}_{-0.036}$$^{+0.019}_{-0.019}$$\phantom{\cdot 10^{-0}}$ &$2.626^{+0.032}_{-0.035}$$^{+0.019}_{-0.019}$$\phantom{\cdot 10^{-0}
}$          \\
 210 &$2.510^{+0.175}_{-0.174}$$^{+0.018}_{-0.018}$$\phantom{\cdot 10^{-0}}$ &$2.438^{+0.080}_{-0.035}$$^{+0.018}_{-0.018}$$\phantom{\cdot 10^{-0}}$ &$2.467^{+0.029}_{-0.038}$$^{+0.018}_{-0.018}$$\phantom{\cdot 10^{-0}
}$          \\
 220 &$2.355^{+0.174}_{-0.172}$$^{+0.016}_{-0.016}$$\phantom{\cdot 10^{-0}}$ &$2.290^{+0.074}_{-0.037}$$^{+0.017}_{-0.017}$$\phantom{\cdot 10^{-0}}$ &$2.316^{+0.030}_{-0.029}$$^{+0.017}_{-0.017}$$\phantom{\cdot 10^{-0}
}$          \\
 230 &$2.213^{+0.176}_{-0.170}$$^{+0.015}_{-0.015}$$\phantom{\cdot 10^{-0}}$ &$2.152^{+0.079}_{-0.028}$$^{+0.016}_{-0.016}$$\phantom{\cdot 10^{-0}}$ &$2.180^{+0.037}_{-0.031}$$^{+0.016}_{-0.016}$$\phantom{\cdot 10^{-0}
}$          \\
 240 &$2.081^{+0.178}_{-0.165}$$^{+0.015}_{-0.015}$$\phantom{\cdot 10^{-0}}$ &$2.028^{+0.069}_{-0.031}$$^{+0.015}_{-0.015}$$\phantom{\cdot 10^{-0}}$ &$2.051^{+0.026}_{-0.021}$$^{+0.015}_{-0.015}$$\phantom{\cdot 10^{-0}
}$          \\
 250 &$1.961^{+0.177}_{-0.164}$$^{+0.014}_{-0.014}$$\phantom{\cdot 10^{-0}}$ &$1.909^{+0.069}_{-0.021}$$^{+0.014}_{-0.014}$$\phantom{\cdot 10^{-0}}$ &$1.935^{+0.026}_{-0.026}$$^{+0.014}_{-0.014}$$\phantom{\cdot 10^{-0}
}$          \\
 260 &$1.847^{+0.177}_{-0.158}$$^{+0.013}_{-0.013}$$\phantom{\cdot 10^{-0}}$ &$1.801^{+0.069}_{-0.025}$$^{+0.013}_{-0.013}$$\phantom{\cdot 10^{-0}}$ &$1.826^{+0.027}_{-0.027}$$^{+0.013}_{-0.013}$$\phantom{\cdot 10^{-0}
}$          \\
 270 &$1.742^{+0.176}_{-0.155}$$^{+0.012}_{-0.012}$$\phantom{\cdot 10^{-0}}$ &$1.701^{+0.070}_{-0.023}$$^{+0.012}_{-0.012}$$\phantom{\cdot 10^{-0}}$ &$1.724^{+0.026}_{-0.018}$$^{+0.013}_{-0.013}$$\phantom{\cdot 10^{-0}
}$          \\
 280 &$1.645^{+0.174}_{-0.151}$$^{+0.012}_{-0.012}$$\phantom{\cdot 10^{-0}}$ &$1.611^{+0.063}_{-0.026}$$^{+0.012}_{-0.012}$$\phantom{\cdot 10^{-0}}$ &$1.634^{+0.023}_{-0.025}$$^{+0.012}_{-0.012}$$\phantom{\cdot 10^{-0}
}$          \\
 290 &$1.558^{+0.171}_{-0.147}$$^{+0.011}_{-0.011}$$\phantom{\cdot 10^{-0}}$ &$1.523^{+0.063}_{-0.022}$$^{+0.011}_{-0.011}$$\phantom{\cdot 10^{-0}}$ &$1.543^{+0.024}_{-0.017}$$^{+0.011}_{-0.011}$$\phantom{\cdot 10^{-0}
}$          \\
 300 &$1.473^{+0.170}_{-0.142}$$^{+0.010}_{-0.010}$$\phantom{\cdot 10^{-0}}$ &$1.443^{+0.059}_{-0.019}$$^{+0.011}_{-0.011}$$\phantom{\cdot 10^{-0}}$ &$1.463^{+0.025}_{-0.018}$$^{+0.011}_{-0.011}$$\phantom{\cdot 10^{-0}
}$          \\
 320 &$1.325^{+0.164}_{-0.133}$$^{+0.009}_{-0.009}$$\phantom{\cdot 10^{-0}}$ &$1.298^{+0.060}_{-0.019}$$^{+0.010}_{-0.010}$$\phantom{\cdot 10^{-0}}$ &$1.317^{+0.020}_{-0.015}$$^{+0.010}_{-0.010}$$\phantom{\cdot 10^{-0}
}$          \\
 340 &$1.194^{+0.157}_{-0.127}$$^{+0.009}_{-0.009}$$\phantom{\cdot 10^{-0}}$ &$1.176^{+0.054}_{-0.021}$$^{+0.009}_{-0.009}$$\phantom{\cdot 10^{-0}}$ &$1.191^{+0.019}_{-0.015}$$^{+0.009}_{-0.009}$$\phantom{\cdot 10^{-0}
}$          \\
 360 &$1.080^{+0.151}_{-0.121}$$^{+0.008}_{-0.008}$$\phantom{\cdot 10^{-0}}$ &$1.062^{+0.054}_{-0.021}$$^{+0.008}_{-0.008}$$\phantom{\cdot 10^{-0}}$ &$1.079^{+0.018}_{-0.012}$$^{+0.008}_{-0.008}$$\phantom{\cdot 10^{-0}
}$          \\
 380 &$9.793^{+1.448}_{-1.136}$$^{+0.072}_{-0.072}$$\cdot 10^{-1}$ &$9.637^{+0.505}_{-0.165}$$^{+0.073}_{-0.073}$$\cdot 10^{-1}$ &$9.810^{+0.160}_{-0.120}$$^{+0.075}_{-0.075}$$\cdot 10^{-1
}$          \\
 400 &$8.904^{+1.384}_{-1.063}$$^{+0.066}_{-0.066}$$\cdot 10^{-1}$ &$8.798^{+0.461}_{-0.158}$$^{+0.067}_{-0.067}$$\cdot 10^{-1}$ &$8.926^{+0.169}_{-0.083}$$^{+0.069}_{-0.069}$$\cdot 10^{-1
}$          \\
 450 &$7.100^{+1.229}_{-0.925}$$^{+0.055}_{-0.055}$$\cdot 10^{-1}$ &$7.032^{+0.384}_{-0.138}$$^{+0.055}_{-0.055}$$\cdot 10^{-1}$ &$7.146^{+0.132}_{-0.075}$$^{+0.056}_{-0.056}$$\cdot 10^{-1
}$          \\
 500 &$5.738^{+1.084}_{-0.801}$$^{+0.046}_{-0.046}$$\cdot 10^{-1}$ &$5.695^{+0.332}_{-0.122}$$^{+0.046}_{-0.046}$$\cdot 10^{-1}$ &$5.794^{+0.117}_{-0.043}$$^{+0.047}_{-0.047}$$\cdot 10^{-1
}$          \\
 550 &$4.683^{+0.955}_{-0.687}$$^{+0.039}_{-0.039}$$\cdot 10^{-1}$ &$4.664^{+0.292}_{-0.112}$$^{+0.039}_{-0.039}$$\cdot 10^{-1}$ &$4.754^{+0.089}_{-0.036}$$^{+0.040}_{-0.040}$$\cdot 10^{-1
}$          \\
 600 &$3.860^{+0.842}_{-0.595}$$^{+0.033}_{-0.033}$$\cdot 10^{-1}$ &$3.863^{+0.243}_{-0.112}$$^{+0.034}_{-0.034}$$\cdot 10^{-1}$ &$3.934^{+0.083}_{-0.027}$$^{+0.034}_{-0.034}$$\cdot 10^{-1
}$          \\
 650 &$3.209^{+0.743}_{-0.516}$$^{+0.029}_{-0.029}$$\cdot 10^{-1}$ &$3.222^{+0.212}_{-0.108}$$^{+0.029}_{-0.029}$$\cdot 10^{-1}$ &$3.286^{+0.064}_{-0.025}$$^{+0.029}_{-0.029}$$\cdot 10^{-1
}$          \\
 700 &$2.687^{+0.660}_{-0.449}$$^{+0.025}_{-0.025}$$\cdot 10^{-1}$ &$2.702^{+0.185}_{-0.090}$$^{+0.025}_{-0.025}$$\cdot 10^{-1}$ &$2.759^{+0.061}_{-0.014}$$^{+0.025}_{-0.025}$$\cdot 10^{-1
}$          \\
 750 &$2.265^{+0.583}_{-0.390}$$^{+0.022}_{-0.022}$$\cdot 10^{-1}$ &$2.285^{+0.164}_{-0.085}$$^{+0.022}_{-0.022}$$\cdot 10^{-1}$ &$2.335^{+0.053}_{-0.015}$$^{+0.022}_{-0.022}$$\cdot 10^{-1
}$          \\
 800 &$1.922^{+0.514}_{-0.345}$$^{+0.019}_{-0.019}$$\cdot 10^{-1}$ &$1.943^{+0.147}_{-0.079}$$^{+0.019}_{-0.019}$$\cdot 10^{-1}$ &$1.982^{+0.051}_{-0.010}$$^{+0.019}_{-0.019}$$\cdot 10^{-1
}$          \\
 850 &$1.637^{+0.458}_{-0.301}$$^{+0.017}_{-0.017}$$\cdot 10^{-1}$ &$1.661^{+0.127}_{-0.070}$$^{+0.017}_{-0.017}$$\cdot 10^{-1}$ &$1.696^{+0.040}_{-0.010}$$^{+0.017}_{-0.017}$$\cdot 10^{-1
}$          \\
 900 &$1.402^{+0.407}_{-0.265}$$^{+0.015}_{-0.015}$$\cdot 10^{-1}$ &$1.424^{+0.114}_{-0.062}$$^{+0.015}_{-0.015}$$\cdot 10^{-1}$ &$1.460^{+0.032}_{-0.013}$$^{+0.015}_{-0.015}$$\cdot 10^{-1
}$          \\
 950 &$1.205^{+0.362}_{-0.233}$$^{+0.013}_{-0.013}$$\cdot 10^{-1}$ &$1.229^{+0.097}_{-0.058}$$^{+0.013}_{-0.013}$$\cdot 10^{-1}$ &$1.260^{+0.030}_{-0.012}$$^{+0.013}_{-0.013}$$\cdot 10^{-1
}$          \\
1000 &$1.040^{+0.323}_{-0.205}$$^{+0.012}_{-0.012}$$\cdot 10^{-1}$ &$1.063^{+0.086}_{-0.054}$$^{+0.012}_{-0.012}$$\cdot 10^{-1}$ &$1.091^{+0.024}_{-0.012}$$^{+0.012}_{-0.012}$$\cdot 10^{-1
}$          \\
\hline
\end{tabular}
\end{center}
\vspace*{-5mm}
\end{footnotesize}
\caption{
\small
\label{tab:table-lhc14-a09-Sc1}
Total VBF cross sections at the
LHC, $\sqrt S = 14 \tev$
at LO, NLO and NNLO in QCD. Errors shown are respectively scale and PDF uncertainities.  Scale uncertainities are evaluated by varying $\mu_r$ and $\mu_f$ in the interval
$\mu_r,\mu_f \in [Q/4,4Q]$.
The ABKM~\cite{Alekhin:2009ni} PDF set has been used.
Numbers are in pb.
}
\end{table}

%***************** 1st table ************************ 

\begin{table}[tb!]
\begin{footnotesize}
\begin{center}
\begin{tabular}{|c|c|c|c|}
\hline
$ \hmass \, [\gev] $ & $ \sigma_{LO} $ & $ \sigma_{NLO} $ & $ \sigma_{NNLO} $ \\
\hline
  90 &$6.192^{+0.000}_{-0.130}$$^{+0.104}_{-0.104}$$\phantom{\cdot 10^{-0}}$ &$5.955^{+0.137}_{-0.076}$$^{+0.103}_{-0.103}$$\phantom{\cdot 10^{-0}}$ &$5.897^{+0.073}_{-0.103}$$^{+0.102}_{-0.102}$$\phantom{\cdot 10^{-0}
}$          \\
  95 &$5.930^{+0.000}_{-0.097}$$^{+0.101}_{-0.101}$$\phantom{\cdot 10^{-0}}$ &$5.701^{+0.108}_{-0.099}$$^{+0.099}_{-0.099}$$\phantom{\cdot 10^{-0}}$ &$5.665^{+0.054}_{-0.100}$$^{+0.099}_{-0.099}$$\phantom{\cdot 10^{-0}
}$          \\
 100 &$5.687^{+0.000}_{-0.074}$$^{+0.097}_{-0.097}$$\phantom{\cdot 10^{-0}}$ &$5.471^{+0.103}_{-0.093}$$^{+0.096}_{-0.096}$$\phantom{\cdot 10^{-0}}$ &$5.448^{+0.025}_{-0.136}$$^{+0.096}_{-0.096}$$\phantom{\cdot 10^{-0}
}$          \\
 105 &$5.438^{+0.013}_{-0.074}$$^{+0.093}_{-0.093}$$\phantom{\cdot 10^{-0}}$ &$5.237^{+0.104}_{-0.059}$$^{+0.092}_{-0.092}$$\phantom{\cdot 10^{-0}}$ &$5.196^{+0.055}_{-0.088}$$^{+0.092}_{-0.092}$$\phantom{\cdot 10^{-0}
}$          \\
 110 &$5.225^{+0.012}_{-0.106}$$^{+0.090}_{-0.090}$$\phantom{\cdot 10^{-0}}$ &$5.026^{+0.072}_{-0.060}$$^{+0.089}_{-0.089}$$\phantom{\cdot 10^{-0}}$ &$4.988^{+0.060}_{-0.081}$$^{+0.089}_{-0.089}$$\phantom{\cdot 10^{-0}
}$          \\
 115 &$5.011^{+0.029}_{-0.096}$$^{+0.087}_{-0.087}$$\phantom{\cdot 10^{-0}}$ &$4.820^{+0.097}_{-0.053}$$^{+0.086}_{-0.086}$$\phantom{\cdot 10^{-0}}$ &$4.786^{+0.039}_{-0.091}$$^{+0.086}_{-0.086}$$\phantom{\cdot 10^{-0}
}$          \\
 120 &$4.816^{+0.031}_{-0.104}$$^{+0.084}_{-0.084}$$\phantom{\cdot 10^{-0}}$ &$4.643^{+0.079}_{-0.074}$$^{+0.083}_{-0.083}$$\phantom{\cdot 10^{-0}}$ &$4.602^{+0.052}_{-0.073}$$^{+0.083}_{-0.083}$$\phantom{\cdot 10^{-0}
}$          \\
 125 &$4.631^{+0.033}_{-0.124}$$^{+0.081}_{-0.081}$$\phantom{\cdot 10^{-0}}$ &$4.450^{+0.095}_{-0.047}$$^{+0.080}_{-0.080}$$\phantom{\cdot 10^{-0}}$ &$4.427^{+0.045}_{-0.077}$$^{+0.080}_{-0.080}$$\phantom{\cdot 10^{-0}
}$          \\
 130 &$4.452^{+0.065}_{-0.133}$$^{+0.079}_{-0.079}$$\phantom{\cdot 10^{-0}}$ &$4.292^{+0.086}_{-0.067}$$^{+0.078}_{-0.078}$$\phantom{\cdot 10^{-0}}$ &$4.264^{+0.037}_{-0.080}$$^{+0.077}_{-0.077}$$\phantom{\cdot 10^{-0}
}$          \\
 135 &$4.283^{+0.078}_{-0.137}$$^{+0.076}_{-0.076}$$\phantom{\cdot 10^{-0}}$ &$4.135^{+0.089}_{-0.056}$$^{+0.075}_{-0.075}$$\phantom{\cdot 10^{-0}}$ &$4.099^{+0.044}_{-0.066}$$^{+0.075}_{-0.075}$$\phantom{\cdot 10^{-0}
}$          \\
 140 &$4.122^{+0.083}_{-0.142}$$^{+0.073}_{-0.073}$$\phantom{\cdot 10^{-0}}$ &$3.980^{+0.088}_{-0.049}$$^{+0.073}_{-0.073}$$\phantom{\cdot 10^{-0}}$ &$3.954^{+0.026}_{-0.077}$$^{+0.072}_{-0.072}$$\phantom{\cdot 10^{-0}
}$          \\
 145 &$3.973^{+0.097}_{-0.148}$$^{+0.071}_{-0.071}$$\phantom{\cdot 10^{-0}}$ &$3.838^{+0.067}_{-0.047}$$^{+0.070}_{-0.070}$$\phantom{\cdot 10^{-0}}$ &$3.823^{+0.020}_{-0.080}$$^{+0.070}_{-0.070}$$\phantom{\cdot 10^{-0}
}$          \\
 150 &$3.830^{+0.099}_{-0.146}$$^{+0.069}_{-0.069}$$\phantom{\cdot 10^{-0}}$ &$3.704^{+0.073}_{-0.050}$$^{+0.068}_{-0.068}$$\phantom{\cdot 10^{-0}}$ &$3.670^{+0.031}_{-0.065}$$^{+0.068}_{-0.068}$$\phantom{\cdot 10^{-0}
}$          \\
 155 &$3.687^{+0.104}_{-0.152}$$^{+0.067}_{-0.067}$$\phantom{\cdot 10^{-0}}$ &$3.565^{+0.071}_{-0.046}$$^{+0.066}_{-0.066}$$\phantom{\cdot 10^{-0}}$ &$3.540^{+0.033}_{-0.067}$$^{+0.066}_{-0.066}$$\phantom{\cdot 10^{-0}
}$          \\
 160 &$3.548^{+0.145}_{-0.149}$$^{+0.064}_{-0.064}$$\phantom{\cdot 10^{-0}}$ &$3.444^{+0.084}_{-0.053}$$^{+0.064}_{-0.064}$$\phantom{\cdot 10^{-0}}$ &$3.413^{+0.033}_{-0.056}$$^{+0.064}_{-0.064}$$\phantom{\cdot 10^{-0}
}$          \\
 165 &$3.432^{+0.123}_{-0.156}$$^{+0.063}_{-0.063}$$\phantom{\cdot 10^{-0}}$ &$3.333^{+0.064}_{-0.057}$$^{+0.062}_{-0.062}$$\phantom{\cdot 10^{-0}}$ &$3.288^{+0.043}_{-0.051}$$^{+0.062}_{-0.062}$$\phantom{\cdot 10^{-0}
}$          \\
 170 &$3.315^{+0.128}_{-0.163}$$^{+0.061}_{-0.061}$$\phantom{\cdot 10^{-0}}$ &$3.205^{+0.078}_{-0.046}$$^{+0.060}_{-0.060}$$\phantom{\cdot 10^{-0}}$ &$3.181^{+0.037}_{-0.050}$$^{+0.060}_{-0.060}$$\phantom{\cdot 10^{-0}
}$          \\
 175 &$3.201^{+0.135}_{-0.158}$$^{+0.059}_{-0.059}$$\phantom{\cdot 10^{-0}}$ &$3.098^{+0.079}_{-0.041}$$^{+0.058}_{-0.058}$$\phantom{\cdot 10^{-0}}$ &$3.079^{+0.040}_{-0.057}$$^{+0.058}_{-0.058}$$\phantom{\cdot 10^{-0}
}$          \\
 180 &$3.090^{+0.139}_{-0.163}$$^{+0.057}_{-0.057}$$\phantom{\cdot 10^{-0}}$ &$2.994^{+0.076}_{-0.047}$$^{+0.057}_{-0.057}$$\phantom{\cdot 10^{-0}}$ &$2.971^{+0.035}_{-0.045}$$^{+0.056}_{-0.056}$$\phantom{\cdot 10^{-0}
}$          \\
 185 &$2.982^{+0.150}_{-0.159}$$^{+0.055}_{-0.055}$$\phantom{\cdot 10^{-0}}$ &$2.899^{+0.073}_{-0.043}$$^{+0.055}_{-0.055}$$\phantom{\cdot 10^{-0}}$ &$2.875^{+0.028}_{-0.049}$$^{+0.055}_{-0.055}$$\phantom{\cdot 10^{-0}
}$          \\
 190 &$2.887^{+0.152}_{-0.160}$$^{+0.054}_{-0.054}$$\phantom{\cdot 10^{-0}}$ &$2.799^{+0.076}_{-0.040}$$^{+0.053}_{-0.053}$$\phantom{\cdot 10^{-0}}$ &$2.786^{+0.023}_{-0.047}$$^{+0.053}_{-0.053}$$\phantom{\cdot 10^{-0}
}$          \\
 195 &$2.792^{+0.149}_{-0.162}$$^{+0.052}_{-0.052}$$\phantom{\cdot 10^{-0}}$ &$2.707^{+0.077}_{-0.036}$$^{+0.052}_{-0.052}$$\phantom{\cdot 10^{-0}}$ &$2.692^{+0.032}_{-0.046}$$^{+0.052}_{-0.052}$$\phantom{\cdot 10^{-0}
}$          \\
 200 &$2.701^{+0.157}_{-0.161}$$^{+0.051}_{-0.051}$$\phantom{\cdot 10^{-0}}$ &$2.623^{+0.070}_{-0.035}$$^{+0.050}_{-0.050}$$\phantom{\cdot 10^{-0}}$ &$2.608^{+0.026}_{-0.039}$$^{+0.050}_{-0.050}$$\phantom{\cdot 10^{-0}
}$          \\
 210 &$2.531^{+0.165}_{-0.160}$$^{+0.048}_{-0.048}$$\phantom{\cdot 10^{-0}}$ &$2.458^{+0.071}_{-0.032}$$^{+0.047}_{-0.047}$$\phantom{\cdot 10^{-0}}$ &$2.449^{+0.026}_{-0.031}$$^{+0.048}_{-0.048}$$\phantom{\cdot 10^{-0}
}$          \\
 220 &$2.376^{+0.161}_{-0.162}$$^{+0.045}_{-0.045}$$\phantom{\cdot 10^{-0}}$ &$2.313^{+0.071}_{-0.031}$$^{+0.045}_{-0.045}$$\phantom{\cdot 10^{-0}}$ &$2.294^{+0.035}_{-0.042}$$^{+0.045}_{-0.045}$$\phantom{\cdot 10^{-0}
}$          \\
 230 &$2.229^{+0.172}_{-0.154}$$^{+0.043}_{-0.043}$$\phantom{\cdot 10^{-0}}$ &$2.175^{+0.065}_{-0.030}$$^{+0.043}_{-0.043}$$\phantom{\cdot 10^{-0}}$ &$2.161^{+0.029}_{-0.036}$$^{+0.042}_{-0.042}$$\phantom{\cdot 10^{-0}
}$          \\
 240 &$2.099^{+0.171}_{-0.150}$$^{+0.040}_{-0.040}$$\phantom{\cdot 10^{-0}}$ &$2.048^{+0.065}_{-0.031}$$^{+0.040}_{-0.040}$$\phantom{\cdot 10^{-0}}$ &$2.034^{+0.029}_{-0.030}$$^{+0.040}_{-0.040}$$\phantom{\cdot 10^{-0}
}$          \\
 250 &$1.984^{+0.161}_{-0.156}$$^{+0.038}_{-0.038}$$\phantom{\cdot 10^{-0}}$ &$1.929^{+0.067}_{-0.028}$$^{+0.038}_{-0.038}$$\phantom{\cdot 10^{-0}}$ &$1.919^{+0.026}_{-0.024}$$^{+0.038}_{-0.038}$$\phantom{\cdot 10^{-0}
}$          \\
 260 &$1.869^{+0.167}_{-0.152}$$^{+0.036}_{-0.036}$$\phantom{\cdot 10^{-0}}$ &$1.821^{+0.066}_{-0.024}$$^{+0.036}_{-0.036}$$\phantom{\cdot 10^{-0}}$ &$1.811^{+0.026}_{-0.032}$$^{+0.036}_{-0.036}$$\phantom{\cdot 10^{-0}
}$          \\
 270 &$1.764^{+0.166}_{-0.149}$$^{+0.034}_{-0.034}$$\phantom{\cdot 10^{-0}}$ &$1.721^{+0.064}_{-0.022}$$^{+0.035}_{-0.035}$$\phantom{\cdot 10^{-0}}$ &$1.710^{+0.026}_{-0.024}$$^{+0.034}_{-0.034}$$\phantom{\cdot 10^{-0}
}$          \\
 280 &$1.667^{+0.165}_{-0.142}$$^{+0.033}_{-0.033}$$\phantom{\cdot 10^{-0}}$ &$1.629^{+0.060}_{-0.023}$$^{+0.033}_{-0.033}$$\phantom{\cdot 10^{-0}}$ &$1.618^{+0.025}_{-0.022}$$^{+0.033}_{-0.033}$$\phantom{\cdot 10^{-0}
}$          \\
 290 &$1.577^{+0.165}_{-0.139}$$^{+0.031}_{-0.031}$$\phantom{\cdot 10^{-0}}$ &$1.541^{+0.059}_{-0.023}$$^{+0.031}_{-0.031}$$\phantom{\cdot 10^{-0}}$ &$1.533^{+0.025}_{-0.022}$$^{+0.031}_{-0.031}$$\phantom{\cdot 10^{-0}
}$          \\
 300 &$1.493^{+0.162}_{-0.137}$$^{+0.030}_{-0.030}$$\phantom{\cdot 10^{-0}}$ &$1.460^{+0.062}_{-0.020}$$^{+0.030}_{-0.030}$$\phantom{\cdot 10^{-0}}$ &$1.454^{+0.022}_{-0.021}$$^{+0.030}_{-0.030}$$\phantom{\cdot 10^{-0}
}$          \\
 320 &$1.344^{+0.155}_{-0.129}$$^{+0.027}_{-0.027}$$\phantom{\cdot 10^{-0}}$ &$1.315^{+0.058}_{-0.019}$$^{+0.027}_{-0.027}$$\phantom{\cdot 10^{-0}}$ &$1.308^{+0.024}_{-0.015}$$^{+0.027}_{-0.027}$$\phantom{\cdot 10^{-0}
}$          \\
 340 &$1.213^{+0.150}_{-0.123}$$^{+0.025}_{-0.025}$$\phantom{\cdot 10^{-0}}$ &$1.191^{+0.052}_{-0.021}$$^{+0.025}_{-0.025}$$\phantom{\cdot 10^{-0}}$ &$1.183^{+0.023}_{-0.016}$$^{+0.025}_{-0.025}$$\phantom{\cdot 10^{-0}
}$          \\
 360 &$1.097^{+0.144}_{-0.116}$$^{+0.022}_{-0.022}$$\phantom{\cdot 10^{-0}}$ &$1.077^{+0.051}_{-0.016}$$^{+0.023}_{-0.023}$$\phantom{\cdot 10^{-0}}$ &$1.073^{+0.018}_{-0.014}$$^{+0.023}_{-0.023}$$\phantom{\cdot 10^{-0}
}$          \\
 380 &$9.950^{+1.400}_{-1.082}$$^{+0.205}_{-0.205}$$\cdot 10^{-1}$ &$9.791^{+0.483}_{-0.149}$$^{+0.207}_{-0.207}$$\cdot 10^{-1}$ &$9.741^{+0.203}_{-0.108}$$^{+0.206}_{-0.206}$$\cdot 10^{-1
}$          \\
 400 &$9.073^{+1.319}_{-1.043}$$^{+0.188}_{-0.188}$$\cdot 10^{-1}$ &$8.928^{+0.448}_{-0.143}$$^{+0.190}_{-0.190}$$\cdot 10^{-1}$ &$8.891^{+0.168}_{-0.103}$$^{+0.190}_{-0.190}$$\cdot 10^{-1
}$          \\
 450 &$7.245^{+1.181}_{-0.896}$$^{+0.153}_{-0.153}$$\cdot 10^{-1}$ &$7.145^{+0.402}_{-0.108}$$^{+0.155}_{-0.155}$$\cdot 10^{-1}$ &$7.123^{+0.150}_{-0.081}$$^{+0.155}_{-0.155}$$\cdot 10^{-1
}$          \\
 500 &$5.859^{+1.053}_{-0.782}$$^{+0.126}_{-0.126}$$\cdot 10^{-1}$ &$5.800^{+0.338}_{-0.113}$$^{+0.128}_{-0.128}$$\cdot 10^{-1}$ &$5.787^{+0.120}_{-0.061}$$^{+0.128}_{-0.128}$$\cdot 10^{-1
}$          \\
 550 &$4.800^{+0.919}_{-0.681}$$^{+0.105}_{-0.105}$$\cdot 10^{-1}$ &$4.764^{+0.289}_{-0.106}$$^{+0.107}_{-0.107}$$\cdot 10^{-1}$ &$4.749^{+0.098}_{-0.052}$$^{+0.107}_{-0.107}$$\cdot 10^{-1
}$          \\
 600 &$3.954^{+0.823}_{-0.584}$$^{+0.088}_{-0.088}$$\cdot 10^{-1}$ &$3.940^{+0.254}_{-0.098}$$^{+0.090}_{-0.090}$$\cdot 10^{-1}$ &$3.931^{+0.091}_{-0.039}$$^{+0.090}_{-0.090}$$\cdot 10^{-1
}$          \\
 650 &$3.299^{+0.726}_{-0.518}$$^{+0.074}_{-0.074}$$\cdot 10^{-1}$ &$3.290^{+0.221}_{-0.091}$$^{+0.076}_{-0.076}$$\cdot 10^{-1}$ &$3.287^{+0.072}_{-0.030}$$^{+0.076}_{-0.076}$$\cdot 10^{-1
}$          \\
 700 &$2.764^{+0.647}_{-0.449}$$^{+0.063}_{-0.063}$$\cdot 10^{-1}$ &$2.775^{+0.184}_{-0.091}$$^{+0.065}_{-0.065}$$\cdot 10^{-1}$ &$2.765^{+0.061}_{-0.023}$$^{+0.065}_{-0.065}$$\cdot 10^{-1
}$          \\
 750 &$2.332^{+0.574}_{-0.392}$$^{+0.054}_{-0.054}$$\cdot 10^{-1}$ &$2.344^{+0.166}_{-0.080}$$^{+0.055}_{-0.055}$$\cdot 10^{-1}$ &$2.340^{+0.057}_{-0.014}$$^{+0.055}_{-0.055}$$\cdot 10^{-1
}$          \\
 800 &$1.981^{+0.509}_{-0.346}$$^{+0.046}_{-0.046}$$\cdot 10^{-1}$ &$1.995^{+0.147}_{-0.072}$$^{+0.048}_{-0.048}$$\cdot 10^{-1}$ &$1.995^{+0.045}_{-0.015}$$^{+0.048}_{-0.048}$$\cdot 10^{-1
}$          \\
 850 &$1.689^{+0.456}_{-0.302}$$^{+0.040}_{-0.040}$$\cdot 10^{-1}$ &$1.709^{+0.129}_{-0.069}$$^{+0.041}_{-0.041}$$\cdot 10^{-1}$ &$1.707^{+0.040}_{-0.010}$$^{+0.041}_{-0.041}$$\cdot 10^{-1
}$          \\
 900 &$1.447^{+0.405}_{-0.267}$$^{+0.035}_{-0.035}$$\cdot 10^{-1}$ &$1.468^{+0.114}_{-0.061}$$^{+0.036}_{-0.036}$$\cdot 10^{-1}$ &$1.467^{+0.036}_{-0.010}$$^{+0.036}_{-0.036}$$\cdot 10^{-1
}$          \\
 950 &$1.246^{+0.363}_{-0.236}$$^{+0.030}_{-0.030}$$\cdot 10^{-1}$ &$1.269^{+0.098}_{-0.058}$$^{+0.032}_{-0.032}$$\cdot 10^{-1}$ &$1.267^{+0.032}_{-0.010}$$^{+0.032}_{-0.032}$$\cdot 10^{-1
}$          \\
1000 &$1.078^{+0.322}_{-0.211}$$^{+0.027}_{-0.027}$$\cdot 10^{-1}$ &$1.099^{+0.088}_{-0.054}$$^{+0.028}_{-0.028}$$\cdot 10^{-1}$ &$1.096^{+0.029}_{-0.008}$$^{+0.028}_{-0.028}$$\cdot 10^{-1
}$          \\
\hline
\end{tabular}
\end{center}
\vspace*{-5mm}
\end{footnotesize}
\caption{
\small
\label{tab:table-lhc14-h15-Sc1}
Total VBF cross sections at the
LHC, $\sqrt S = 14 \tev$
at LO, NLO and NNLO in QCD. Errors shown are respectively scale and PDF uncertainities.  Scale uncertainities are evaluated by varying $\mu_r$ and $\mu_f$ in the interval
$\mu_r,\mu_f \in [Q/4,4Q]$.
The HERAPDF1.5~\cite{herapdf:2009wt,herapdfgrid:2010} PDF set has been used.
Numbers are in pb.
}
\end{table}

%***************** 1st table ************************ 

\begin{table}[tb!]
\begin{footnotesize}
\begin{center}
\begin{tabular}{|c|c|c|c|}
\hline
$ \hmass \, [\gev] $ & $ \sigma_{LO} $ & $ \sigma_{NLO} $ & $ \sigma_{NNLO} $ \\
\hline
  90 &$5.930^{+0.018}_{-0.184}$$^{+0.065}_{-0.065}$$\phantom{\cdot 10^{-0}}$ &$5.691^{+0.122}_{-0.071}$$^{+0.065}_{-0.065}$$\phantom{\cdot 10^{-0}}$ &$5.724^{+0.048}_{-0.098}$$^{+0.060}_{-0.060}$$\phantom{\cdot 10^{-0}
}$          \\
  95 &$5.689^{+0.021}_{-0.178}$$^{+0.063}_{-0.063}$$\phantom{\cdot 10^{-0}}$ &$5.401^{+0.181}_{-0.058}$$^{+0.062}_{-0.062}$$\phantom{\cdot 10^{-0}}$ &$5.473^{+0.056}_{-0.098}$$^{+0.057}_{-0.057}$$\phantom{\cdot 10^{-0}
}$          \\
 100 &$5.452^{+0.002}_{-0.146}$$^{+0.060}_{-0.060}$$\phantom{\cdot 10^{-0}}$ &$5.236^{+0.111}_{-0.072}$$^{+0.060}_{-0.060}$$\phantom{\cdot 10^{-0}}$ &$5.250^{+0.041}_{-0.095}$$^{+0.055}_{-0.055}$$\phantom{\cdot 10^{-0}
}$          \\
 105 &$5.227^{+0.000}_{-0.109}$$^{+0.058}_{-0.058}$$\phantom{\cdot 10^{-0}}$ &$5.018^{+0.109}_{-0.057}$$^{+0.057}_{-0.057}$$\phantom{\cdot 10^{-0}}$ &$5.029^{+0.062}_{-0.067}$$^{+0.053}_{-0.053}$$\phantom{\cdot 10^{-0}
}$          \\
 110 &$5.025^{+0.000}_{-0.080}$$^{+0.056}_{-0.056}$$\phantom{\cdot 10^{-0}}$ &$4.828^{+0.085}_{-0.060}$$^{+0.055}_{-0.055}$$\phantom{\cdot 10^{-0}}$ &$4.852^{+0.041}_{-0.078}$$^{+0.051}_{-0.051}$$\phantom{\cdot 10^{-0}
}$          \\
 115 &$4.824^{+0.000}_{-0.064}$$^{+0.054}_{-0.054}$$\phantom{\cdot 10^{-0}}$ &$4.642^{+0.092}_{-0.056}$$^{+0.053}_{-0.053}$$\phantom{\cdot 10^{-0}}$ &$4.656^{+0.046}_{-0.066}$$^{+0.049}_{-0.049}$$\phantom{\cdot 10^{-0}
}$          \\
 120 &$4.636^{+0.004}_{-0.072}$$^{+0.052}_{-0.052}$$\phantom{\cdot 10^{-0}}$ &$4.461^{+0.084}_{-0.050}$$^{+0.051}_{-0.051}$$\phantom{\cdot 10^{-0}}$ &$4.473^{+0.047}_{-0.070}$$^{+0.047}_{-0.047}$$\phantom{\cdot 10^{-0}
}$          \\
 125 &$4.464^{+0.004}_{-0.087}$$^{+0.050}_{-0.050}$$\phantom{\cdot 10^{-0}}$ &$4.293^{+0.077}_{-0.047}$$^{+0.049}_{-0.049}$$\phantom{\cdot 10^{-0}}$ &$4.305^{+0.035}_{-0.068}$$^{+0.046}_{-0.046}$$\phantom{\cdot 10^{-0}
}$          \\
 130 &$4.293^{+0.017}_{-0.084}$$^{+0.048}_{-0.048}$$\phantom{\cdot 10^{-0}}$ &$4.136^{+0.068}_{-0.051}$$^{+0.048}_{-0.048}$$\phantom{\cdot 10^{-0}}$ &$4.152^{+0.026}_{-0.072}$$^{+0.044}_{-0.044}$$\phantom{\cdot 10^{-0}
}$          \\
 135 &$4.132^{+0.021}_{-0.089}$$^{+0.047}_{-0.047}$$\phantom{\cdot 10^{-0}}$ &$3.986^{+0.081}_{-0.046}$$^{+0.046}_{-0.046}$$\phantom{\cdot 10^{-0}}$ &$3.996^{+0.040}_{-0.054}$$^{+0.043}_{-0.043}$$\phantom{\cdot 10^{-0}
}$          \\
 140 &$3.985^{+0.028}_{-0.100}$$^{+0.045}_{-0.045}$$\phantom{\cdot 10^{-0}}$ &$3.840^{+0.074}_{-0.041}$$^{+0.045}_{-0.045}$$\phantom{\cdot 10^{-0}}$ &$3.850^{+0.038}_{-0.050}$$^{+0.042}_{-0.042}$$\phantom{\cdot 10^{-0}
}$          \\
 145 &$3.838^{+0.034}_{-0.102}$$^{+0.044}_{-0.044}$$\phantom{\cdot 10^{-0}}$ &$3.707^{+0.063}_{-0.049}$$^{+0.043}_{-0.043}$$\phantom{\cdot 10^{-0}}$ &$3.706^{+0.033}_{-0.052}$$^{+0.040}_{-0.040}$$\phantom{\cdot 10^{-0}
}$          \\
 150 &$3.703^{+0.037}_{-0.111}$$^{+0.042}_{-0.042}$$\phantom{\cdot 10^{-0}}$ &$3.562^{+0.079}_{-0.037}$$^{+0.042}_{-0.042}$$\phantom{\cdot 10^{-0}}$ &$3.578^{+0.033}_{-0.049}$$^{+0.039}_{-0.039}$$\phantom{\cdot 10^{-0}
}$          \\
 155 &$3.568^{+0.058}_{-0.112}$$^{+0.041}_{-0.041}$$\phantom{\cdot 10^{-0}}$ &$3.450^{+0.063}_{-0.061}$$^{+0.041}_{-0.041}$$\phantom{\cdot 10^{-0}}$ &$3.452^{+0.032}_{-0.050}$$^{+0.038}_{-0.038}$$\phantom{\cdot 10^{-0}
}$          \\
 160 &$3.445^{+0.065}_{-0.114}$$^{+0.040}_{-0.040}$$\phantom{\cdot 10^{-0}}$ &$3.331^{+0.071}_{-0.034}$$^{+0.040}_{-0.040}$$\phantom{\cdot 10^{-0}}$ &$3.332^{+0.028}_{-0.046}$$^{+0.037}_{-0.037}$$\phantom{\cdot 10^{-0}
}$          \\
 165 &$3.332^{+0.066}_{-0.124}$$^{+0.039}_{-0.039}$$\phantom{\cdot 10^{-0}}$ &$3.216^{+0.073}_{-0.051}$$^{+0.038}_{-0.038}$$\phantom{\cdot 10^{-0}}$ &$3.222^{+0.028}_{-0.049}$$^{+0.037}_{-0.037}$$\phantom{\cdot 10^{-0}
}$          \\
 170 &$3.216^{+0.077}_{-0.121}$$^{+0.038}_{-0.038}$$\phantom{\cdot 10^{-0}}$ &$3.112^{+0.070}_{-0.045}$$^{+0.037}_{-0.037}$$\phantom{\cdot 10^{-0}}$ &$3.114^{+0.025}_{-0.049}$$^{+0.036}_{-0.036}$$\phantom{\cdot 10^{-0}
}$          \\
 175 &$3.111^{+0.070}_{-0.132}$$^{+0.037}_{-0.037}$$\phantom{\cdot 10^{-0}}$ &$3.005^{+0.069}_{-0.031}$$^{+0.036}_{-0.036}$$\phantom{\cdot 10^{-0}}$ &$3.010^{+0.027}_{-0.041}$$^{+0.035}_{-0.035}$$\phantom{\cdot 10^{-0}
}$          \\
 180 &$3.004^{+0.090}_{-0.122}$$^{+0.036}_{-0.036}$$\phantom{\cdot 10^{-0}}$ &$2.912^{+0.063}_{-0.039}$$^{+0.035}_{-0.035}$$\phantom{\cdot 10^{-0}}$ &$2.907^{+0.041}_{-0.037}$$^{+0.034}_{-0.034}$$\phantom{\cdot 10^{-0}
}$          \\
 185 &$2.905^{+0.100}_{-0.123}$$^{+0.035}_{-0.035}$$\phantom{\cdot 10^{-0}}$ &$2.813^{+0.056}_{-0.034}$$^{+0.034}_{-0.034}$$\phantom{\cdot 10^{-0}}$ &$2.815^{+0.024}_{-0.037}$$^{+0.033}_{-0.033}$$\phantom{\cdot 10^{-0}
}$          \\
 190 &$2.813^{+0.101}_{-0.127}$$^{+0.034}_{-0.034}$$\phantom{\cdot 10^{-0}}$ &$2.724^{+0.069}_{-0.034}$$^{+0.033}_{-0.033}$$\phantom{\cdot 10^{-0}}$ &$2.723^{+0.027}_{-0.037}$$^{+0.033}_{-0.033}$$\phantom{\cdot 10^{-0}
}$          \\
 195 &$2.719^{+0.109}_{-0.123}$$^{+0.033}_{-0.033}$$\phantom{\cdot 10^{-0}}$ &$2.638^{+0.066}_{-0.035}$$^{+0.032}_{-0.032}$$\phantom{\cdot 10^{-0}}$ &$2.637^{+0.027}_{-0.032}$$^{+0.032}_{-0.032}$$\phantom{\cdot 10^{-0}
}$          \\
 200 &$2.633^{+0.113}_{-0.128}$$^{+0.032}_{-0.032}$$\phantom{\cdot 10^{-0}}$ &$2.555^{+0.065}_{-0.031}$$^{+0.032}_{-0.032}$$\phantom{\cdot 10^{-0}}$ &$2.557^{+0.038}_{-0.037}$$^{+0.031}_{-0.031}$$\phantom{\cdot 10^{-0}
}$          \\
 210 &$2.473^{+0.119}_{-0.130}$$^{+0.030}_{-0.030}$$\phantom{\cdot 10^{-0}}$ &$2.407^{+0.057}_{-0.035}$$^{+0.030}_{-0.030}$$\phantom{\cdot 10^{-0}}$ &$2.401^{+0.025}_{-0.039}$$^{+0.030}_{-0.030}$$\phantom{\cdot 10^{-0}
}$          \\
 220 &$2.327^{+0.119}_{-0.135}$$^{+0.029}_{-0.029}$$\phantom{\cdot 10^{-0}}$ &$2.257^{+0.065}_{-0.029}$$^{+0.029}_{-0.029}$$\phantom{\cdot 10^{-0}}$ &$2.258^{+0.023}_{-0.028}$$^{+0.029}_{-0.029}$$\phantom{\cdot 10^{-0}
}$          \\
 230 &$2.188^{+0.124}_{-0.128}$$^{+0.028}_{-0.028}$$\phantom{\cdot 10^{-0}}$ &$2.126^{+0.062}_{-0.021}$$^{+0.027}_{-0.027}$$\phantom{\cdot 10^{-0}}$ &$2.129^{+0.019}_{-0.035}$$^{+0.028}_{-0.028}$$\phantom{\cdot 10^{-0}
}$          \\
 240 &$2.063^{+0.121}_{-0.133}$$^{+0.026}_{-0.026}$$\phantom{\cdot 10^{-0}}$ &$2.005^{+0.055}_{-0.025}$$^{+0.026}_{-0.026}$$\phantom{\cdot 10^{-0}}$ &$2.002^{+0.026}_{-0.025}$$^{+0.027}_{-0.027}$$\phantom{\cdot 10^{-0}
}$          \\
 250 &$1.946^{+0.129}_{-0.133}$$^{+0.025}_{-0.025}$$\phantom{\cdot 10^{-0}}$ &$1.895^{+0.055}_{-0.026}$$^{+0.025}_{-0.025}$$\phantom{\cdot 10^{-0}}$ &$1.890^{+0.020}_{-0.022}$$^{+0.026}_{-0.026}$$\phantom{\cdot 10^{-0}
}$          \\
 260 &$1.840^{+0.128}_{-0.129}$$^{+0.024}_{-0.024}$$\phantom{\cdot 10^{-0}}$ &$1.790^{+0.053}_{-0.022}$$^{+0.024}_{-0.024}$$\phantom{\cdot 10^{-0}}$ &$1.788^{+0.018}_{-0.022}$$^{+0.025}_{-0.025}$$\phantom{\cdot 10^{-0}
}$          \\
 270 &$1.739^{+0.129}_{-0.128}$$^{+0.023}_{-0.023}$$\phantom{\cdot 10^{-0}}$ &$1.697^{+0.048}_{-0.026}$$^{+0.023}_{-0.023}$$\phantom{\cdot 10^{-0}}$ &$1.692^{+0.018}_{-0.022}$$^{+0.024}_{-0.024}$$\phantom{\cdot 10^{-0}
}$          \\
 280 &$1.645^{+0.131}_{-0.123}$$^{+0.022}_{-0.022}$$\phantom{\cdot 10^{-0}}$ &$1.603^{+0.051}_{-0.021}$$^{+0.022}_{-0.022}$$\phantom{\cdot 10^{-0}}$ &$1.600^{+0.021}_{-0.019}$$^{+0.023}_{-0.023}$$\phantom{\cdot 10^{-0}
}$          \\
 290 &$1.558^{+0.131}_{-0.122}$$^{+0.021}_{-0.021}$$\phantom{\cdot 10^{-0}}$ &$1.520^{+0.050}_{-0.017}$$^{+0.021}_{-0.021}$$\phantom{\cdot 10^{-0}}$ &$1.518^{+0.018}_{-0.019}$$^{+0.022}_{-0.022}$$\phantom{\cdot 10^{-0}
}$          \\
 300 &$1.478^{+0.130}_{-0.119}$$^{+0.020}_{-0.020}$$\phantom{\cdot 10^{-0}}$ &$1.442^{+0.048}_{-0.018}$$^{+0.020}_{-0.020}$$\phantom{\cdot 10^{-0}}$ &$1.439^{+0.019}_{-0.017}$$^{+0.022}_{-0.022}$$\phantom{\cdot 10^{-0}
}$          \\
 320 &$1.331^{+0.130}_{-0.114}$$^{+0.019}_{-0.019}$$\phantom{\cdot 10^{-0}}$ &$1.297^{+0.052}_{-0.013}$$^{+0.018}_{-0.018}$$\phantom{\cdot 10^{-0}}$ &$1.299^{+0.017}_{-0.016}$$^{+0.020}_{-0.020}$$\phantom{\cdot 10^{-0}
}$          \\
 340 &$1.204^{+0.124}_{-0.110}$$^{+0.017}_{-0.017}$$\phantom{\cdot 10^{-0}}$ &$1.180^{+0.042}_{-0.017}$$^{+0.017}_{-0.017}$$\phantom{\cdot 10^{-0}}$ &$1.174^{+0.018}_{-0.013}$$^{+0.019}_{-0.019}$$\phantom{\cdot 10^{-0}
}$          \\
 360 &$1.092^{+0.122}_{-0.103}$$^{+0.016}_{-0.016}$$\phantom{\cdot 10^{-0}}$ &$1.071^{+0.041}_{-0.015}$$^{+0.016}_{-0.016}$$\phantom{\cdot 10^{-0}}$ &$1.067^{+0.017}_{-0.012}$$^{+0.018}_{-0.018}$$\phantom{\cdot 10^{-0}
}$          \\
 380 &$9.938^{+1.175}_{-0.990}$$^{+0.150}_{-0.150}$$\cdot 10^{-1}$ &$9.766^{+0.391}_{-0.159}$$^{+0.148}_{-0.148}$$\cdot 10^{-1}$ &$9.713^{+0.139}_{-0.102}$$^{+0.166}_{-0.166}$$\cdot 10^{-1
}$          \\
 400 &$9.057^{+1.142}_{-0.938}$$^{+0.140}_{-0.140}$$\cdot 10^{-1}$ &$8.907^{+0.377}_{-0.128}$$^{+0.138}_{-0.138}$$\cdot 10^{-1}$ &$8.861^{+0.137}_{-0.089}$$^{+0.156}_{-0.156}$$\cdot 10^{-1
}$          \\
 450 &$7.266^{+1.039}_{-0.826}$$^{+0.118}_{-0.118}$$\cdot 10^{-1}$ &$7.165^{+0.326}_{-0.104}$$^{+0.117}_{-0.117}$$\cdot 10^{-1}$ &$7.126^{+0.109}_{-0.072}$$^{+0.134}_{-0.134}$$\cdot 10^{-1
}$          \\
 500 &$5.901^{+0.934}_{-0.722}$$^{+0.101}_{-0.101}$$\cdot 10^{-1}$ &$5.831^{+0.288}_{-0.091}$$^{+0.100}_{-0.100}$$\cdot 10^{-1}$ &$5.797^{+0.095}_{-0.049}$$^{+0.116}_{-0.116}$$\cdot 10^{-1
}$          \\
 550 &$4.847^{+0.828}_{-0.637}$$^{+0.088}_{-0.088}$$\cdot 10^{-1}$ &$4.804^{+0.244}_{-0.088}$$^{+0.086}_{-0.086}$$\cdot 10^{-1}$ &$4.765^{+0.087}_{-0.040}$$^{+0.100}_{-0.100}$$\cdot 10^{-1
}$          \\
 600 &$4.011^{+0.745}_{-0.549}$$^{+0.076}_{-0.076}$$\cdot 10^{-1}$ &$3.989^{+0.227}_{-0.082}$$^{+0.075}_{-0.075}$$\cdot 10^{-1}$ &$3.958^{+0.069}_{-0.033}$$^{+0.088}_{-0.088}$$\cdot 10^{-1
}$          \\
 650 &$3.351^{+0.661}_{-0.483}$$^{+0.066}_{-0.066}$$\cdot 10^{-1}$ &$3.343^{+0.190}_{-0.076}$$^{+0.066}_{-0.066}$$\cdot 10^{-1}$ &$3.313^{+0.057}_{-0.024}$$^{+0.076}_{-0.076}$$\cdot 10^{-1
}$          \\
 700 &$2.816^{+0.593}_{-0.426}$$^{+0.058}_{-0.058}$$\cdot 10^{-1}$ &$2.818^{+0.168}_{-0.071}$$^{+0.058}_{-0.058}$$\cdot 10^{-1}$ &$2.788^{+0.055}_{-0.014}$$^{+0.067}_{-0.067}$$\cdot 10^{-1
}$          \\
 750 &$2.382^{+0.531}_{-0.374}$$^{+0.051}_{-0.051}$$\cdot 10^{-1}$ &$2.393^{+0.145}_{-0.074}$$^{+0.051}_{-0.051}$$\cdot 10^{-1}$ &$2.363^{+0.048}_{-0.011}$$^{+0.059}_{-0.059}$$\cdot 10^{-1
}$          \\
 800 &$2.027^{+0.471}_{-0.328}$$^{+0.046}_{-0.046}$$\cdot 10^{-1}$ &$2.038^{+0.132}_{-0.061}$$^{+0.045}_{-0.045}$$\cdot 10^{-1}$ &$2.014^{+0.041}_{-0.008}$$^{+0.052}_{-0.052}$$\cdot 10^{-1
}$          \\
 850 &$1.733^{+0.424}_{-0.292}$$^{+0.041}_{-0.041}$$\cdot 10^{-1}$ &$1.746^{+0.116}_{-0.058}$$^{+0.040}_{-0.040}$$\cdot 10^{-1}$ &$1.727^{+0.036}_{-0.008}$$^{+0.046}_{-0.046}$$\cdot 10^{-1
}$          \\
 900 &$1.488^{+0.380}_{-0.259}$$^{+0.036}_{-0.036}$$\cdot 10^{-1}$ &$1.507^{+0.099}_{-0.058}$$^{+0.036}_{-0.036}$$\cdot 10^{-1}$ &$1.486^{+0.030}_{-0.008}$$^{+0.041}_{-0.041}$$\cdot 10^{-1
}$          \\
 950 &$1.283^{+0.338}_{-0.229}$$^{+0.032}_{-0.032}$$\cdot 10^{-1}$ &$1.300^{+0.090}_{-0.050}$$^{+0.032}_{-0.032}$$\cdot 10^{-1}$ &$1.284^{+0.025}_{-0.008}$$^{+0.036}_{-0.036}$$\cdot 10^{-1
}$          \\
1000 &$1.110^{+0.304}_{-0.204}$$^{+0.029}_{-0.029}$$\cdot 10^{-1}$ &$1.128^{+0.081}_{-0.046}$$^{+0.029}_{-0.029}$$\cdot 10^{-1}$ &$1.112^{+0.023}_{-0.007}$$^{+0.032}_{-0.032}$$\cdot 10^{-1
}$          \\
\hline
\end{tabular}
\end{center}
\vspace*{-5mm}
\end{footnotesize}
\caption{
\small
\label{tab:table-lhc14-gjr-Sc1}
Total VBF cross sections at the
LHC, $\sqrt S = 14 \tev$
at LO, NLO and NNLO in QCD. Errors shown are respectively scale and PDF uncertainities.  Scale uncertainities are evaluated by varying $\mu_r$ and $\mu_f$ in the interval
$\mu_r,\mu_f \in [Q/4,4Q]$.
The JR09~\cite{JimenezDelgado:2008hf,JimenezDelgado:2009tv} PDF set has been used.
Numbers are in pb.
}
\end{table}

%***************** 1st table ************************ 

\begin{table}[tb!]
\begin{footnotesize}
\begin{center}
\begin{tabular}{|c|c|c|c|}
\hline
$ \hmass \, [\gev] $ & $ \sigma_{LO} $ & $ \sigma_{NLO} $ & $ \sigma_{NNLO} $ \\
\hline
  90 &$5.508^{+0.000}_{-0.151}$$^{+0.059}_{-0.059}$$\phantom{\cdot 10^{-0}}$ &$5.809^{+0.147}_{-0.072}$$^{+0.105}_{-0.105}$$\phantom{\cdot 10^{-0}}$ &$5.857^{+0.076}_{-0.098}$$^{+0.094}_{-0.094}$$\phantom{\cdot 10^{-0}
}$          \\
  95 &$5.266^{+0.000}_{-0.094}$$^{+0.057}_{-0.057}$$\phantom{\cdot 10^{-0}}$ &$5.583^{+0.124}_{-0.068}$$^{+0.101}_{-0.101}$$\phantom{\cdot 10^{-0}}$ &$5.644^{+0.045}_{-0.146}$$^{+0.091}_{-0.091}$$\phantom{\cdot 10^{-0}
}$          \\
 100 &$5.040^{+0.000}_{-0.079}$$^{+0.055}_{-0.055}$$\phantom{\cdot 10^{-0}}$ &$5.352^{+0.110}_{-0.069}$$^{+0.097}_{-0.097}$$\phantom{\cdot 10^{-0}}$ &$5.382^{+0.052}_{-0.104}$$^{+0.086}_{-0.086}$$\phantom{\cdot 10^{-0}
}$          \\
 105 &$4.837^{+0.000}_{-0.070}$$^{+0.053}_{-0.053}$$\phantom{\cdot 10^{-0}}$ &$5.141^{+0.101}_{-0.066}$$^{+0.093}_{-0.093}$$\phantom{\cdot 10^{-0}}$ &$5.147^{+0.073}_{-0.075}$$^{+0.083}_{-0.083}$$\phantom{\cdot 10^{-0}
}$          \\
 110 &$4.645^{+0.001}_{-0.092}$$^{+0.051}_{-0.051}$$\phantom{\cdot 10^{-0}}$ &$4.939^{+0.081}_{-0.061}$$^{+0.090}_{-0.090}$$\phantom{\cdot 10^{-0}}$ &$4.961^{+0.027}_{-0.119}$$^{+0.080}_{-0.080}$$\phantom{\cdot 10^{-0}
}$          \\
 115 &$4.456^{+0.007}_{-0.145}$$^{+0.049}_{-0.049}$$\phantom{\cdot 10^{-0}}$ &$4.748^{+0.068}_{-0.072}$$^{+0.086}_{-0.086}$$\phantom{\cdot 10^{-0}}$ &$4.752^{+0.054}_{-0.089}$$^{+0.077}_{-0.077}$$\phantom{\cdot 10^{-0}
}$          \\
 120 &$4.285^{+0.019}_{-0.113}$$^{+0.048}_{-0.048}$$\phantom{\cdot 10^{-0}}$ &$4.548^{+0.095}_{-0.062}$$^{+0.083}_{-0.083}$$\phantom{\cdot 10^{-0}}$ &$4.562^{+0.082}_{-0.044}$$^{+0.074}_{-0.074}$$\phantom{\cdot 10^{-0}
}$          \\
 125 &$4.115^{+0.021}_{-0.119}$$^{+0.046}_{-0.046}$$\phantom{\cdot 10^{-0}}$ &$4.375^{+0.077}_{-0.052}$$^{+0.080}_{-0.080}$$\phantom{\cdot 10^{-0}}$ &$4.391^{+0.040}_{-0.087}$$^{+0.072}_{-0.072}$$\phantom{\cdot 10^{-0}
}$          \\
 130 &$3.961^{+0.040}_{-0.131}$$^{+0.044}_{-0.044}$$\phantom{\cdot 10^{-0}}$ &$4.219^{+0.071}_{-0.062}$$^{+0.077}_{-0.077}$$\phantom{\cdot 10^{-0}}$ &$4.224^{+0.091}_{-0.075}$$^{+0.069}_{-0.069}$$\phantom{\cdot 10^{-0}
}$          \\
 135 &$3.802^{+0.058}_{-0.128}$$^{+0.043}_{-0.043}$$\phantom{\cdot 10^{-0}}$ &$4.064^{+0.073}_{-0.076}$$^{+0.075}_{-0.075}$$\phantom{\cdot 10^{-0}}$ &$4.076^{+0.030}_{-0.081}$$^{+0.067}_{-0.067}$$\phantom{\cdot 10^{-0}
}$          \\
 140 &$3.670^{+0.061}_{-0.137}$$^{+0.042}_{-0.042}$$\phantom{\cdot 10^{-0}}$ &$3.906^{+0.087}_{-0.043}$$^{+0.072}_{-0.072}$$\phantom{\cdot 10^{-0}}$ &$3.926^{+0.030}_{-0.084}$$^{+0.064}_{-0.064}$$\phantom{\cdot 10^{-0}
}$          \\
 145 &$3.531^{+0.075}_{-0.142}$$^{+0.040}_{-0.040}$$\phantom{\cdot 10^{-0}}$ &$3.764^{+0.080}_{-0.044}$$^{+0.070}_{-0.070}$$\phantom{\cdot 10^{-0}}$ &$3.781^{+0.039}_{-0.071}$$^{+0.062}_{-0.062}$$\phantom{\cdot 10^{-0}
}$          \\
 150 &$3.405^{+0.085}_{-0.145}$$^{+0.039}_{-0.039}$$\phantom{\cdot 10^{-0}}$ &$3.630^{+0.084}_{-0.063}$$^{+0.067}_{-0.067}$$\phantom{\cdot 10^{-0}}$ &$3.642^{+0.042}_{-0.077}$$^{+0.060}_{-0.060}$$\phantom{\cdot 10^{-0}
}$          \\
 155 &$3.283^{+0.100}_{-0.149}$$^{+0.038}_{-0.038}$$\phantom{\cdot 10^{-0}}$ &$3.498^{+0.081}_{-0.044}$$^{+0.065}_{-0.065}$$\phantom{\cdot 10^{-0}}$ &$3.514^{+0.042}_{-0.068}$$^{+0.058}_{-0.058}$$\phantom{\cdot 10^{-0}
}$          \\
 160 &$3.165^{+0.107}_{-0.156}$$^{+0.037}_{-0.037}$$\phantom{\cdot 10^{-0}}$ &$3.377^{+0.087}_{-0.050}$$^{+0.063}_{-0.063}$$\phantom{\cdot 10^{-0}}$ &$3.385^{+0.039}_{-0.068}$$^{+0.056}_{-0.056}$$\phantom{\cdot 10^{-0}
}$          \\
 165 &$3.053^{+0.104}_{-0.149}$$^{+0.035}_{-0.035}$$\phantom{\cdot 10^{-0}}$ &$3.258^{+0.079}_{-0.044}$$^{+0.061}_{-0.061}$$\phantom{\cdot 10^{-0}}$ &$3.275^{+0.037}_{-0.067}$$^{+0.055}_{-0.055}$$\phantom{\cdot 10^{-0}
}$          \\
 170 &$2.947^{+0.123}_{-0.149}$$^{+0.034}_{-0.034}$$\phantom{\cdot 10^{-0}}$ &$3.150^{+0.079}_{-0.046}$$^{+0.059}_{-0.059}$$\phantom{\cdot 10^{-0}}$ &$3.160^{+0.042}_{-0.060}$$^{+0.053}_{-0.053}$$\phantom{\cdot 10^{-0}
}$          \\
 175 &$2.851^{+0.115}_{-0.159}$$^{+0.033}_{-0.033}$$\phantom{\cdot 10^{-0}}$ &$3.046^{+0.076}_{-0.047}$$^{+0.057}_{-0.057}$$\phantom{\cdot 10^{-0}}$ &$3.075^{+0.016}_{-0.081}$$^{+0.052}_{-0.052}$$\phantom{\cdot 10^{-0}
}$          \\
 180 &$2.756^{+0.124}_{-0.160}$$^{+0.032}_{-0.032}$$\phantom{\cdot 10^{-0}}$ &$2.938^{+0.079}_{-0.040}$$^{+0.055}_{-0.055}$$\phantom{\cdot 10^{-0}}$ &$2.950^{+0.040}_{-0.062}$$^{+0.050}_{-0.050}$$\phantom{\cdot 10^{-0}
}$          \\
 185 &$2.663^{+0.129}_{-0.158}$$^{+0.031}_{-0.031}$$\phantom{\cdot 10^{-0}}$ &$2.842^{+0.076}_{-0.040}$$^{+0.054}_{-0.054}$$\phantom{\cdot 10^{-0}}$ &$2.853^{+0.049}_{-0.054}$$^{+0.048}_{-0.048}$$\phantom{\cdot 10^{-0}
}$          \\
 190 &$2.571^{+0.136}_{-0.152}$$^{+0.031}_{-0.031}$$\phantom{\cdot 10^{-0}}$ &$2.749^{+0.070}_{-0.035}$$^{+0.052}_{-0.052}$$\phantom{\cdot 10^{-0}}$ &$2.760^{+0.041}_{-0.048}$$^{+0.047}_{-0.047}$$\phantom{\cdot 10^{-0}
}$          \\
 195 &$2.491^{+0.139}_{-0.154}$$^{+0.030}_{-0.030}$$\phantom{\cdot 10^{-0}}$ &$2.650^{+0.088}_{-0.030}$$^{+0.050}_{-0.050}$$\phantom{\cdot 10^{-0}}$ &$2.668^{+0.041}_{-0.045}$$^{+0.045}_{-0.045}$$\phantom{\cdot 10^{-0}
}$          \\
 200 &$2.412^{+0.144}_{-0.158}$$^{+0.029}_{-0.029}$$\phantom{\cdot 10^{-0}}$ &$2.574^{+0.077}_{-0.036}$$^{+0.049}_{-0.049}$$\phantom{\cdot 10^{-0}}$ &$2.581^{+0.040}_{-0.047}$$^{+0.044}_{-0.044}$$\phantom{\cdot 10^{-0}
}$          \\
 210 &$2.262^{+0.148}_{-0.156}$$^{+0.027}_{-0.027}$$\phantom{\cdot 10^{-0}}$ &$2.416^{+0.074}_{-0.035}$$^{+0.046}_{-0.046}$$\phantom{\cdot 10^{-0}}$ &$2.423^{+0.040}_{-0.045}$$^{+0.041}_{-0.041}$$\phantom{\cdot 10^{-0}
}$          \\
 220 &$2.120^{+0.156}_{-0.150}$$^{+0.026}_{-0.026}$$\phantom{\cdot 10^{-0}}$ &$2.267^{+0.075}_{-0.029}$$^{+0.044}_{-0.044}$$\phantom{\cdot 10^{-0}}$ &$2.276^{+0.037}_{-0.039}$$^{+0.039}_{-0.039}$$\phantom{\cdot 10^{-0}
}$          \\
 230 &$1.998^{+0.154}_{-0.152}$$^{+0.025}_{-0.025}$$\phantom{\cdot 10^{-0}}$ &$2.132^{+0.074}_{-0.027}$$^{+0.041}_{-0.041}$$\phantom{\cdot 10^{-0}}$ &$2.142^{+0.033}_{-0.033}$$^{+0.037}_{-0.037}$$\phantom{\cdot 10^{-0}
}$          \\
 240 &$1.880^{+0.156}_{-0.149}$$^{+0.023}_{-0.023}$$\phantom{\cdot 10^{-0}}$ &$2.010^{+0.072}_{-0.031}$$^{+0.039}_{-0.039}$$\phantom{\cdot 10^{-0}}$ &$2.018^{+0.033}_{-0.032}$$^{+0.035}_{-0.035}$$\phantom{\cdot 10^{-0}
}$          \\
 250 &$1.773^{+0.155}_{-0.147}$$^{+0.022}_{-0.022}$$\phantom{\cdot 10^{-0}}$ &$1.894^{+0.074}_{-0.028}$$^{+0.037}_{-0.037}$$\phantom{\cdot 10^{-0}}$ &$1.903^{+0.028}_{-0.037}$$^{+0.033}_{-0.033}$$\phantom{\cdot 10^{-0}
}$          \\
 260 &$1.673^{+0.156}_{-0.143}$$^{+0.021}_{-0.021}$$\phantom{\cdot 10^{-0}}$ &$1.785^{+0.068}_{-0.025}$$^{+0.035}_{-0.035}$$\phantom{\cdot 10^{-0}}$ &$1.796^{+0.032}_{-0.031}$$^{+0.032}_{-0.032}$$\phantom{\cdot 10^{-0}
}$          \\
 270 &$1.580^{+0.156}_{-0.141}$$^{+0.020}_{-0.020}$$\phantom{\cdot 10^{-0}}$ &$1.691^{+0.061}_{-0.030}$$^{+0.033}_{-0.033}$$\phantom{\cdot 10^{-0}}$ &$1.695^{+0.032}_{-0.028}$$^{+0.030}_{-0.030}$$\phantom{\cdot 10^{-0}
}$          \\
 280 &$1.496^{+0.153}_{-0.139}$$^{+0.019}_{-0.019}$$\phantom{\cdot 10^{-0}}$ &$1.598^{+0.064}_{-0.025}$$^{+0.032}_{-0.032}$$\phantom{\cdot 10^{-0}}$ &$1.604^{+0.028}_{-0.028}$$^{+0.029}_{-0.029}$$\phantom{\cdot 10^{-0}
}$          \\
 290 &$1.416^{+0.155}_{-0.135}$$^{+0.018}_{-0.018}$$\phantom{\cdot 10^{-0}}$ &$1.512^{+0.062}_{-0.020}$$^{+0.030}_{-0.030}$$\phantom{\cdot 10^{-0}}$ &$1.519^{+0.027}_{-0.023}$$^{+0.027}_{-0.027}$$\phantom{\cdot 10^{-0}
}$          \\
 300 &$1.342^{+0.150}_{-0.132}$$^{+0.017}_{-0.017}$$\phantom{\cdot 10^{-0}}$ &$1.430^{+0.067}_{-0.020}$$^{+0.029}_{-0.029}$$\phantom{\cdot 10^{-0}}$ &$1.442^{+0.024}_{-0.025}$$^{+0.026}_{-0.026}$$\phantom{\cdot 10^{-0}
}$          \\
 320 &$1.208^{+0.149}_{-0.124}$$^{+0.016}_{-0.016}$$\phantom{\cdot 10^{-0}}$ &$1.288^{+0.063}_{-0.018}$$^{+0.026}_{-0.026}$$\phantom{\cdot 10^{-0}}$ &$1.298^{+0.024}_{-0.021}$$^{+0.024}_{-0.024}$$\phantom{\cdot 10^{-0}
}$          \\
 340 &$1.093^{+0.143}_{-0.121}$$^{+0.014}_{-0.014}$$\phantom{\cdot 10^{-0}}$ &$1.165^{+0.054}_{-0.016}$$^{+0.024}_{-0.024}$$\phantom{\cdot 10^{-0}}$ &$1.172^{+0.023}_{-0.019}$$^{+0.022}_{-0.022}$$\phantom{\cdot 10^{-0}
}$          \\
 360 &$9.925^{+1.344}_{-1.156}$$^{+0.133}_{-0.133}$$\cdot 10^{-1}$ &$1.055^{+0.054}_{-0.015}$$^{+0.022}_{-0.022}$$\phantom{\cdot 10^{-0}}$ &$1.061^{+0.021}_{-0.015}$$^{+0.020}_{-0.020}$$\phantom{\cdot 10^{-0}
}$          \\
 380 &$8.999^{+1.354}_{-1.073}$$^{+0.122}_{-0.122}$$\cdot 10^{-1}$ &$9.599^{+0.490}_{-0.142}$$^{+0.199}_{-0.199}$$\cdot 10^{-1}$ &$9.656^{+0.190}_{-0.141}$$^{+0.181}_{-0.181}$$\cdot 10^{-1
}$          \\
 400 &$8.192^{+1.294}_{-0.997}$$^{+0.112}_{-0.112}$$\cdot 10^{-1}$ &$8.746^{+0.478}_{-0.157}$$^{+0.183}_{-0.183}$$\cdot 10^{-1}$ &$8.780^{+0.205}_{-0.110}$$^{+0.166}_{-0.166}$$\cdot 10^{-1
}$          \\
 450 &$6.573^{+1.137}_{-0.878}$$^{+0.092}_{-0.092}$$\cdot 10^{-1}$ &$7.008^{+0.394}_{-0.129}$$^{+0.149}_{-0.149}$$\cdot 10^{-1}$ &$7.045^{+0.156}_{-0.101}$$^{+0.136}_{-0.136}$$\cdot 10^{-1
}$          \\
 500 &$5.330^{+1.018}_{-0.761}$$^{+0.077}_{-0.077}$$\cdot 10^{-1}$ &$5.677^{+0.342}_{-0.117}$$^{+0.123}_{-0.123}$$\cdot 10^{-1}$ &$5.714^{+0.127}_{-0.067}$$^{+0.113}_{-0.113}$$\cdot 10^{-1
}$          \\
 550 &$4.372^{+0.908}_{-0.663}$$^{+0.064}_{-0.064}$$\cdot 10^{-1}$ &$4.654^{+0.302}_{-0.115}$$^{+0.102}_{-0.102}$$\cdot 10^{-1}$ &$4.678^{+0.118}_{-0.046}$$^{+0.094}_{-0.094}$$\cdot 10^{-1
}$          \\
 600 &$3.621^{+0.809}_{-0.580}$$^{+0.054}_{-0.054}$$\cdot 10^{-1}$ &$3.852^{+0.253}_{-0.106}$$^{+0.086}_{-0.086}$$\cdot 10^{-1}$ &$3.877^{+0.091}_{-0.042}$$^{+0.079}_{-0.079}$$\cdot 10^{-1
}$          \\
 650 &$3.024^{+0.713}_{-0.504}$$^{+0.046}_{-0.046}$$\cdot 10^{-1}$ &$3.211^{+0.220}_{-0.100}$$^{+0.073}_{-0.073}$$\cdot 10^{-1}$ &$3.232^{+0.079}_{-0.028}$$^{+0.067}_{-0.067}$$\cdot 10^{-1
}$          \\
 700 &$2.539^{+0.640}_{-0.442}$$^{+0.039}_{-0.039}$$\cdot 10^{-1}$ &$2.694^{+0.201}_{-0.089}$$^{+0.062}_{-0.062}$$\cdot 10^{-1}$ &$2.716^{+0.070}_{-0.022}$$^{+0.057}_{-0.057}$$\cdot 10^{-1
}$          \\
 750 &$2.148^{+0.569}_{-0.388}$$^{+0.034}_{-0.034}$$\cdot 10^{-1}$ &$2.283^{+0.166}_{-0.087}$$^{+0.053}_{-0.053}$$\cdot 10^{-1}$ &$2.297^{+0.063}_{-0.017}$$^{+0.049}_{-0.049}$$\cdot 10^{-1
}$          \\
 800 &$1.827^{+0.511}_{-0.340}$$^{+0.029}_{-0.029}$$\cdot 10^{-1}$ &$1.937^{+0.150}_{-0.075}$$^{+0.045}_{-0.045}$$\cdot 10^{-1}$ &$1.955^{+0.054}_{-0.013}$$^{+0.042}_{-0.042}$$\cdot 10^{-1
}$          \\
 850 &$1.563^{+0.456}_{-0.300}$$^{+0.025}_{-0.025}$$\cdot 10^{-1}$ &$1.656^{+0.130}_{-0.068}$$^{+0.039}_{-0.039}$$\cdot 10^{-1}$ &$1.673^{+0.041}_{-0.015}$$^{+0.037}_{-0.037}$$\cdot 10^{-1
}$          \\
 900 &$1.342^{+0.408}_{-0.265}$$^{+0.022}_{-0.022}$$\cdot 10^{-1}$ &$1.421^{+0.116}_{-0.063}$$^{+0.034}_{-0.034}$$\cdot 10^{-1}$ &$1.436^{+0.037}_{-0.014}$$^{+0.032}_{-0.032}$$\cdot 10^{-1
}$          \\
 950 &$1.158^{+0.363}_{-0.234}$$^{+0.019}_{-0.019}$$\cdot 10^{-1}$ &$1.226^{+0.101}_{-0.057}$$^{+0.030}_{-0.030}$$\cdot 10^{-1}$ &$1.237^{+0.035}_{-0.012}$$^{+0.028}_{-0.028}$$\cdot 10^{-1
}$          \\
1000 &$1.002^{+0.326}_{-0.208}$$^{+0.017}_{-0.017}$$\cdot 10^{-1}$ &$1.062^{+0.090}_{-0.053}$$^{+0.026}_{-0.026}$$\cdot 10^{-1}$ &$1.070^{+0.030}_{-0.011}$$^{+0.024}_{-0.024}$$\cdot 10^{-1
}$          \\
\hline
\end{tabular}
\end{center}
\vspace*{-5mm}
\end{footnotesize}
\caption{
\small
\label{tab:table-lhc14-m68-Sc1}
Total VBF cross sections at the
LHC, $\sqrt S = 14 \tev$
at LO, NLO and NNLO in QCD. Errors shown are respectively scale and PDF uncertainities.  Scale uncertainities are evaluated by varying $\mu_r$ and $\mu_f$ in the interval
$\mu_r,\mu_f \in [Q/4,4Q]$.
The MSTW2008~\cite{Martin:2009iq} PDF set (68\% CL) has been used.
Numbers are in pb.
}
\end{table}

%***************** 1st table ************************ 

\begin{table}[tb!]
\begin{footnotesize}
\begin{center}
\begin{tabular}{|c|c|c|c|}
\hline
$ \hmass \, [\gev] $ & $ \sigma_{LO} $ & $ \sigma_{NLO} $ & $ \sigma_{NNLO} $ \\
\hline
  90 &$5.980^{+0.022}_{-0.186}$$^{+0.080}_{-0.080}$$\phantom{\cdot 10^{-0}}$ &$5.712^{+0.163}_{-0.068}$$^{+0.078}_{-0.078}$$\phantom{\cdot 10^{-0}}$ &$5.810^{+0.056}_{-0.132}$$^{+0.072}_{-0.072}$$\phantom{\cdot 10^{-0}
}$          \\
  95 &$5.733^{+0.011}_{-0.182}$$^{+0.076}_{-0.076}$$\phantom{\cdot 10^{-0}}$ &$5.498^{+0.128}_{-0.069}$$^{+0.075}_{-0.075}$$\phantom{\cdot 10^{-0}}$ &$5.565^{+0.062}_{-0.111}$$^{+0.069}_{-0.069}$$\phantom{\cdot 10^{-0}
}$          \\
 100 &$5.497^{+0.015}_{-0.147}$$^{+0.073}_{-0.073}$$\phantom{\cdot 10^{-0}}$ &$5.265^{+0.122}_{-0.088}$$^{+0.072}_{-0.072}$$\phantom{\cdot 10^{-0}}$ &$5.344^{+0.049}_{-0.115}$$^{+0.066}_{-0.066}$$\phantom{\cdot 10^{-0}
}$          \\
 105 &$5.262^{+0.000}_{-0.101}$$^{+0.070}_{-0.070}$$\phantom{\cdot 10^{-0}}$ &$5.065^{+0.105}_{-0.071}$$^{+0.069}_{-0.069}$$\phantom{\cdot 10^{-0}}$ &$5.106^{+0.072}_{-0.085}$$^{+0.063}_{-0.063}$$\phantom{\cdot 10^{-0}
}$          \\
 110 &$5.055^{+0.000}_{-0.080}$$^{+0.067}_{-0.067}$$\phantom{\cdot 10^{-0}}$ &$4.862^{+0.100}_{-0.077}$$^{+0.066}_{-0.066}$$\phantom{\cdot 10^{-0}}$ &$4.913^{+0.045}_{-0.108}$$^{+0.060}_{-0.060}$$\phantom{\cdot 10^{-0}
}$          \\
 115 &$4.848^{+0.012}_{-0.114}$$^{+0.065}_{-0.065}$$\phantom{\cdot 10^{-0}}$ &$4.668^{+0.090}_{-0.067}$$^{+0.064}_{-0.064}$$\phantom{\cdot 10^{-0}}$ &$4.716^{+0.061}_{-0.085}$$^{+0.058}_{-0.058}$$\phantom{\cdot 10^{-0}
}$          \\
 120 &$4.684^{+0.000}_{-0.107}$$^{+0.062}_{-0.062}$$\phantom{\cdot 10^{-0}}$ &$4.490^{+0.090}_{-0.069}$$^{+0.061}_{-0.061}$$\phantom{\cdot 10^{-0}}$ &$4.542^{+0.040}_{-0.093}$$^{+0.055}_{-0.055}$$\phantom{\cdot 10^{-0}
}$          \\
 125 &$4.479^{+0.017}_{-0.089}$$^{+0.060}_{-0.060}$$\phantom{\cdot 10^{-0}}$ &$4.317^{+0.083}_{-0.061}$$^{+0.059}_{-0.059}$$\phantom{\cdot 10^{-0}}$ &$4.363^{+0.047}_{-0.081}$$^{+0.053}_{-0.053}$$\phantom{\cdot 10^{-0}
}$          \\
 130 &$4.307^{+0.024}_{-0.100}$$^{+0.058}_{-0.058}$$\phantom{\cdot 10^{-0}}$ &$4.148^{+0.078}_{-0.055}$$^{+0.057}_{-0.057}$$\phantom{\cdot 10^{-0}}$ &$4.218^{+0.019}_{-0.104}$$^{+0.051}_{-0.051}$$\phantom{\cdot 10^{-0}
}$          \\
 135 &$4.148^{+0.026}_{-0.110}$$^{+0.056}_{-0.056}$$\phantom{\cdot 10^{-0}}$ &$4.002^{+0.071}_{-0.057}$$^{+0.055}_{-0.055}$$\phantom{\cdot 10^{-0}}$ &$4.035^{+0.045}_{-0.071}$$^{+0.049}_{-0.049}$$\phantom{\cdot 10^{-0}
}$          \\
 140 &$3.996^{+0.029}_{-0.122}$$^{+0.054}_{-0.054}$$\phantom{\cdot 10^{-0}}$ &$3.854^{+0.068}_{-0.050}$$^{+0.053}_{-0.053}$$\phantom{\cdot 10^{-0}}$ &$3.890^{+0.031}_{-0.081}$$^{+0.047}_{-0.047}$$\phantom{\cdot 10^{-0}
}$          \\
 145 &$3.846^{+0.044}_{-0.116}$$^{+0.052}_{-0.052}$$\phantom{\cdot 10^{-0}}$ &$3.714^{+0.070}_{-0.052}$$^{+0.051}_{-0.051}$$\phantom{\cdot 10^{-0}}$ &$3.747^{+0.039}_{-0.067}$$^{+0.046}_{-0.046}$$\phantom{\cdot 10^{-0}
}$          \\
 150 &$3.725^{+0.042}_{-0.138}$$^{+0.051}_{-0.051}$$\phantom{\cdot 10^{-0}}$ &$3.561^{+0.093}_{-0.038}$$^{+0.049}_{-0.049}$$\phantom{\cdot 10^{-0}}$ &$3.616^{+0.035}_{-0.075}$$^{+0.044}_{-0.044}$$\phantom{\cdot 10^{-0}
}$          \\
 155 &$3.575^{+0.068}_{-0.113}$$^{+0.049}_{-0.049}$$\phantom{\cdot 10^{-0}}$ &$3.449^{+0.081}_{-0.051}$$^{+0.048}_{-0.048}$$\phantom{\cdot 10^{-0}}$ &$3.487^{+0.032}_{-0.082}$$^{+0.042}_{-0.042}$$\phantom{\cdot 10^{-0}
}$          \\
 160 &$3.449^{+0.067}_{-0.134}$$^{+0.047}_{-0.047}$$\phantom{\cdot 10^{-0}}$ &$3.325^{+0.086}_{-0.035}$$^{+0.046}_{-0.046}$$\phantom{\cdot 10^{-0}}$ &$3.362^{+0.039}_{-0.068}$$^{+0.041}_{-0.041}$$\phantom{\cdot 10^{-0}
}$          \\
 165 &$3.333^{+0.081}_{-0.142}$$^{+0.046}_{-0.046}$$\phantom{\cdot 10^{-0}}$ &$3.212^{+0.084}_{-0.037}$$^{+0.045}_{-0.045}$$\phantom{\cdot 10^{-0}}$ &$3.257^{+0.026}_{-0.070}$$^{+0.040}_{-0.040}$$\phantom{\cdot 10^{-0}
}$          \\
 170 &$3.213^{+0.088}_{-0.138}$$^{+0.044}_{-0.044}$$\phantom{\cdot 10^{-0}}$ &$3.109^{+0.076}_{-0.044}$$^{+0.044}_{-0.044}$$\phantom{\cdot 10^{-0}}$ &$3.135^{+0.034}_{-0.061}$$^{+0.038}_{-0.038}$$\phantom{\cdot 10^{-0}
}$          \\
 175 &$3.105^{+0.095}_{-0.142}$$^{+0.043}_{-0.043}$$\phantom{\cdot 10^{-0}}$ &$3.001^{+0.082}_{-0.042}$$^{+0.042}_{-0.042}$$\phantom{\cdot 10^{-0}}$ &$3.033^{+0.034}_{-0.060}$$^{+0.037}_{-0.037}$$\phantom{\cdot 10^{-0}
}$          \\
 180 &$2.993^{+0.113}_{-0.137}$$^{+0.041}_{-0.041}$$\phantom{\cdot 10^{-0}}$ &$2.900^{+0.071}_{-0.036}$$^{+0.041}_{-0.041}$$\phantom{\cdot 10^{-0}}$ &$2.932^{+0.029}_{-0.058}$$^{+0.036}_{-0.036}$$\phantom{\cdot 10^{-0}
}$          \\
 185 &$2.902^{+0.107}_{-0.150}$$^{+0.040}_{-0.040}$$\phantom{\cdot 10^{-0}}$ &$2.810^{+0.074}_{-0.041}$$^{+0.040}_{-0.040}$$\phantom{\cdot 10^{-0}}$ &$2.830^{+0.033}_{-0.048}$$^{+0.035}_{-0.035}$$\phantom{\cdot 10^{-0}
}$          \\
 190 &$2.804^{+0.120}_{-0.143}$$^{+0.039}_{-0.039}$$\phantom{\cdot 10^{-0}}$ &$2.714^{+0.076}_{-0.039}$$^{+0.039}_{-0.039}$$\phantom{\cdot 10^{-0}}$ &$2.742^{+0.033}_{-0.049}$$^{+0.034}_{-0.034}$$\phantom{\cdot 10^{-0}
}$          \\
 195 &$2.707^{+0.126}_{-0.142}$$^{+0.038}_{-0.038}$$\phantom{\cdot 10^{-0}}$ &$2.625^{+0.071}_{-0.034}$$^{+0.037}_{-0.037}$$\phantom{\cdot 10^{-0}}$ &$2.651^{+0.034}_{-0.051}$$^{+0.033}_{-0.033}$$\phantom{\cdot 10^{-0}
}$          \\
 200 &$2.628^{+0.117}_{-0.149}$$^{+0.037}_{-0.037}$$\phantom{\cdot 10^{-0}}$ &$2.542^{+0.076}_{-0.033}$$^{+0.036}_{-0.036}$$\phantom{\cdot 10^{-0}}$ &$2.565^{+0.037}_{-0.046}$$^{+0.032}_{-0.032}$$\phantom{\cdot 10^{-0}
}$          \\
 210 &$2.461^{+0.131}_{-0.147}$$^{+0.035}_{-0.035}$$\phantom{\cdot 10^{-0}}$ &$2.387^{+0.069}_{-0.031}$$^{+0.034}_{-0.034}$$\phantom{\cdot 10^{-0}}$ &$2.406^{+0.036}_{-0.042}$$^{+0.030}_{-0.030}$$\phantom{\cdot 10^{-0}
}$          \\
 220 &$2.311^{+0.141}_{-0.152}$$^{+0.033}_{-0.033}$$\phantom{\cdot 10^{-0}}$ &$2.244^{+0.071}_{-0.034}$$^{+0.033}_{-0.033}$$\phantom{\cdot 10^{-0}}$ &$2.263^{+0.032}_{-0.037}$$^{+0.028}_{-0.028}$$\phantom{\cdot 10^{-0}
}$          \\
 230 &$2.174^{+0.141}_{-0.151}$$^{+0.031}_{-0.031}$$\phantom{\cdot 10^{-0}}$ &$2.112^{+0.064}_{-0.031}$$^{+0.031}_{-0.031}$$\phantom{\cdot 10^{-0}}$ &$2.128^{+0.031}_{-0.038}$$^{+0.026}_{-0.026}$$\phantom{\cdot 10^{-0}
}$          \\
 240 &$2.045^{+0.143}_{-0.144}$$^{+0.029}_{-0.029}$$\phantom{\cdot 10^{-0}}$ &$1.989^{+0.066}_{-0.030}$$^{+0.029}_{-0.029}$$\phantom{\cdot 10^{-0}}$ &$2.000^{+0.032}_{-0.033}$$^{+0.025}_{-0.025}$$\phantom{\cdot 10^{-0}
}$          \\
 250 &$1.927^{+0.148}_{-0.141}$$^{+0.028}_{-0.028}$$\phantom{\cdot 10^{-0}}$ &$1.874^{+0.065}_{-0.024}$$^{+0.028}_{-0.028}$$\phantom{\cdot 10^{-0}}$ &$1.890^{+0.028}_{-0.032}$$^{+0.024}_{-0.024}$$\phantom{\cdot 10^{-0}
}$          \\
 260 &$1.821^{+0.142}_{-0.143}$$^{+0.027}_{-0.027}$$\phantom{\cdot 10^{-0}}$ &$1.772^{+0.061}_{-0.027}$$^{+0.027}_{-0.027}$$\phantom{\cdot 10^{-0}}$ &$1.784^{+0.028}_{-0.028}$$^{+0.022}_{-0.022}$$\phantom{\cdot 10^{-0}
}$          \\
 270 &$1.718^{+0.146}_{-0.138}$$^{+0.025}_{-0.025}$$\phantom{\cdot 10^{-0}}$ &$1.673^{+0.059}_{-0.025}$$^{+0.025}_{-0.025}$$\phantom{\cdot 10^{-0}}$ &$1.685^{+0.027}_{-0.027}$$^{+0.021}_{-0.021}$$\phantom{\cdot 10^{-0}
}$          \\
 280 &$1.624^{+0.147}_{-0.135}$$^{+0.024}_{-0.024}$$\phantom{\cdot 10^{-0}}$ &$1.585^{+0.056}_{-0.022}$$^{+0.024}_{-0.024}$$\phantom{\cdot 10^{-0}}$ &$1.595^{+0.029}_{-0.026}$$^{+0.020}_{-0.020}$$\phantom{\cdot 10^{-0}
}$          \\
 290 &$1.538^{+0.145}_{-0.132}$$^{+0.023}_{-0.023}$$\phantom{\cdot 10^{-0}}$ &$1.501^{+0.060}_{-0.023}$$^{+0.023}_{-0.023}$$\phantom{\cdot 10^{-0}}$ &$1.511^{+0.029}_{-0.024}$$^{+0.019}_{-0.019}$$\phantom{\cdot 10^{-0}
}$          \\
 300 &$1.457^{+0.144}_{-0.129}$$^{+0.022}_{-0.022}$$\phantom{\cdot 10^{-0}}$ &$1.424^{+0.053}_{-0.022}$$^{+0.022}_{-0.022}$$\phantom{\cdot 10^{-0}}$ &$1.436^{+0.021}_{-0.027}$$^{+0.018}_{-0.018}$$\phantom{\cdot 10^{-0}
}$          \\
 320 &$1.310^{+0.142}_{-0.123}$$^{+0.020}_{-0.020}$$\phantom{\cdot 10^{-0}}$ &$1.278^{+0.055}_{-0.018}$$^{+0.020}_{-0.020}$$\phantom{\cdot 10^{-0}}$ &$1.291^{+0.023}_{-0.022}$$^{+0.017}_{-0.017}$$\phantom{\cdot 10^{-0}
}$          \\
 340 &$1.186^{+0.135}_{-0.121}$$^{+0.018}_{-0.018}$$\phantom{\cdot 10^{-0}}$ &$1.158^{+0.053}_{-0.016}$$^{+0.018}_{-0.018}$$\phantom{\cdot 10^{-0}}$ &$1.167^{+0.021}_{-0.018}$$^{+0.015}_{-0.015}$$\phantom{\cdot 10^{-0}
}$          \\
 360 &$1.073^{+0.132}_{-0.113}$$^{+0.017}_{-0.017}$$\phantom{\cdot 10^{-0}}$ &$1.051^{+0.047}_{-0.014}$$^{+0.017}_{-0.017}$$\phantom{\cdot 10^{-0}}$ &$1.056^{+0.022}_{-0.015}$$^{+0.014}_{-0.014}$$\phantom{\cdot 10^{-0}
}$          \\
 380 &$9.737^{+1.290}_{-1.059}$$^{+0.155}_{-0.155}$$\cdot 10^{-1}$ &$9.551^{+0.467}_{-0.151}$$^{+0.155}_{-0.155}$$\cdot 10^{-1}$ &$9.612^{+0.184}_{-0.153}$$^{+0.129}_{-0.129}$$\cdot 10^{-1
}$          \\
 400 &$8.859^{+1.226}_{-1.002}$$^{+0.143}_{-0.143}$$\cdot 10^{-1}$ &$8.711^{+0.421}_{-0.141}$$^{+0.144}_{-0.144}$$\cdot 10^{-1}$ &$8.764^{+0.169}_{-0.125}$$^{+0.119}_{-0.119}$$\cdot 10^{-1
}$          \\
 450 &$7.079^{+1.118}_{-0.868}$$^{+0.118}_{-0.118}$$\cdot 10^{-1}$ &$6.985^{+0.368}_{-0.115}$$^{+0.119}_{-0.119}$$\cdot 10^{-1}$ &$7.024^{+0.152}_{-0.102}$$^{+0.098}_{-0.098}$$\cdot 10^{-1
}$          \\
 500 &$5.733^{+1.011}_{-0.750}$$^{+0.098}_{-0.098}$$\cdot 10^{-1}$ &$5.675^{+0.321}_{-0.106}$$^{+0.099}_{-0.099}$$\cdot 10^{-1}$ &$5.704^{+0.114}_{-0.077}$$^{+0.082}_{-0.082}$$\cdot 10^{-1
}$          \\
 550 &$4.696^{+0.899}_{-0.664}$$^{+0.083}_{-0.083}$$\cdot 10^{-1}$ &$4.660^{+0.277}_{-0.100}$$^{+0.084}_{-0.084}$$\cdot 10^{-1}$ &$4.679^{+0.108}_{-0.050}$$^{+0.069}_{-0.069}$$\cdot 10^{-1
}$          \\
 600 &$3.877^{+0.794}_{-0.573}$$^{+0.070}_{-0.070}$$\cdot 10^{-1}$ &$3.860^{+0.246}_{-0.097}$$^{+0.071}_{-0.071}$$\cdot 10^{-1}$ &$3.881^{+0.087}_{-0.042}$$^{+0.059}_{-0.059}$$\cdot 10^{-1
}$          \\
 650 &$3.236^{+0.699}_{-0.506}$$^{+0.060}_{-0.060}$$\cdot 10^{-1}$ &$3.228^{+0.210}_{-0.086}$$^{+0.061}_{-0.061}$$\cdot 10^{-1}$ &$3.241^{+0.077}_{-0.031}$$^{+0.051}_{-0.051}$$\cdot 10^{-1
}$          \\
 700 &$2.711^{+0.626}_{-0.440}$$^{+0.051}_{-0.051}$$\cdot 10^{-1}$ &$2.717^{+0.184}_{-0.084}$$^{+0.052}_{-0.052}$$\cdot 10^{-1}$ &$2.724^{+0.070}_{-0.021}$$^{+0.044}_{-0.044}$$\cdot 10^{-1
}$          \\
 750 &$2.289^{+0.564}_{-0.385}$$^{+0.044}_{-0.044}$$\cdot 10^{-1}$ &$2.305^{+0.156}_{-0.081}$$^{+0.046}_{-0.046}$$\cdot 10^{-1}$ &$2.310^{+0.057}_{-0.021}$$^{+0.038}_{-0.038}$$\cdot 10^{-1
}$          \\
 800 &$1.945^{+0.497}_{-0.340}$$^{+0.039}_{-0.039}$$\cdot 10^{-1}$ &$1.959^{+0.141}_{-0.071}$$^{+0.040}_{-0.040}$$\cdot 10^{-1}$ &$1.967^{+0.048}_{-0.014}$$^{+0.033}_{-0.033}$$\cdot 10^{-1
}$          \\
 850 &$1.662^{+0.442}_{-0.300}$$^{+0.034}_{-0.034}$$\cdot 10^{-1}$ &$1.681^{+0.120}_{-0.067}$$^{+0.035}_{-0.035}$$\cdot 10^{-1}$ &$1.684^{+0.043}_{-0.012}$$^{+0.029}_{-0.029}$$\cdot 10^{-1
}$          \\
 900 &$1.425^{+0.396}_{-0.265}$$^{+0.029}_{-0.029}$$\cdot 10^{-1}$ &$1.442^{+0.111}_{-0.058}$$^{+0.030}_{-0.030}$$\cdot 10^{-1}$ &$1.446^{+0.038}_{-0.010}$$^{+0.026}_{-0.026}$$\cdot 10^{-1
}$          \\
 950 &$1.227^{+0.354}_{-0.234}$$^{+0.026}_{-0.026}$$\cdot 10^{-1}$ &$1.246^{+0.096}_{-0.056}$$^{+0.027}_{-0.027}$$\cdot 10^{-1}$ &$1.250^{+0.032}_{-0.010}$$^{+0.023}_{-0.023}$$\cdot 10^{-1
}$          \\
1000 &$1.060^{+0.317}_{-0.207}$$^{+0.023}_{-0.023}$$\cdot 10^{-1}$ &$1.080^{+0.086}_{-0.049}$$^{+0.024}_{-0.024}$$\cdot 10^{-1}$ &$1.083^{+0.030}_{-0.010}$$^{+0.020}_{-0.020}$$\cdot 10^{-1
}$          \\
\hline
\end{tabular}
\end{center}
\vspace*{-5mm}
\end{footnotesize}
\caption{
\small
\label{tab:table-lhc14-nnn-Sc1}
Total VBF cross sections at the
LHC, $\sqrt S = 14 \tev$
at LO, NLO and NNLO in QCD. Errors shown are respectively scale and PDF uncertainities.  Scale uncertainities are evaluated by varying $\mu_r$ and $\mu_f$ in the interval
$\mu_r,\mu_f \in [Q/4,4Q]$.
The NNPDF2.1~\cite{Ball:2011uy} PDF set has been used.
Numbers are in pb.
}
\end{table}

%%
%% ---------------------------------------------------------------------
%%
\cleardoublepage
\newpage
{\footnotesize

\bibliography{vbfbib}
\bibliographystyle{h-physrev5.bst}
}

\end{document}